\documentclass[12pt]{article}
\usepackage{amsmath,amsthm,amssymb,epsfig,times}
\usepackage[utf8]{inputenc}
\usepackage{float}
\usepackage{subfigure}
\usepackage{natbib}
\usepackage{times}
\usepackage[usenames]{color}
\usepackage{enumerate}
\usepackage{arydshln}
\usepackage[title]{appendix}
\usepackage{enumitem}
\usepackage{tikz}
\usetikzlibrary{decorations.pathreplacing,calligraphy}
\usepackage[ruled]{algorithm2e}
\usepackage{setspace}
\usepackage{hyperref}
\usepackage{caption}
\bibpunct{(}{)}{;}{a}{,}{,}

\newcommand{\btheta}{\boldsymbol{\theta}}

\newcommand{\by}{\mathbf{y}}
\newcommand{\bk}{\mathbf{k}}
\newcommand{\bx}{\mathbf{x}}
\newcommand{\bz}{\mathbf{z}}

\newcommand{\bR}{\mathbf{R}}
\newcommand{\bH}{\mathbf{H}}

\newtheorem{theorem}{Theorem}
\newtheorem{definition}{Definition}

\newtheorem{proposition}{Proposition}

\title{\bf Bayesian Copula Density Estimation Using Bernstein Yett-Uniform Priors}
\author{{\sc  Nicolás Kuschinski, Richard Warr, and Alejandro Jara}}
\begin{document}
\date{\today}
\maketitle

\footnotetext[1]{ Nicolás Kuschinski is a Postdoctoral Researcher at the Department of Statistics, Pontificia Universidad
Cat\'olica de Chile and the Center for the Discovery of Structures in Complex Data, Casilla 306, Correo 22, Santiago, Chile (E-mail: nicolas.kuschinski@mat.uc.cl). 
Richard Warr is an Associate Professor, Department of Statistics, Brigham Young University, Provo, UT 84602, USA (warr@stat.byu.edu). Alejandro Jara is 
an Associate Professor, Department of Statistics, Pontificia Universidad
Cat\'olica de Chile and Center for the Discovery of Structures in Complex Data Casilla 306, Correo 22, Santiago, Chile (E-mail: atjara@uc.cl).}

\begin{abstract}
Probability density estimation is a central task in statistics. Copula-based models provide a great deal of flexibility in modelling
multivariate distributions, allowing for the specifications of models for the marginal
distributions separately from the dependence structure (copula) that links them to form a joint distribution. Choosing a class of copula models is not a trivial
task and its misspecification can lead to wrong conclusions. We introduce a novel class of random Bernstein copula functions, and studied its support and the behavior of its posterior distribution. The proposal is based on a particular class of random grid-uniform copulas, referred to as yett-uniform copulas. Alternative Markov chain Monte Carlo algorithms for exploring the  posterior distribution under the proposed model are also studied. 
The methodology is illustrated by means of simulated  and real data.

\end{abstract}

\section{Introduction}
The marginals-copula representation of a joint distribution is a flexible tool for understanding complex structures of dependence for multivariate data and for incorporating a prior on a given functional of an unknown multivariate distribution generating the data or latent variables in a hierarchical model. Under this representation  the associations among the random variables are parameterized separately from their univariate marginal distributions, where the copula $C$ is a function that contains all the information about the association structure. For any $d$-variate distribution $H:\mathbb{R}^d\rightarrow[0,1]$ with $d>1$ and marginals $F_1,$ $F_2,$ $\ldots,$ $F_d$, a copula is a function $C:\mathbb{R}^d\rightarrow[0,1]$ such that
$$
H(\boldsymbol{x})=C(F_1(x_1),F_2(x_2), \ldots, F_d(x_d)),
$$
where $\boldsymbol{x}=(x_1,\ldots,x_d) \in \mathbb{R}^d$. Sklar's theorem \citep{sklarthm1, sklarthm, librocop} is a classical result which states that this function $C$ exists for any multivariate distribution $H$, and that if $H$ is continuous, $C$ is unique. Copula functions are themselves multivariate probability distributions supported on the unit hyper-cube and are such that the single variate marginals are uniform \citep[see, e.g.,][]{librocop, jimenezvaron2023visualization}.

There is a large body of literature on copula estimation in parametric contexts  \citep{bayesgausscop, lambert_2007, librocop,  librocop1, summaryfreqcop, summarycop, archimcop1, archimcop, bayesarchim}. However, this does not apply to nonparametric methods, which happens for at least three reasons: 1) the univariate uniform marginals constraints are difficult to accommodate in available nonparametric procedures, 2) copulas have the potential to work in high-dimensional contexts, and nonparametric techniques tend to suffer from the curse of dimensionality problem, and 3) it is not possible to find the induced 
prior distribution on the copula function  or the expression of the copula function for the Bayesian nonparametric methods commonly used for multivariate density estimation \citep[see, e.g.,][]{bnpbook}, even if finite-dimensional approximations are considered.

Non-parametric approaches for discrete data have been attempted in numerous ways, as described in \cite{countcops} and \cite{discretecops}. In the context of continuous data, classical nonparametric approaches can be traced back to \cite {empcop}, and can be found in \cite{semiparamfreq}, \cite{freqsplinecop}, \cite{momentcop}, \cite{factorcop}, and \cite{freqnpcop}. The classical methods commonly rely on the use of a partial- or a pseudo-likelihood, and do not allow for a proper quantification of the uncertainties associated to the lack of knowledge of the marginal distributions. Furthermore, these approaches cannot be employed for modelling the association structure of latent variables in the context of hierarchical models. 

Flexible Bayesian approaches for copula functions can be found  in \cite{bnpjeffreys}, \cite{gausmixcop}, \cite{abccop}, \cite{skewmixcop}, \cite{polyacop}, and \cite{kuschinski;jara;2021}. \cite{bnpjeffreys} proposed an interesting semi-parametric Bayesian approach for bivariate copulas based on a finite-dimensional approximation. This approximation is structured from a partition of the unit interval based on intervals of the same length and indicator functions for the corresponding intervals. The density of the copula is constructed via a mixture of the cross products of the indicator functions, resulting in a locally uniform distribution over the unit square, parameterized by a doubly stochastic matrix. Taking advantage of the properties of doubly stochastic matrices, the authors proceed to develop both a conjugate Jeffreys prior and a Markov chain Monte Carlo (MCMC) algorithm to sample from the posterior distribution. The approach is flexible but is difficult to generalize to nonequally-spaced partitions and  higher dimensions because of its reliance on the properties of doubly stochastic matrices. 

\cite{gausmixcop} and \cite{skewmixcop} proposed a model based on mixtures of Gaussian and skew-normal copulas, respectively. Mixture models are being increasingly used as a way to arbitrarily approximate any probability distribution. However, this approximation property depends on the kernel and the mixing distribution, and the Gaussian copula kernel is not rich enough to form a dense class. For instance, the density function of a bivariate Gaussian copula has the property that $c_{\boldsymbol{R}}(u_1,u_2) = c_{\boldsymbol{R}} (u_2,u_1)$. Therefore, the density of a mixture of arbitrarily many Gaussian copulas also has this feature and, thus, cannot approximate an asymmetrical copula function, such as the  asymmetrical $t$ copula described by \cite{asymtcop}. 

 \cite{abccop} proposed an estimation method for a functional of a multivariate distribution using a copula representation. The prior on the copula is constructed by eliciting the prior on functionals of interest and it is implied for the remaining elements of the copula function by the use of the exponentially tilted empirical likelihood.  The main advantage of their approach is the ease of elicitation, because it does not require the prior elicitation about all aspects of the multivariate dependence structure. However, it does not allow for posterior inferences about all aspects of the multivariate dependence structure.  Finally, \cite{polyacop} employed Dirichlet-based Polya trees models  to propose a fully non-parametric Bayesian approach to modeling copula functions in any number of dimensions and a method for conjugate posterior simulation from the resulting posterior. This attractive result is, however, significantly marred by the flaw that the simulations from the posterior distribution are not themselves copula functions. 

\cite{kuschinski;jara;2021} introduced the class of grid-uniform copula functions and showed that the class is dense in the space of all continuous copula functions in a Hellinger sense. They proposed a hierarchically centered prior distribution based on the class of grid-uniform copula functions, borrowing ideas from spatial statistics, and that have appealing support and posterior consistency properties. They also described a class of transformations on grid-uniform copulas which is closed in the space of grid-uniform copula functions and that is able to span the complete space of grid-uniform copula functions in finite number of steps. This family of transformations, referred to as rectangle exchanges, was also employed to develop an automatic MCMC algorithm for exploring the corresponding posterior distribution. 

In contrast to classical methods, the hierarchically centered grid-uniform prior of \cite{kuschinski;jara;2021}  allows for a proper quantification of
the uncertainties associated to the lack of knowledge of the marginal distributions and can be employed for modelling the association structure of latent variables in the context of hierarchical models. However, important disadvantages are that grid-uniform copulas are not smooth and may not be sensitive enough toward the local properties of the true copula model. In this paper, we introduce a novel prior model for  continuous copula functions that overcomes the disadvantages of grid uniform copulas. The prior model corresponds to the induced probability distribution that arises from the computation of the multivariate Bernstein polynomial (MBP) of a particular class of random grid-uniform copulas, referred to as yett-uniform copulas,
which has appealing smoothness, support and posterior consistency properties.

MBP have been also used to proposed inference procedures for copulas in the classical statistical literature \citep[see, e.g.,][]{berncops, asymbernmix, asymbern2, asymbern3, embern, bernindep, bernmaxcorrel}. Most of the classical procedures are based on the empirical Bernstein copula estimator proposed by \cite{berncops}, which has appealing asymptotic properties \citep[see, e.g.,][]{asymbernmix, asymbern2, asymbern3}. However, there is no guarantee that point estimates are copula functions for finite sample sizes. \cite{embern} proposed an estimator based on MBP and using an EM algorithm applied to a pseudo-likelihood function. However, this technique also has no guarantees that the resulting estimate is a copula function. \cite{posquaddep} proposed an estimator based on Bernstein copulas while assuring that the resulting estimates are in fact copula functions. This is done by finding the Bernstein copula that is closest to the empirical Bernstein copula estimator. While this is an improvement over the previous classical approaches, it only provides a point estimator which is has not been proven to be optimal. The approach also does not allow for quantification of the uncertainties associated with the lack of knowledge of the marginal distributions and cannot be employed for modelling the association structure of latent variables in the context of hierarchical models.

This paper is organized as follows. In Section \ref{sec2} we briefly describe the properties of multivariate Bernstein copulas, grid-uniform, and yett-uniform copulas, so as to make the discussion self contained. In Section \ref{sec3} we introduce the Bernstein yett-uniform process prior and state its main properties. In Section \ref{sec4} we propose and compare alternative MCMC algorithms for exploring the corresponding posterior distribution under the proposed model.  In Section \ref{sec5} we illustrate the behavior of the model by means of analyses of simulated and real data. A final discussion concludes the article. The proofs of the theoretical results are provided in the online supplementary material.

\section{Bernstein polynomials and grid-uniform copulas}\label{sec2}

\subsection{Multivariate Bernstein polynomials and Bernstein copulas}

For a given function  $G: [0,1]^d \longrightarrow [0,1]$,  the associated MBP of degree $\bk =(k_1,\ldots,k_d) \in \mathbb{N}^d$ is defined by
\begin{eqnarray}\label{MBP2}
B_{\bk,G}(\bz) 
= \sum_{\nu_1=0}^{k_1} \cdots    \sum_{\nu_d=0}^{k_d} G\left(\frac{\nu_1}{k_1}, \ldots, \frac{\nu_d}{k_d}\right) M\left(\nu_1, \ldots, \nu_d  \mid \bk, \bz\right), 
\end{eqnarray}
where  
$$
M\left(\nu_1, \ldots, \nu_d  \mid \bk, \bz\right) = \mathcal{P}_{\nu_1, k_1} \left(z_1\right) \cdots  \mathcal{P}_{\nu_d, k_d} \left(z_d \right),
$$
 with $ \mathcal{P}_{\nu_j, k_j}  \left(z_j \right) = \binom{k_j}{\nu_j} z_j^{\nu_j} \left(1-z_j \right)^{k_j - \nu_j}$. MBPs have appealing approximation properties.  In particular, the relation
$$
\lim_{\bk \rightarrow  \infty} B_{\bk,G}(\bz) = G(\bz),
$$
holds at each point of continuity $\bz$ of $G$. The relation holds uniformly on $[0,1]^d$ if $G$ is a continuous function. It has been shown that if $G$ is the restriction of the CDF of a probability measure defined on $[0,1]^d$, then $B_{\bk,G}(\cdot)$ is also the restriction of the CDF of a probability measure defined on $[0,1]^d$. It has also been shown that if $G$ is the restriction of the CDF of a probability measure defined on $[0,1]^d$, such that $G(\bz)=0$ if $\bz \in \{\by \in [0,1]^d : y_l>0, l=1,\ldots,d \}^C$, then $B_{\bk,G}(\cdot)$ has a density w.r.t. Lebesgue measure \citep[see, e.g.,][]{petrone;99a, petrone;99b, barrientos;etal:2015}. 

It can be shown that if $G$ is the restriction of the CDF of an absolutely continuous probability measure defined on $[0,1]^d$ such that the single variate marginal distributions are uniform, then $B_{\bk,G}(\bz)$ is also the restriction of the CDF of a copula. Furthermore, if $G$ is the restriction of the CDF of a particular discrete copula function, then $B_{\bk,G}(\bz)$ is also the restriction of the CDF of a copula. In both cases, we referred to $B_{\bk,G}(\bz)$ to as a Bernstein copula.

\begin{theorem} For any integer $d \geq 2$,  let $\bk \in \mathbb{N}^d$. Let $G$ be the restriction of the CDF of either:
\begin{itemize}
\item[(i)] an absolutely continuous probability distribution defined on $[0,1]^d$ such that the single variate marginal distributions are uniform, or 

\item[(ii)] a  discrete probability measure defined on $(0,1]^d$, with uniform marginals having $p_1 \times \cdots \times p_d$ support points, which are such that $\frac{p_i}{k_i}$ is an integer for all $i \in \{1,\ldots,d\}$.
\end{itemize}
Then $B_{\bk,G}(\bz)$ is also the restriction of the CDF of an absolutely  continuous probability measure defined on $[0,1]^d$, with uniform marginal distributions, and that admits a density w.r.t. Lebesgue measure given by the following mixture of products of beta distributions
\begin{eqnarray}\label{densityMBP}
b_{\bk,G}(\bz) & = & \sum_{\nu_1=1}^{k_1} \cdots    \sum_{\nu_d=1}^{k_d} 
W_{\nu_1, \ldots, \nu_d}  \prod_{j=1}^d \beta(z_j \mid \nu_j, k_j - \nu_j +1),
\end{eqnarray}
where $W_{\nu_1, \ldots, \nu_d}  =
G\left( 
\left(\frac{\nu_1-1}{k_1},\frac{\nu_1}{k_1}\right]
\times \ldots \times 
\left(\frac{\nu_d-1}{k_d},\frac{\nu_d}{k_d}\right]
\right)$, and $\beta(\cdot \mid a, b)$ stands for the density of beta distribution with parameters $a$ and $b$. 
\end{theorem}

Bernstein copula densities share the approximation properties of the original MBP $B_{\bk,G}$. In addition to proving the uniform approximation property, we derive the approximation rate of $b_{\bk,G}$ for $\alpha$-smooth $d$-dimensional copula densities functions, $\alpha \in (0,2]$. A function $g$ on $[0,1]^d$ is called $\alpha$-smooth if for some $\alpha \in (0,1]$ there exists a constant $L_{\alpha}(g)$ such that, for every pair of points $\bz_1, \bz_2 \in [0,1]^d$, $\left|g(\bz_1) - g(\bz_2)\right| \leq L_{\alpha}(g) \Vert\bz_1- \bz_2 \Vert^{\alpha}$ or if for $\alpha \in (1,2]$ its first order partial derivatives are $(\alpha-1)$--smooth. 
\begin{theorem}\label{approxTeo}
Let $G$  be the CDF of an absolutely continuous $d$-dimensional copula, w.r.t. Lebesgue measure, and  with continuous density function, $g$. Then,  $b_{\bk,G}(\cdot)$ converges uniformly to $g$ as  $k_1,\ldots,k_d \rightarrow  \infty$. In addition, if $k = \min\{k_1, \ldots, k_d\}$ and  $g$ is $\alpha$-smooth, $\alpha \in (0,1]$, then
\begin{eqnarray}\nonumber
\Vert g -  b_{\bk,G} \Vert_{\infty}  & \leq & d L_{\alpha}(g)\left(\frac{1}{2k}+\frac{1}{k^2}\right)^{\alpha/2}
 +
\left|1 -  
\frac{(k+d-1)!}{(k-1)!k^d}
\right|
\left\Vert 
g
\right\Vert_{\infty}.
\end{eqnarray}
Finally, if $g$ is $\alpha$-smooth, $\alpha \in (1,2]$, then
\begin{eqnarray}\nonumber
\Vert g -  b_{\bk,G} \Vert_{\infty}  & \leq & 
\frac{d}{k}\max_{l \in \{1,\ldots,d \}}
\left\Vert D_{1}^{z_l}g \right\Vert_{\infty} 
+
\sum_{l=1}^d
\frac{1}{k^l} 
\binom {d} {l}
\max_{ \tiny
(j_1,\ldots, j_d) \in  \mathcal{D}_l^d
}
\left\Vert
D_{j_d}^{y_d} \ldots D_{j_1}^{z_1} G
\right\Vert_{\infty}
\\\nonumber & & 
+
 \sqrt{d} \max_{l \in \{1,\ldots,d\}}L_{\alpha}(D_1^{z_l}g)
\left(\frac{d(1+d)}{2k}  \right)^{\alpha/2}
+
\frac{d^2(d-1)!}{k}\left\Vert g \right\Vert_{\infty},
\end{eqnarray}
where $D_{j_i}^{z_i}$ denotes the partial derivative operator 
$
D_{j_i}^{z_i} = \partial^{j_i} / \partial z_i^{j_i}
$
and
$
\mathcal{D}_l^d := \left\{
(j_1,\ldots,j_d) : j_{t}=2, \right.$ $\left. \mbox{ if } t\in\left\{ t_1, \ldots, t_l \right\} \subseteq {\left\{ 1, \ldots, d \right\}}, \mbox{ and }, j_t=1, \mbox{ if } t\in\left\{ t_1, \ldots, t_l \right\}^C
\right\}.
$
\end{theorem}

\subsection{Grid-uniform copulas and rectangle exchanges}

For any integer $d \geq 2$, let $\rho$ be an orthogonal grid of $[0,1]^d$. Specifically, let $\rho_i$ be an ordered collection of points in $[0,1]$, and set $\rho = \rho_1 \times \cdots \times \rho_d$, such that $\boldsymbol{1}_d \in \rho$. Let  $\nu^\rho$  be the collection of sets formed by $\rho$, which are indexed by their upper right (or higher dimensional equivalent) corner. Now, let $F$ be a probability measure defined on an appropriate space and $B$ a measurable set such that $F(B) > 0$. We denote $F\mid_B$ to be the restriction of $F$ to $B$ defined by $F_{|_B}(A)=F(A\mid B)=F(A \cap B)/F(B)$. A probability distribution $F$ on $[0,1]^d$ is said to be \emph{$\rho$-uniform}, if for each set  $ B \in \nu^\rho$, such that $F(B)>0$, the restriction of $F$ to the set $B$, $F_{|_B}$, is uniform on $B$. A distribution $C$ on $[0,1]^d$ is a $\rho$-uniform copula if it is $\rho$-uniform and if all of its one-dimensional marginal distributions are uniform.

\cite{kuschinski;jara;2021} showed that if $C$ is an arbitrary copula that is absolutely continuous with respect to Lebesgue measure, then for every $\epsilon>0$, there exists a grid-uniform copula $D$ such that the Hellinger distance between $D$ and $C$ is smaller than $\epsilon$. \cite{kuschinski;jara;2021} also introduced a class of transformations on grid-uniform copulas, referred to as rectangle exchanges, which have the following important properties: (i) a rectangle exchange on a $\rho$-uniform copula produces another $\rho$-uniform copula, and (ii) given a grid $\rho$, and two $\rho$-uniform copulas $C$ and $D$, there is a finite sequence of rectangle exchanges which can transform $C$ into $D$.

\begin{definition} \label{def:rectExchange}
For any integer $d \geq 2$ let $\rho$ be a grid on $[0,1]^d$ and  $C$ be a grid $\rho$-uniform copula function. The function $C^*$ is the result of a rectangle exchange of  $C$,
if  $C^*$ is constructed using the following steps:
\begin{itemize}
    \item[(1)] Set $C^* = C$, and pick $i$ and $j$ in the set $\{1,\ldots, d\}$, such that $i <j$ and the cardinality of $\rho_i$ and $\rho_j$ is greater than or equal to 2. Also, for all $k \in \{1,\ldots,d\} \setminus \{i,j\}$ pick point $x_k \in \rho_k$.
    
    \item[(2)] Pick $a_1,a_2 \in \rho_i$ and $b_1,b_2 \in \rho_j$. 

    \item[(3)] Set 
    $$
    \boldsymbol{p}_{(a_l,b_m)} = (x_1,\ldots,x_{i-1},a_l,,x_{i+1},\ldots,x_{j-1},b_m, x_{j+1}, \ldots,x_d), 
    $$
    where $l,m \in \{1,2\}$.  
        
    \item[(4)] Pick some $\epsilon$ in the interval 
        $$\left[\max\left\{-C\left(\nu^{\rho}_{\boldsymbol{p}_{(a_1,b_2)}} \right) ,-C\left(\nu^{\rho}_{\boldsymbol{p}_{(a_2,b_1)}} \right) \right\}, 
                \min\left\{ C\left(\nu^{\rho}_{\boldsymbol{p}_{(a_1,b_1)}} \right),C\left(\nu^{\rho}_{\boldsymbol{p}_{(a_2,b_2)}} \right) \right\}
        \right].$$
        
            \item[(5)] Set
        \begin{itemize} 
            \item[] $C^*\left(\nu^{\rho}_{\boldsymbol{p}_{(a_1,b_1)}} \right) = C\left(\nu^{\rho}_{\boldsymbol{p}_{(a_1,b_1)}} \right)  - \epsilon,$  $C^*\left(\nu^{\rho}_{\boldsymbol{p}_{(a_1,b_2)}} \right) = C\left(\nu^{\rho}_{\boldsymbol{p}_{(a_1,b_2)}} \right) + \epsilon,$
            \item[]  $C^*\left(\nu^{\rho}_{\boldsymbol{p}_{(a_2,b_1)}} \right) = C\left(\nu^{\rho}_{\boldsymbol{p}_{(a_2,b_1)}} \right) +\epsilon,$ and $C^*\left(\nu^{\rho}_{\boldsymbol{p}_{(a_2,b_2)}}\right) =  C\left(\nu^{\rho}_{\boldsymbol{p}_{(a_2,b_2)}}\right)  - \epsilon.$
        \end{itemize}
\end{itemize}
\end{definition}

Rectangle exchanges are a closed operation on grid-uniform copulas since they conserve the uniformity of all marginals. 
Another interesting property of grid-uniform copula functions is that starting from an uniform distribution on $[0,1]^d$, it is possible to reach any given grid-uniform copula, say $C_0$, by doing certain types of operations. Specifically for a given grid-uniform copula $C$, 
\cite{kuschinski;jara;2021} referred to as a \textit{grid division} on $C$, to the addition of a division along any of the coordinates, such that the sets that are not divided retain their probabilities, and those that are divided distribute their probability in proportion to their volume. In other words, the resulting copula function arising from a grid division of $C$ is identical to $C$, but is mapped onto a more refined grid. Finally, via rectangle exchanges it is possible to generate the full space of grid-uniform copula functions. More specifically, 
let $C_1$ and $C_2$ be two $\rho$-uniform copulas. Then, there is a finite sequence of rectangle exchanges to transform $C_1$ into $C_2$.

\subsection{Yett-uniform copulas and its connection with Bernstein copulas}

Recall that a Bernstein copula density $b_{\bk,G}$, given by expression (\ref{densityMBP}), depends only on the degree $\bk$ and on weights $W_{\nu_1, \ldots, \nu_d}$, which are established by the copula function $G$, 
$$W_{\nu_1, \ldots, \nu_d} =
G\left( 
\left(\frac{\nu_1-1}{k_1},\frac{\nu_1}{k_1}\right]
\times \ldots \times 
\left(\frac{\nu_d-1}{k_d},\frac{\nu_d}{k_d}\right]
\right).$$ 
This implies that any two copula functions $G$ and $G'$ having the same weights $W_{\nu_1, \ldots, \nu_d}$, have the same Bernstein copula density, $b_{\bk,G} = b_{\bk,G'}$. Thus, the mapping induced by the multivariate Bernstein polynomial of degree $\bk$ from the space of $d$-dimensional copula functions to the space of Bernstein copulas, is many-to-few.  If we restrict the domain to a particular class of grid-uniform copulas, this is no longer the case. For a given $\bk \in \mathbb{N}^d$, set  
$$\rho^{\bk} = \{0/k_1,1/k_1,2/k_1,\ldots, k_1/k_1\} \times \cdots \times \{0/k_d,1/k_d,2/k_d,\ldots, k_d/k_d\}.$$
As before, let $\nu^{\rho^{\bk}}$ be the collection of sets formed by $\rho^{\bk}$, of the form
$$
\left(\frac{l_1-1}{k_1}, \frac{l_1}{k_1} \right] \times \cdots \times \left(\frac{l_d-1}{k_d}, \frac{l_d}{k_d}\right],
$$
where $l_j \in \{1,\ldots,k_j\}$,  which are indexed by their upper right (or higher dimensional equivalent).

\begin{definition}{\bf (Yett-uniform copula)}  
A distribution $G$ on $[0,1]^d$ is said to be a 
yett-uniform copula of degree $\bk \in \mathbb{N}^d$, if it is $\rho^{\bk}$-uniform and its single-dimensional marginal distributions are uniform.
\end{definition}

Let $\widetilde{\mathcal{C}}_{d,\bk}$ be the space of yett-uniform $d$-dimensional copulas functions of degree $\bk$. Let $\mathcal{BC}_{d,\bk}$ be the space of probability distributions induced by Bernstein copulas of degree $\bk$. Let $BP_{\bk, G}$ be the mapping induced by its multivariate Bernstein polynomial of degree $\bk$, $B_{\bk, G}$, which sends a yett-uniform copula $G \in \mathcal{C}_{d,\rho^{\bk}}$ to the
restriction of its CDF to $[0,1]^d$.  Let $T_{[0,1]^d}$ be the mapping that sends the restriction of a CDF to $[0,1]^d$, 
to the corresponding probability measure defined on $[0,1]^d$ and having uniform marginal distributions.
 
\begin{proposition}
For every $\bk \in \mathbb{N}^d$, the mapping $T_{[0,1]^d}
\circ 
BP_{\rho^\bk,G}$ is bijective from  $\widetilde{\mathcal{C}}_{d,\bk}$ to 
$\mathcal{BC}_{d,\bk}$.\end{proposition}

\section{Bernstein yett-uniform processes}\label{sec3}

\subsection{General definition} \label{sec:def}

Let $\mathcal{C}_d$ be the space of absolutely continuous $d$-dimensional copulas. We induce a prior probability model on  $\mathcal{C}_d$ by considering the multivariate Bernstein polynomial of degree $\bk$, $B_{\bk,G}$, of the restriction of the CDF of a random yett-uniform copula, $G$.

\begin{definition}{\bf } Let $T_{[0,1]^d}$ be the mapping that sends the restriction of a CDF to $[0,1]^d$, 
to the corresponding probability measure defined on $[0,1]^d$ and having uniform marginal distributions. Let $P_{\bk,\alpha, G_0}$ be the prior probability measure on yett-uniform copulas of degree $\bk$,
where $ \alpha > 0$ and $G_0$ is a reference copula. Let $BP_{\bk, G}$ be the mapping induced by its multivariate Bernstein polynomial of degree $\bk$, which sends a yett-uniform copula $G \in \widetilde{\mathcal{C}}_{d,\bk}$ to the
restriction of its CDF to $[0,1]^d$. The Bernstein yett-uniform process $C$ is the random probability measure having probability law given by $P_{\bk, \alpha, G_0} \circ BP^{-1}_{\bk,  G} \circ T^{-1}_{[0,1]^d}$ on $\mathcal{C}_d$. We will use the notation 
$$C \mid \bk,  \alpha, G_0 \sim BYUP(\bk,  \alpha, G_0),$$
to refer to the random Bernstein yett-uniform copula $C$, which is referred to as the Bernstein yett-uniform process (BYUP) and is parameterized by $(\bk,  \alpha, G_0)$.
\end{definition}

The BYUPs are well defined stochastic processes. 
Specifically, the mappings defining the trajectories of the process is Borel 
under standard topologies. 
 
\begin{theorem}
Let $\mathcal{C}_d$ be the space of absolutely continuous probability distributions defined on $[0,1]^d$ and with uniform marginals. Let $\mathcal{C}^*_{d}$ be the space of yett-uniform 
$d$-dimensional probability distributions 
with uniform marginals. Let $T_{[0,1]^d}
\circ 
BP_{\bk,G} :\mathbb{N}^d \times \mathcal{C}^*_{d} \longrightarrow \mathcal{C}_d$ be the mapping  induced by multivariate Bernstein polynomials.
The mapping 
$T_{[0,1]^d}
\circ 
BP_{\rho^\bk,G}
$ is continuous under the Hellinger topology, that is,  for every Hellinger neighborhood $\mathcal{A}\subset\mathcal{C}_d$ either $ 
BP_{\rho^\bk,G}^{-1}
\circ 
T_{[0,1]^d}^{-1}
(\mathcal{A})$ is empty or there is a $\bk \in \mathbb{N}^d $ and a nonempty  open Hellinger ball of radius $\delta>0$, $\mathcal{B}_\delta \subset 
\widetilde{\mathcal{C}}_{d,\bk}$, such that $\{T_{[0,1]^d}
\circ 
BP_{\rho^\bk,G}: G \in  \mathcal{B}_\delta \} \subset \mathcal{A}$.
\end{theorem}

\begin{theorem}
Let $\mathcal{C}_d$ be the space of absolutely continuous probability distributions defined on $[0,1]^d$ with uniform marginals and with strictly positive continuous density and finite entropy. Let $\mathcal{C}^*_{d}$ be the space of yett-uniform 
$d$-dimensional probability distributions 
with uniform marginals and which assign strictly positive probability to each cell in $\nu^{\rho^k}$. Let $T_{[0,1]^d}
\circ 
BP_{\rho^\bk,G} :\mathbb{N}^d \times \mathcal{C}^*_{d} \longrightarrow \mathcal{C}_d$ be the mapping  induced by multivariate Bernstein polynomials.
The mapping 
$T_{[0,1]^d}
\circ 
BP_{\rho^\bk,G}
$ is continuous under the Kulllback-Lebiler topology, that is,  for every Kullback-Leibler neighborhood $\mathcal{A}\subset\mathcal{C}_d$ either $ 
BP_{\rho^\bk,G}^{-1}
\circ 
T_{[0,1]^d}^{-1}
(\mathcal{A})$ is empty or there is a $\bk \in \mathbb{N}^d $ and a nonempty  open Kullback-Leibler ball of radius $\delta>0$, $\mathcal{B}_\delta \subset 
\widetilde{\mathcal{C}}_{d,\bk}$, such that $\{T_{[0,1]^d}
\circ 
BP_{\rho^\bk,G}: G \in  \mathcal{B}_\delta \} \subset \mathcal{A}$.
\end{theorem}

\subsection{A specific construction} \label{sec:spec}

We build the BYUP on probability models for yett-uniform copulas of the form
$$
p(G \mid \bk, \alpha, G_0 )\propto \exp\left\{ -\frac{1}{2} \alpha \times \mathcal{D}\left(G,G_0 \right)\right\} \times I_{\widetilde{\mathcal{C}}_{d,\bk}}(G),
$$
where $ \alpha > 0$, $G_0$ is a reference copula, and $\mathcal{D}$ is a suitable distance for probability distributions. Many choices for $\mathcal{D}$ could be considered. One choice is the squared-$L_2$ distance. Let $g_0$ and $g$ be densities for $g_0$ and $g$, respectively. Let $B_1, \ldots, B_p$ be the sets included in $\nu^{\rho^{\bk}}$.  Under the squared-$L_2$ distance the yett-uniform prior model is given by 
\begin{eqnarray}\nonumber  \label{prior1}
 p(G \mid \bk, \alpha, G_0 ) &\propto& \exp \left\{  -\frac{\alpha}{2}  \times \int_{[0,1]^d} \left(g(\boldsymbol{x}) - g_0(\boldsymbol{x}) \right)^2 d \boldsymbol{x} \right\}  \times I_{\widetilde{\mathcal{C}}_{d,\bk}}(G),\\\nonumber
     &=& \exp \left\{ -\frac{\alpha}{2}  \times 
    \sum_{l=1}^{\left|\nu^{\rho} \right|} \left[  \int_{B_l} (g(\boldsymbol{x})-g_0(\boldsymbol{x}))^2 d \boldsymbol{x}  \right] \right\} \times I_{\widetilde{\mathcal{C}}_{d,\bk}}(G),\\\nonumber
        &\propto& \exp \left \{ -\frac{\alpha}{2}  \times \sum_{l=1}^{\left|\nu^{\rho^{\bk}} \right|} \left[-2g_l \int_{B_l} g_0(\boldsymbol{x}) d \boldsymbol{x}  + g_l^2 \right] \right\} \times I_{\widetilde{\mathcal{C}}_{d,\bk}}(G),
\end{eqnarray}
where $g_l= \frac{\int_{B_l} g(\boldsymbol{x}) d \boldsymbol{x} }{\lambda (B_l)}$,  with $\lambda(A)$ being the  Lebesgue measure of the set $A$. The prior takes the form of a truncated $\left|\nu^{\rho^{\bk}} \right|$-variate normal random distribution, centered at the yett-uniform version of $G_0$, $G_{0,\rho^{\bk}}$, and precision matrix given by  $\alpha \times \mathbf{I}_{\left|\nu^{\rho^{\bk}} \right|}$. $G_{0,\rho^{\bk}}$ plays the role a centering parameter under the proposed prior and corresponds to the prior mode. On the other hand,  $\alpha$ plays the role of a precision parameter, since as $\alpha$ goes to $+\infty$, the prior variance for $G$, $\mbox{var}(G|\bk,\alpha,G_0)$ goes to $0$.

Assigning $\mathcal{D}$ to the squared-$L_2$ norm provides easy to interpret parameters. However, it does not allow for the incorporation of prior information on the degree of smoothness of the copula function. Let $\mathcal{V}^{\rho^{\bk}}$ be a set containing  the elements of $\nu^{\rho^{\bk}}$ in a given order. Let $\boldsymbol{A}$ be a symmetric matrix in which each entry $\boldsymbol{A}_{i,j}$ encodes information about the spatial relationship of the sets $\mathcal{V}^{\rho^{\bk}}_i$ and $\mathcal{V}^{\rho^{\bk}}_j$. Finally, let $\boldsymbol{D}_A$ be a diagonal matrix with $\boldsymbol{D}_{A_{i,i}}= \sum_{j=1}^{|\nu^{\rho^{\bk}}|} \boldsymbol{A}_{i,j}$. Borrowing ideas from models commonly used in spatial statistics and the nature of the grid-uniform model, and following \cite{kuschinski;jara;2021}, we take
\begin{equation*}
\label{eq:dist}
 \mathcal{D}\left(G,G_0 \right ) = \sum_{i=1}^{|\nu^\rho|}\sum_{j=1}^{|\nu^{\rho^{\bk}}|} \left( \boldsymbol{D}_A- \gamma \boldsymbol{A} \right)_{i,j} \int_{\mathcal{V}^{\rho^{\bk}}_i}(g(\boldsymbol{x})-g_0(\boldsymbol{x}) d\boldsymbol{x}   \int_{\mathcal{V}^{\rho^{\bk}}_j}(g(\boldsymbol{y}) - g_0(\boldsymbol{y}) )d \boldsymbol{y},
\end{equation*}
where $\gamma >0$. Under this distance, the grid-uniform prior is given by
\begin{eqnarray}\nonumber  \label{prior2}
 p(G \mid \rho^{\bk},\alpha,\boldsymbol{A}, \gamma, G_0) &\propto& \exp \left\{ - \frac{\alpha}{2}  \mathcal{D} \left(G, G_0 \right)  \right\} \times  I_{\widetilde{\mathcal{C}}_{d,\bk}}(G),\\\nonumber
  &=&  \exp\left\{ - \frac{\alpha}{2}  \overrightarrow{G-G_0}^T \left(\boldsymbol{D}_A- \gamma \boldsymbol{A} \right) \overrightarrow{G-G_0}\ \right\} \times   I_{\widetilde{\mathcal{C}}_{d,\bk}}(G),
\end{eqnarray}
where $\overrightarrow{G-G_0}$ is the vector representation of $\left\{\int_B g( \boldsymbol{x})-g_0(\boldsymbol{x})d \boldsymbol{x} : B\in\nu^{\rho^{\bk}} \right \}$ corresponding to the order induced by $\mathcal{V}^{\rho^{\bk}}$. The prior corresponds to a truncated Gaussian conditional autoregressive (CAR) model, which allows for correlation of nearby regions, as specified by a smoothing parameter $\gamma >0$. Under the ICAR, $\mathcal{D}$ is given by
$$
\mathcal{D}(G,G_0)=\sum_{i=1}^{|V^{\rho^{\bk}}|}\sum_{j=1}^{|\mathcal{V}^{\rho^{\bk}}|}A_{i,j}\left(\int_{\mathcal{V}^{\rho^{\bk}}_i} (g(\boldsymbol{x}))-g_0(\boldsymbol{x}))d\boldsymbol{x}-\int_{V^{\rho^{\bk}}_j}(c(\boldsymbol{y})-g_0(\boldsymbol{y}))d \boldsymbol{y}\right)^2.
$$

In addition, when $\boldsymbol{A}$ is the adjacency matrix, the distance reduces to
$$
\mathcal{D}(G,G_0)=\sum_{B\in\nu^{^{\rho^{\bk}}}}\sum_{B'\in N_B}\left(\int_B (g(\boldsymbol{x})-g_0(\boldsymbol{x}))d \boldsymbol{x}-\int_{B'} (g(\boldsymbol{y})-g_0(\boldsymbol{y}))d\boldsymbol{y}\right)^2.
$$
Please notice that, in general, when $\gamma=1$ the latter expression is not a distance, but only a pseudo-metric, since adding a constant to either $c$ or $g_0$ does not change the value of $\mathcal{D}(G,G_0)$. However, since $G$ and $G_0$ are both restricted to the space of yett-uniform copulas, $\mathcal{D}$ does define a distance on the corresponding domain.

\subsection{Support properties}

The large topological support of a prior on an infinite-dimensional space is an important property that any Bayesian model should ideally possess. In fact, assigning positive prior probability mass to neighborhoods of any probability distribution is a minimum requirement for a Bayesian model to be considered nonparametric. The following theorem provides sufficient conditions for the space of $d$-dimensional absolutely continuous copulas, $\mathcal{C}_d$, to be the Hellinger support of a BYUP with random degree $\bk$.

\begin{theorem}\label{L1support} Assume the hierarchical model given by
$C \mid \bk,\alpha, G_0 \sim BYUP(\bk,  \alpha, G_0)$ and $\bk \sim p_{\bk}$. Let $\Pi\left( \cdot \mid  \alpha, G_0,  p_{\bk}\right)$ be the probability law of the mixture of BYUPs induced by it. Assume that $p_{\bk}$ is such that for every $\nu\in\mathbb{N}$, there is a $\mathbf{l} \in\mathbb{N}^d$, such that each $l_i>\nu$ and $p_{\bk}(\mathbf{l})>0$. Assume also that $\alpha > 0$ and $G_0 \in \mathcal{C}_d$. 
Then,  for every $M \in \mathcal{C}_d$
and for every $\epsilon>0$,
    $$
    \Pi \left( A^M_{\epsilon} \mid  \alpha, G_0, p_{\bk} \right) > 0,
    $$
were $A^H_\epsilon=\{N \in \mathcal{C}_d: \mathcal{H}(H||N)<\epsilon\}$, with $\mathcal{H}(\cdot||\cdot)$ being the Hellinger distance.
\end{theorem}

If stronger assumptions on the space of copula functions are imposed, a stronger support property can be obtained.

\begin{theorem}\label{KLsupport}
Assume the hierarchical model given by
$C \mid \bk,\alpha, G_0 \sim BYUP(\bk,  \alpha, G_0)$ and $\bk \sim p_{\bk}$. Let $\Pi\left( \cdot \mid  \alpha, G_0,  p_{\bk}\right)$ be the probability law of the mixture of BYUPs induced by it. Assume that $p_{\bk}$ is such that for every $\nu\in\mathbb{N}$, there is a $\mathbf{l} \in\mathbb{N}^d$, such that each $l_i>\nu$ and $p_{\bk}(\mathbf{l})>0$. Assume also that $\alpha > 0$ and $G_0 \in \mathcal{C}_d$. 
Let $\widetilde{\mathcal{C}}_d \subset \mathcal{C}_d$ be the space of absolutely continuous 
copulas that admit a positive density $h$ fulfilling the following conditions:    
\begin{enumerate}
\item[(i)] $-\infty < \int h(\bx)\log(h(\bx))d\bx<\infty$, and
        
\item[(ii)] there is a $B\in\mathcal{R}$ and a $\gamma>0$ such that if $h(\bx)>B$ and $L_2(\bx,\bx')<\gamma$ then $h(\bx')>1$ for all $\bx' \in [0,1]^d$.
\end{enumerate}
Then, for every $M \in \widetilde{\mathcal{C}}_d$
and for every $\epsilon>0$
 $$
    \Pi \left( A^M_{\epsilon} \mid  \alpha, G_0, p_{\bk} \right) > 0,
    $$
where  $A^H_\epsilon=\{N \in \widetilde{\mathcal{C}}_d: \mathcal{KL}(H||N)<\epsilon\}$, with $\mathcal{KL}(\cdot||\cdot)$ being he Kullback-Leibler divergence.
\end{theorem}

\subsection{The asymptotic behavior of the posterior distribution}

We study the asymptotic behavior of the posterior distribution of BYUP, by assuming that we observe a random sample of size $n$, $\bx_1, \ldots, \bx_n \mid H \overset{i.i.d.}{\sim} H$,
where 
$$
H(\bx)=C(F_1(x_1),F_2(x_2), \ldots, F_d(x_d)),$$
with $C$ being the corresponding copula
function of $H$. The following theorem shows the weak consistency of the posterior distribution induced by an BYUP prior.

\begin{theorem}\label{weak_pc}
Assume that $\bx_1, \ldots, \bx_n \mid H \overset{i.i.d.}{\sim} H$, where $H(\bx)=C(F_1(x_1),F_2(x_2), \ldots, F_d(x_d))$, and $F_1, \ldots, F_d$ being known marginal distributions. Consider the mixture of BYUPs prior for $C$ induced by the following hierarchical model
$C \mid \bk,\alpha, G_0 \sim BYUP(\bk,  \alpha, G_0)$ and $\bk \sim p_{\bk}$.  Suppose that $p_{\bk}$ is such that for every $\nu\in\mathbb{N}$, there is a $\mathbf{l} \in\mathbb{N}^d$, such that each $l_i>\nu$ and $p_{\bk}(\mathbf{l})>0$. Assume also that $\alpha > 0$ and $G_0 \in \mathcal{C}_d$. 
Then,  for every true copula model 
$M \in \widetilde{\mathcal{C}}_d$ and for every weak neighborhood of $M$, $A^M$,
    $$
    \Pi \left( A^M \mid  \alpha, G_0, p_{\bk}, \bx_1, \ldots, \bx_n\right) 
\xrightarrow[n \longrightarrow +\infty]{} 1,$$
$P$-almost surely, where
$P(\bx)=M(F_1(x_1),F_2(x_2), \ldots, F_d(x_d))$ and 
 $
    \Pi \left( \cdot \mid  \alpha, G_0, p_{\bk}, \bx_1, \ldots, \bx_n\right)$ denotes the posterior distribution under the mixture of BYUPs.
\end{theorem}

By considering a prior on the degree of the MBP with a particular tail behavior, the posterior distribution of the proposed model is strongly consistent at
any continuous copula function under simple random sampling.

\begin{theorem}\label{l1_pc}
Assume that $\bx_1, \ldots, \bx_n \mid H \overset{i.i.d.}{\sim} H$, where $H(\bx)=C(F_1(x_1),F_2(x_2), \ldots, F_d(x_d))$, and $F_1, \ldots, F_d$ being known marginal distributions. Consider the mixture of BYUPs prior for $C$ induced by the following hierarchical model
$C \mid \bk,\alpha, G_0 \sim BYUP(\bk,  \alpha, G_0)$, where $k=k_1=k_2=\ldots=k_d$, 
and $k \sim p_{k}$.  Now let $\bar{P}_l=\sum_{r=l}^\infty p_{k}(r)$, and assume $p_k$ satisfies $\bar{P}_l<p^{l^d+a}$ for some $0<p<1$, some $a>0$ and large enough $l$,
Then,  for every true copula model $M \in \widetilde{\mathcal{C}}_d$
and for every $\epsilon>0$,
    $$
    \Pi \left( A^M_{\epsilon} \mid  \alpha, G_0, p_{\bk}, \bx_1, \ldots, \bx_n\right) 
\xrightarrow[n \longrightarrow +\infty]{} 1,$$
$P$-almost surely, where
$P(\bx)=M(F_1(x_1),F_2(x_2), \ldots, F_d(x_d))$  and $A^H_\epsilon=\{N \in \widetilde{\mathcal{C}}_d: \mathcal{H}(H||N)<\epsilon\}$, with $\mathcal{H}(\cdot||\cdot)$ being the Hellinger distance.
\end{theorem}

\section{Posterior computation for finite Bernstein yett-uniform process priors}\label{sec4}

We propose a Metropolis-within Gibbs algorithm for exploring the posterior distribution of the copula functions and the parameters associated with the marginal distributions. Assume that we observe an independent and identically distributed (i.i.d.) sample of size $n$ from a $d$-variate continuous distribution $H$, $\bx_1, \ldots, \bx_n \mid H \overset{i.i.d.}{\sim} H$, 
where  $H(\bx)=C(F_1(x_1),F_2(x_2), \ldots, F_d(x_d))$, with $F_1,\ldots, F_n$ being the marginal distributions of $H$, and $C$ is the corresponding copula function. We model $C$ using a finite BYUP prior,
$C \mid \bk,\alpha, G_0 \sim BYUP(\bk,  \alpha, G_0)$, where $\bk$ is fixed. Under the BYUP prior, the hierarchical representation of the model is given by
\begin{eqnarray}\label{like}\nonumber
p(\bx_1,\ldots,  \bx_n \mid H) &=& 
\prod_{i=1}^n \left( \prod_{j=1}^d   f_j\left(x_{ij}\right) \right) c\left(F_j(x_{i1}, \ldots, F_j(x_{id})
\right),
\\\nonumber
&=&  \prod_{i=1}^n \left( \prod_{j=1}^d   f_j\left(x_{ij}\right) \right) \times  \\
&&
 \left(
\sum_{\nu_1=1}^{k_1} \cdots    \sum_{\nu_d=1}^{k_d} 
W_{\nu_1, \ldots, \nu_d}  \prod_{j=1}^d \beta(
F_j(x_{ij}) \mid \nu_j, k_j - \nu_j +1)
\right),
\end{eqnarray}
where $W_{\nu_1, \ldots, \nu_d}  =
G\left( 
\left(\frac{\nu_1-1}{k_1},\frac{\nu_1}{k_1}\right]
\times \ldots \times 
\left(\frac{\nu_d-1}{k_d},\frac{\nu_d}{k_d}\right]
\right)$, and 
\begin{eqnarray}
p(G \mid \bk, \alpha, G_0 )\propto \exp\left\{ -\frac{1}{2} \alpha \times \mathcal{D}\left(G,G_0 \right)\right\} \times I_{\widetilde{\mathcal{C}}_{d,\bk}}(G),
\end{eqnarray}
where $\alpha > 0$, $G_0$ is a reference copula, and $\mathcal{D}$ is a suitable distance for probability distributions. 

\subsection{Updating the marginal distributions}

There is no single technique which will work efficiently for the updating of the parameters of all possible marginal distributions, and tuning the posterior sampler may be difficult. However, there are some algorithms which are effective for a broad scope of distributions, and can reasonably explore the posterior without having to worry about tuning. A good starting point when parametric marginal distributions are considered is the $t$-walk \citep{twalk}, which is a general purpose sampler for parametric continuous distributions. The $t$-walk is a MH algorithm which adapts to the scale of the target distribution and can sample well from most finite dimensional continuous distributions without tuning.

The $t$-walk requires a support function for the parameters, an energy function, and two starting points within the support. The energy function as required by the algorithm is the negative log-density of the posterior distribution (up to a proportionality constant). We write the marginal distributions corresponding to some parametric family as $F_1(x_1\mid\btheta_1),\, F_2(x_2\mid\btheta_2), \ldots F_d(x_d\mid \btheta_d)$, where $\btheta_j$ is a set of parameters which defines the $j$th marginal, with prior $\pi_j(\btheta_j)$, $j=1,\ldots,d$. The energy function is, then, given by
\begin{eqnarray}\nonumber
E(\btheta_1,\ldots,\btheta_d) &\propto& -
\sum_{i=1}^n \left( \sum_{j=1}^d   \log f_j\left(x_{ij}\mid \btheta_j\right) \right),\\\nonumber
&& -
\sum_{i=1}^n 
 \log 
 \left(
\sum_{\nu_1=1}^{k_1} \cdots    \sum_{\nu_d=1}^{k_d} 
W_{\nu_1, \ldots, \nu_d}  \prod_{j=1}^d \beta(
F_j(x_{ij}) \mid \nu_j, k_j - \nu_j +1)
\right),\\\nonumber
&& - \sum_{j=1}^{d} \log \pi_j(\btheta_j).
\end{eqnarray}
If Dirichlet process mixture models \citep[see, e.g.,][]{bnpbook} are employed for modeling the marginal distributions, the marginal algorithm of \cite{ishwaran01}, which assume a finite-dimensional approximation to the nonparametric marginals, can be easily adapted to our context. 

\subsection{Updating the parameters of $G_0$} \label{hierarchicalmcmc}

Selecting a single default $G_0$ around which to center the prior is difficult because, once specified, a single centering distribution may affect inferences. Rather than selecting a single centering copula function, one option is to consider a hierarchical prior by allowing the parameters of the centering copula function $G_0$ to be random. One possible choice  is to use the Gaussian copula family given by
$$
G_{0, \bR}( \boldsymbol{x})=\Phi_{\bR}(\Phi^{-1}(x_1), \Phi^{-1}(x_2), \ldots, \Phi^{-1}(x_d)),
$$
parametrized by the correlation matrix $\bR$. Choosing a prior for $\bR$ is delicate since $p(G|\bk,\alpha,G_0)$ is known only up to a proportionality term, and this term depends on $G_0$ (and hence on $\bR$). We can write out the full prior for $G$ as
$$
p(G|\bk, \alpha,\bR)=N(\bR) \exp\left\{-\frac{1}{2}\alpha \times \mathcal{D}(G,G_{0,\bR})\right\},
$$
where $N(\bR)$ is a normalizing constant.  By default, we consider a conditional prior for $\bR$, such that 
$p(\bR \mid \bk, \alpha,  G_0)\propto \frac{1}{N(\bR)}$. 

When working with a hierarchical prior that establishes a prior distribution for $G_0$ which depends on $\bR$, updating $\bR$ can be done by adding a kernel to the MCMC chain.
To update $G_0$, we use a variation of the metropolized hit-and-run algorithm \citep{hitandrun}, which makes proposals that are always valid correlation matrices.

In our specific case, we allow $\delta$ to be a pre-specified tuning parameter. We consider values between 0.3 and 1.0, as discussed by \cite {hitandrun}. To update $\bR$, we propose a move from $\bR^{(i)}$ to $\bR^{(i+1)}=\bR^{(i)}+\bH$, by picking $\bH$ as follows:
\begin{enumerate}
    \item[(1)]{Let $\xi^{(i)}$ be the least eigenvalue of $\bR^{(i)}$.}
    \item[(2)]{Pick a sequence of i.i.d. standard normal variables $z_{1,2}, z_{1,3}, \ldots z_{d-1,d}$.}
    \item[(3)]{Pick $\delta\sim N(0,\boldsymbol{r}^2)$ truncated to $(-\frac{\xi^{(i)}}{\sqrt{2}},\frac{\xi^{(i)}}{\sqrt{2}})$.}
    \item[(4)]{For $i<j$ set $h_{i,j}=\frac{\delta z_{i,j}}{\sum_{j=1}^{d-1}\sum_{l=j}^D z_{j,l}^2}$. Also set $h_{i,i}=0$ for all $i$, and for $i>j$ set $h_{i,j}=h_{j,i}$. Set the matrix $\bH=[h_{i,j}]$.}
\end{enumerate}
The acceptance probability is given by 
$$
\max\{0,r(\bR^{(i+1)},\bR^{(i)})\},
$$
where 
\begin{eqnarray}\nonumber
    \log r(\bR^{(i+1)},\bR^{(i)})&=& \log p(G^{(i)}\mid \bk, \alpha^{(i)},\bR^{(i+1)}) -\log p(G^{(i)}\mid \bk, \alpha^{(i)}, \bR^{(i)}), \\\nonumber
    &=& \frac{1}{2}\alpha^{(i)}\left(\mathcal{D}(G,G_{0,\bR^{(i)}})-\mathcal{D}(G, G_{0,\bR^{(i+1)}})\right).
\end{eqnarray}

\subsection{Updating $G$}\label{updatingG}

Proposing a move in the copula space is not trivial.  Previously, \cite{kuschinski;jara;2021} introduced the rectangular exchange random walk proposal scheme which ensures a copula, relatively near the current state, is proposed.  While this proposal mechanism is useful, a drawback is that it moves slowly around the space of copulas. In each move only
4 blocks in the grid are possibly updated. In this section we introduce three alternative proposal schemes that can explore the space in every dimension at each step. For sampling notation we define $G^{(i)}$ to be the current value of $G$ in the Markov chain and $G^*$ to be a proposed value for $G$.
We point out that with $\bk$ fixed, $G$ is fully specified if we know the weights $W_{\nu_1, \ldots, \nu_d}$ in expression (\ref{densityMBP}).  The reverse is also true, given $\bk$, if $G$ is specified then $W_{\nu_1, \ldots, \nu_d}$ is also known.  Thus in our MCMC we could update $G$ or $W_{\nu_1, \ldots, \nu_d}$ and the result would be equivalent.  We opt to update $G$, in which we rely on some geometrical aspects of the yett-uniform copula space.

\subsubsection{Iterated rectangular exchanges (IRE) proposal}
This proposal is a simple extension of the rectangular exchanges (RE) introduced in \cite{kuschinski;jara;2021} and given in Definition \ref{def:rectExchange}.  The main idea of this proposal is to iterate multiple rectangular exchanges before actually computing the Metropolis-Hastings (M-H) acceptance ratio. Since  the RE and IRE proposals are symmetric, the M-H acceptance probability is given by 
$\min \left\{r, 1 \right\}$, where  
\begin{equation}
\label{eq:mh-ratio}
r =
\frac{  
\prod_{i=1}^n 
 \left(
\sum_{\nu_1=1}^{k_1} \cdots    \sum_{\nu_d=1}^{k_d} 
W^*_{\nu_1, \ldots, \nu_d}  \prod_{j=1}^d \beta(
F_j(x_{ij}) \mid \nu_j, k_j - \nu_j +1)
\right)
 \ p(G^* \mid \bk,\alpha,G_0)}{ 
 \prod_{i=1}^n 
 \left(
 \sum_{\nu_1=1}^{k_1} \cdots  
 \sum_{\nu_d=1}^{k_d} 
W^{(i)}_{\nu_1, \ldots, \nu_d}  \prod_{j=1}^d \beta(F_j(x_{ij}) \mid \nu_j, k_j - \nu_j +1)\right)
 \ p(G^{(i)} \mid \bk,\alpha,G_0)},
\end{equation}
with
$$W^*_{\nu_1, \ldots, \nu_d}  =
G^*\left( 
\left(\frac{\nu_1-1}{k_1},\frac{\nu_1}{k_1}\right]
\times \ldots \times 
\left(\frac{\nu_d-1}{k_d},\frac{\nu_d}{k_d}\right]
\right),$$ 
and
$$W^{(i)}_{\nu_1, \ldots, \nu_d}  =
G^{(i)}\left( 
\left(\frac{\nu_1-1}{k_1},\frac{\nu_1}{k_1}\right]
\times \ldots \times 
\left(\frac{\nu_d-1}{k_d},\frac{\nu_d}{k_d}\right]
\right).$$ 
The algorithm to implement the iterated rectangular exchange proposal is given in Algorithm \ref{alg:iter-rec-exchanges}.

\begin{algorithm}
\setstretch{1}
\caption{The iterated rectangular exchange proposal.} \label{alg:iter-rec-exchanges}
\KwIn{$G^{(i)}$, $d$, $\bk$, and a positive integer tuning parameter $u$.}
\KwSty{Step 1:} Let $G^* = G^{(i)}$\;
\KwSty{Step 2:} Modify $G^*$ by performing a rectangular exchange (as given in Definition \ref{def:rectExchange})\;
\KwSty{Step 3:} Repeat Step 2 $u$ times\;
\KwOut{$G^*$, the proposed yett-uniform copula.}
\end{algorithm}

In implementation, the primary issue to resolve is tuning the number of RE iterations so that the probability of acceptance is neither too high nor too low.  If too many REs are iterated, the acceptance probability will be low and result in poor mixing of the MCMC chain.  Thus the discrete tuning parameter for this proposal scheme is simply the number of iterations.  If the acceptance probability is too high, increase the number of iterations, and conversely, if the acceptance probability is too low reduce the number of iterations.  

\subsubsection{Generalized rectangular exchange proposal}
The next proposal can be considered a generalization of rectangular exchanges.  This method can simultaneously propose rectangular exchanges in multiple dimensions of the yett.  The steps to generate a yett-uniform copula proposal is given in Algorithm \ref{alg:gen-rec-exchanges}.  It is worth noting that the structures $Z$, $Z_1$, and $Z_2$ in the algorithm in general are not copulas, but multivariate non-continuous probability density functions in the $d$-dimensional hypercube.  When generating $Z_2$ in Step 6 of the algorithm, it is critical that only indices from $Z_1$ with positive probability mass are permuted, the others must remain fixed.  In Steps 7 and 8 in the algorithm, to test that a copula is still contained in $\tilde{\mathcal{C}}_{d,\bk}$, one only needs to ensure that each cell is nonnegative.  In other words, $\varepsilon_m$ and $\varepsilon_M$ are selected when the cell (or cells) with the minimum probability mass is exactly zero.

\begin{algorithm}
\setstretch{1}
\caption{The generalized rectangular exchange proposal.} \label{alg:gen-rec-exchanges}
\KwIn{$G^{(i)}$, $d$, $\bk$, and a positive integer tuning parameter $u$.}
\KwSty{Step 1:} Define $k_m = \min\{k_1,\ldots,k_d\}$ and $G^* = G^{(i)}$\;
\KwSty{Step 2:} Define a yett-uniform $Z$ with the same yett as $G^{*}$, but no assigned probability\;
\KwSty{Step 3:} Find the cell in $Z$ that has index 1 in each dimension and assign it probability $1/k_m$\;
\KwSty{Step 4:} Repeat Step 3 for indices $2, 3, \ldots, k_m$\;
\KwSty{Step 5:} Randomly permute the indices of $Z$ to generate $Z_1$\;
\KwSty{Step 6:} Randomly permute the indices of $Z_1$ that have nonzero mass to generate $Z_2$\;
\KwSty{Step 7:} Find $\varepsilon_m$, the min value of $\varepsilon$ such that $G^{*}+\varepsilon \, (Z_1-Z_2)$ is still contained in $\tilde{\mathcal{C}}_{d,\bk}$\;
\KwSty{Step 8:} Find $\varepsilon_M$, the max value of $\varepsilon$ such that $G^{*}+\varepsilon \, (Z_1-Z_2)$ is still contained in $\tilde{\mathcal{C}}_{d,\bk}$\;
\KwSty{Step 9:} Randomly sample $\varepsilon^* \sim \text{uniform}\left(\varepsilon_m,\varepsilon_M\right)$\;
\KwSty{Step 10:} Define $G^* := G^{*} + \varepsilon^* \left(Z_1-Z_2 \right)$\;
\KwSty{Step 11:} $u := u-1$\;
\KwSty{Step 12:} If $u \geq 1$ go to Step 5\;
\KwOut{$G^*$, the proposed yett-uniform copula.}
\end{algorithm}

This proposal is symmetric, and the Metropolis acceptance probability is given by
$\min \left\{r, 1 \right\}$, where $r$ is defined in Equation \ref{eq:mh-ratio}. This proposal scheme always produces a yett-uniform copula.  It can ambitiously explore the copula space in one move, but that ambitiousness can be tempered with lower values of $u$. A simple two dimensional illustration follows.  The inputs are 
\begin{equation*}
 G^{(i)} = \begin{bmatrix}
1/24 &|& 1/24 &|& 4/24 &|& 2/24 \\ \hline
2/24 &|& 4/24 &|& 1/24 &|& 1/24 \\ \hline
3/24 &|& 1/24 &|& 1/24 &|& 3/24 \\
\end{bmatrix}, 
\end{equation*}
$d=2$, $\bk=(3,4)$, and $u=1$.  In Step 1 we have $k_m=3$.  For Steps 2-4 we define
\begin{equation*}
 Z = \begin{bmatrix}
1/3 &|& 0 &|& 0 &|& 0 \\ \hline
0 &|& 1/3 &|& 0 &|& 0 \\ \hline
0 &|& 0 &|& 1/3 &|& 0 \\
\end{bmatrix}. 
\end{equation*}
In Step 5 we randomly permute the rows and columns of $Z$ and obtain
\begin{equation*}
 Z_1 = \begin{bmatrix}
0 &|& 0 &|& 1/3 &|& 0 \\ \hline
0 &|& 0 &|& 0 &|& 1/3 \\ \hline
0 &|& 1/3 &|& 0 &|& 0 \\
\end{bmatrix}. 
\end{equation*}
Next, in Step 6 we randomly permute the nonzero rows and columns of $Z_1$ and get
\begin{equation*}
 Z_2 = \begin{bmatrix}
0 &|& 1/3 &|& 0 &|& 0 \\ \hline
0 &|& 0 &|& 1/3 &|& 0 \\ \hline
0 &|& 0 &|& 0 &|& 1/3 \\
\end{bmatrix}.
\end{equation*}
For Steps 7 and 8 we find $\varepsilon_m$ and $\varepsilon_M$ to be $\varepsilon_m=-3/24$ and $\varepsilon_2=3/24$.  Now in Step 9 we sample $\varepsilon^*$ from a uniform$(-3/24,3/24)$ and in one instance get $\varepsilon^* \approx 9/800$.  
Finally, in Step 10 we get the proposed copula
\begin{equation*}
 G^{*} \approx \begin{bmatrix}
1/24 &|& 1/24-3/800 &|& 4/24+3/800 &|& 2/24 \\ \hline
2/24 &|& 4/24 &|& 1/24-3/800 &|& 1/24+3/800 \\ \hline
3/24 &|& 1/24+3/800 &|& 1/24 &|& 3/24-3/800 \\
\end{bmatrix}.
\end{equation*}
It is clear that the same amount of probability mass is being added and subtracted from each row and column.  If some mass of $Z_1$ and $Z_2$ aligned in one of their cells, then this example would reduce to a rectangular exchange.  If $u$ was greater than 1 this process would be repeated. 

\subsubsection{Vertex-line proposal}
Before introducing the next proposal scheme we discuss some geometrical aspects of the yett-uniform copula space of degree $\bk$, denoted by 
$\tilde{\mathcal{C}}_{d,\bk}$.  We can describe $\tilde{\mathcal{C}}_{d,\bk}$ as a convex polytope in $\mathbb{R}^d$.  This polytope is defined by a finite number of vertices that bound the space. We denote the set of vertices of $\tilde{\mathcal{C}}_{d,\bk}$ as $\mathcal{E}_{d,\bk}$.  Any copula $G \in \tilde{\mathcal{C}}_{d,\bk}$ can be (non-uniquely) represented as a mixture of the vertices.  Thus for any yett-uniform copula, $G$, we can express it as  $G = \sum_{E_i \in \mathcal{E}_{d,\bk}} w_i E_i$, where $w_i$ are non-negative values that sum to one.
An issue with the vertex representation is that the number of vertices grows at a factorial rate as the dimension and the yett resolution increases.  Therefore it is not practical to enumerate all possible vertices even for relatively small values of $d$ and $k_i$, $i=1,\ldots,d$.

Using the fact that the vertices bound the yett-uniform copula space, we use them to navigate inside that space for MCMC proposals.  In this proposal we use the vertices of the polytope to propose a move $G^* \in \tilde{\mathcal{C}}_{d,\bk}$. Additionally, it can be shown that a finite sequence of these proposals can produce a Markov chain from any arbitrary point in the yett-uniform polytope to any other.  It is a simple process to sample a vertex (uniformly) from $\mathcal{E}_{d,\bk} \in \tilde{\mathcal{C}}_{d,\bk}$. An algorithm to uniformly sample a vertex from $\tilde{\mathcal{C}}_{d,\bk}$ (when $k_i=k_j$ for all $i$ and $j$ in $\{1,\ldots,d\}$) is given in Appendix 10 of the online supplementary material. 

\begin{figure}
    \centering
    \subfigure[Part of the polytope showing $G^{(i)}$.]{\resizebox{.23\textwidth}{!}{\begin{tikzpicture} 
\draw[ultra thick](6,1)--(6,-2);
\draw[gray,ultra thick,dashed](1.5,-2.77)--(6,-2); 
\draw[gray,ultra thick,dashed](2.25,2.63)--(6,1);
\draw[ultra thick](0,0) arc(20:35:-12cm); 
\draw[ultra thick](0,0) arc(-20:-60:-5cm);
\filldraw[black] (4,-.667) circle (2pt) node[anchor=north]{$G^{(i)}$};
\end{tikzpicture}}} \qquad 
    \subfigure[Step 1, sampling $E$.]{\resizebox{.25\textwidth}{!}{\begin{tikzpicture} 
\draw[ultra thick](6,1)--(6,-2); 
\draw[gray,ultra thick,dashed](1.5,-2.77)--(6,-2); 
\draw[gray,ultra thick,dashed](2.25,2.63)--(6,1);
\draw[ultra thick](0,0) arc(20:35:-12cm); 
\draw[ultra thick](0,0) arc(-20:-60:-5cm);
\draw[fill] (0,0) circle [radius=1pt];
\filldraw[black] (4,-.667) circle (2pt) node[anchor=north]{$G^{(i)}$};
\filldraw[black] (0,0) circle (2pt) node[anchor=east]{$E$};
\end{tikzpicture}}} \qquad
    \subfigure[Draw line through $E$ and $G^{(i)}$ and find the polytope boundary.]{\resizebox{.30\textwidth}{!}{\begin{tikzpicture} 
\draw[ultra thick](0,0)--(6,-1) node[pos=0,left]{$E$} node[right]{boundary};
\draw[ultra thick](6,1)--(6,-2); 
\draw[gray,ultra thick,dashed](1.5,-2.77)--(6,-2); 
\draw[gray,ultra thick,dashed](2.25,2.63)--(6,1);
\draw[ultra thick](0,0) arc(20:35:-12cm); 
\draw[ultra thick](0,0) arc(-20:-60:-5cm);
\draw[fill] (0,0) circle [radius=2pt];
\draw[fill] (6,-1) circle [radius=2pt];
\filldraw[black] (4,-.667) circle (2pt) node[anchor=north]{$G^{(i)}$};
\end{tikzpicture}}}\qquad 
    \caption{A graphical depiction of some steps in the vertex-line proposal algorithm.  $G^{(i)}$ is the current state of the Markov chain.  $E$ is the randomly selected vertex of the polytope.  The line segment from $E$ to ``boundary" shows the possible yett-uniform copula proposals.}
    \label{fig:vertexLineProp}
\end{figure}
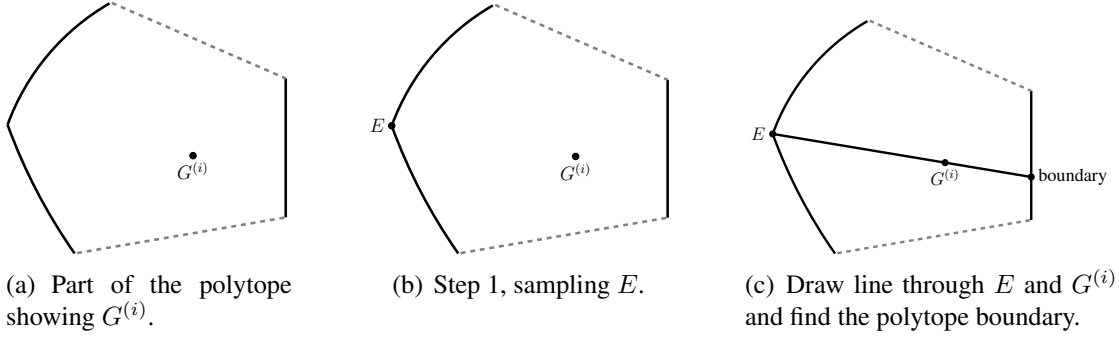

To generate a proposal, first sample one vertex $E \in \mathcal{E}_{d,\bk}$.  Then a line is defined, given by the function
$$f(\varepsilon) = \varepsilon G^{(i)} + (1-\varepsilon) E,$$ which passes through both points $G^{(i)}$ and $E$. A segment of this line exists in the polytope $\tilde{\mathcal{C}}_{d,\bk}$ and the proposal will lie on this segment.  A geometrical depiction of the proposal is shown in Figure \ref{fig:vertexLineProp}.  The steps to generate a copula proposal $G^*$ are outlined in Algorithm \ref{alg:vertex-line-prop}. 
\begin{algorithm}
\setstretch{1}
\caption{The vertex-line proposal.} \label{alg:vertex-line-prop}
\KwIn{$G^{(i)}$, $d$, $\bk$, and a positive tuning parameter $\tau$.}
\KwSty{Step 1:} Uniformly sample a vertex from $\mathcal{E}_{d,\bk}$, labeled as $E$\;
\KwSty{Step 2:} Find $\varepsilon^M$, the maximum value of $\varepsilon$ such that $\varepsilon G^{(i)} + (1-\varepsilon) E$ is still in $\tilde{\mathcal{C}}_{d,\bk}$\;
\KwSty{Step 3:} Sample $\varepsilon^* \sim \text{Truncated-Normal}\left(\mu=1,\sigma=1/\sqrt{\tau},min=0,max=\varepsilon_M\right)$\;
\KwSty{Step 4:} Define $G^* = \varepsilon^* \,G^{(i)} + (1-\varepsilon^* ) E$\;
\KwOut{$G^*$, the proposed yett-uniform copula.}
\end{algorithm}

The acceptance rates can be controlled with $\tau$; larger values of $\tau$ will, on average, produce proposals closer to $G^{(i)}$ and increase the proposal's acceptance probability.  
This proposal is not symmetric. The M-H acceptance probability is: $\min \left\{r \times \frac{q(G^{(i)}\mid G^*)
}{q( G^*\mid G^{(i)})}, 1 \right\}$, where $r$ is given in Equation \ref{eq:mh-ratio} and 
\begin{equation} \label{eqn:hastingsRatio}
\frac{q(G^{(i)}\mid G^*)
}{q( G^*\mid G^{(i)})} =
\frac{\phi\left( \sqrt{\tau} \left( \varepsilon^{(i)} - \varepsilon^* \right) \right)
\left[ \Phi\left( \sqrt{\tau} \left( \varepsilon_M - 1\right) \right) - \Phi\left( -\sqrt{\tau} \right) \right]
}
{\phi\left( \sqrt{\tau} \left( \varepsilon^{*} - 1 \right) \right)
\left[ \Phi\left( \sqrt{\tau} \left( \varepsilon_M - \varepsilon^*\right) \right) - \Phi\left( -\varepsilon^*\sqrt{\tau} \right) \right]}
\end{equation}
where, respectively, $\phi$ and $\Phi$ represent the pdf and cdf of the standard normal distribution. This proposal can also ambitiously explore the copula space in one move, but that ambitiousness is controlled by adjusting $\tau$. 

As in the previous subsection a simple two dimensional illustration is given.  The inputs are 
\begin{equation*}
 G^{(i)} = \begin{bmatrix}
1/24 &|& 1/24 &|& 4/24 &|& 2/24 \\ \hline
2/24 &|& 4/24 &|& 1/24 &|& 1/24 \\ \hline
3/24 &|& 1/24 &|& 1/24 &|& 3/24 \\
\end{bmatrix}, 
\end{equation*}
$d=2$, $\bk=(3,4)$, and $\tau=6$.  In Step 1 we randomly sample the vertex
\begin{equation*}
 E = \begin{bmatrix}
0 &|& 1/12 &|& 0 &|& 1/4 \\ \hline
1/4 &|& 1/12 &|& 0 &|& 0 \\ \hline
0 &|& 1/12 &|& 1/4 &|& 0 \\
\end{bmatrix}.
\end{equation*}
We obtain $\varepsilon_M = 1.2$, to complete Step 2.  In the next step we sample $\varepsilon^*$ from a Truncated-Normal$(\mu=1,\sigma=1/\sqrt{6},min=0,max=1.2)$ and in one instance get $\varepsilon^* \approx 0.936$.

Step 4 produces the proposal and output of  
\begin{equation*}
G^{*} \approx \begin{bmatrix}
0.039 &|& 0.039+0.005\bar{3} &|& 0.156 &|& 0.078+0.016 \\ \hline
0.078+0.016 &|& 0.156+0.005\bar{3} &|& 0.039 &|& 0.039 \\ \hline
0.117 &|& 0.039+0.005\bar{3} &|& 0.039+0.016 &|& 0.117 \\
\end{bmatrix}. 
\end{equation*}
Additional details regarding the generalized rectangular exchange proposal and the vertex-line proposal are provided in Appendix 11 of the online supplementary material.

\section{Illustrations}\label{sec5}

We illustrate the behavior of the proposed model and compare the proposed MCMC algorithms by means of the analysis of simulated and real data. Functions implementing the MCMC algorithms employed in these analyses were written in Julia. The codes are available upon request to the authors.

\subsection{Comparison of proposal distributions for $G$}

\begin{figure}[!h]
\centering
\subfigure[M1 - $n = 2,500$.] 
{
\includegraphics[width=3.5cm]{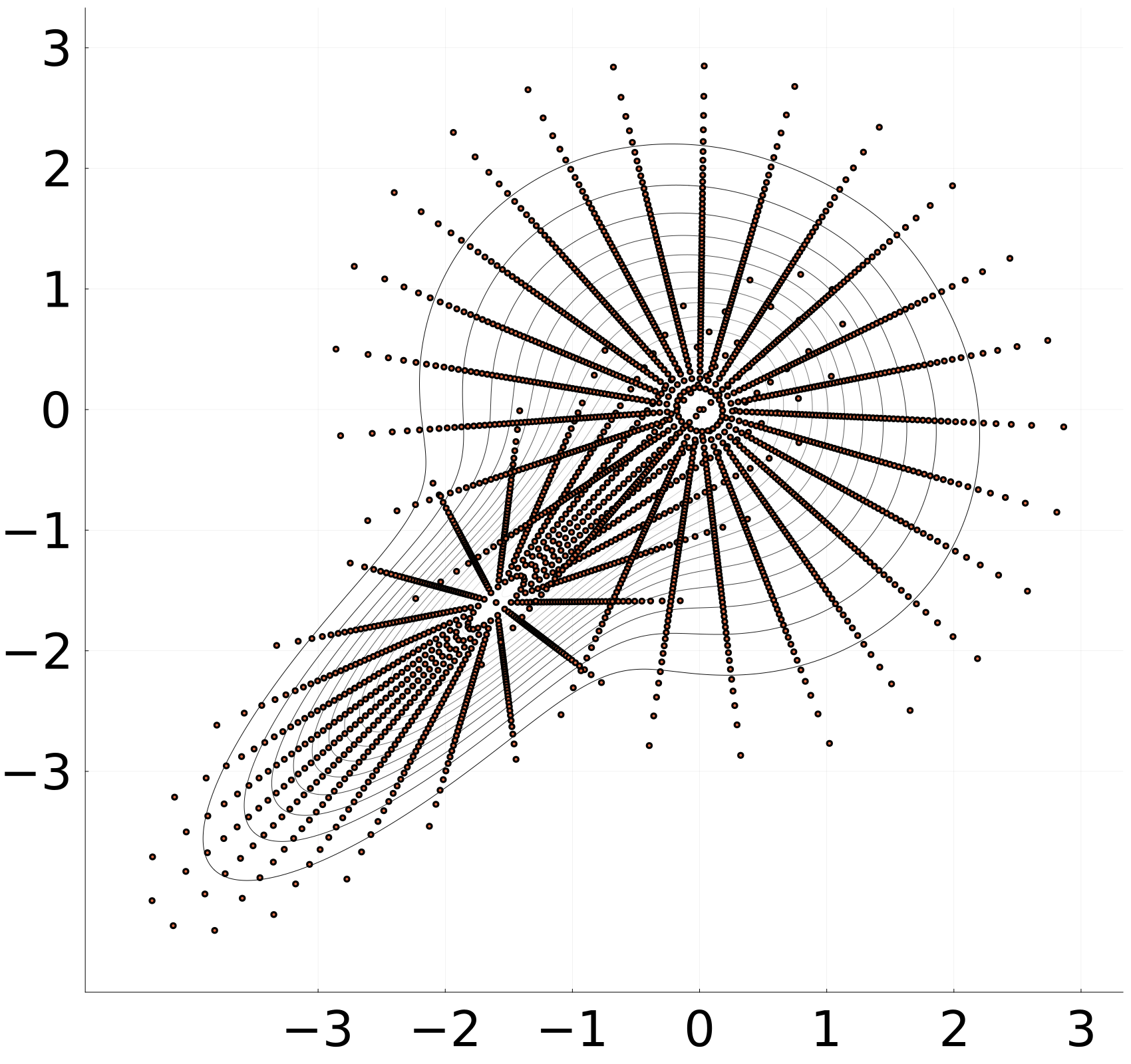}
 }      
\subfigure[M1  - $n = 10,000$.] 
{
\includegraphics[width=3.5cm]{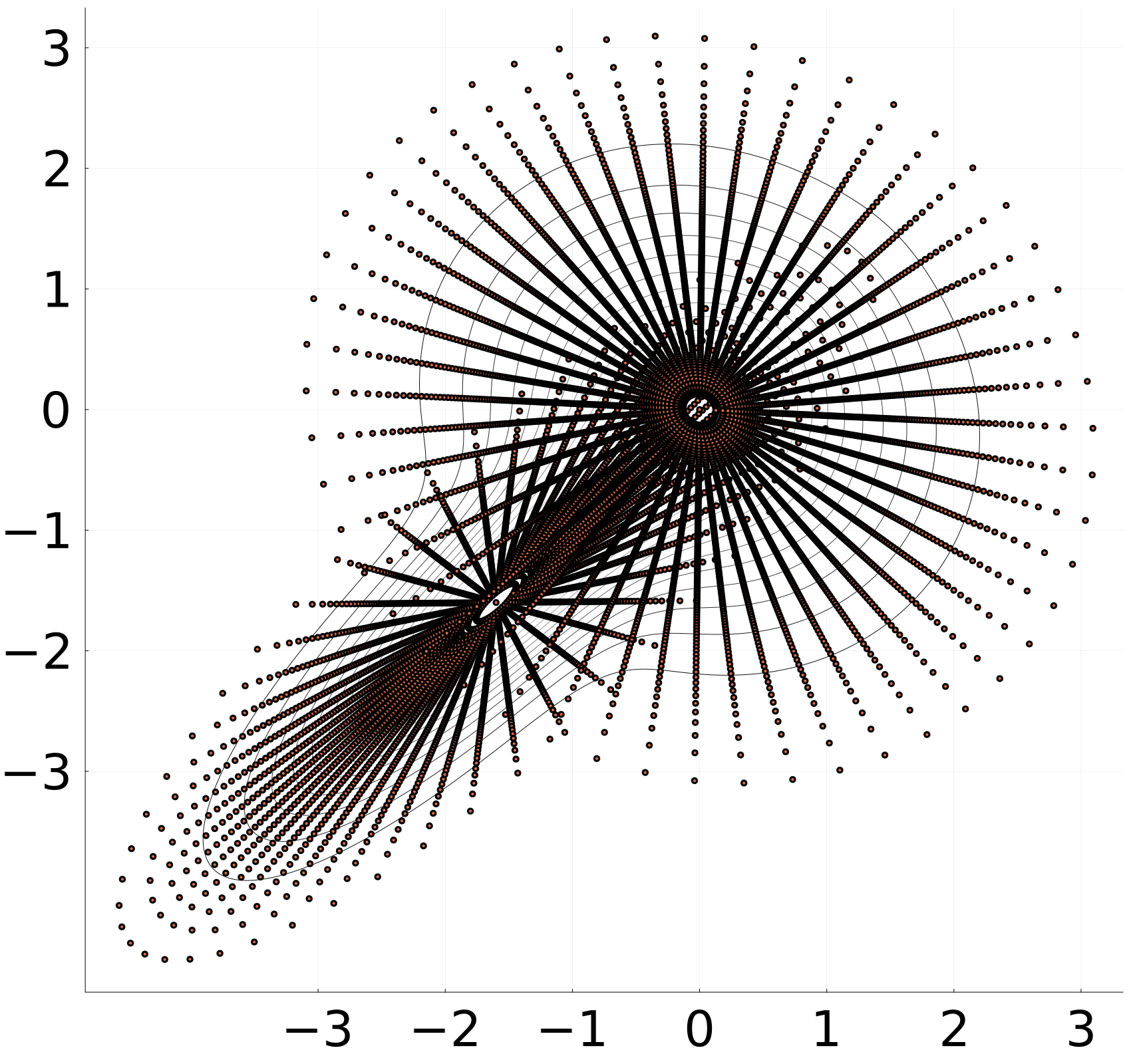}
 }
\subfigure[M2 - $n = 2,500$.] 
{
\includegraphics[width=3.5cm]{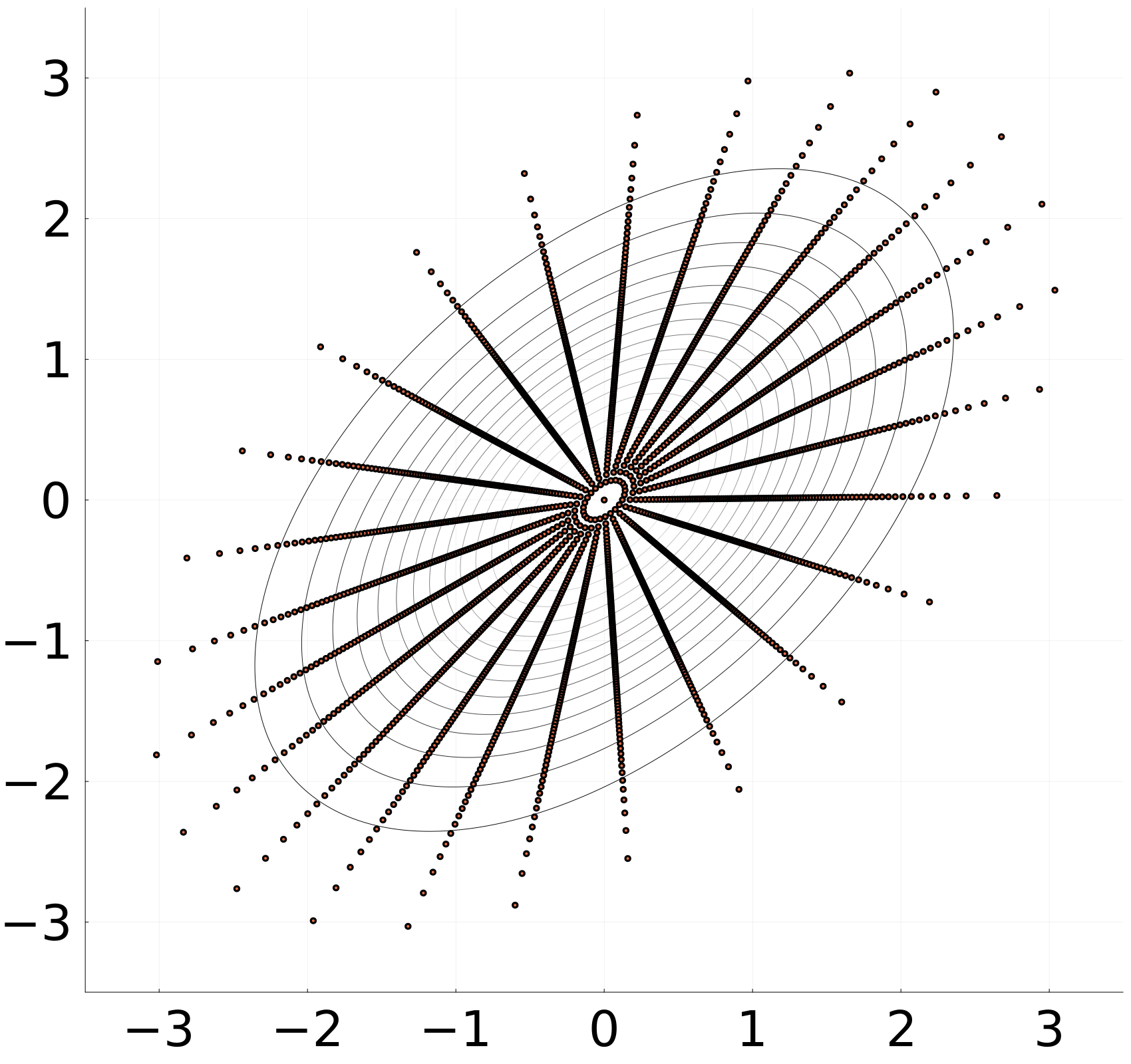}
 }      
\subfigure[M2 - $n = 10,000$.] 
{
\includegraphics[width=3.5cm]{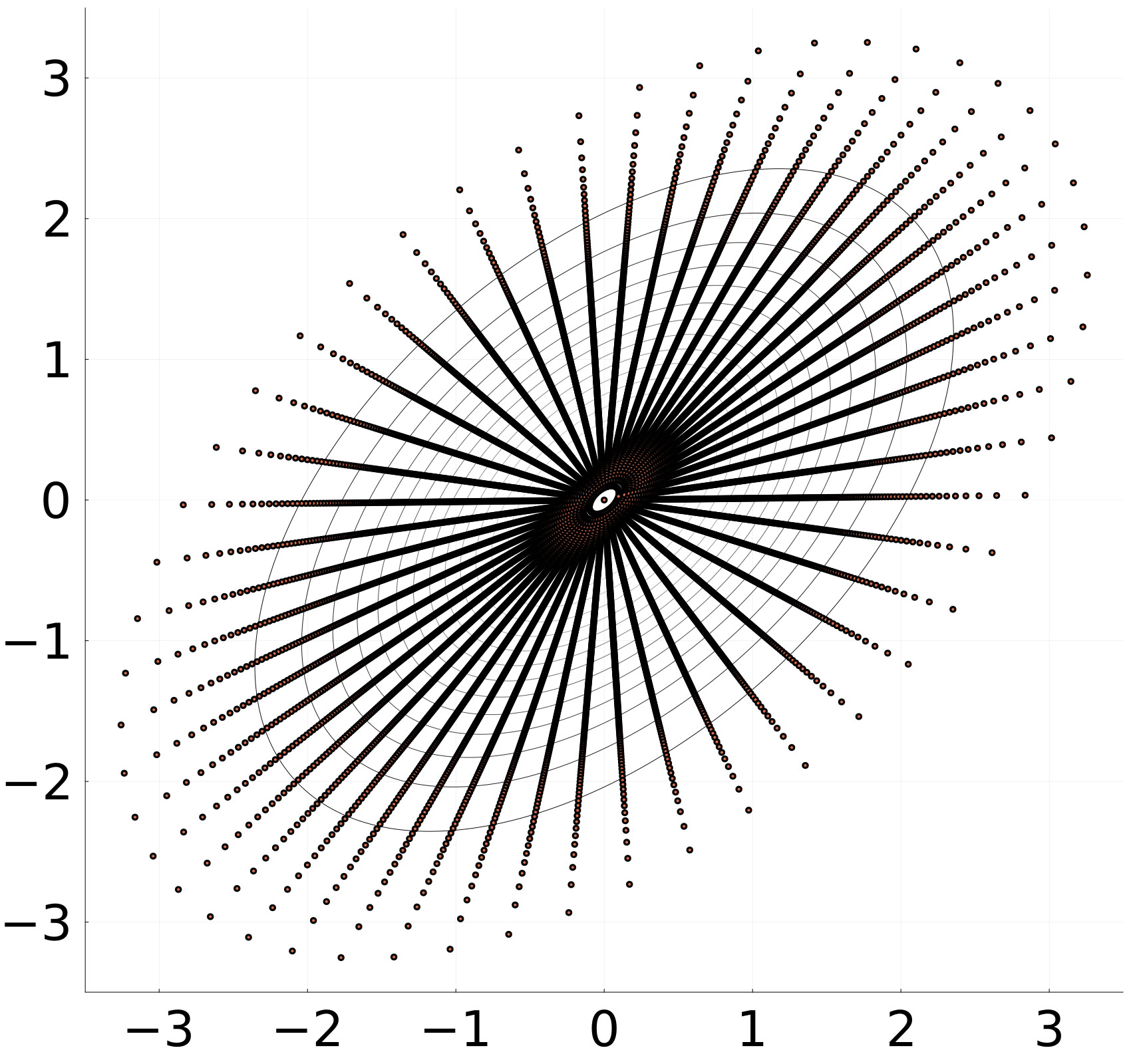}
 }
\\ 
 \subfigure[M3 - $n = 2,500$.] 
{
\includegraphics[width=3.5cm]{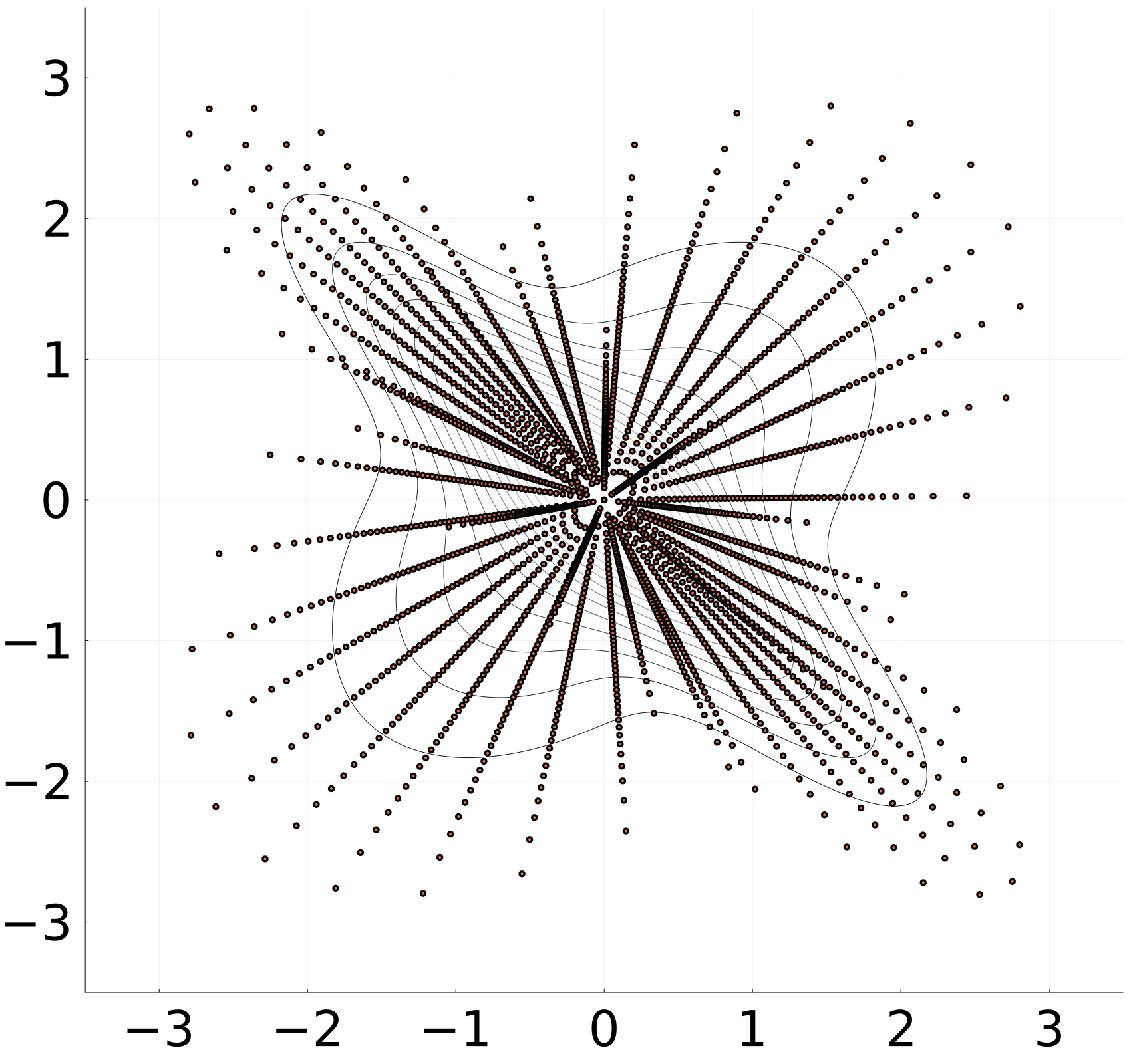}
 }      
\subfigure[M3 - $n = 10,000$.] 
{
\includegraphics[width=3.5cm]{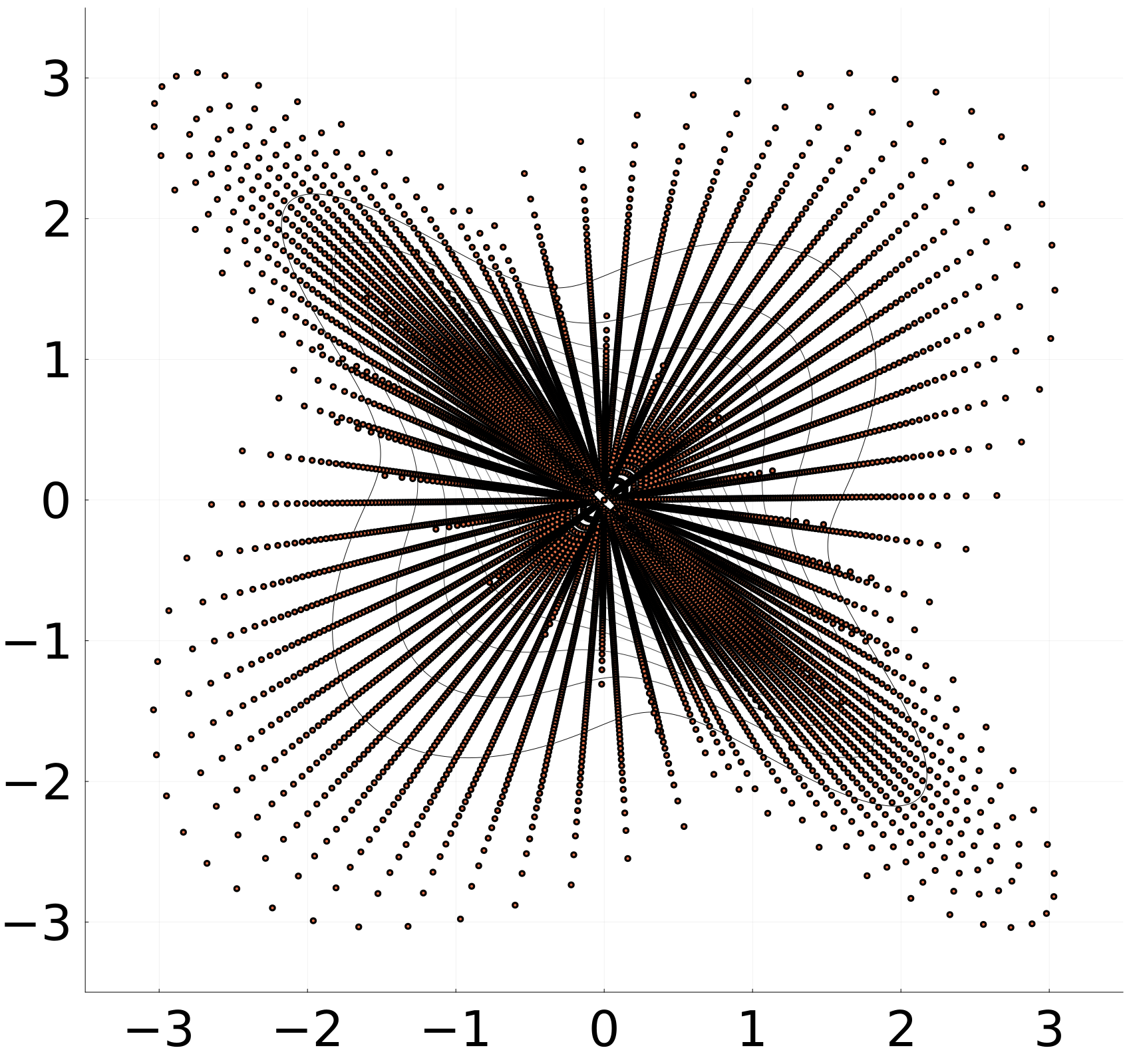}
 }
 \subfigure[M4 - $n = 2,500$.] 
{
\includegraphics[width=3.5cm]{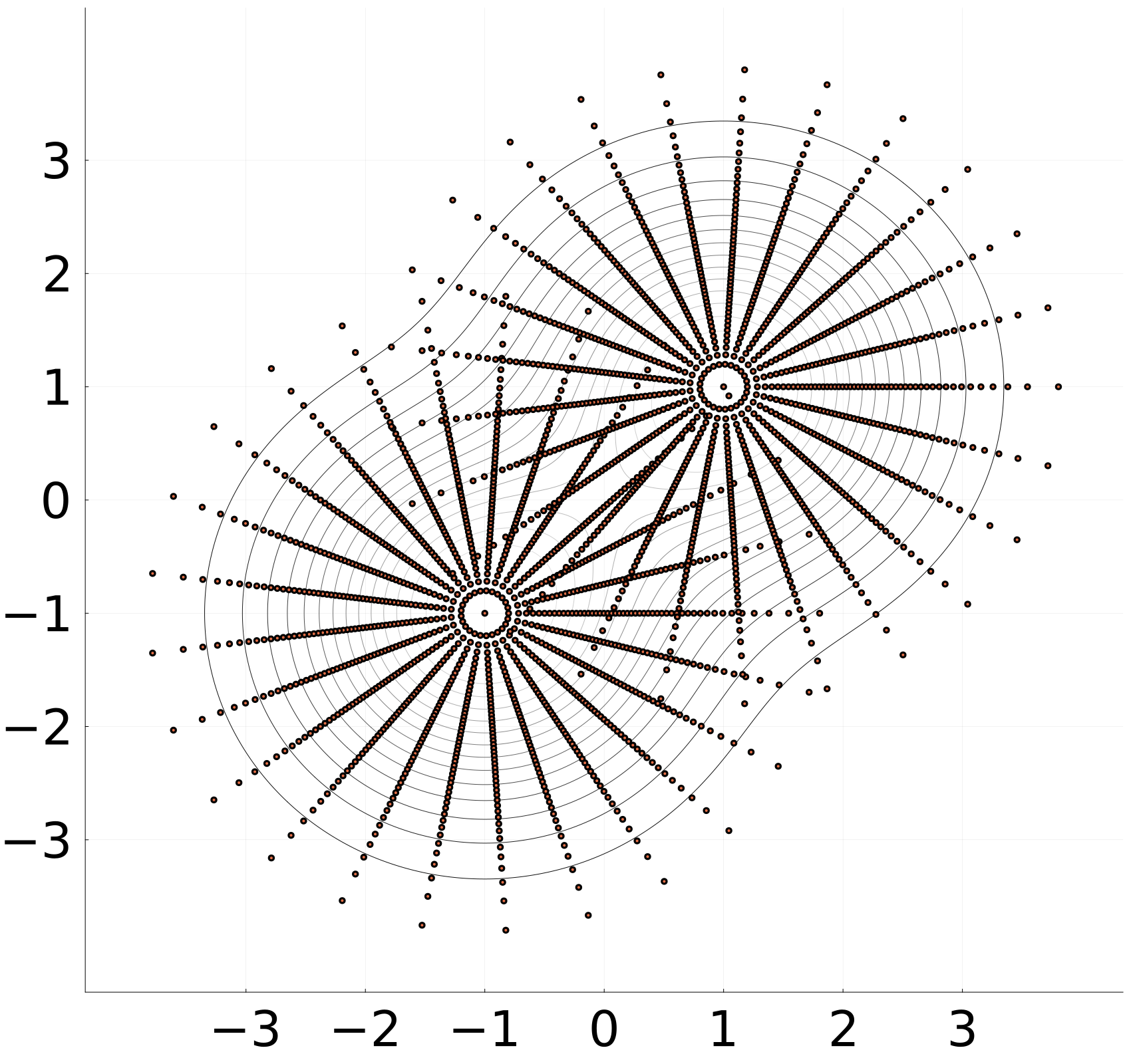}
 }      
\subfigure[M4 - $n = 10,000$.] 
{
\includegraphics[width=3.5cm]{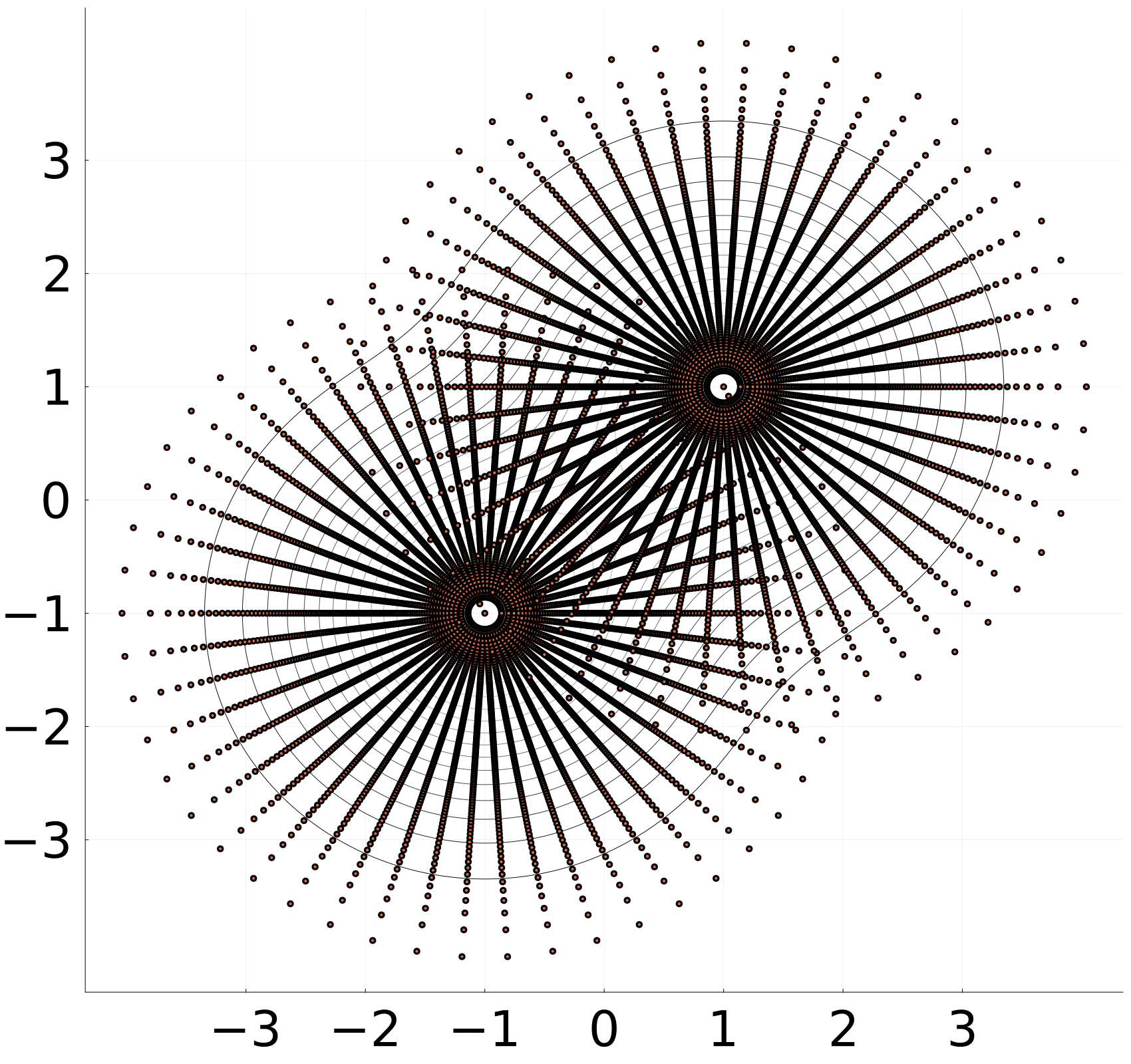}
 }      
 \caption{Simulated data - True models: Panels (a) and (b), (c) and (d),  (e) and (f), and (g) and (h) display the true models 1 (M1), 2 (M2), 3 (M3) and 4 (M4), respectively. Panels (a), (c), (d), and (g) display the perfect samples for $n= 2,500$. Panels (b), (d), (e), and (h) display the perfect samples for $n= 10,500$.}
\label{fig1:simulation1}
\end{figure}

The aims of this section are (i) to provide information on the optimal tuning parameters for each of the three updating schemes discussed in Section~\ref{updatingG} for $G$, and (ii) to compare the efficiency of the different updating schemes. We consider four true bivariate models, with different levels of complexity. 
Model 1 is a mixture of two Gaussian distributions that emulate the behavior of a bivariate model, with Gaussian $(0,1)$ marginals and a Clayton copula with parameter $3$. The true model in this case is given by:
$$
0.4 \times N_2\left(
\left(
\begin{array}{c}
-1.6 \\
-1.6
\end{array}\right),
\left(
\begin{array}{cc}
1.00 &  0.85 \\
 0.85 & 1.00
\end{array}
\right)
\right) +
0.6 \times 
 N_2\left(
\left(
\begin{array}{c}
0.0 \\
0.0
\end{array}\right),
\left(
\begin{array}{cc}
1.00 &  -0.10 \\
 -0.10 & 1.00
\end{array}
\right)
\right).
$$
Model 2 is a bivariate Gaussian model with standard marginal distributions and correlation of 0.5.
Model 3 is the following mixture of Gaussian distributions:
$$
0.5 \times N_2\left(
\left(
\begin{array}{c}
0.0 \\
0.0
\end{array}\right),
\left(
\begin{array}{cc}
1.00 &  0.50 \\
 0.50 & 1.00
\end{array}
\right)
\right) +
0.5 \times 
 N_2\left(
\left(
\begin{array}{c}
0.0 \\
0.0
\end{array}\right),
\left(
\begin{array}{cc}
1.00 &  -0.90 \\
 -0.90 & 1.00
\end{array}
\right)
\right).
$$
Finally, Model 4 is the following mixture of Gaussian distributions:
$$
0.5 \times N_2\left(
\left(
\begin{array}{c}
-1.0 \\
-1.0
\end{array}\right),
\left(
\begin{array}{cc}
1.00 &  0.00 \\
 0.00 & 1.00
\end{array}
\right)
\right) +
0.5 \times 
 N_2\left(
\left(
\begin{array}{c}
1.0 \\
1.0
\end{array}\right),
\left(
\begin{array}{cc}
1.00 &  0.00 \\
 0.00 & 1.00
\end{array}
\right)
\right).
$$
For each model, we simulate a perfect sample of size $N = $ 2,500 and 10,000. By a perfect sample we mean a deterministic selection of points in the corresponding sample space that try reproduce the main characteristics of the target distribution. Since our true models are either bi-variate Gaussian or mixture of bi-variate Gaussian distributions, we describe a generation mechanism of perfect samples for $2$-dimensional Gaussian  distributions only. For a $2$-dimensional standard Gaussian distribution, a perfect sample of size $a_1 \times a_2$ is achieved 
by generating a vector $\mathbf{q}$ of $a_1$ quantiles of a $\chi$ distribution with 2 degrees of freedom; please note that this is the distribution of a distance from a point distributed as a $2$-dimensional standard Gaussian distribution to the origin of the Euclidean space. Set the vector $\mathbf{r}$ to contain $a_2$ evenly spaced numbers in the interval $[0, 2\pi]$. Then the $i\times a_1 +j^{th}$ element of the perfect sample is given by $(q_j \times \cos(r_i), q_j \times \sin(r_i))$. Finally, to define a perfect sample of an arbitrary bivariate gaussian distribution with mean $\boldsymbol{\mu}$ and covariance matrix $\boldsymbol{\Sigma}$, $\mathbf{y}_P$,
we set $\mathbf{y}_P=\boldsymbol{\mu} + \mathbf{U} \times \mathbf{s}_P$, where $\mathbf{s}_P$ is the perfect sample of a bivariate standard gaussian and $\mathbf{U}$ is a symmetric square root of $\boldsymbol{\Sigma}$. Figure~\ref{fig1:simulation1} displays the true models under consideration, with the perfect samples for the two sample sizes under consideration. 

We consider 6 settings for each of the three  candidate generating kernels for $G$. Using each of these algorithms, we create a Markov chain of size 2,400,000, a burn-in period of 400,000, and a thin-in of 200 to explore the posterior distribution under the proposed model for each simulated data set, by considering $k$ = 10, the hierarchically centered prior with the ICAR correlation structure, and with $G_0$ being an independent copula. In these analyses we assume the marginal distributions to be known. 

\begin{figure}[!h]
\centering
\subfigure[] 
{
\includegraphics[width=3.8cm]{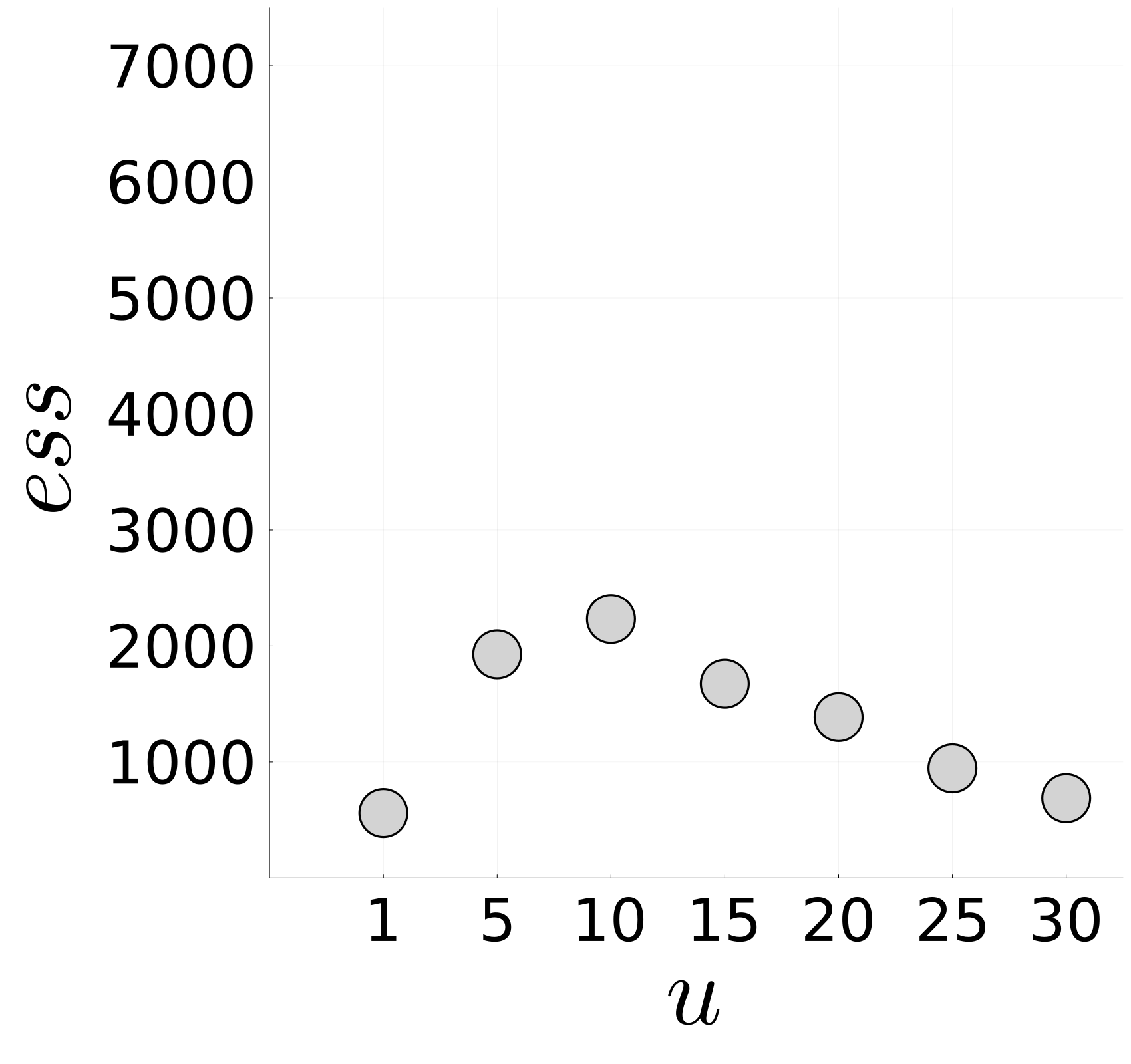}
 }
\subfigure[] 
{
\includegraphics[width=3.8cm]{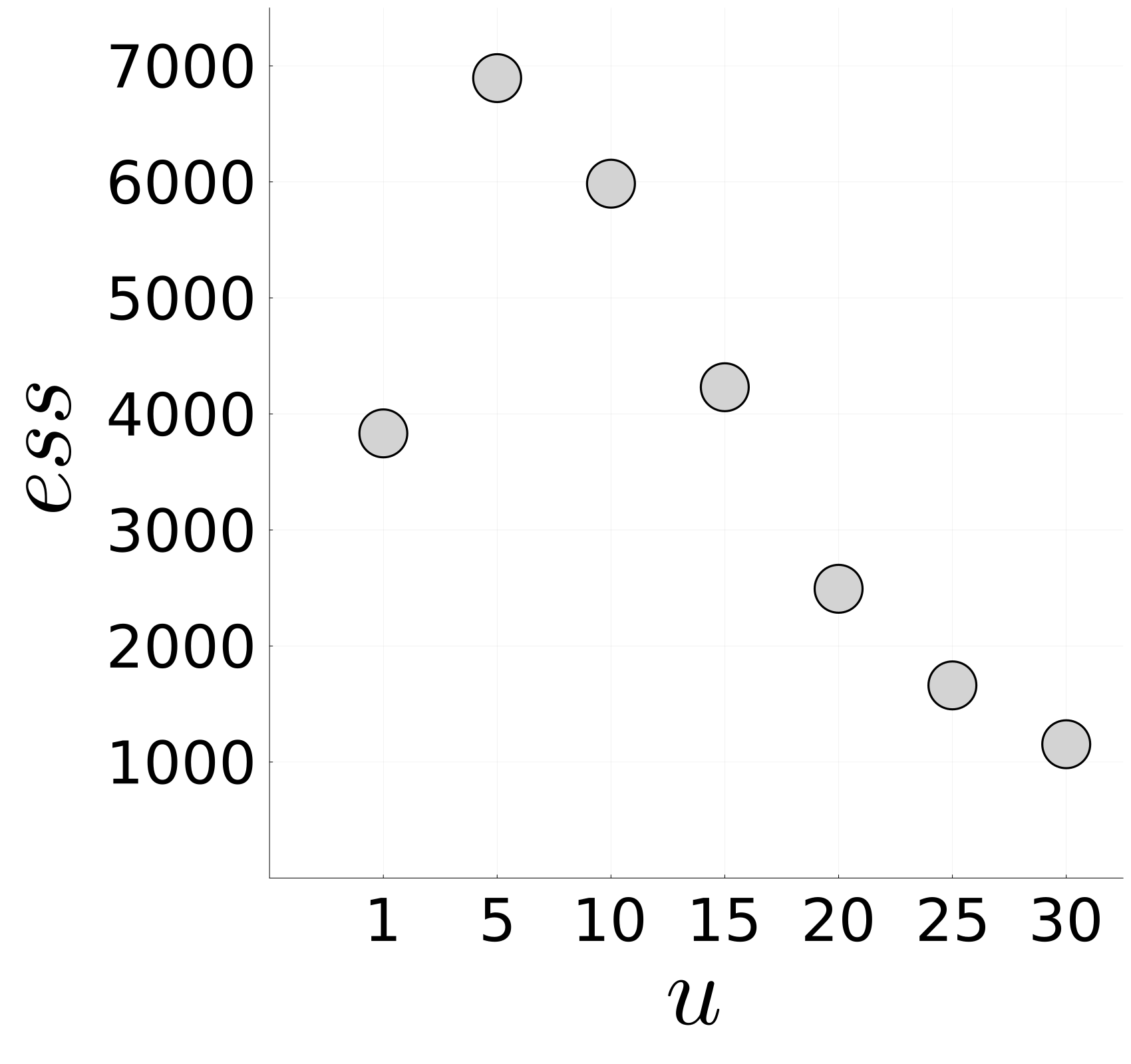}
 }
\subfigure[] 
{
\includegraphics[width=3.8cm]{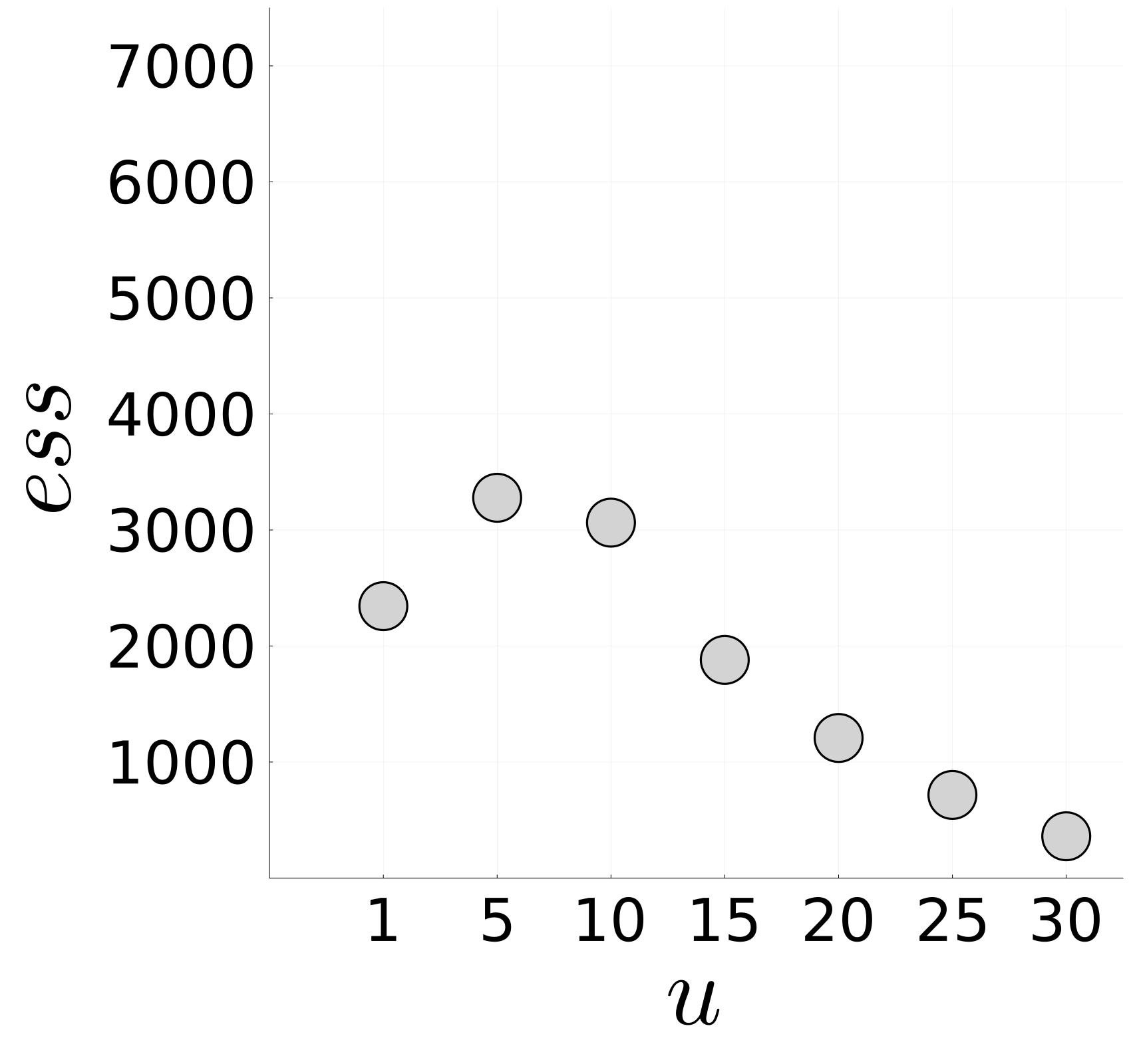}
 }      
\subfigure[] 
{
\includegraphics[width=3.8cm]{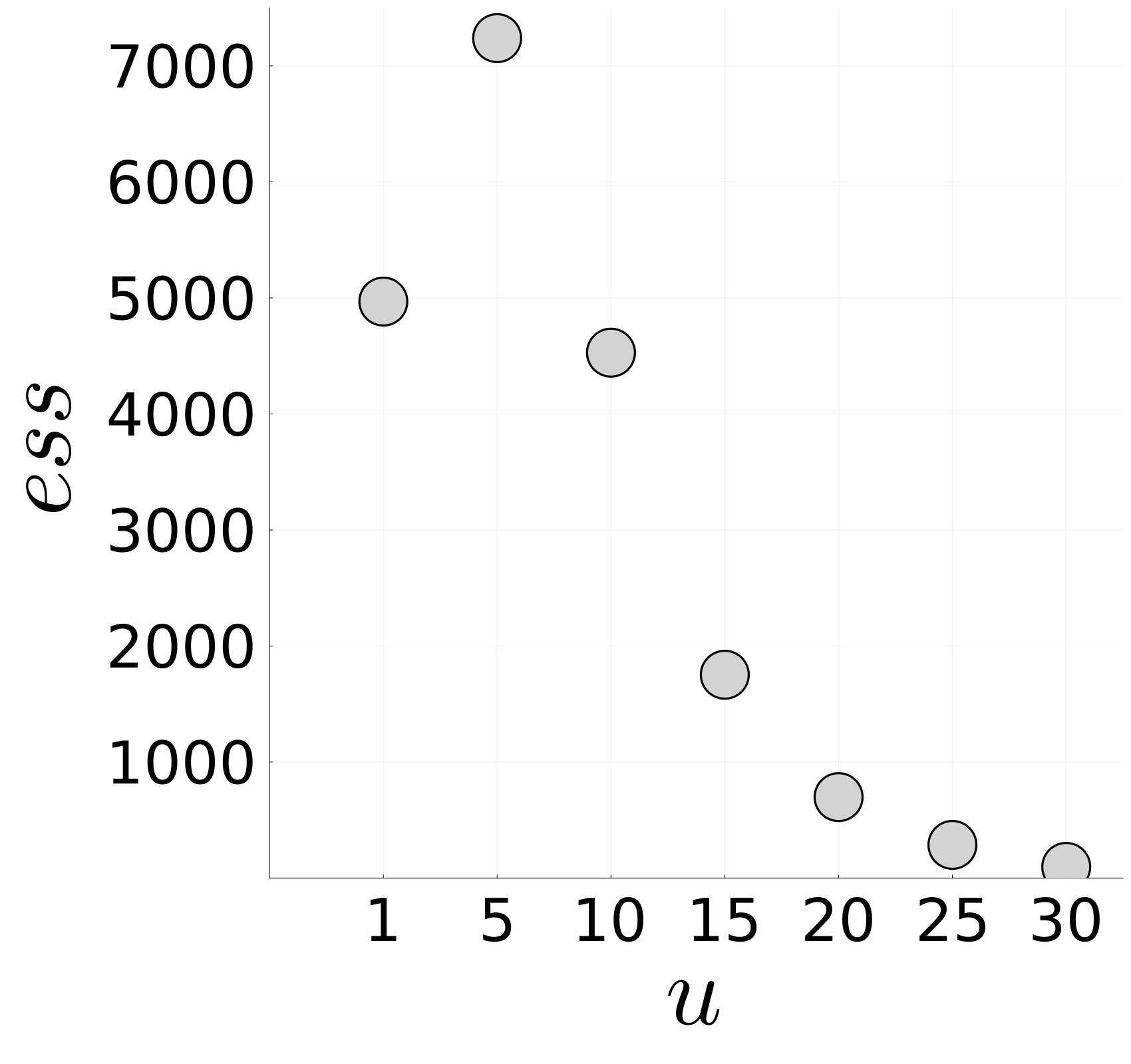}
 }      
\\
\subfigure[]
{
       \includegraphics[width=3.8cm]{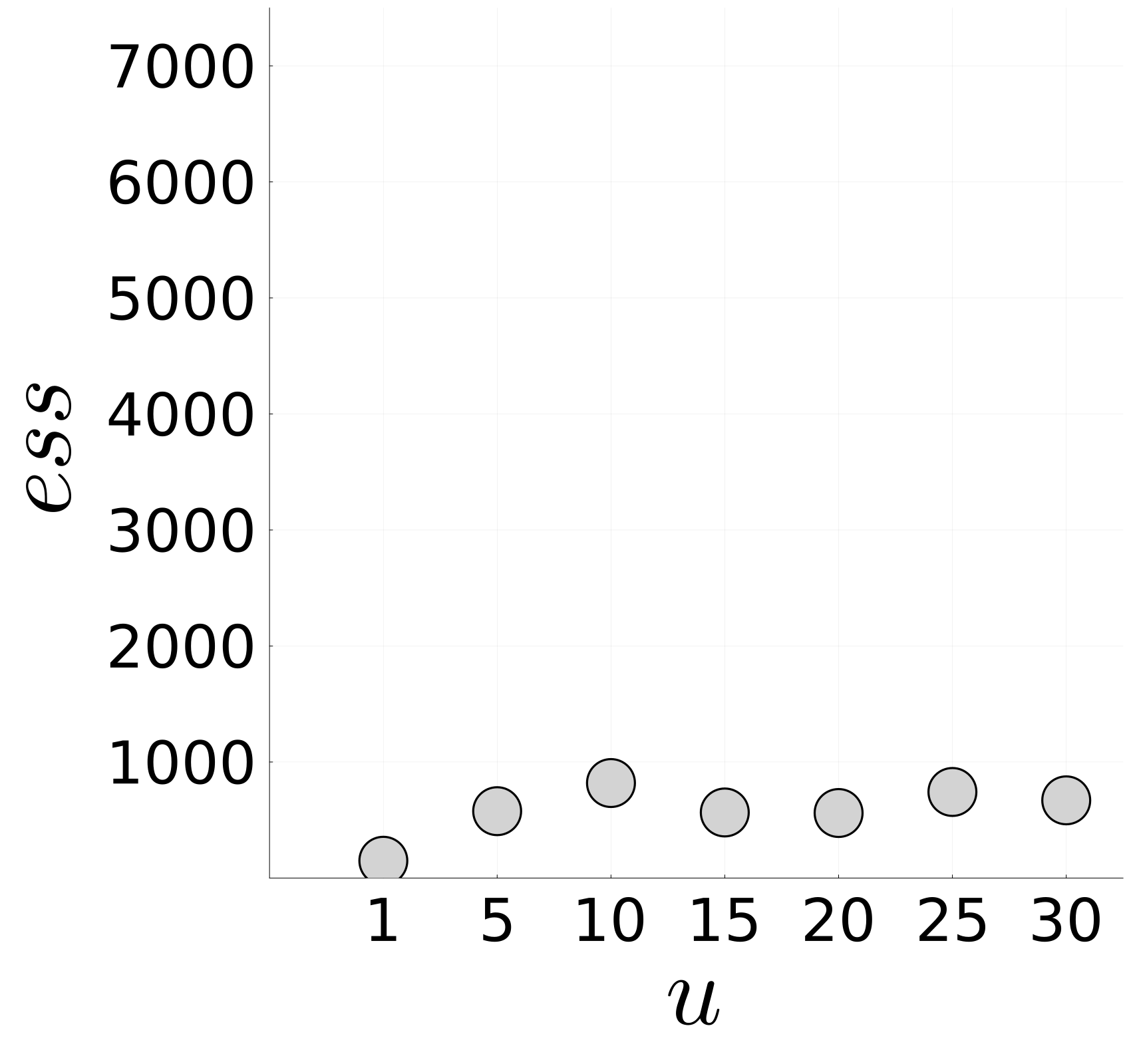}
 }
\subfigure[] 
{
       \includegraphics[width=3.8cm]{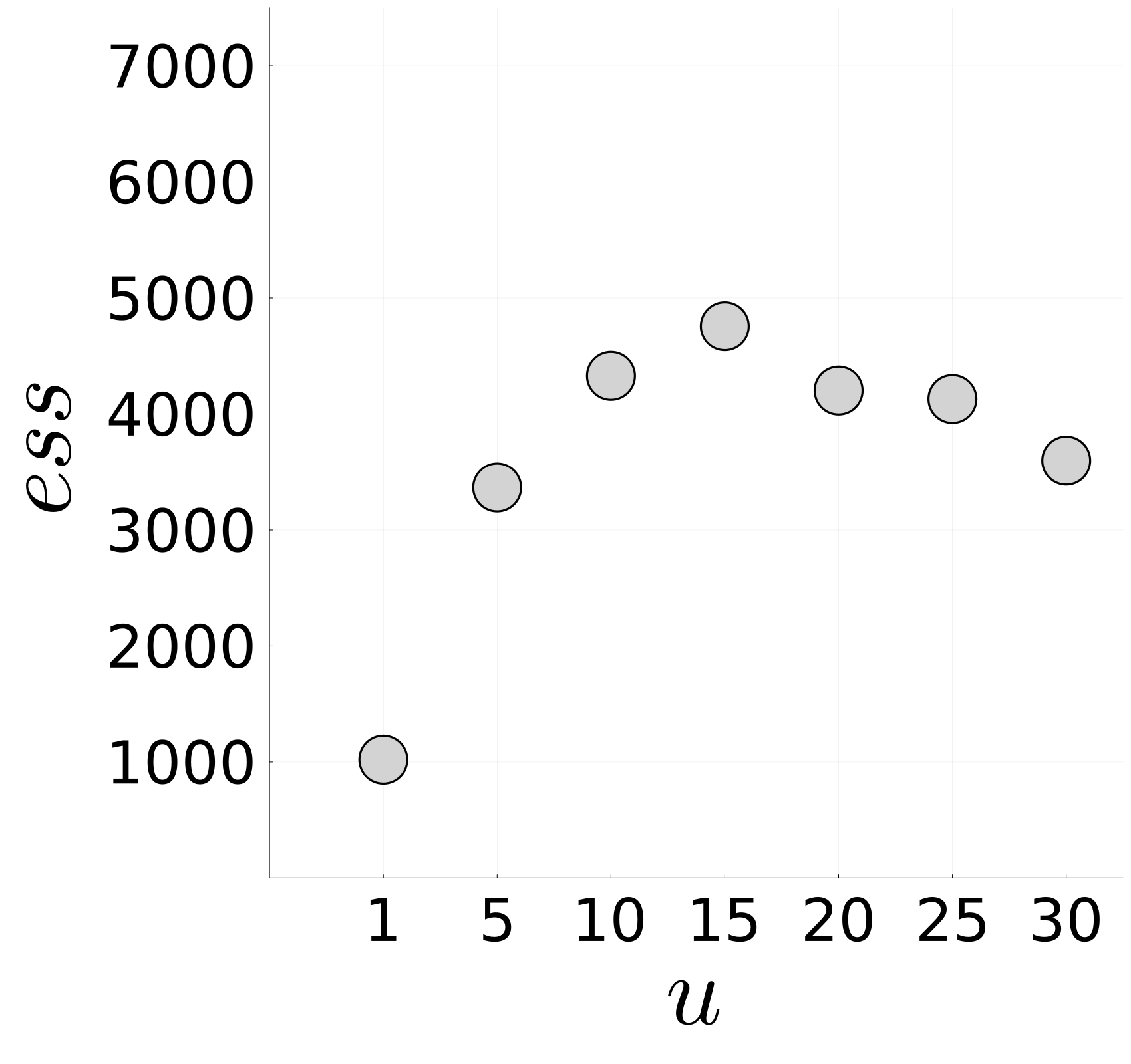}
 }
\subfigure[] 
{
       \includegraphics[width=3.8cm]{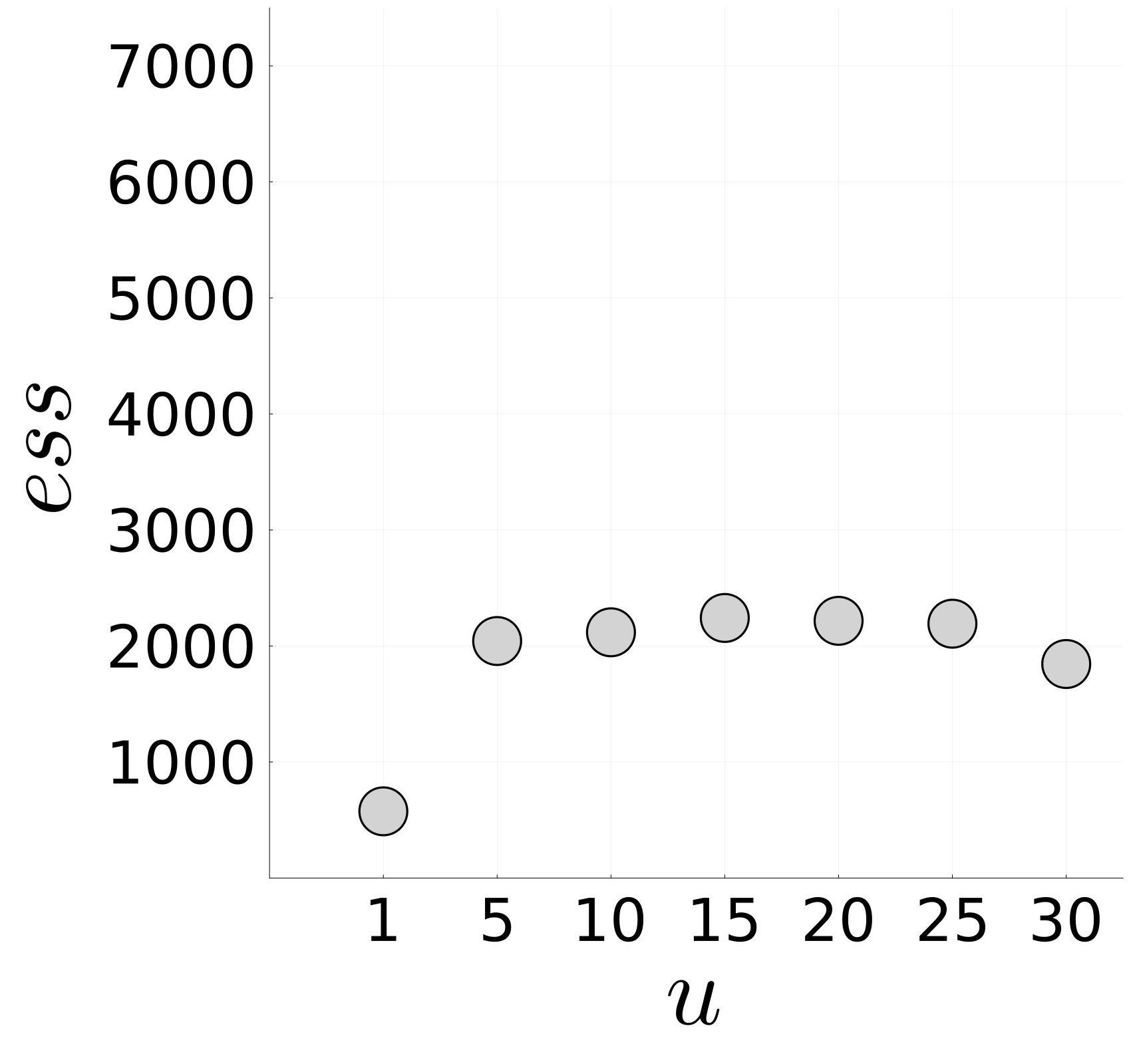}
 }      
\subfigure[] 
{
       \includegraphics[width=3.8cm]{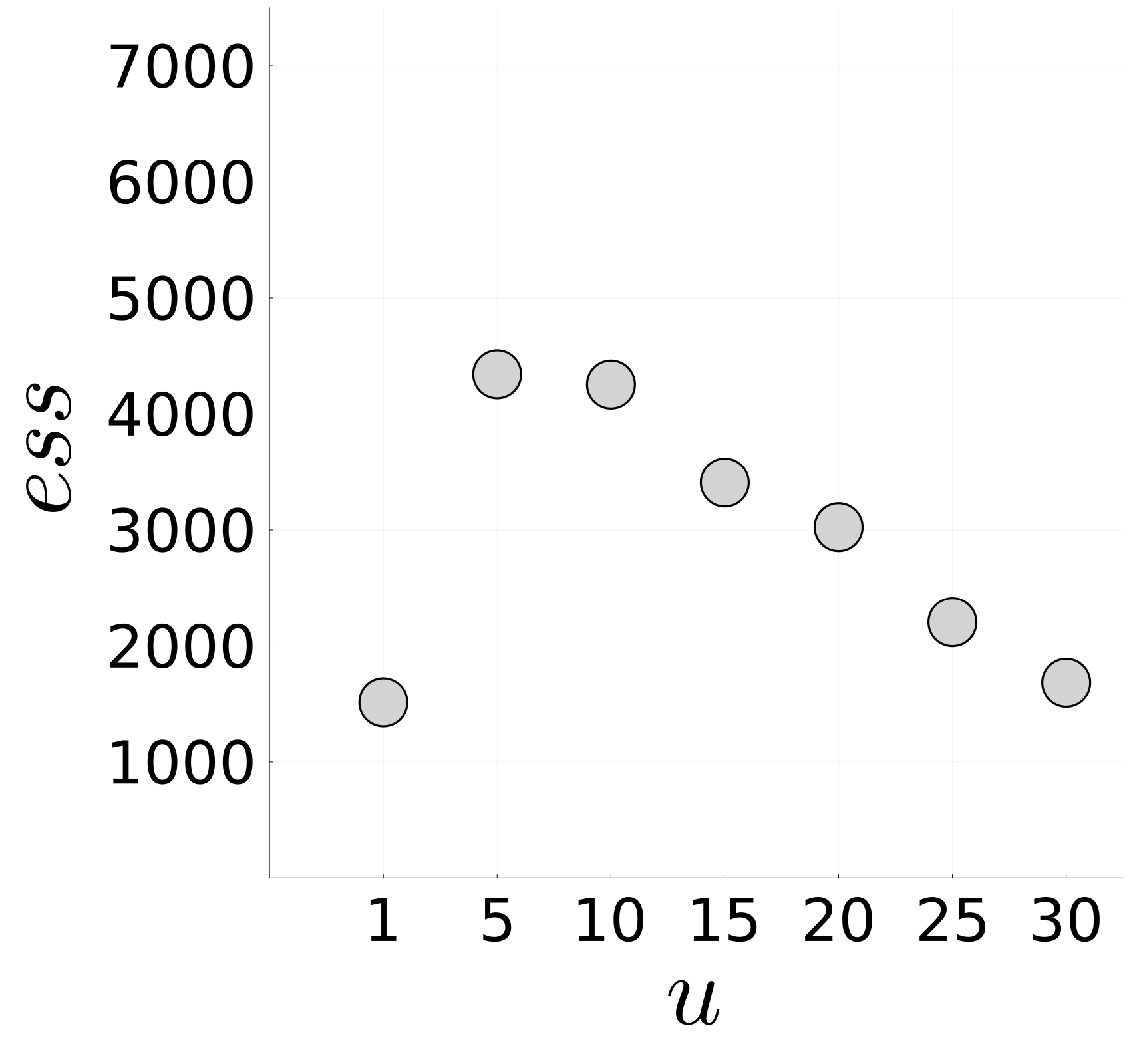}
 }  
\\ 
\subfigure[] 
{
       \includegraphics[width=3.8cm]{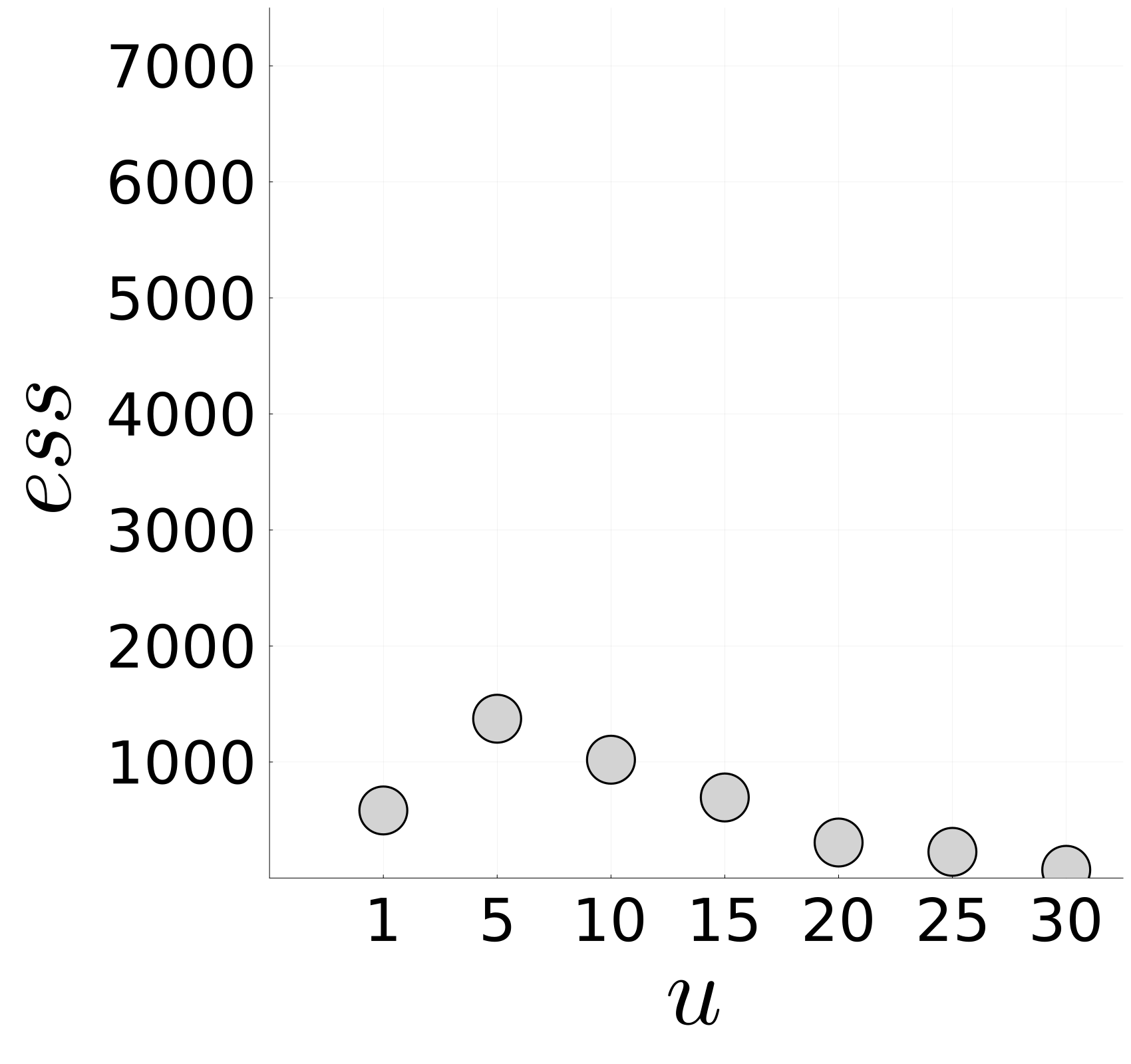}
 }
\subfigure[] 
{
       \includegraphics[width=3.8cm]{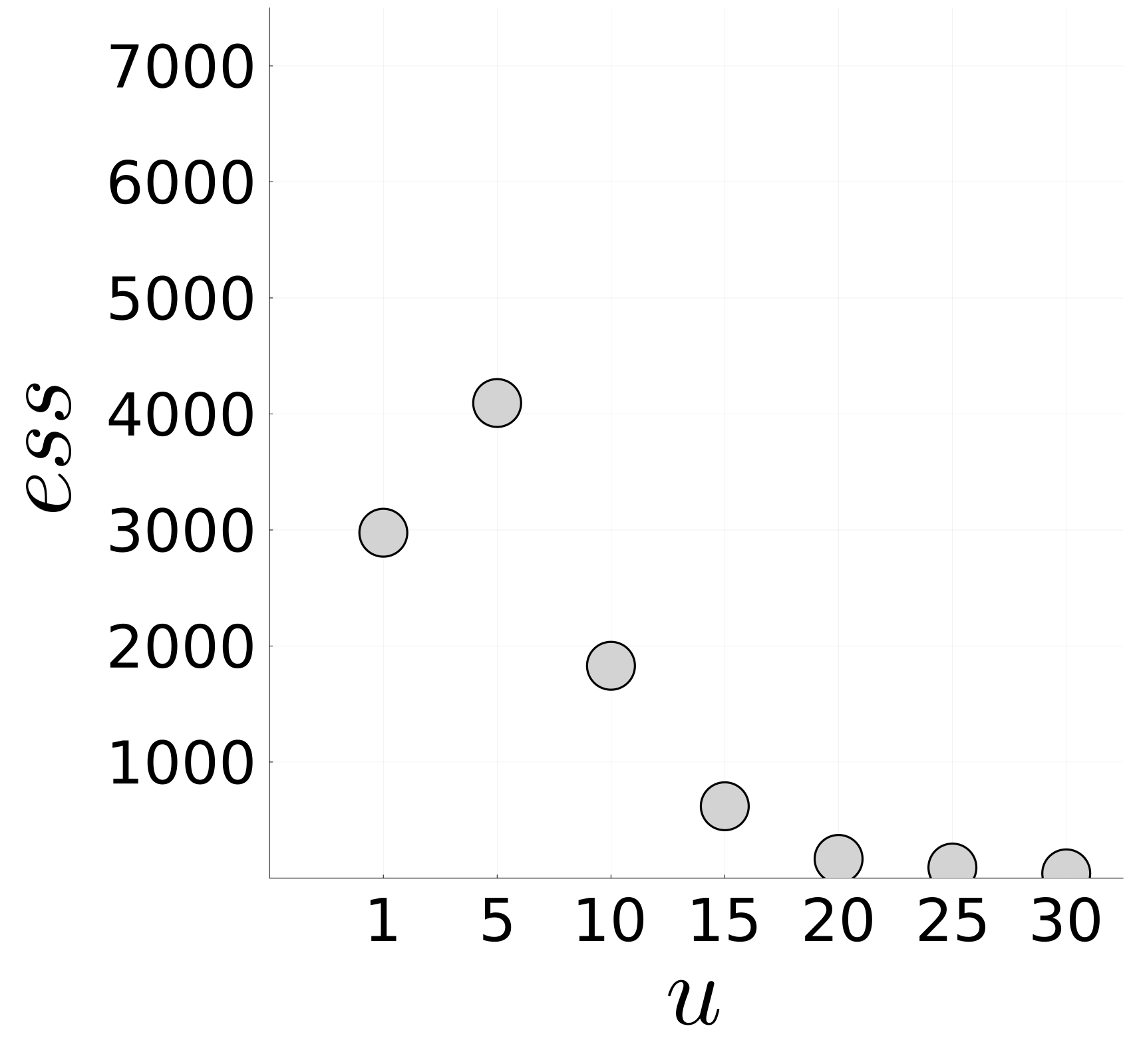}
 }
\subfigure[] 
{
       \includegraphics[width=3.8cm]{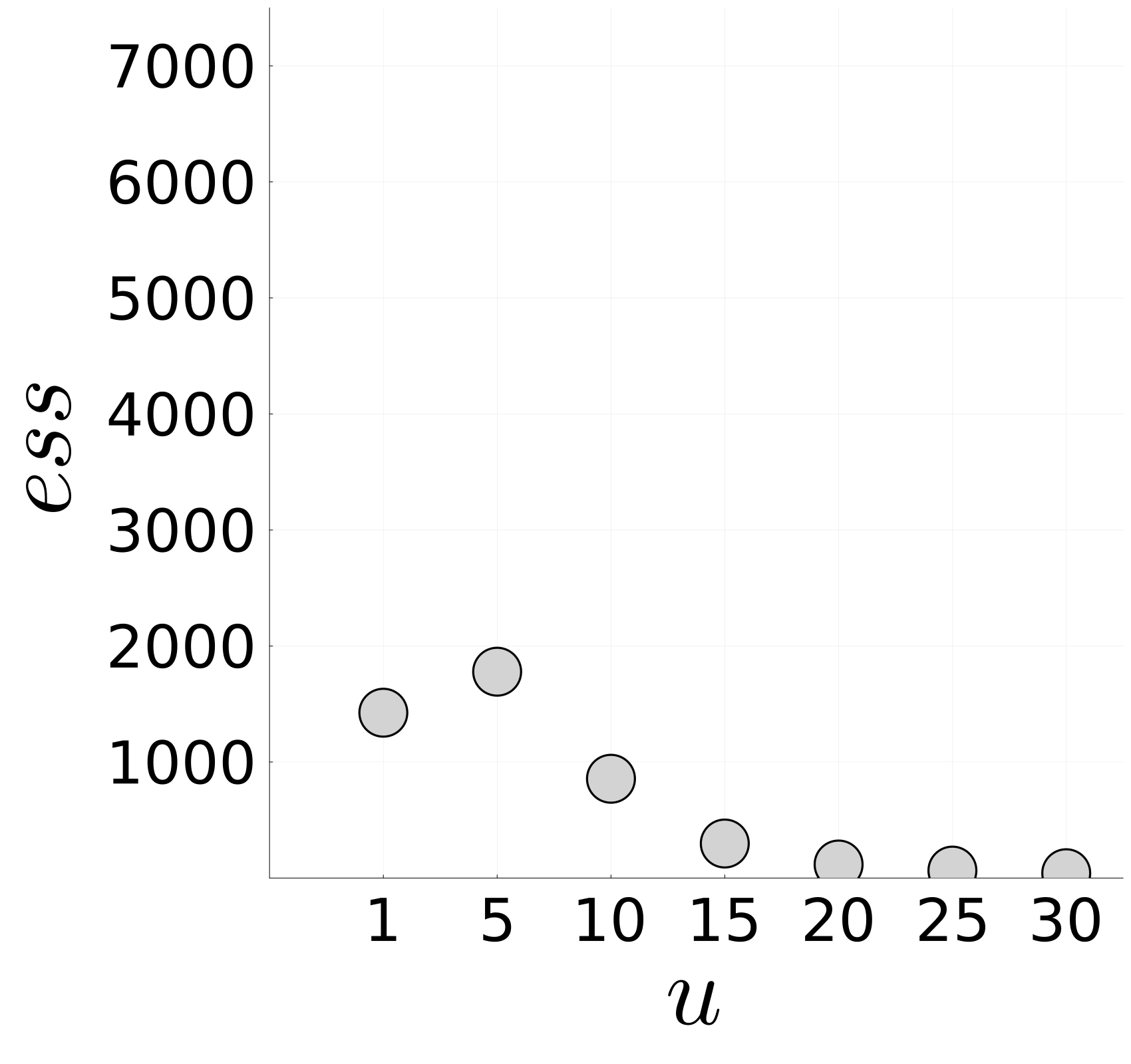}
 }      
\subfigure[] 
{
       \includegraphics[width=3.8cm]{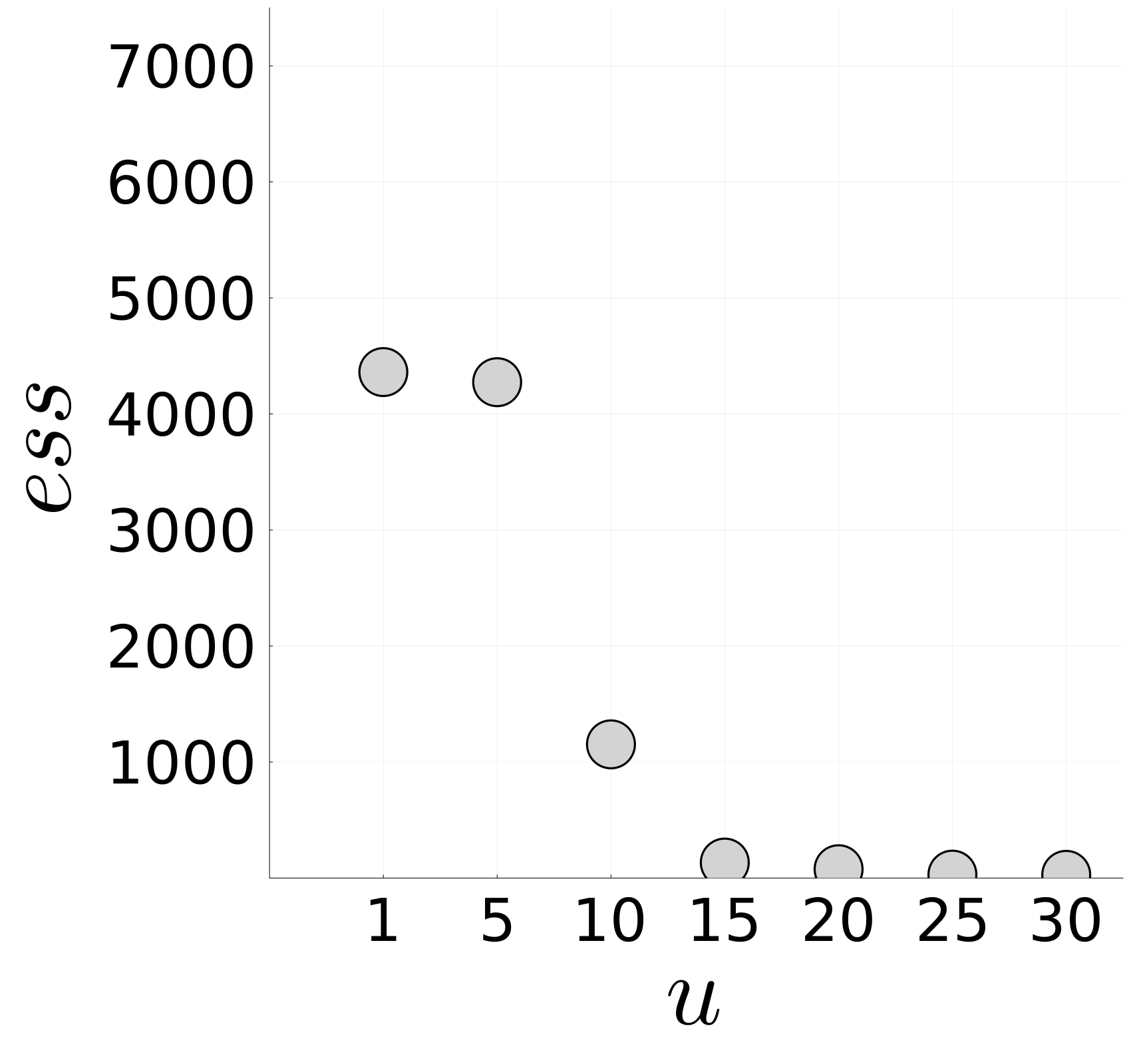}
 }      
\\ 
\subfigure[] 
{
       \includegraphics[width=3.8cm]{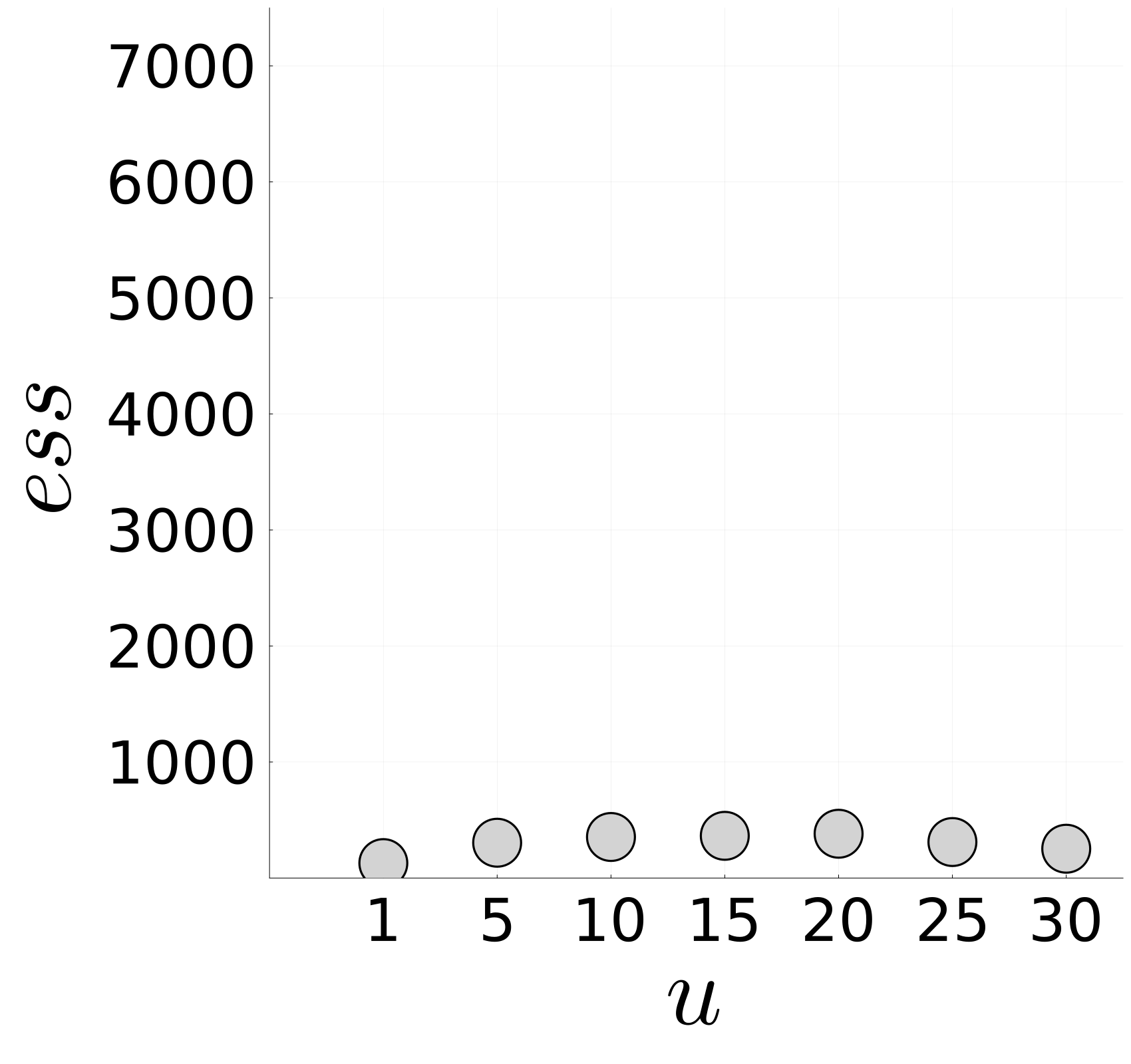}
 }
\subfigure[] 
{
       \includegraphics[width=3.8cm]{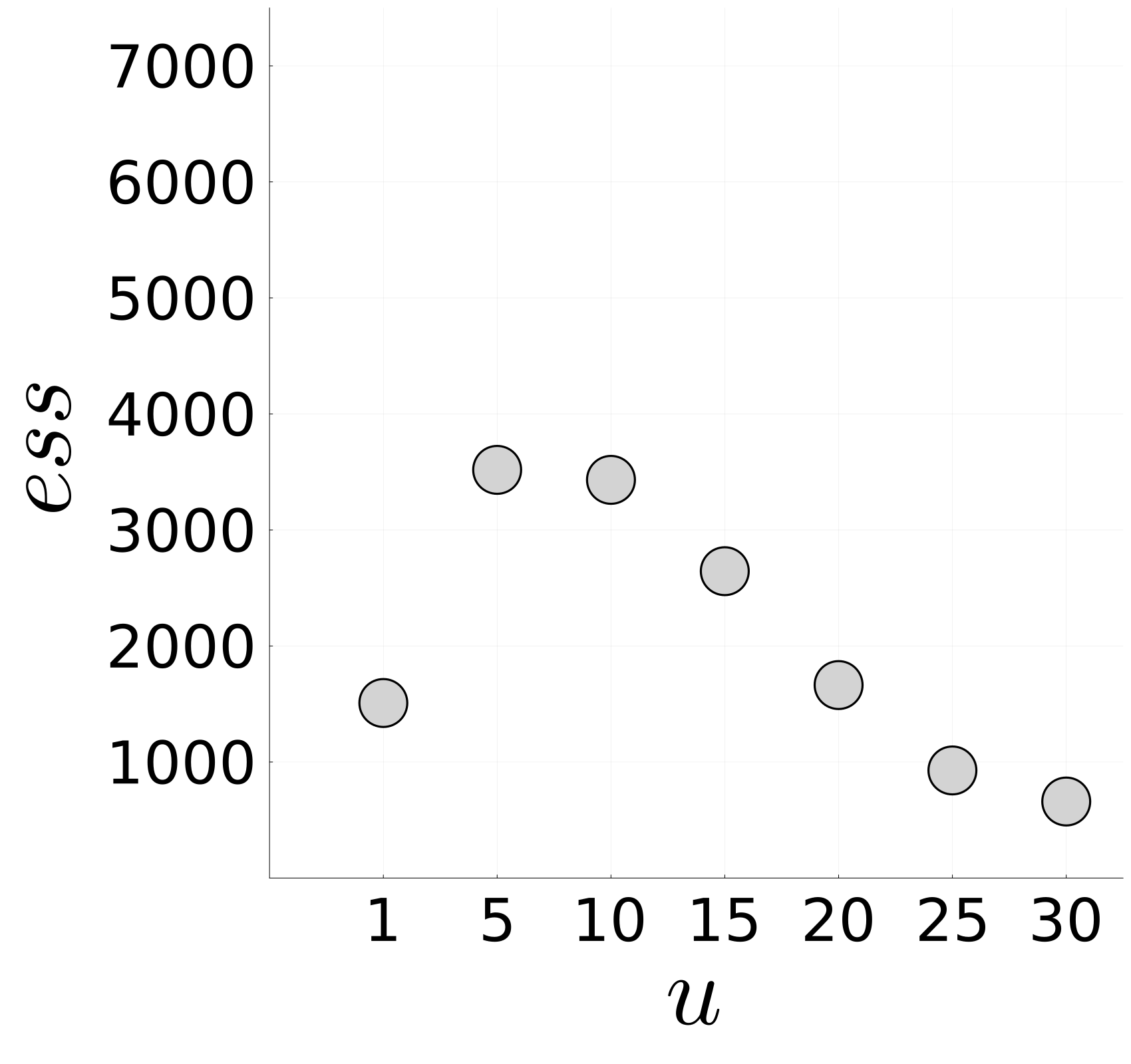}
 }
\subfigure[] 
{
       \includegraphics[width=3.8cm]{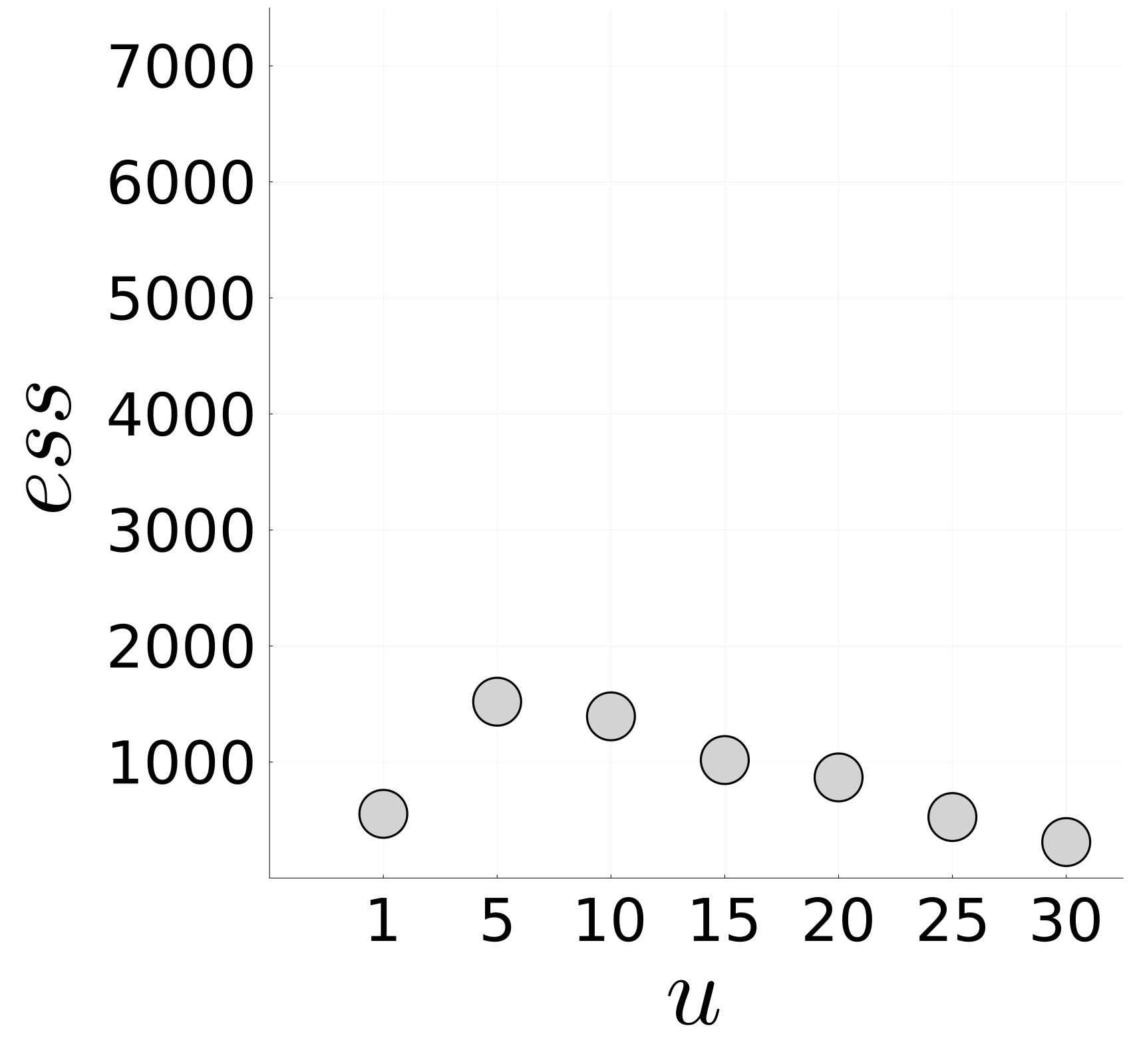}
 }      
\subfigure[] 
{
       \includegraphics[width=3.8cm]{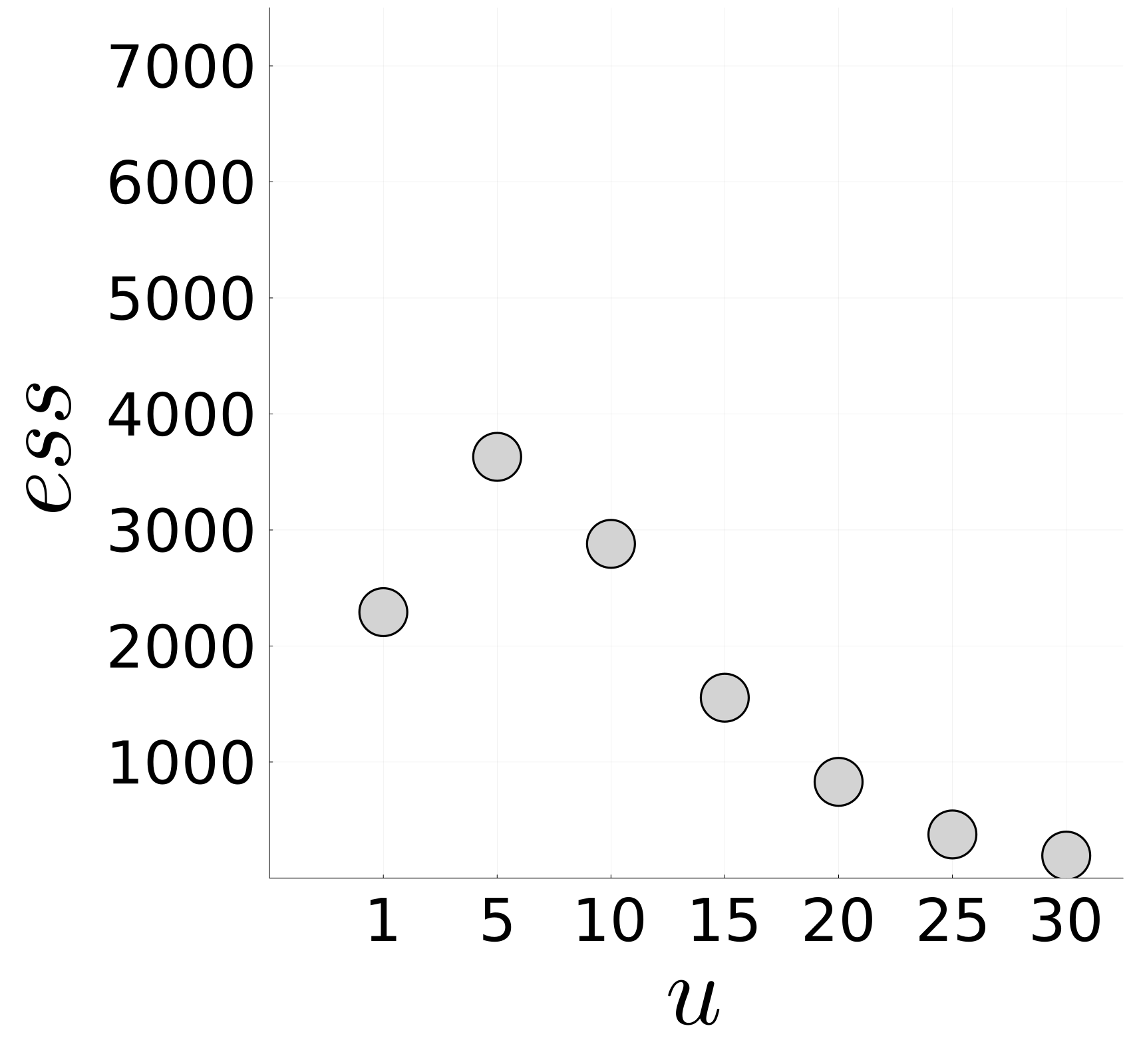}
 }  
 \caption{Simulated data -  Algorithm 1: Effective sample size ($ess$) for different values of the number of rectangle exchanges ($u$).   Panels (a), (e), (i), and (m) display the results for Model 1. Panels (b), (f), (j), and (n)  display the results for Model 2. 
 Panels (c), (g), (k), and (o)  display the results for Model 3.
 Panels (d), (h), (l), and (p)  display the results for Model 4. Panels (a) - (d), (e) - (h), (i) - (l), and (m) - (p) display the results for $k = 10$ and $n = 2,500$, $k = 20$ and $n = 2,500$, $k = 10$ and $n = 10,000$, and $k = 20$ and $n = 10,000$, respectively.}
\label{fig2:simulation1}
\end{figure}

\begin{figure}[!h]
\centering
\subfigure[] 
{
\includegraphics[width=3.8cm]{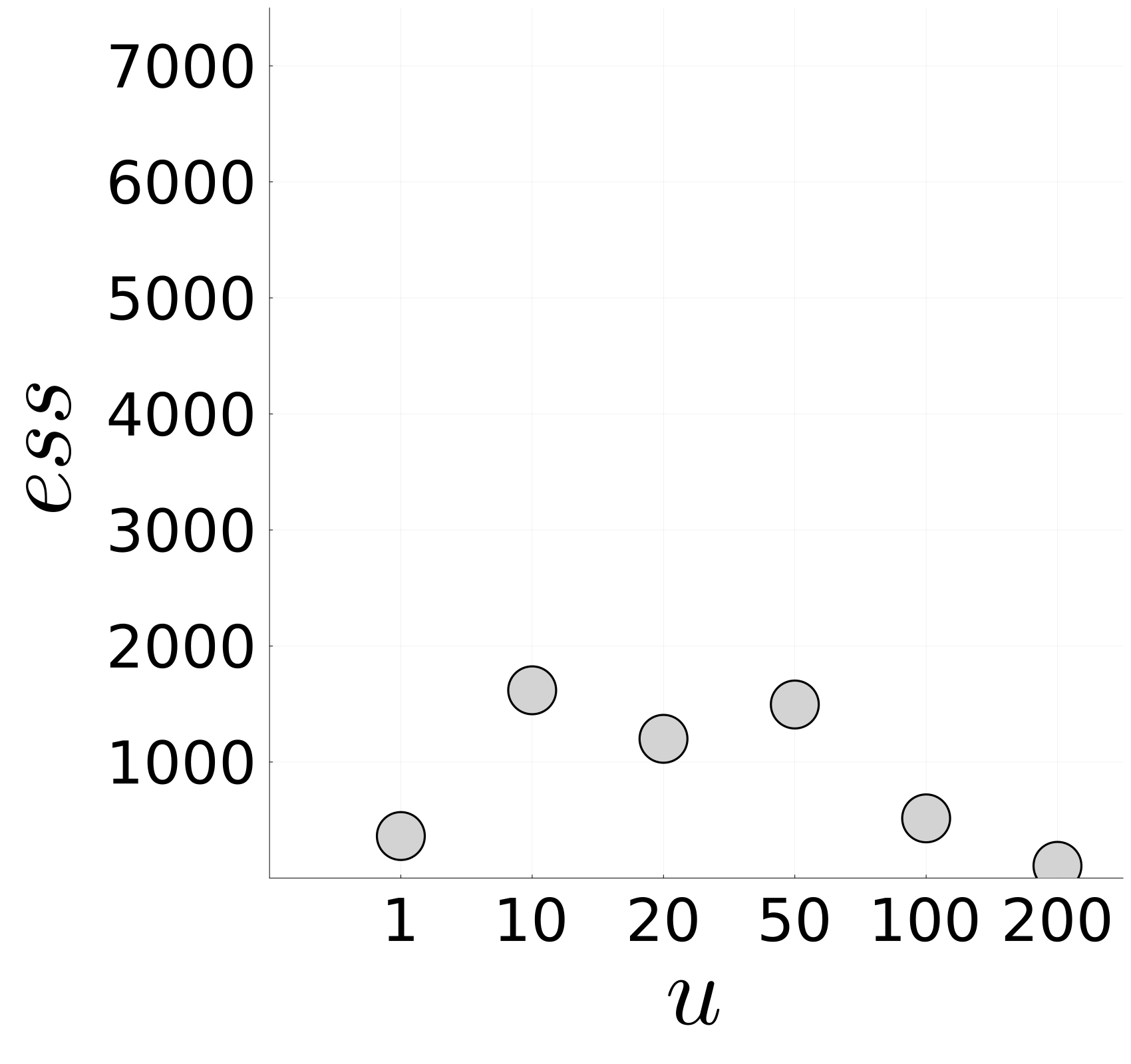}
 }
\subfigure[] 
{
\includegraphics[width=3.8cm]{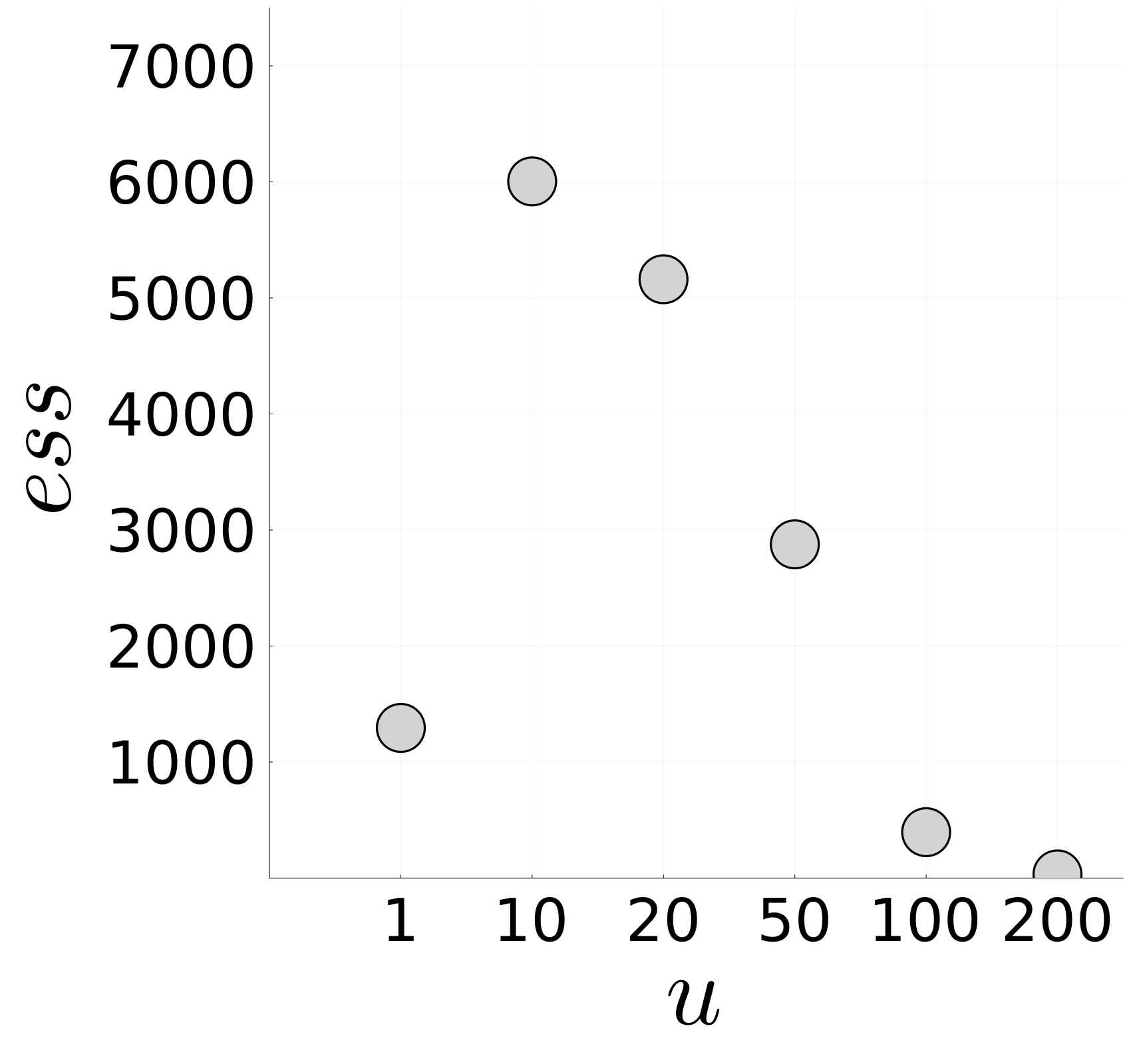}
 }
\subfigure[] 
{
\includegraphics[width=3.8cm]{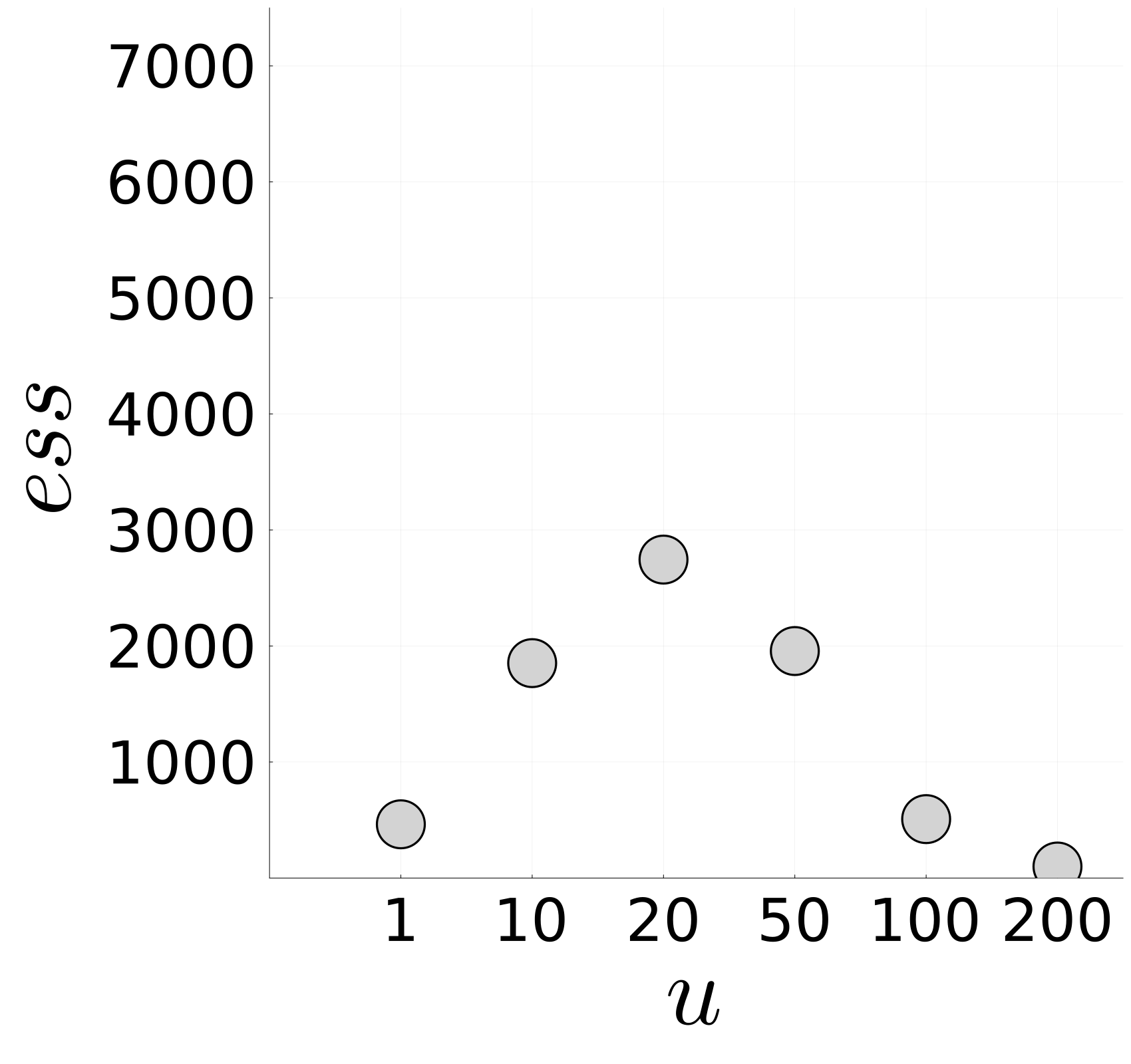}
 }      
\subfigure[] 
{
\includegraphics[width=3.8cm]{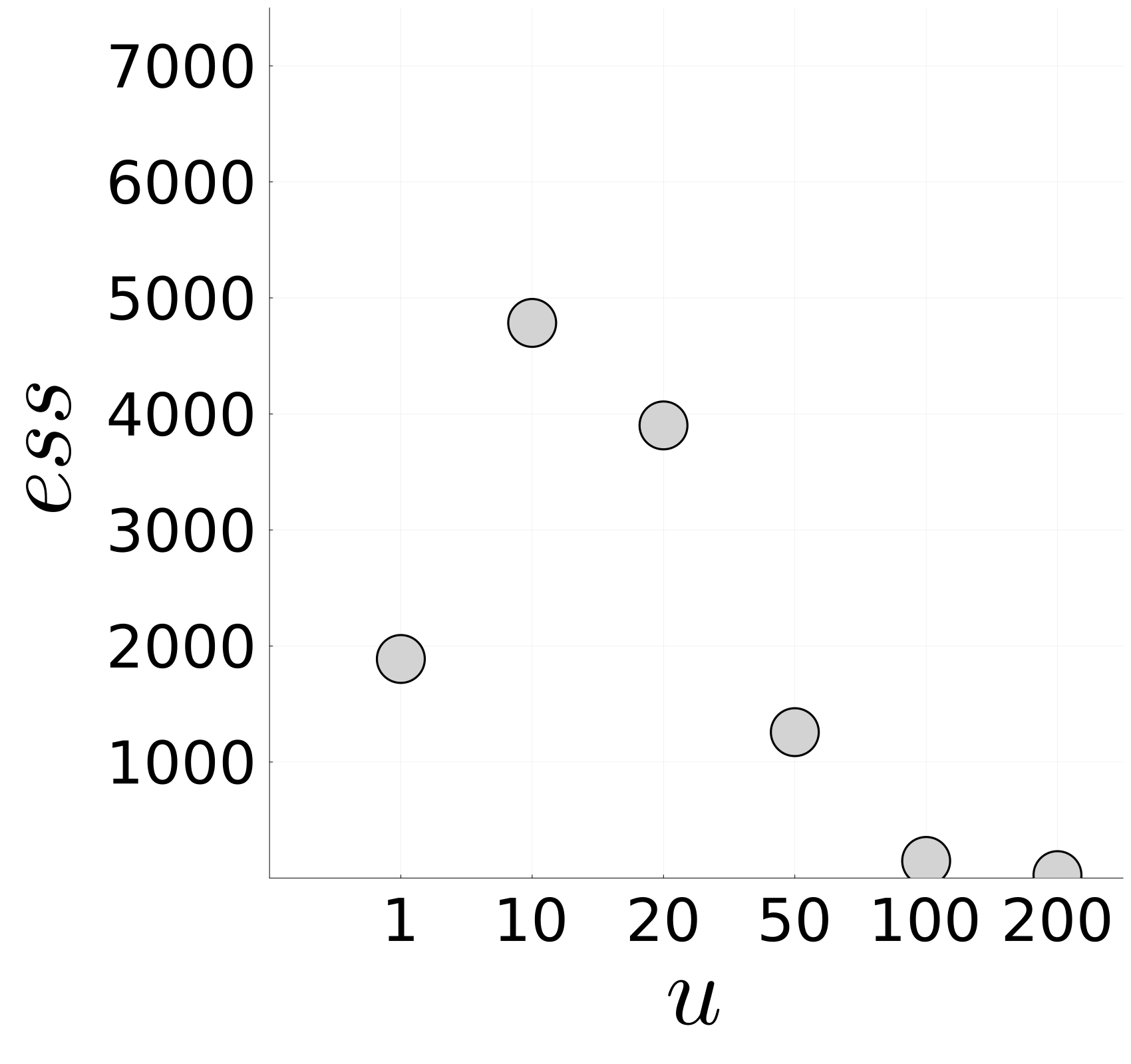}
 }      
\\
\subfigure[]
{
\includegraphics[width=3.8cm]{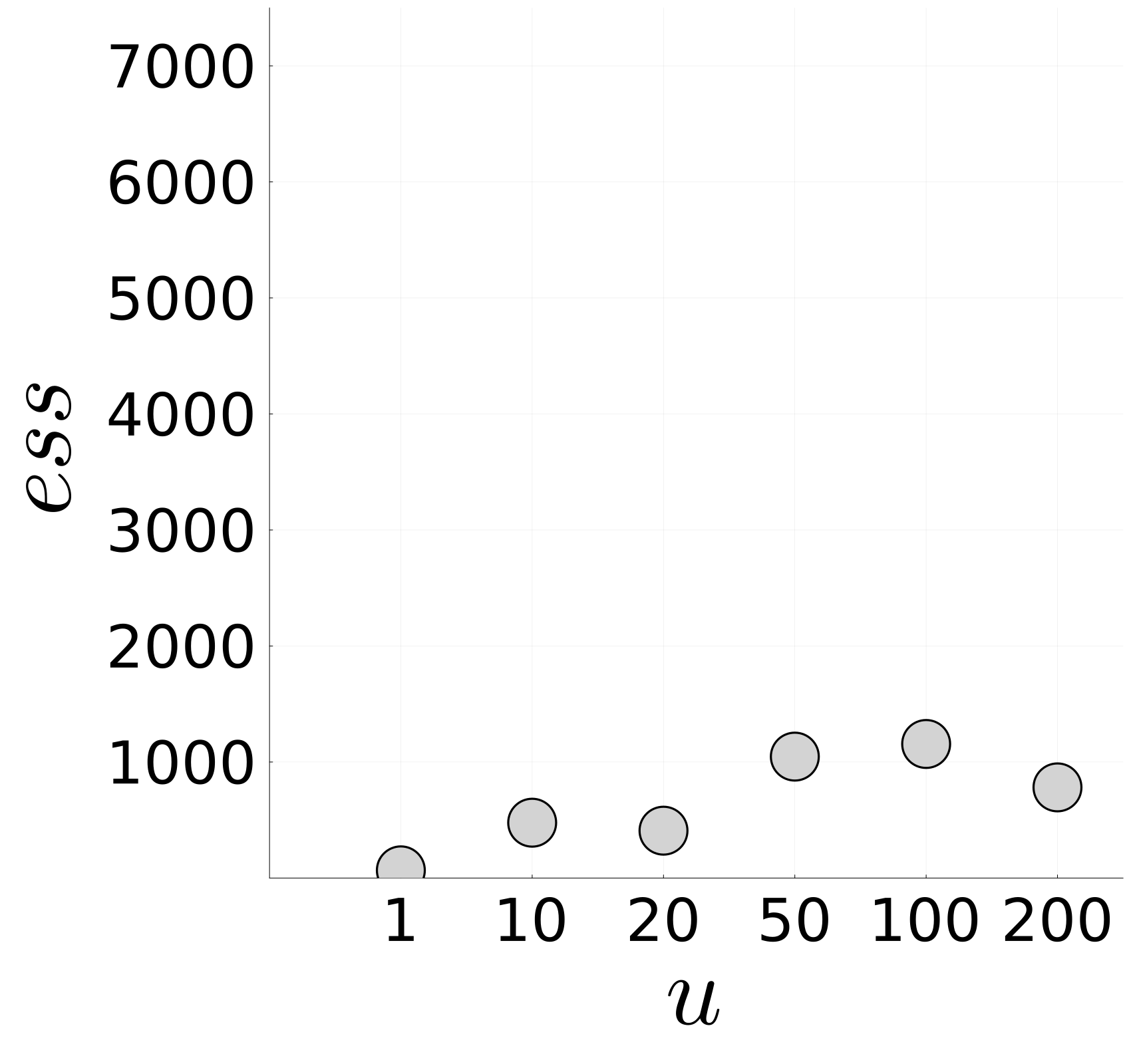}
 }
\subfigure[] 
{
\includegraphics[width=3.8cm]{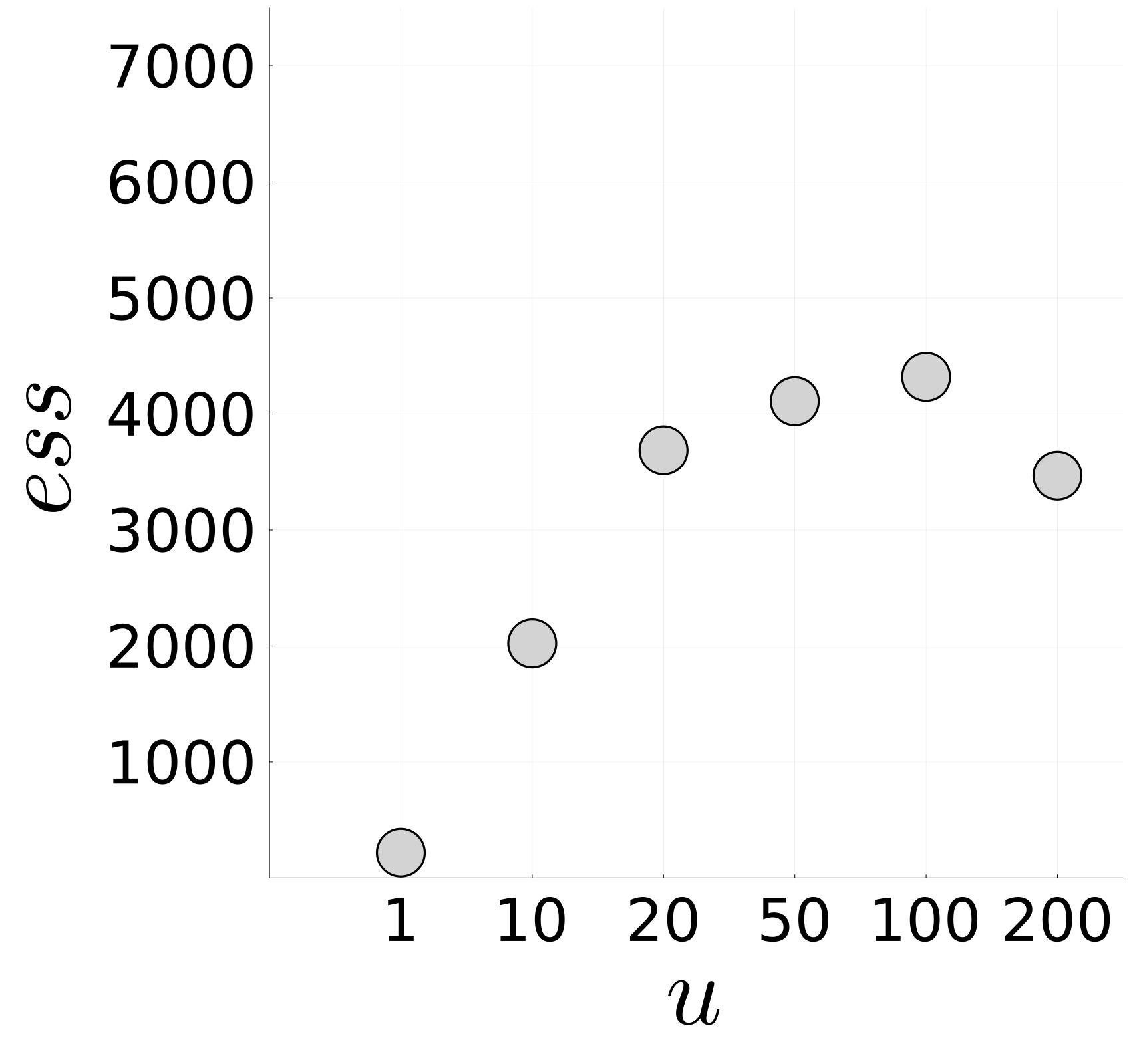}
 }
\subfigure[] 
{
\includegraphics[width=3.8cm]{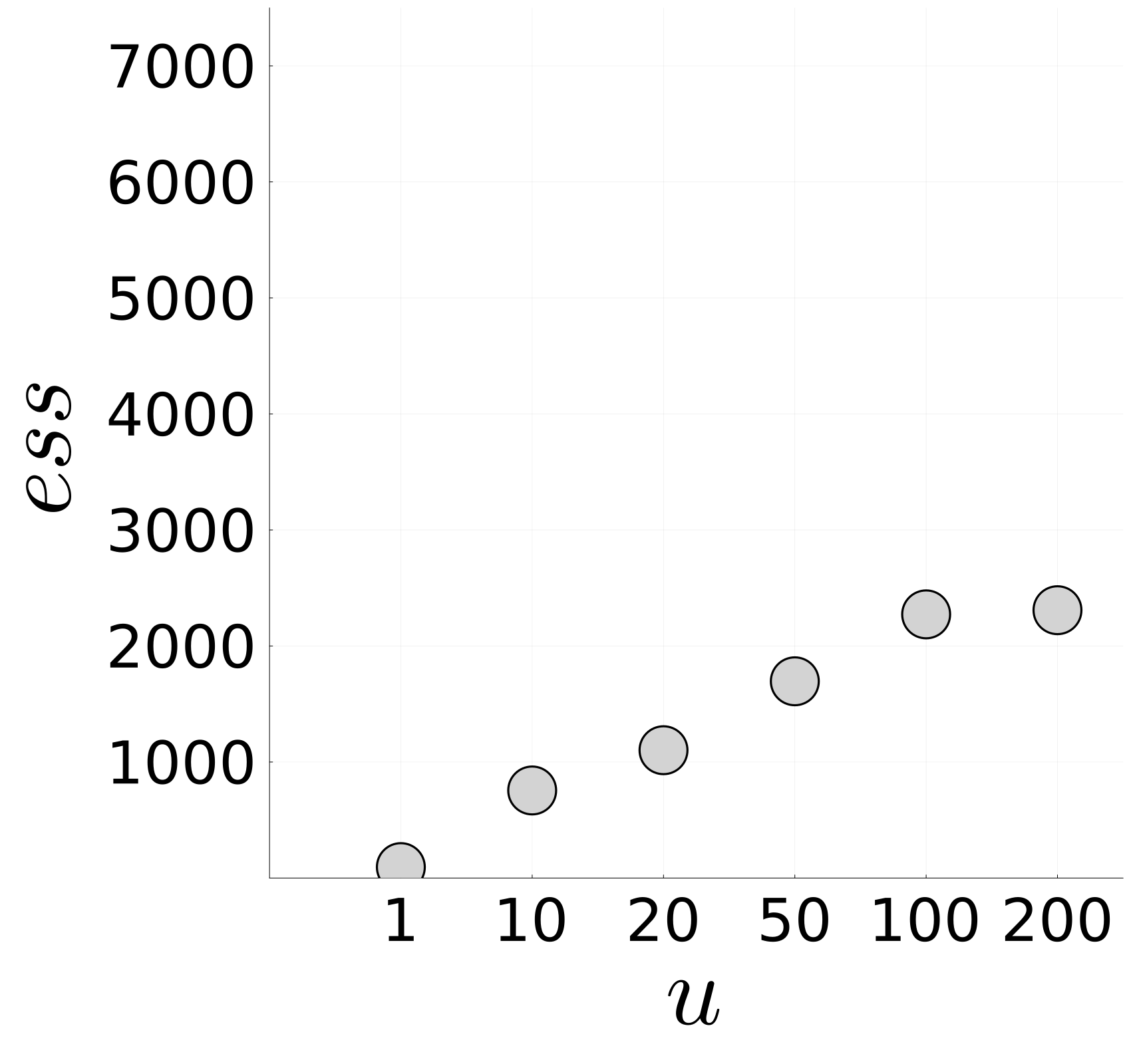}
 }      
\subfigure[] 
{
\includegraphics[width=3.8cm]{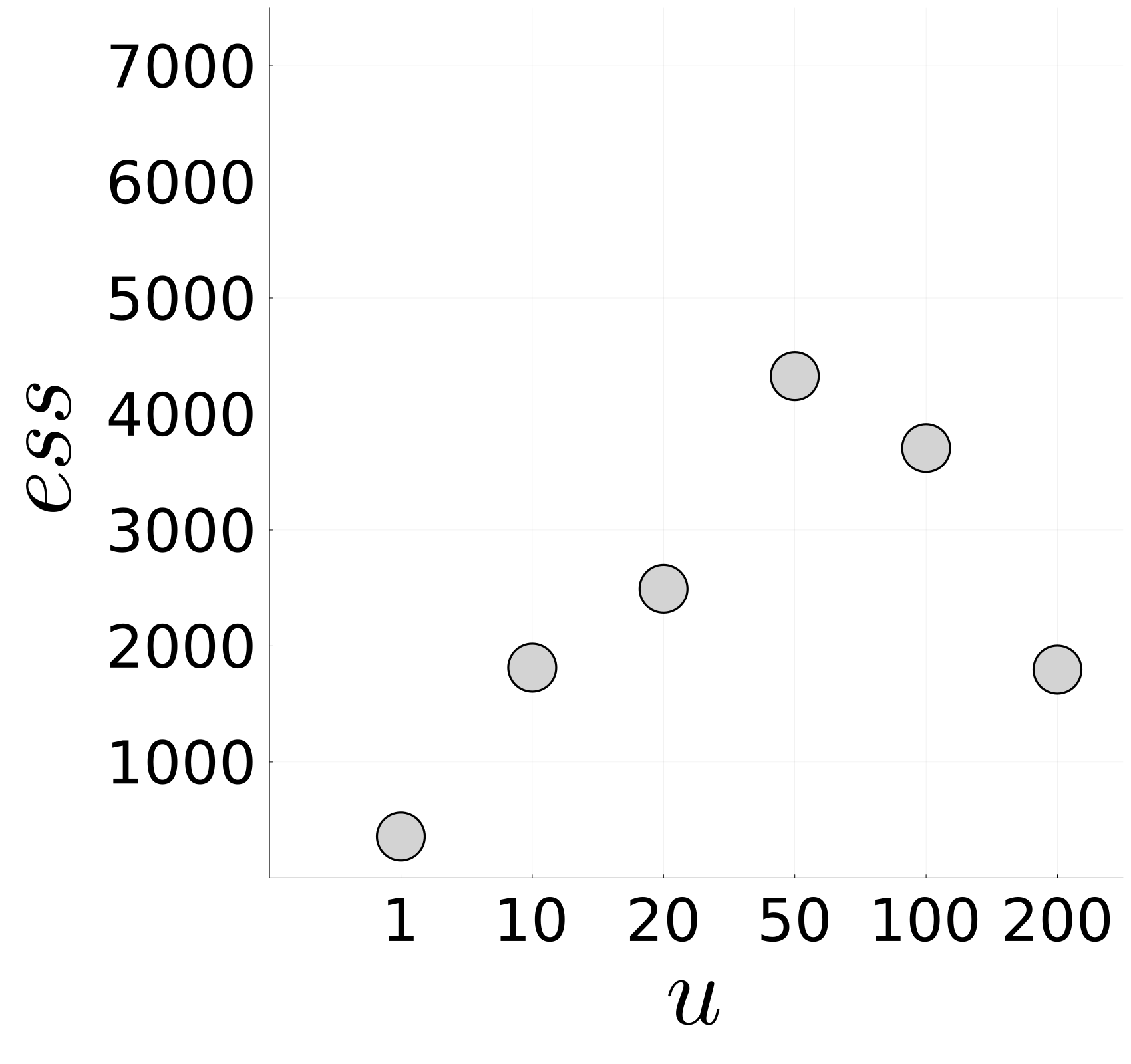}
 }  
\\ 
\subfigure[] 
{
\includegraphics[width=3.8cm]{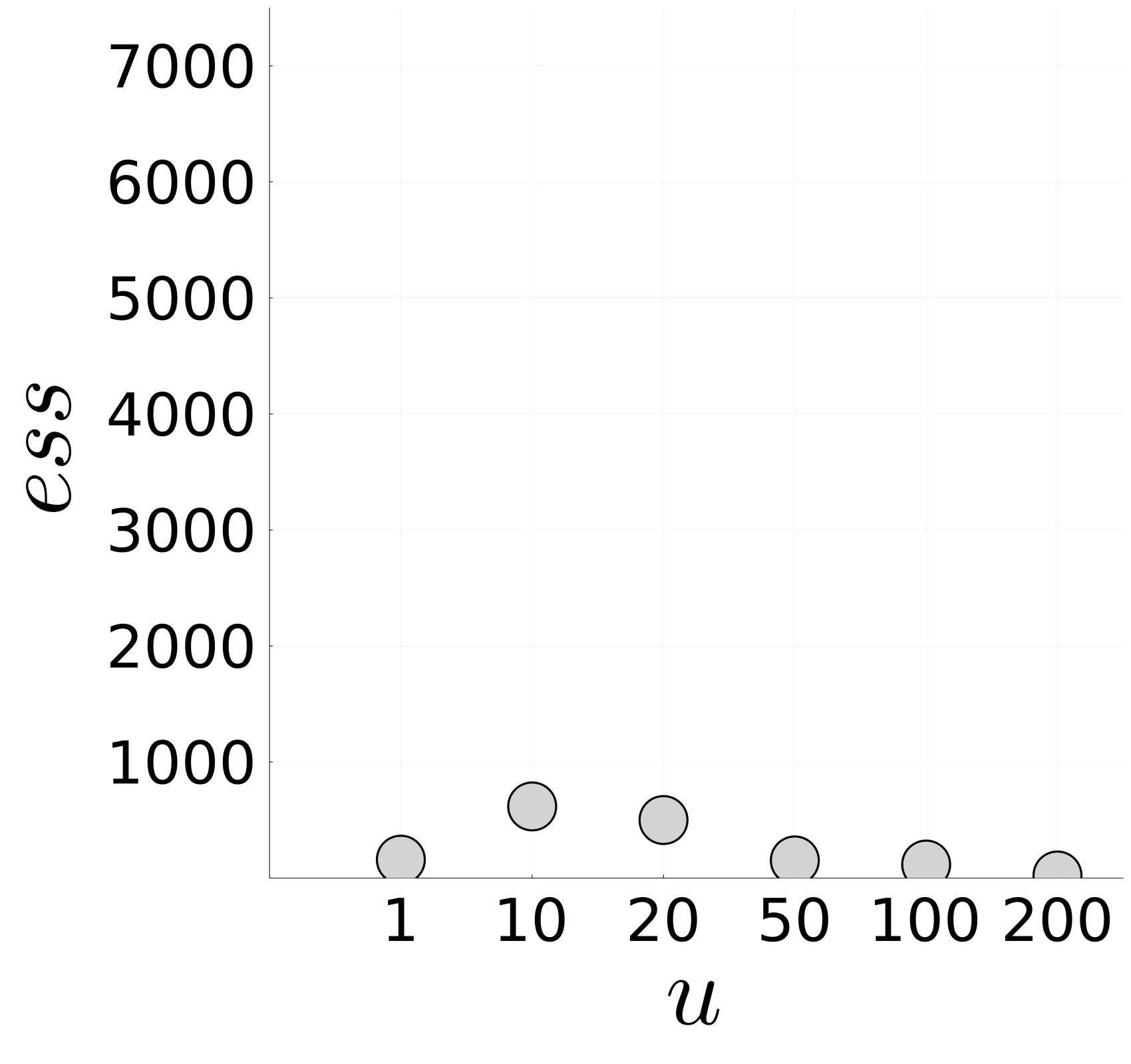}
 }
\subfigure[] 
{
\includegraphics[width=3.8cm]{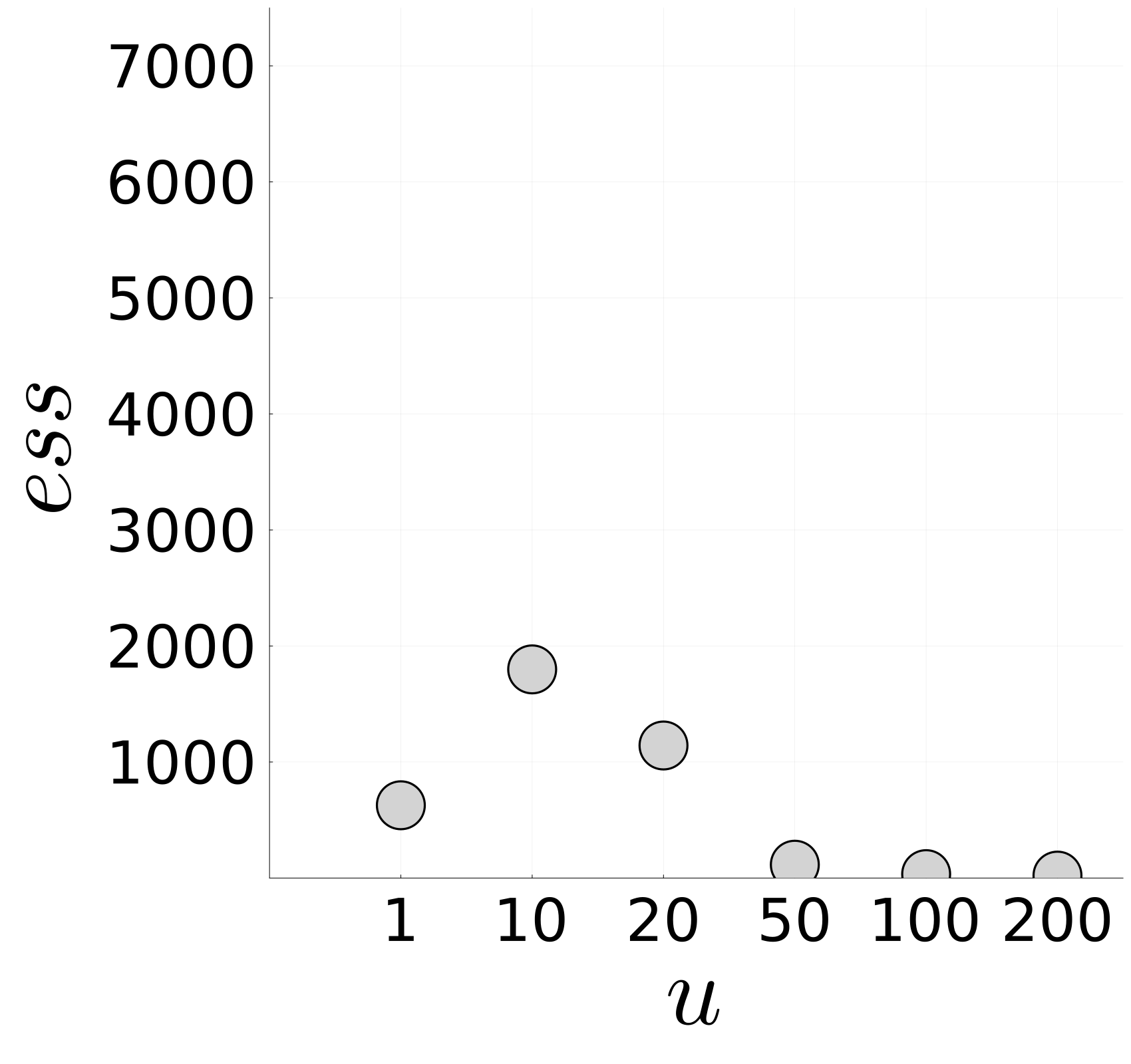}
 }
\subfigure[] 
{
\includegraphics[width=3.8cm]{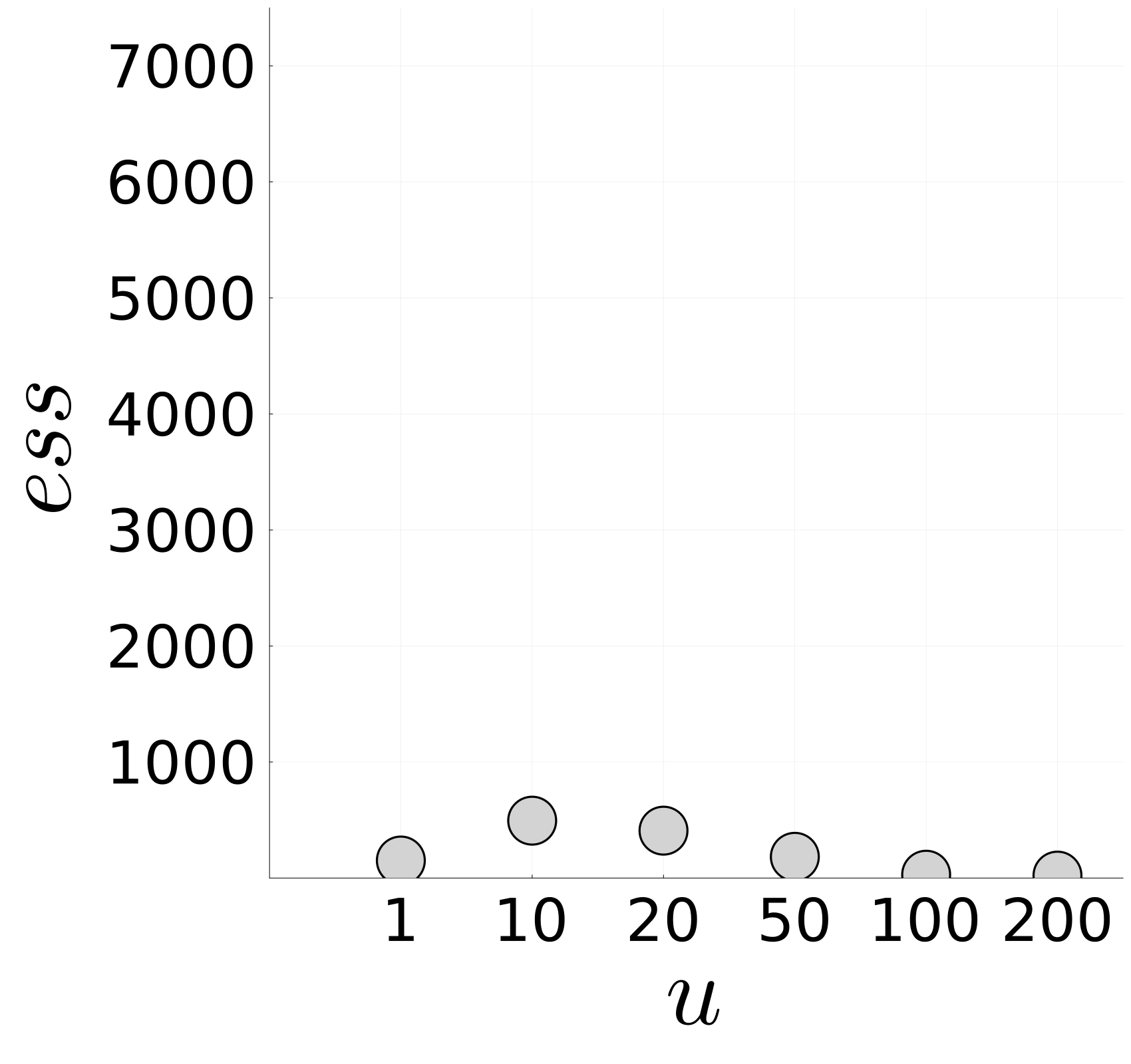}
 }      
\subfigure[] 
{
\includegraphics[width=3.8cm]{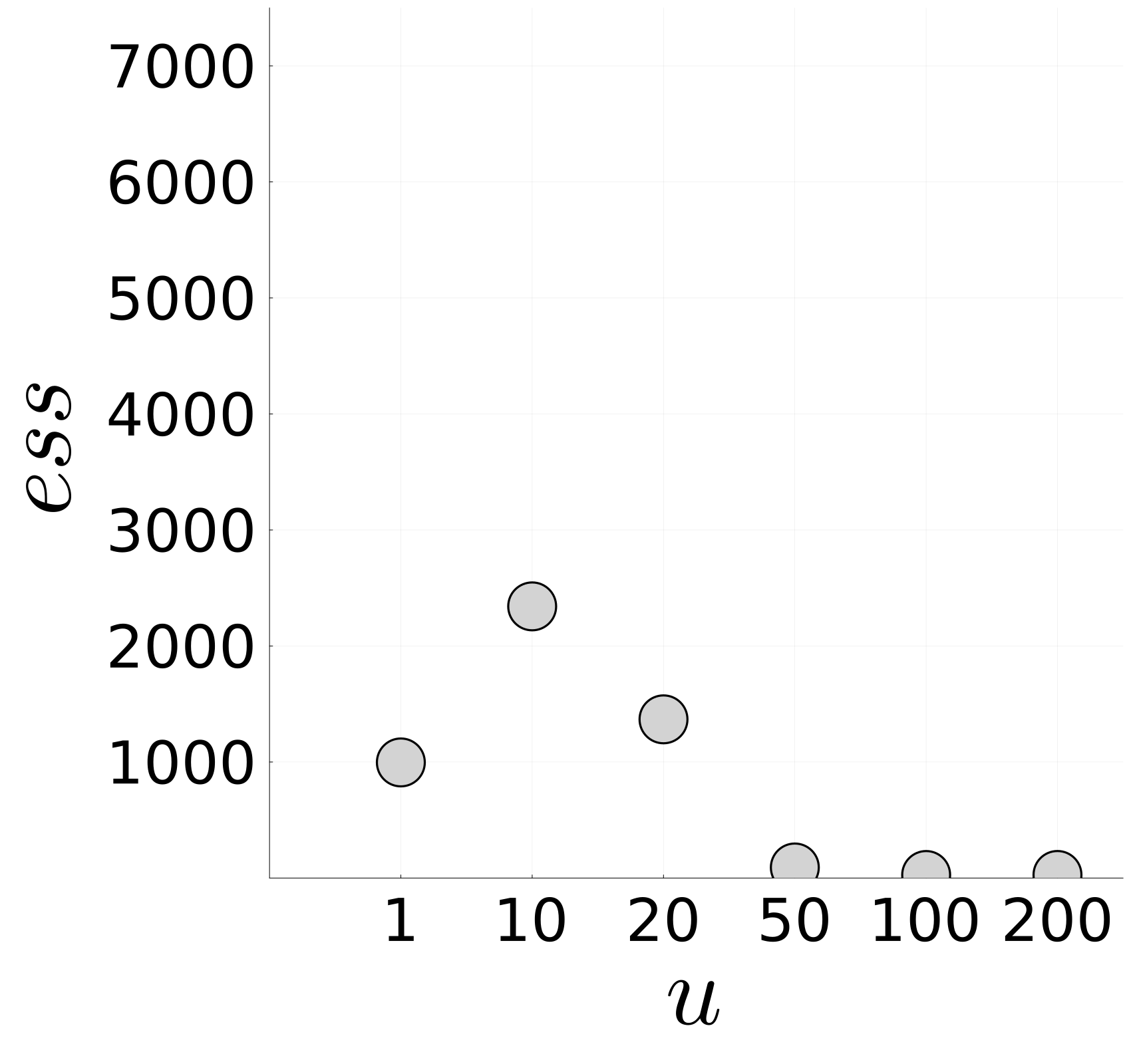}
 }      
\\ 
\subfigure[] 
{
\includegraphics[width=3.8cm]{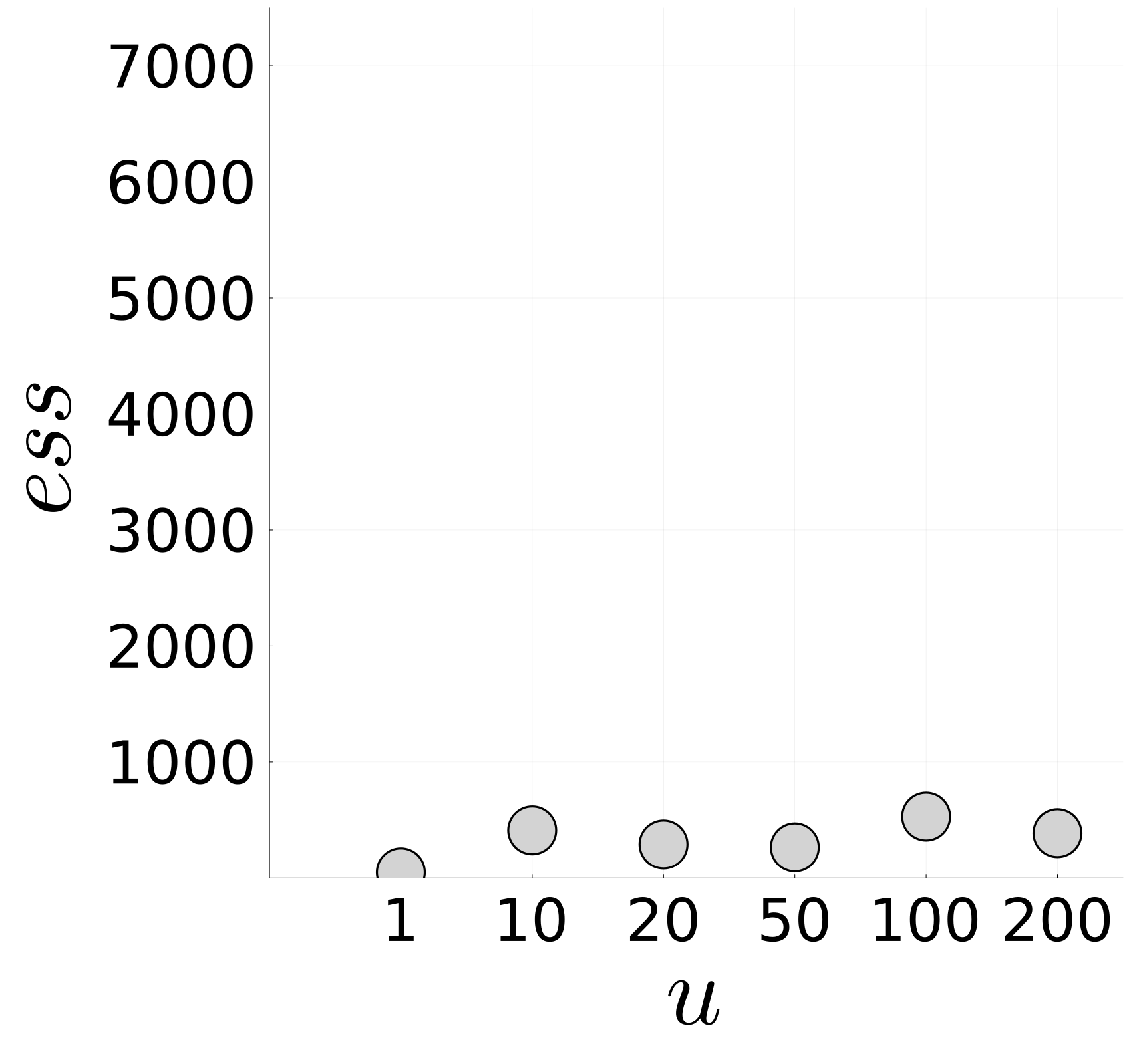}
 }
\subfigure[] 
{
\includegraphics[width=3.8cm]{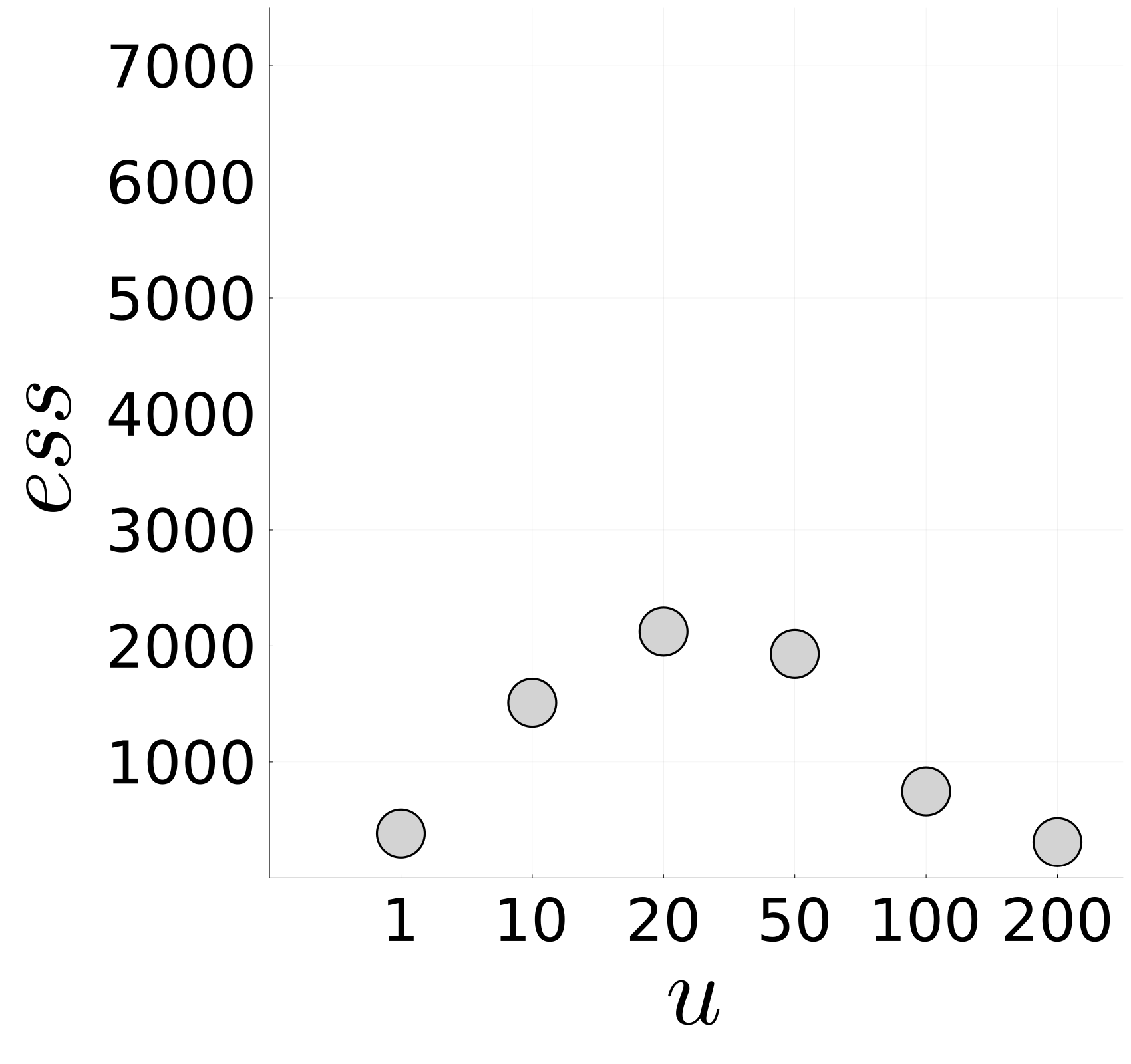}
 }
\subfigure[] 
{
\includegraphics[width=3.8cm]{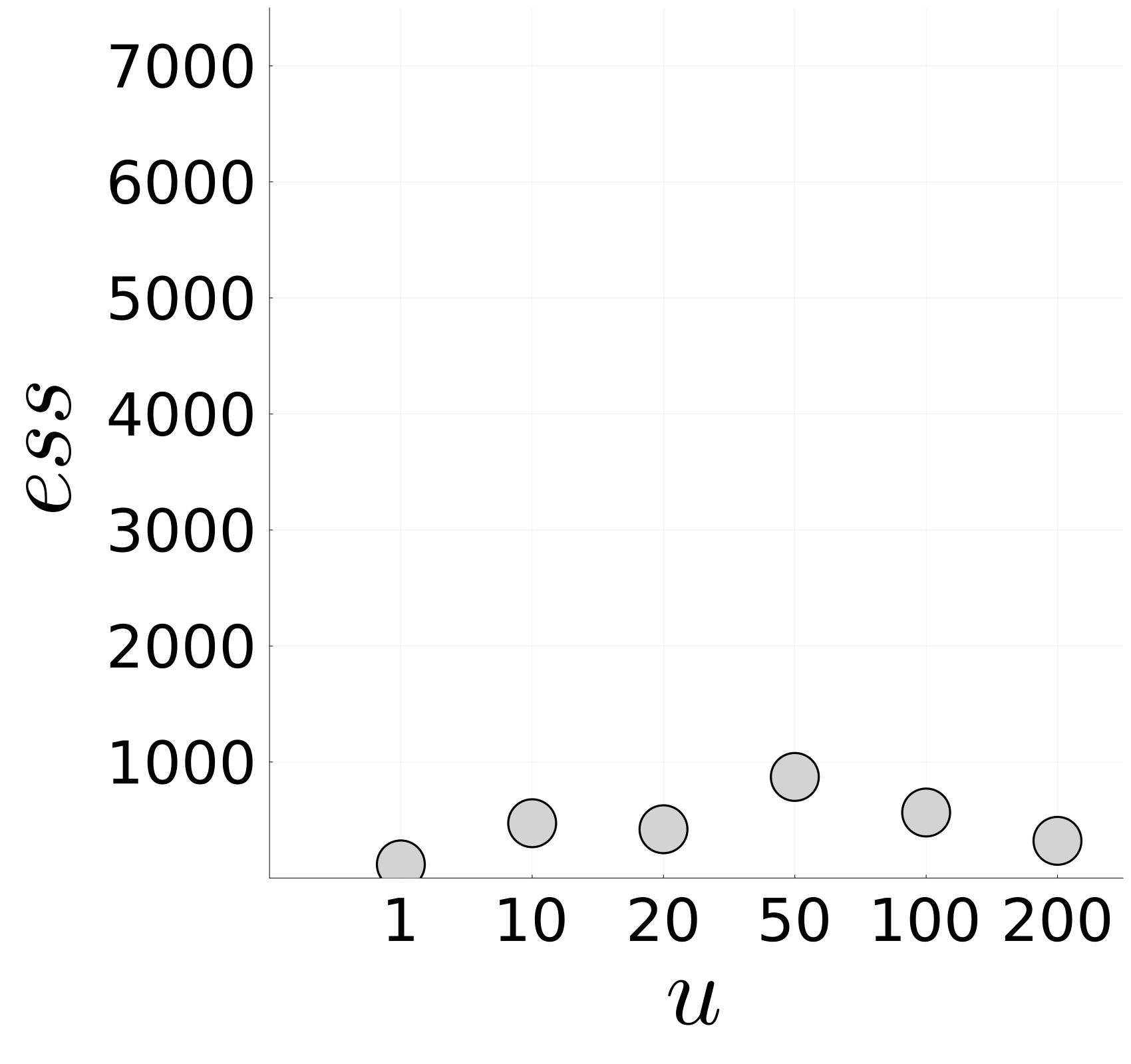}
 }      
\subfigure[] 
{
\includegraphics[width=3.8cm]{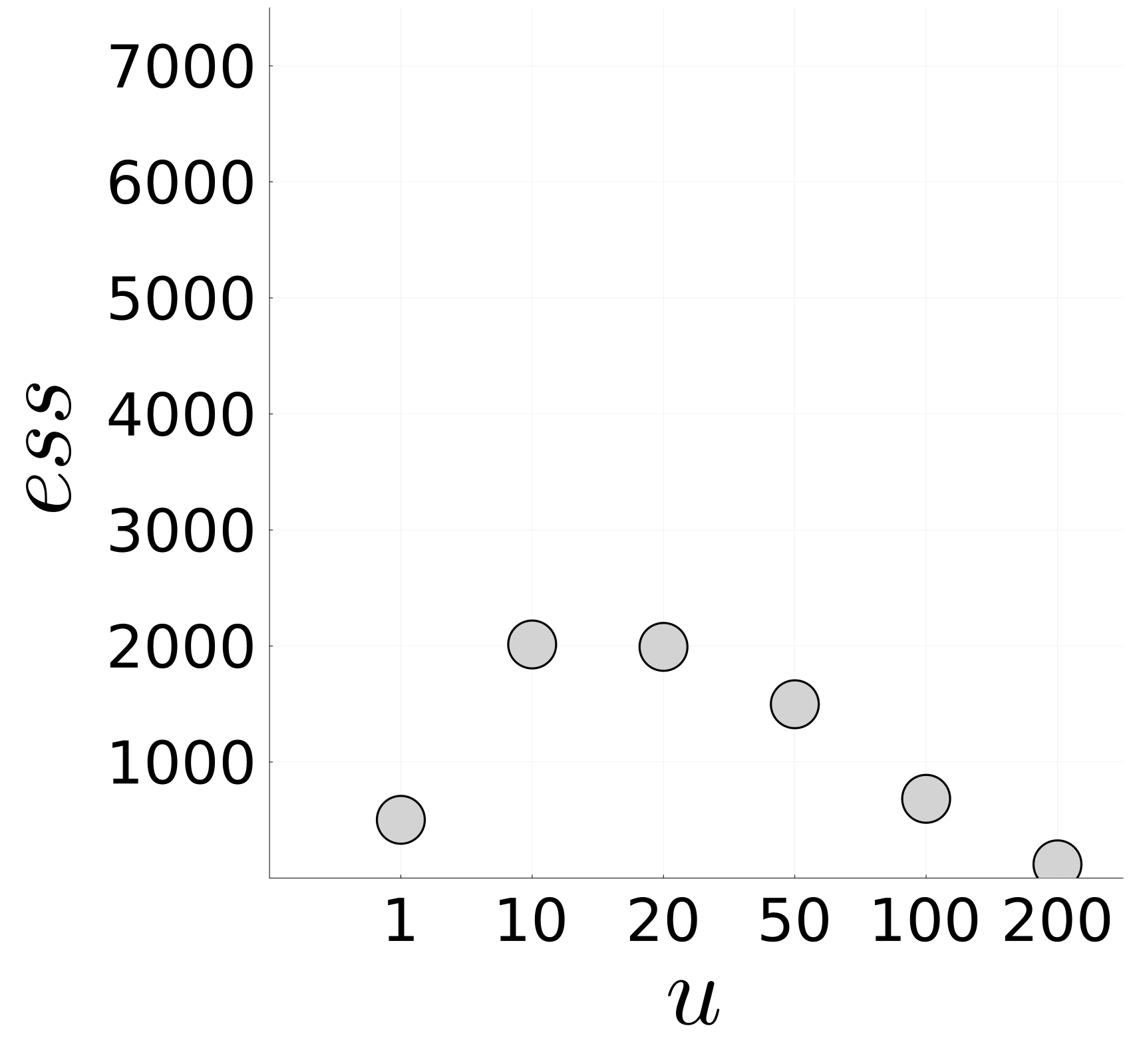}
 }  
 \caption{Simulated data -  Algorithm 2: Effective sample size ($ess$) for different values of $\tau$.   Panels (a), (e), (i), and (m) display the results for Model 1. Panels (b), (f), (j), and (n)  display the results for Model 2. 
 Panels (c), (g), (k), and (o)  display the results for Model 3.
 Panels (d), (h), (l), and (p)  display the results for Model 4. Panels (a) - (d), (e) - (h), (i) - (l), and (m) - (p) display the results for $k = 10$ and $n = 2,500$, $k = 20$ and $n = 2,500$, $k = 10$ and $n = 10,000$, and $k = 20$ and $n = 10,000$, respectively.}
\label{fig3:simulation1}
\end{figure}

\begin{figure}[!h]
\centering
\subfigure[] 
{
\includegraphics[width=3.8cm]{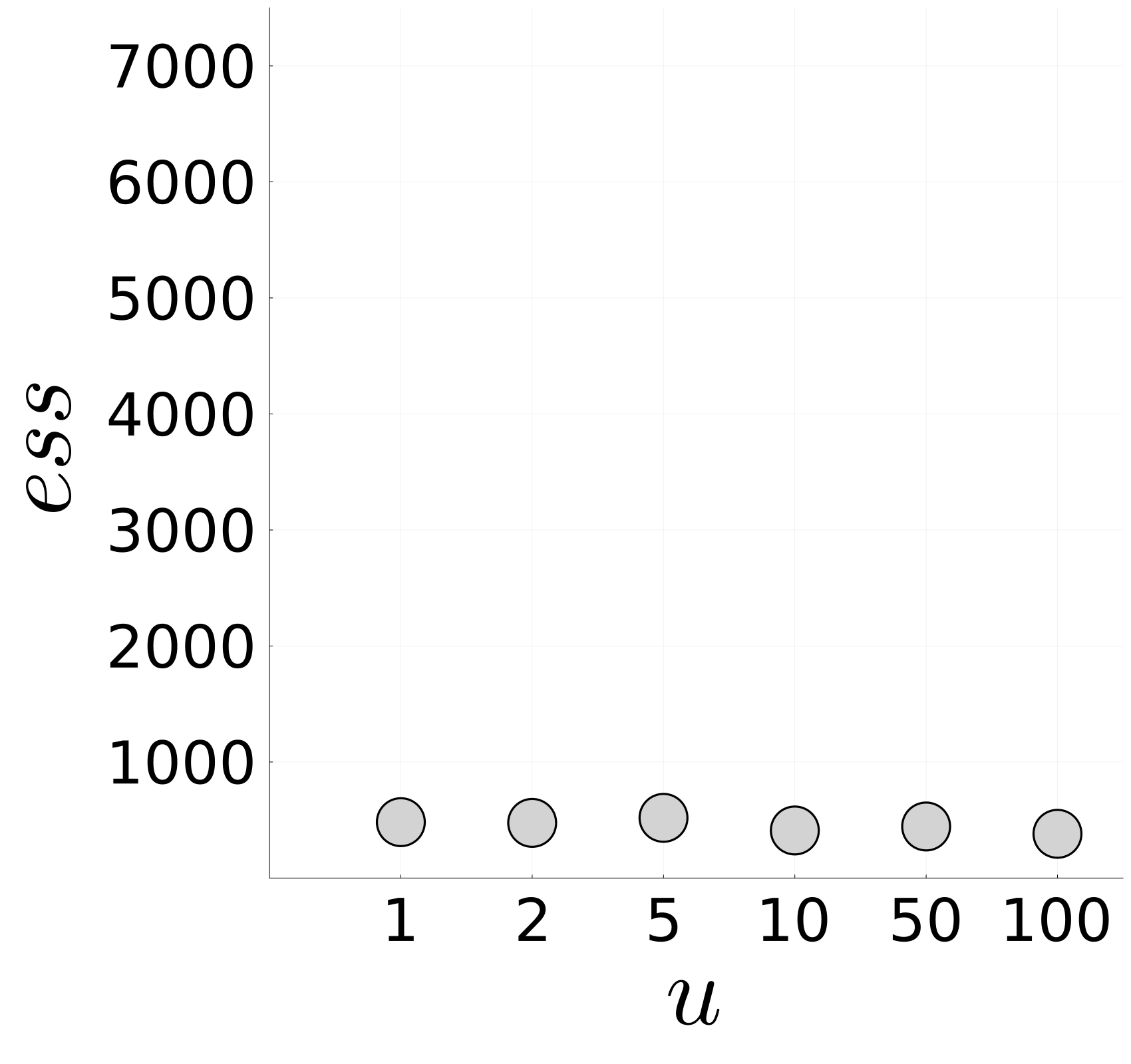}
 }
\subfigure[] 
{
\includegraphics[width=3.8cm]{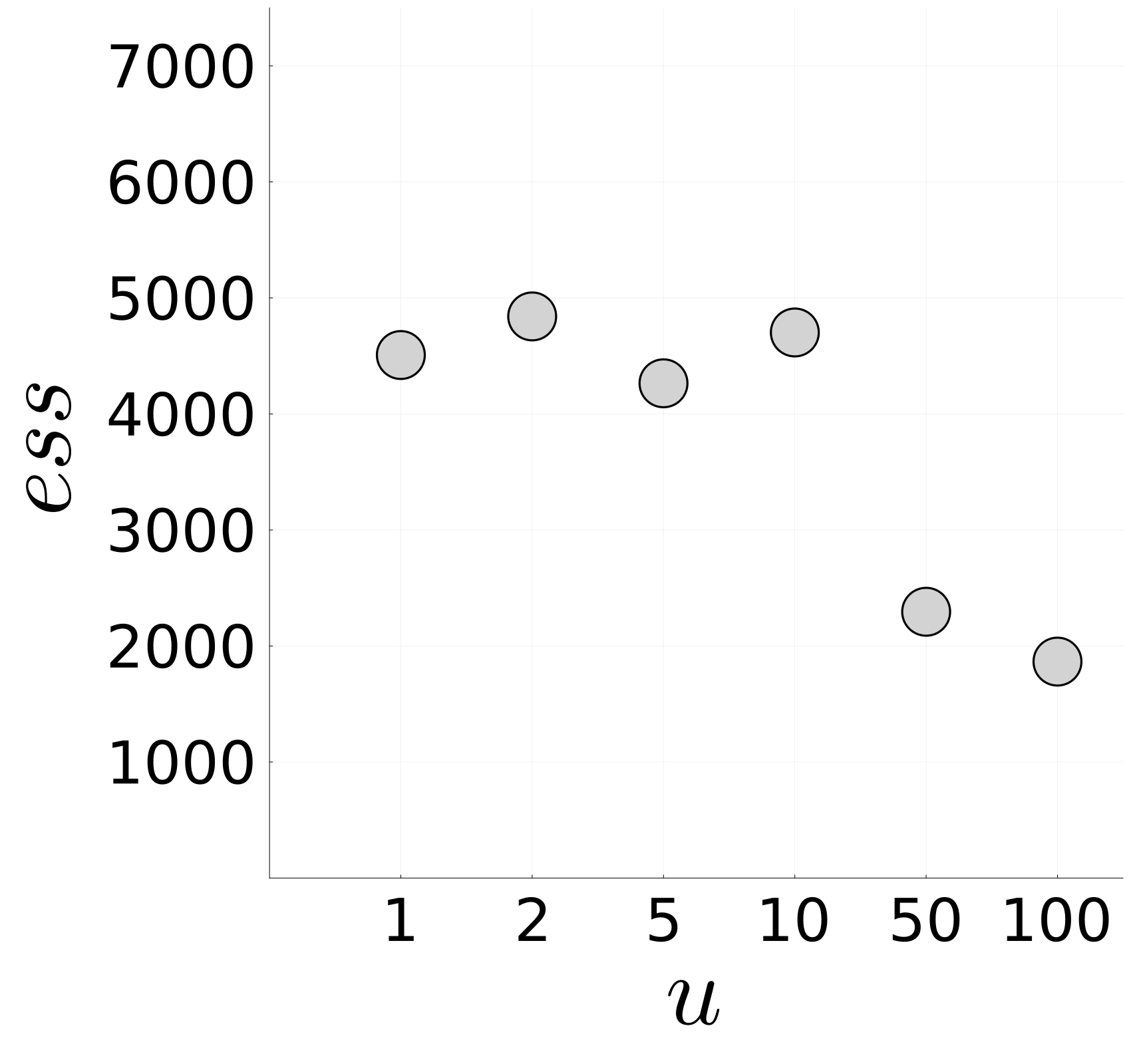}
 }
\subfigure[] 
{
\includegraphics[width=3.8cm]{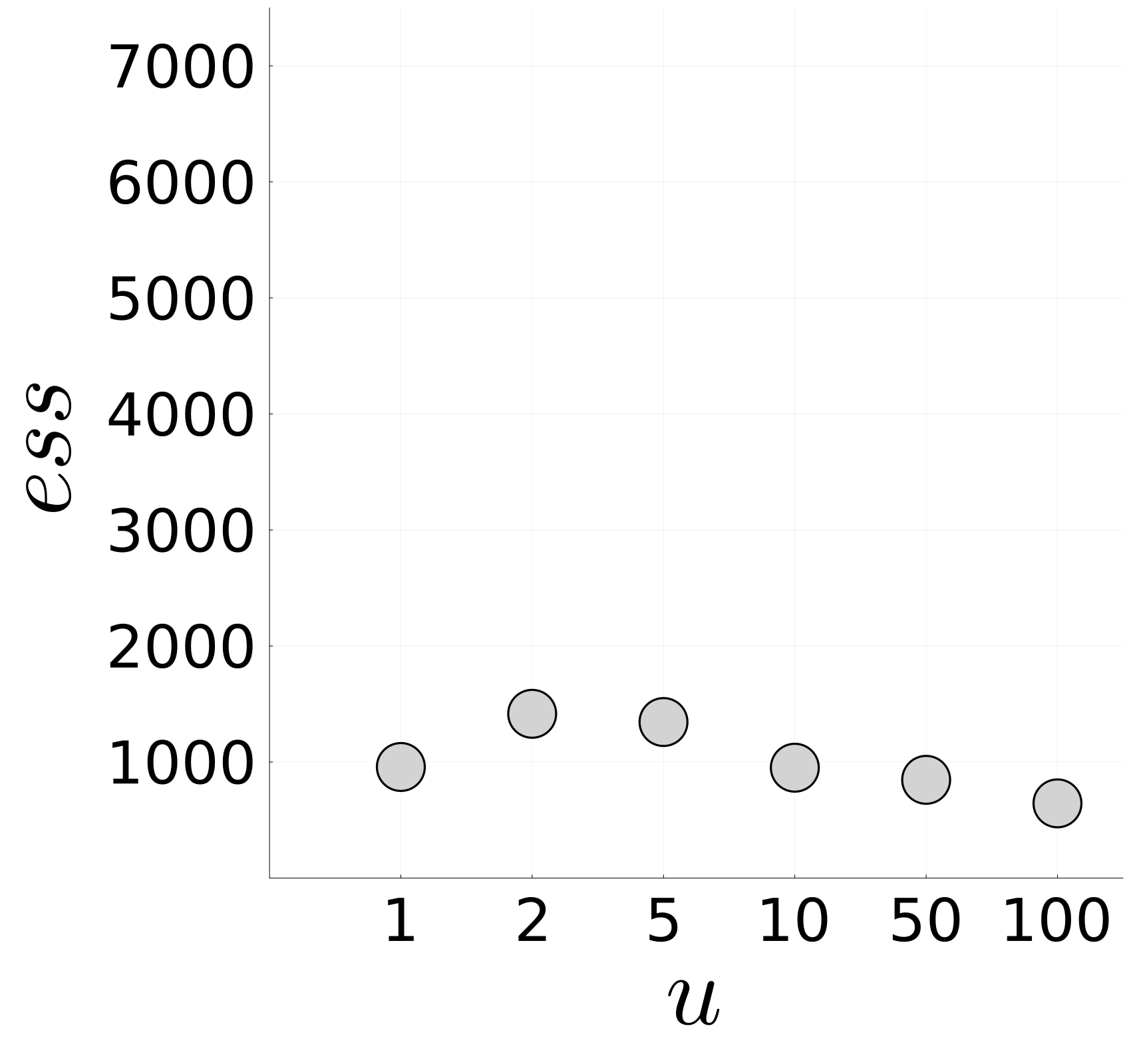}
 }      
\subfigure[] 
{
\includegraphics[width=3.8cm]{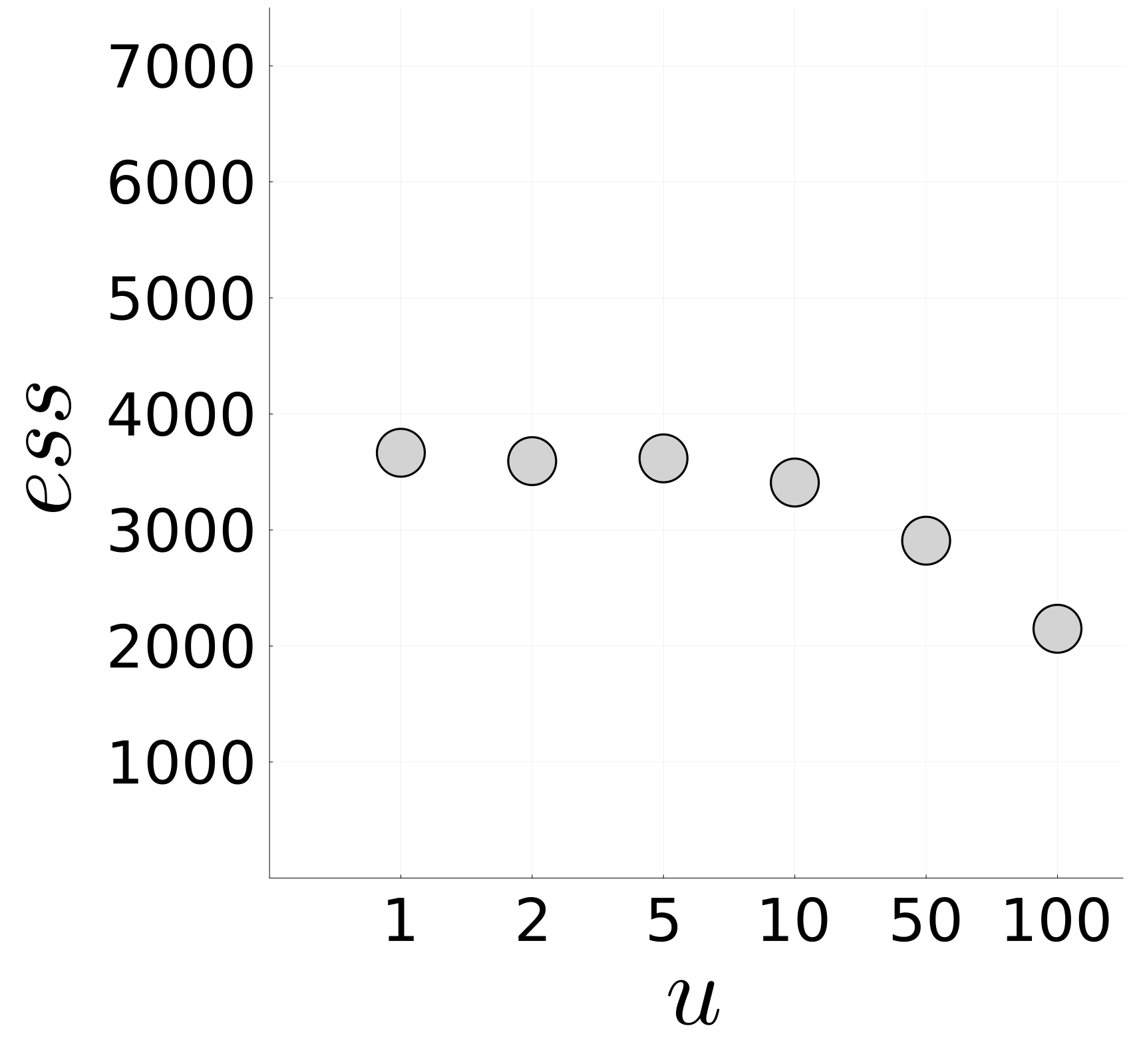}
 }      
\\
\subfigure[]
{
       \includegraphics[width=3.8cm]{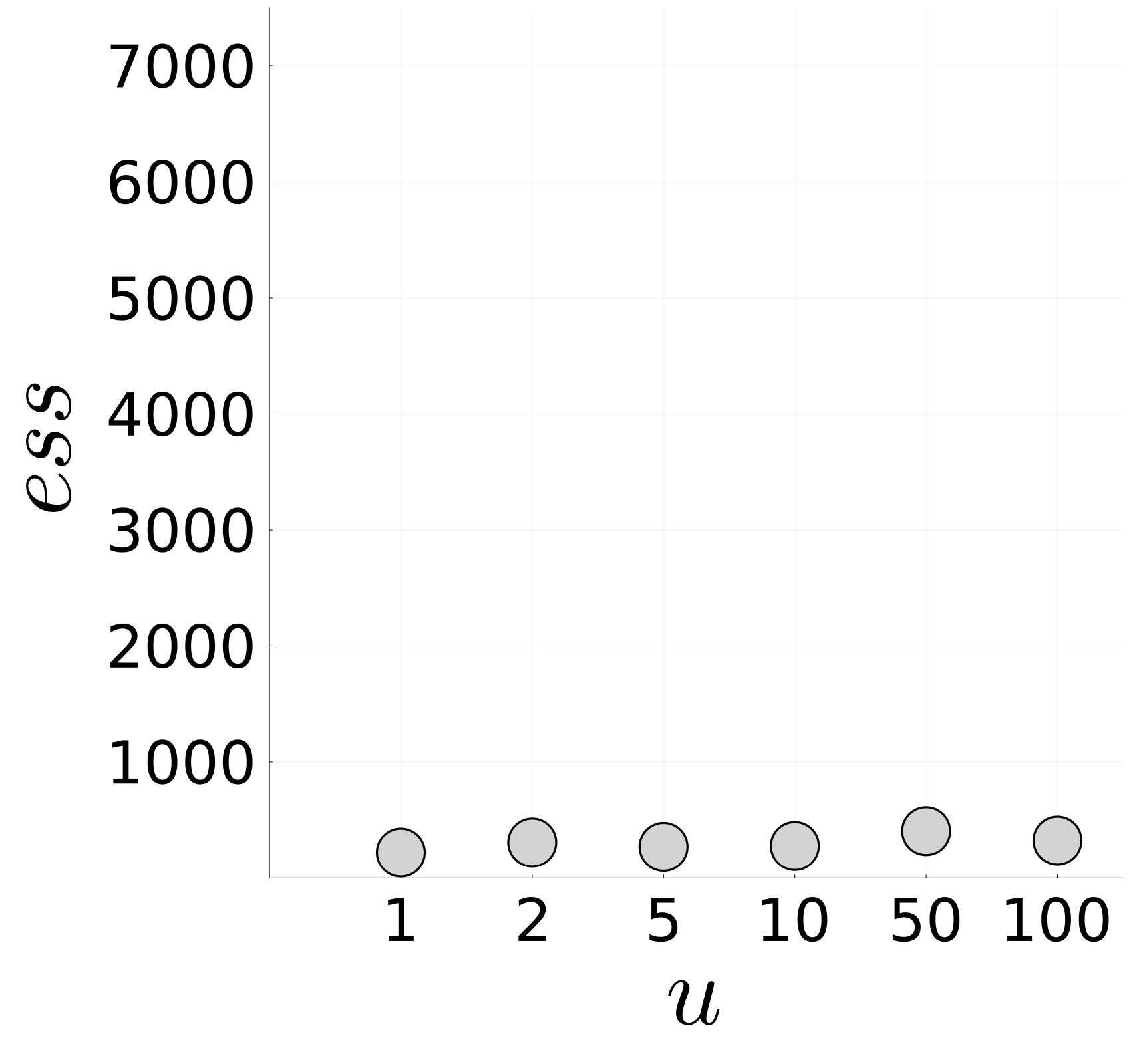}
 }
\subfigure[] 
{
\includegraphics[width=3.8cm]{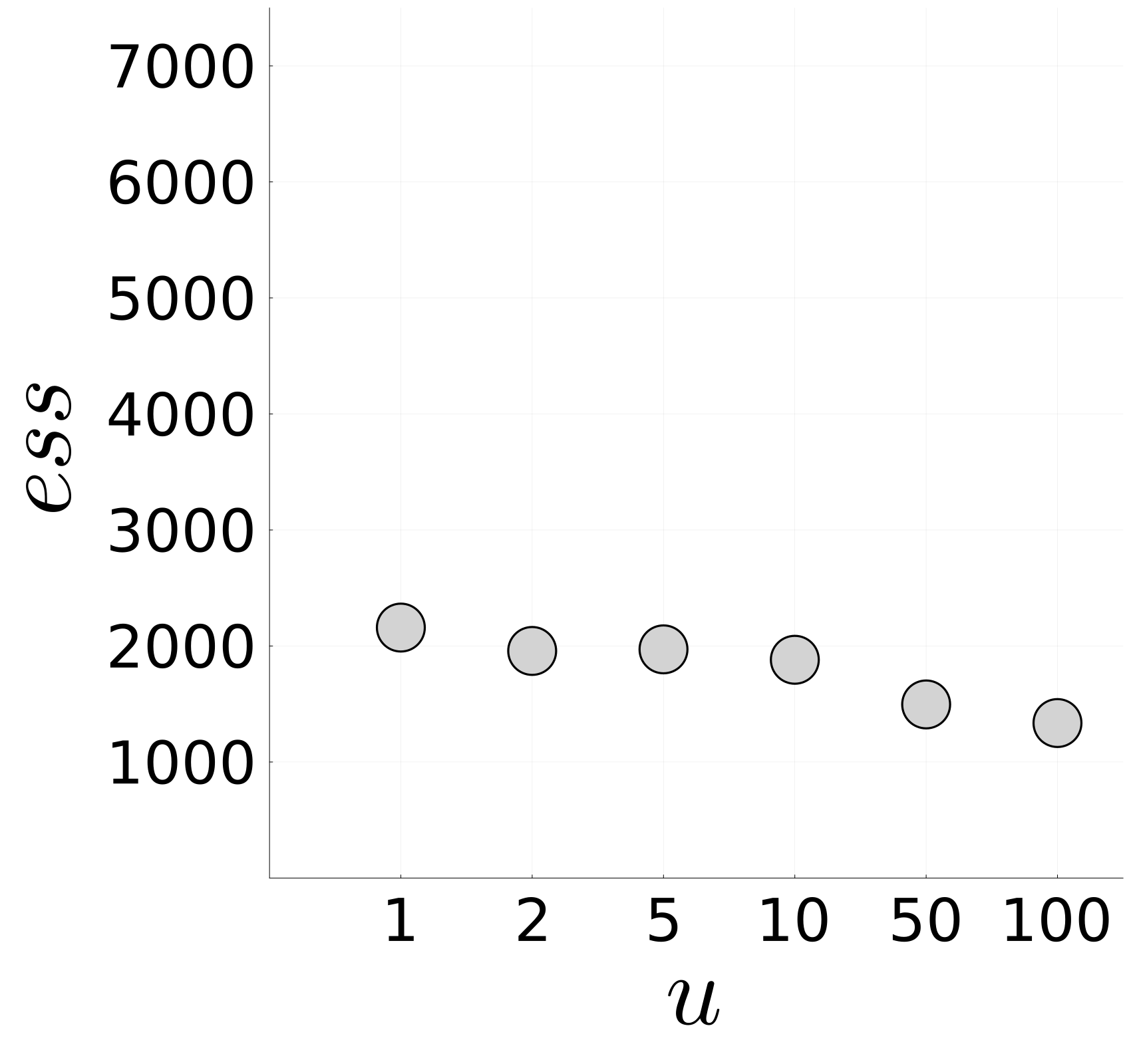}
 }
\subfigure[] 
{
       \includegraphics[width=3.8cm]{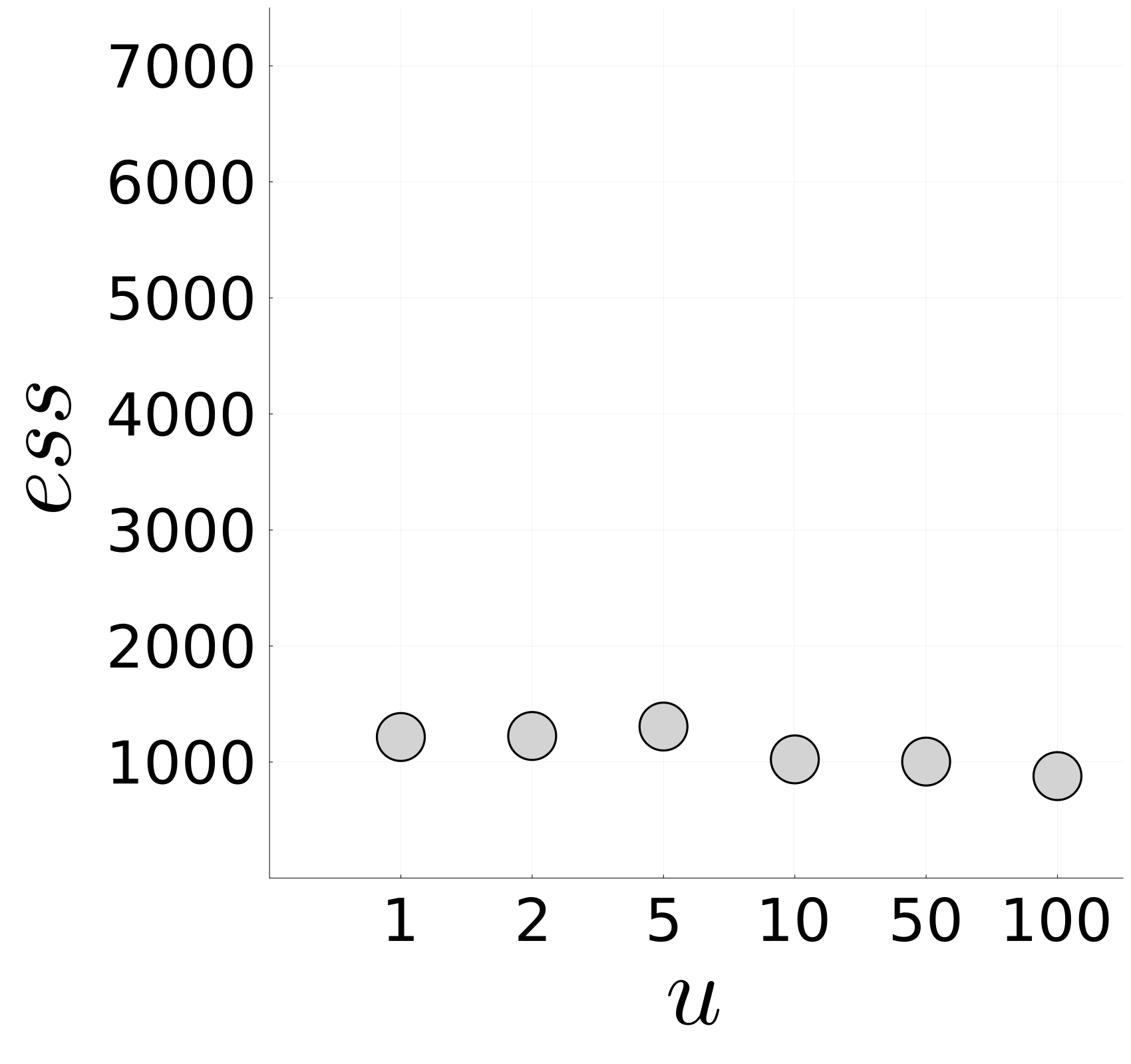}
 }      
\subfigure[] 
{
       \includegraphics[width=3.8cm]{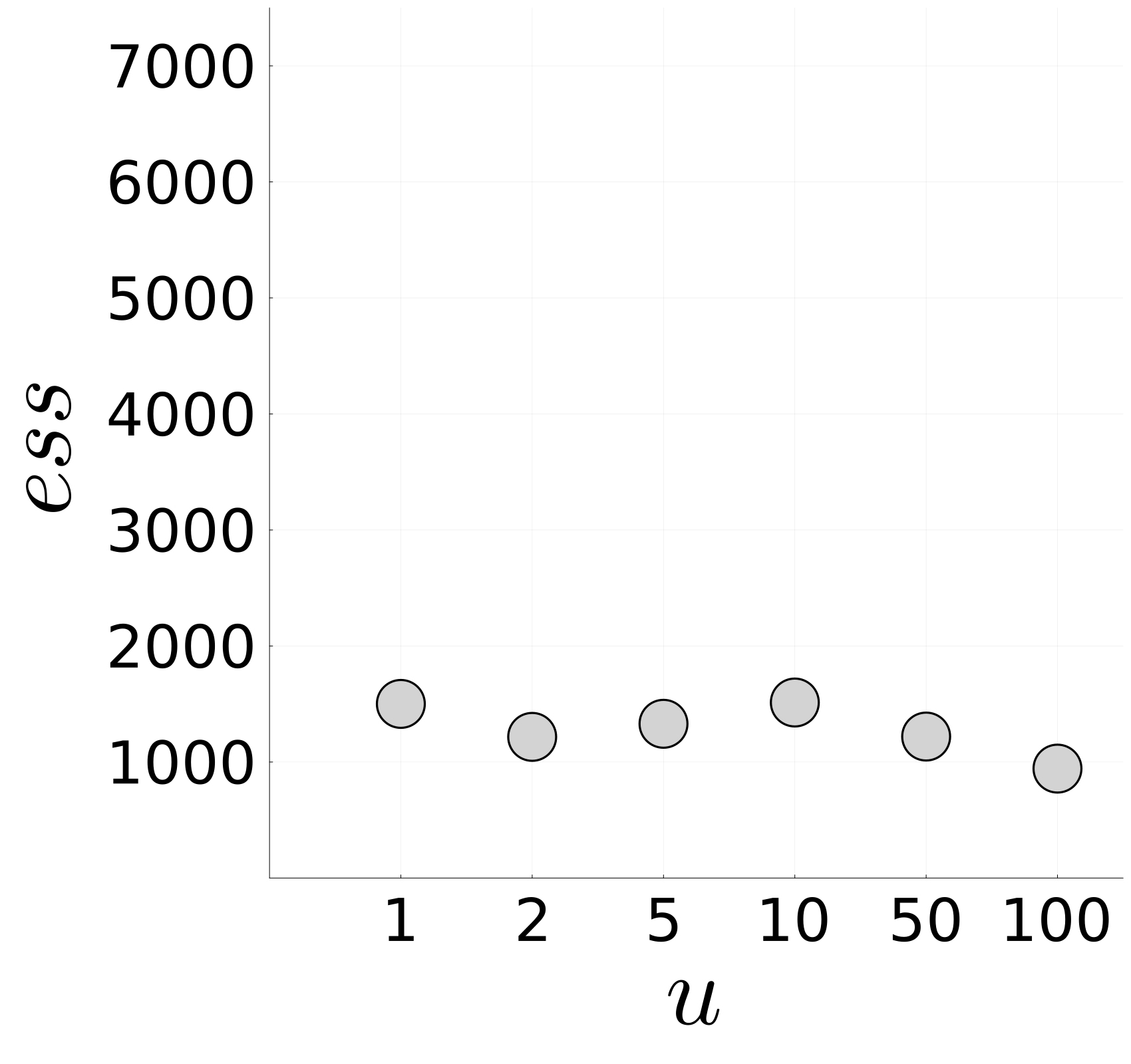}
 }  
\\ 
\subfigure[] 
{
       \includegraphics[width=3.8cm]{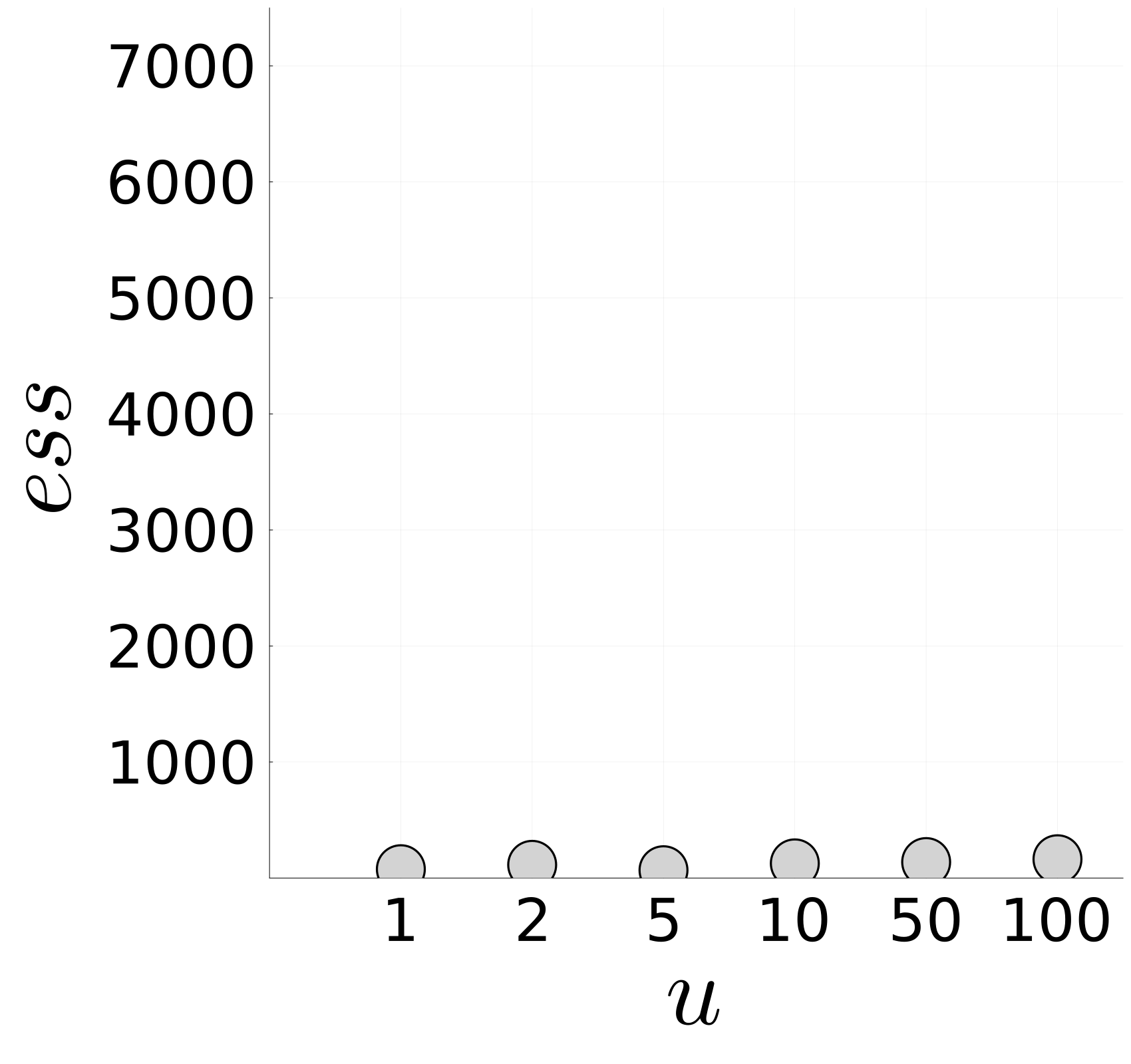}
 }
\subfigure[] 
{
       \includegraphics[width=3.8cm]{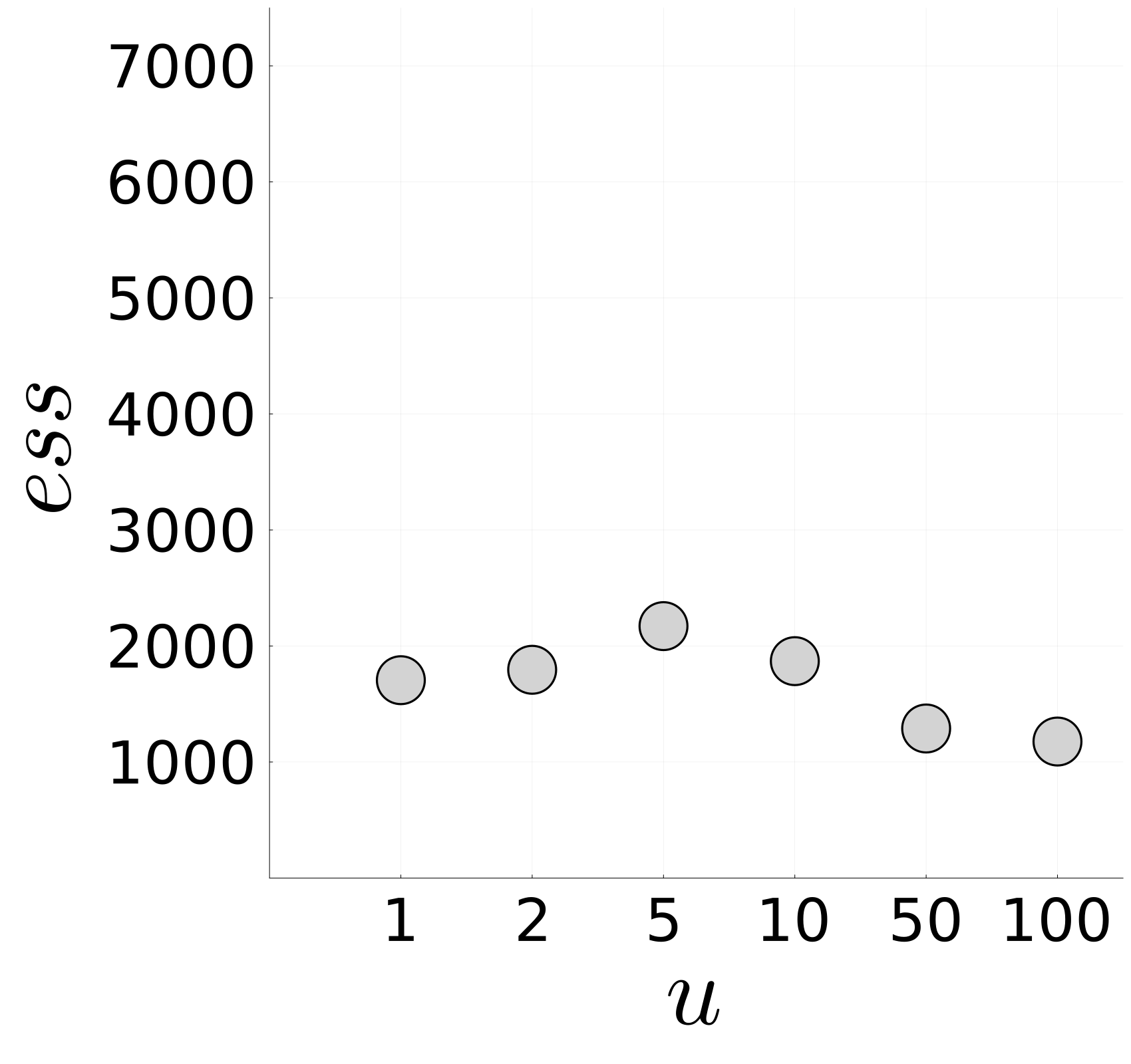}
 }
\subfigure[] 
{
       \includegraphics[width=3.8cm]{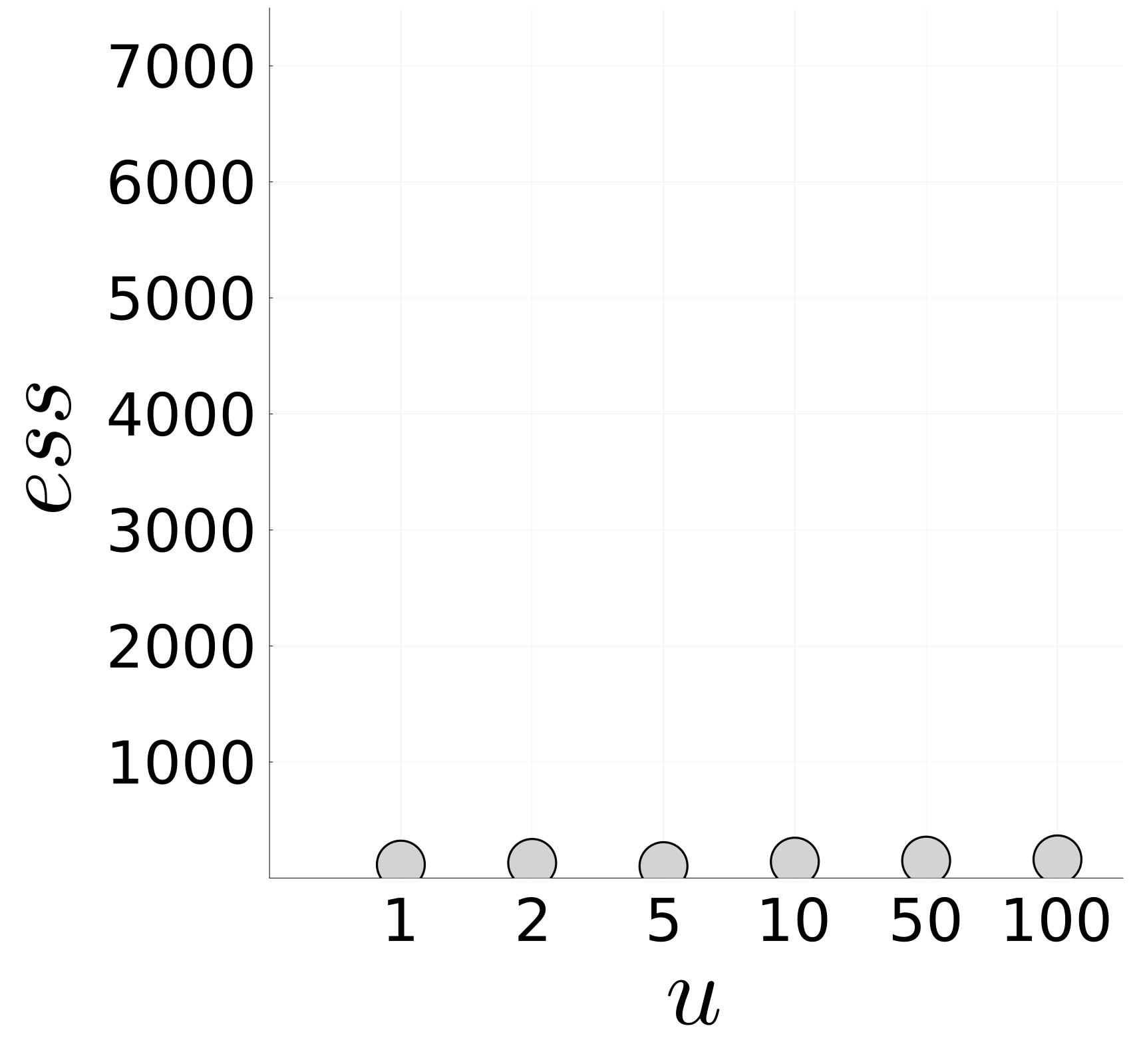}
 }      
\subfigure[] 
{
       \includegraphics[width=3.8cm]{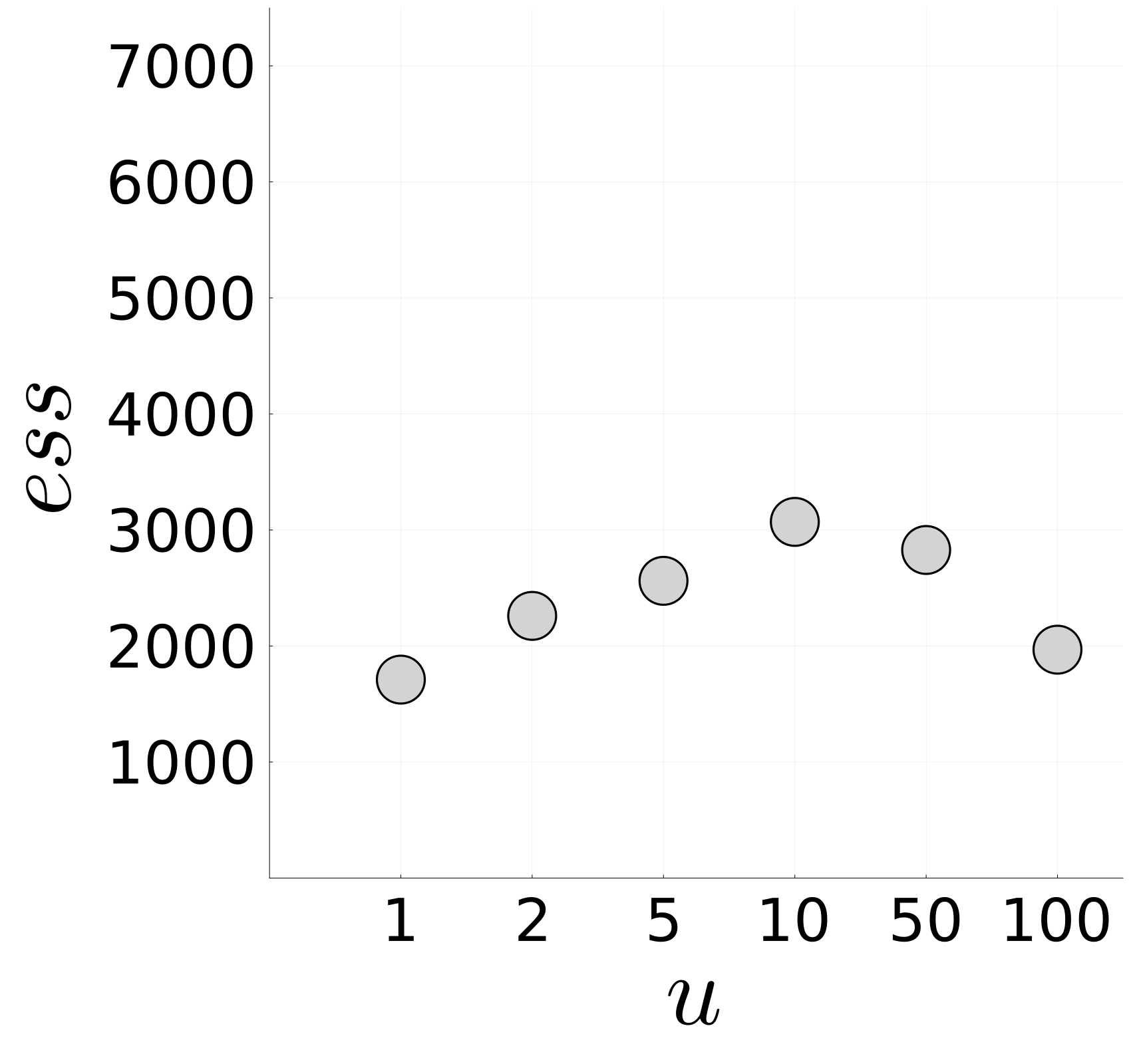}
 }      
\\ 
\subfigure[] 
{
       \includegraphics[width=3.8cm]{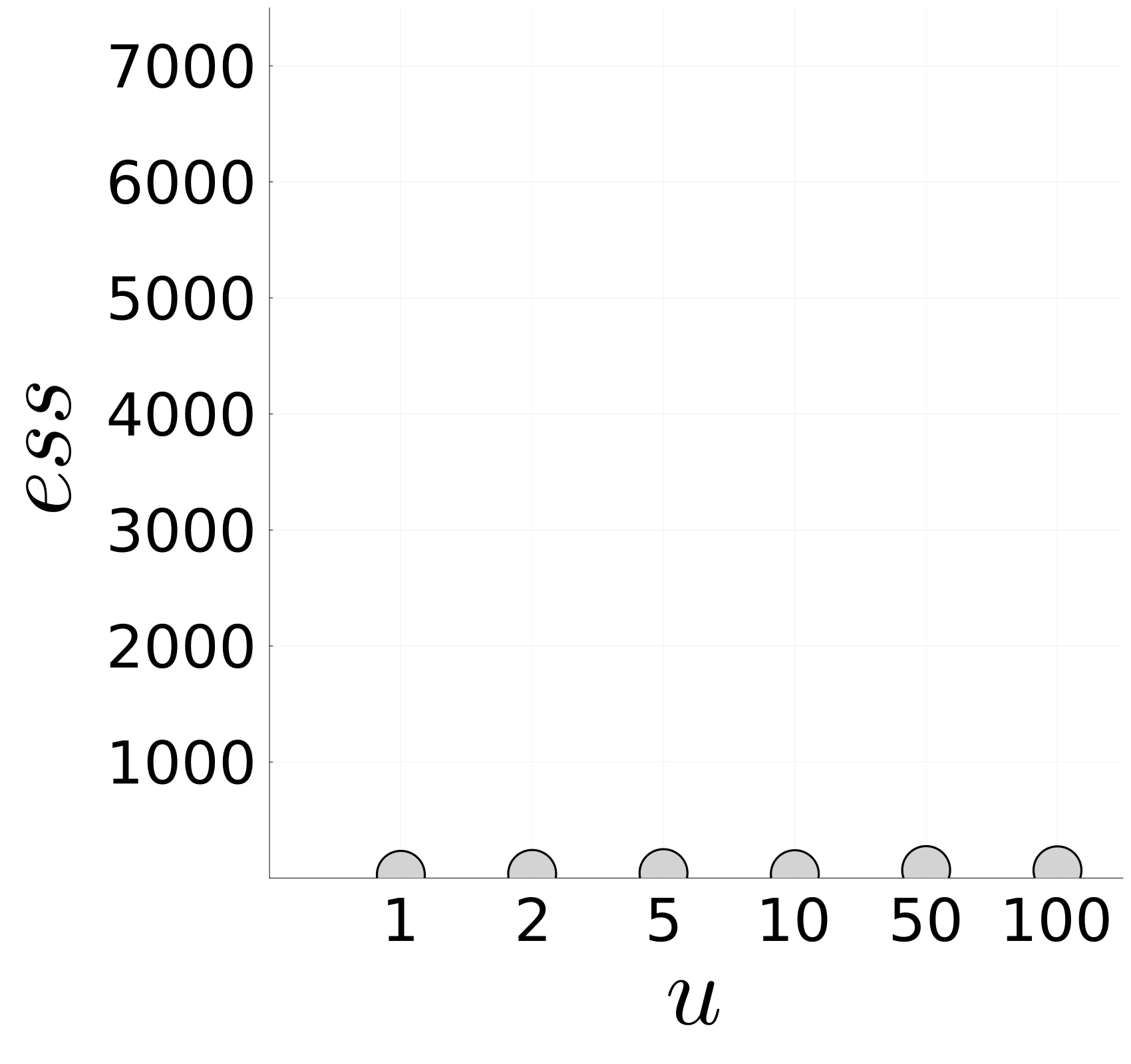}
 }
\subfigure[] 
{
       \includegraphics[width=3.8cm]{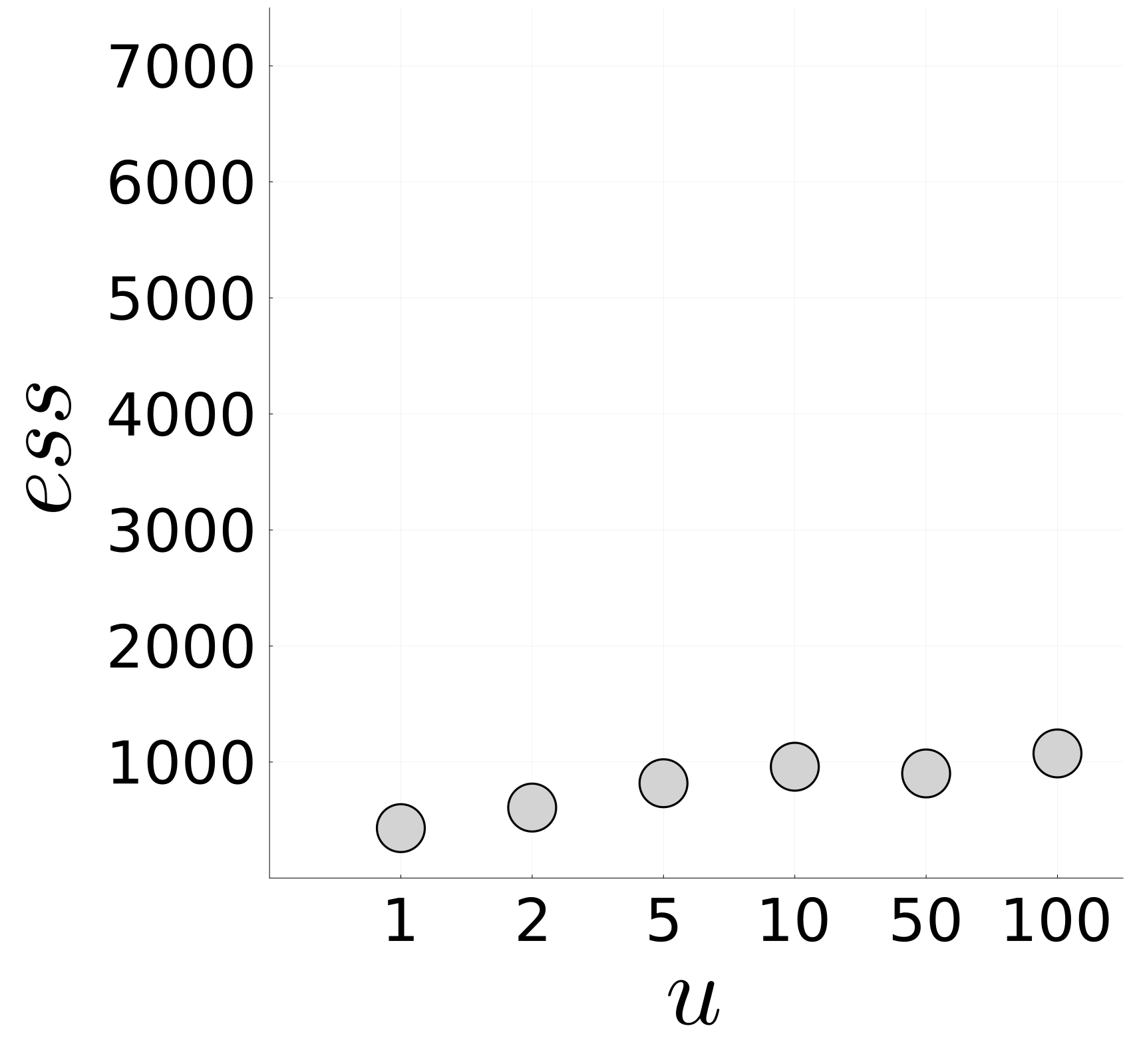}
 }
\subfigure[] 
{
       \includegraphics[width=3.8cm]{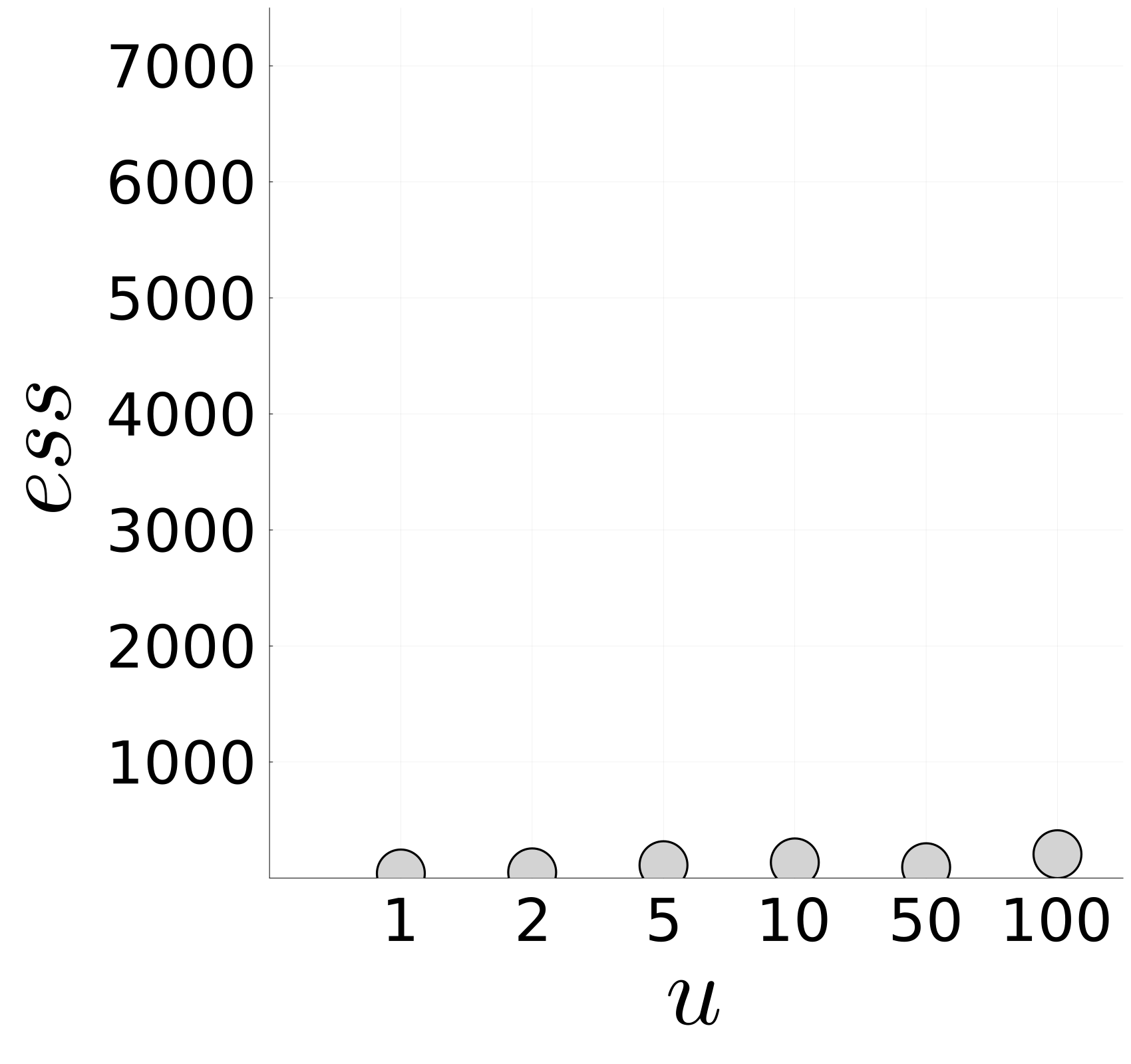}
 }      
\subfigure[] 
{
       \includegraphics[width=3.8cm]{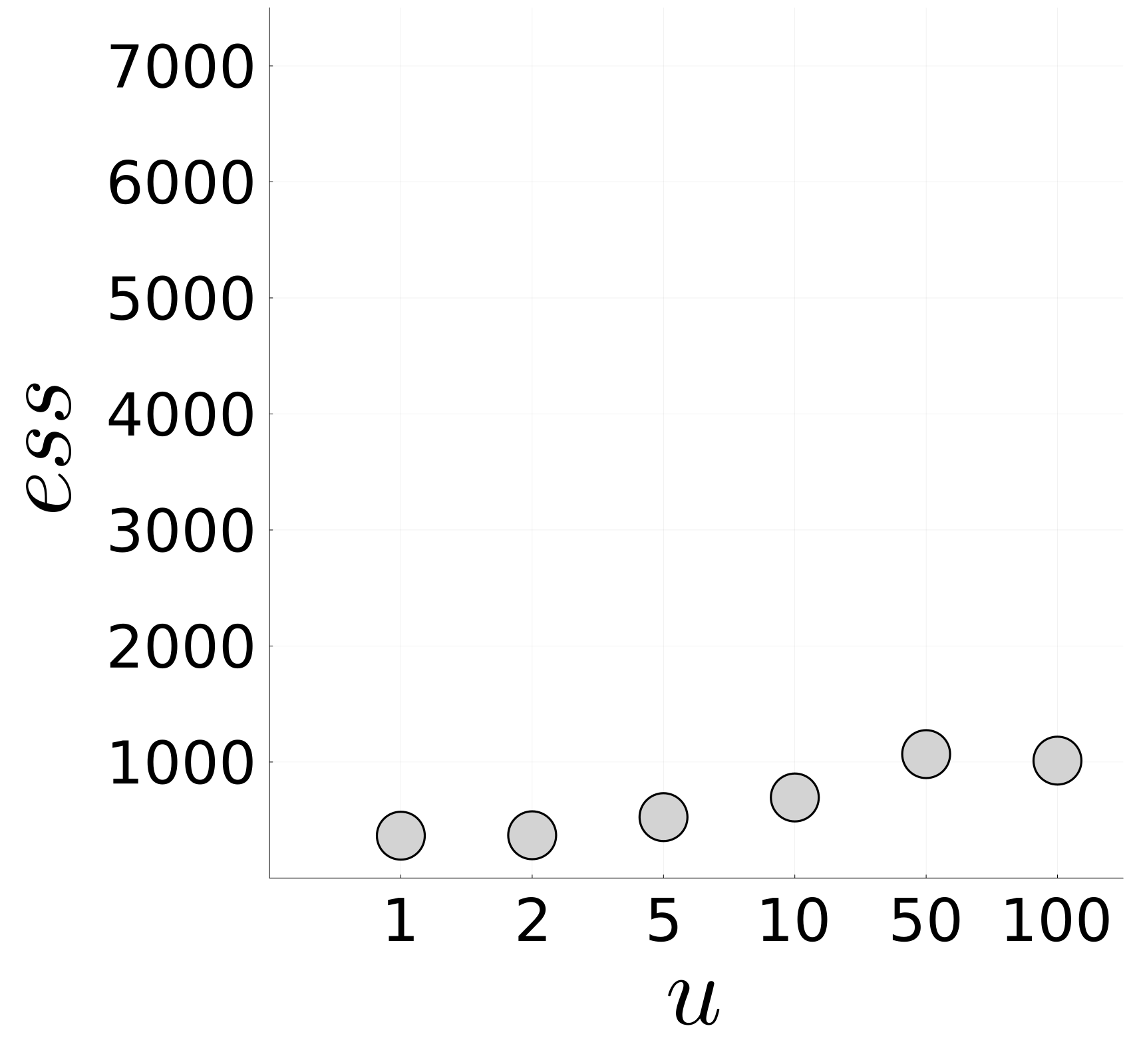}
 }  
 \caption{Simulated data -  Algorithm 3: Effective sample size ($ess$) for different values of $u=\tau/100$.   Panels (a), (e), (i), and (m) display the results for Model 1. Panels (b), (f), (j), and (n)  display the results for Model 2. 
 Panels (c), (g), (k), and (o)  display the results for Model 3.
 Panels (d), (h), (l), and (p)  display the results for Model 4. Panels (a) - (d), (e) - (h), (i) - (l), and (m) - (p) display the results for $k = 10$ and $n = 2,500$, $k = 20$ and $n = 2,500$, $k = 10$ and $n = 10,000$, and $k = 20$ and $n = 10,000$, respectively.}
\label{fig4:simulation1}
\end{figure}

We compare the algorithms by evaluating the effective sample size (ESS) for the model parameters. We do not consider the ESS per second as a comparison criteria because the computational effort of the updating scheme is similar. Figures \ref{fig2:simulation1} -\ref{fig4:simulation1} show the distribution of the ESS across the model parameters for the different updating schemes, tuning parameter, true model, and sample size.
\clearpage

\clearpage

\clearpage

We can see from Figures \ref{fig2:simulation1} -- \ref{fig4:simulation1}  that the complexity of the true model, the sample size, grid resolution, and the choice of the tuning parameter of the MCMC sampling scheme impact the mixing of the chain. The amount of data influences the complexity of the posterior, and as the amount of data increases, we generally see the mixing degrades. As a general rule we also see the mixing worsens as the grid resolution increases from 10 to 20.  The simulations clearly show that the dependency structure of the model also has a dramatic influence on the mixing.  Model 2 appears to have the best mixing properties followed closely by Model 4.  Both of these models' contour plots are nearly convex and appear to allow for easier posterior exploration.  Generally, Model 1 appears to have the worst mixing and Model 3 is not much better.  These models have irregular shapes which could complicate the MCMC mixing properties.

The primary purpose of this simulation study was to help determine which MCMC algorithm is best.  The results for these scenarios nearly always favors the iterated rectangular exchanges proposal and the generalized rectangular exchange proposal typically outperforms the vertex-line proposal. Tuning the proposal for a specific problem will yield best results. 
For the iterated rectangular exchanges, a number of rectangle exchanges ranging from 5 to 10 can be considered as a good starting point in this process.

Finally, an important aspect of the iterated rectangular exchanges and the generalized rectangular exchange proposal is that they can be implemented on non-yett uniform copulas regardless of the grid spacing.  Whereas the vertex-line proposal relies on the equal spacing of the grid in any dimension to generate vertices.

\subsection{Estimation of parametric and non-standard copula functions with known marginals}

The aim of this section is to illustrate the behavior of the model and its ability to properly estimate simple and complex true copula functions. For this we consider the same four true models of the previous section and assume that the marginal distributions are known.  For each model, we simulate a perfect sample of size $n = $ 600, 1,400, 2,500, 4,500, 10,000 and 18,000.  For each simulated dataset we fit our proposed model by considering $k$ = 10 and 20, the hierarchically centered prior with the ICAR correlation structure, and with $G_0$ being an independent copula. We create a Markov chain of (conservative) size of 2,400,000 using the iterated rectangular exchanges algorithm. We considered a burn-in period of 400,000 and a thinning of 1,000. 

\begin{figure}[!h]
\begin{centering}
\subfigure[M1 - $k$ = 10.]{%
\includegraphics[width=3.8cm]{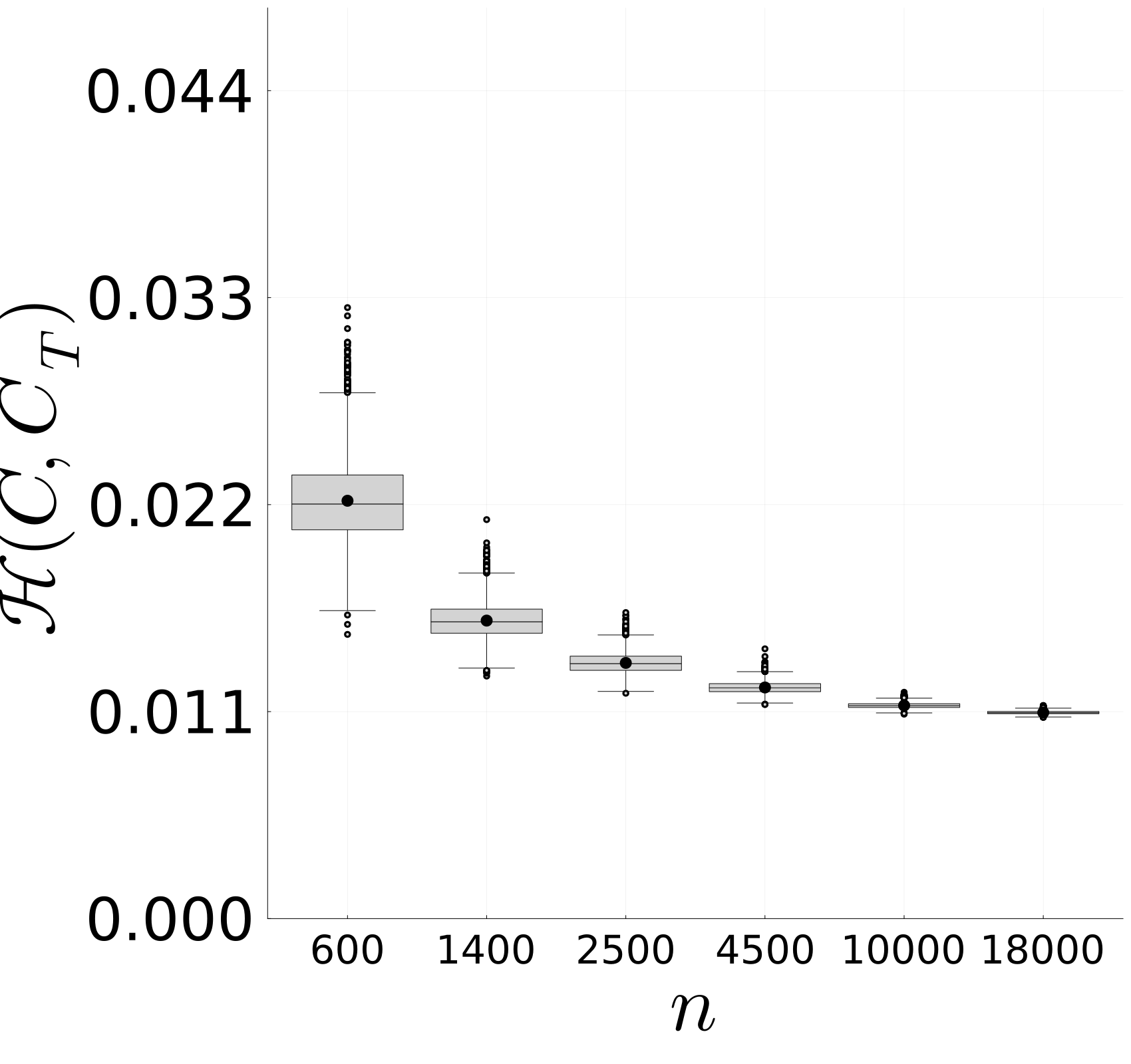} 
}      
\subfigure[M1 - $k$ = 20.]{%
\includegraphics[width=3.8cm]{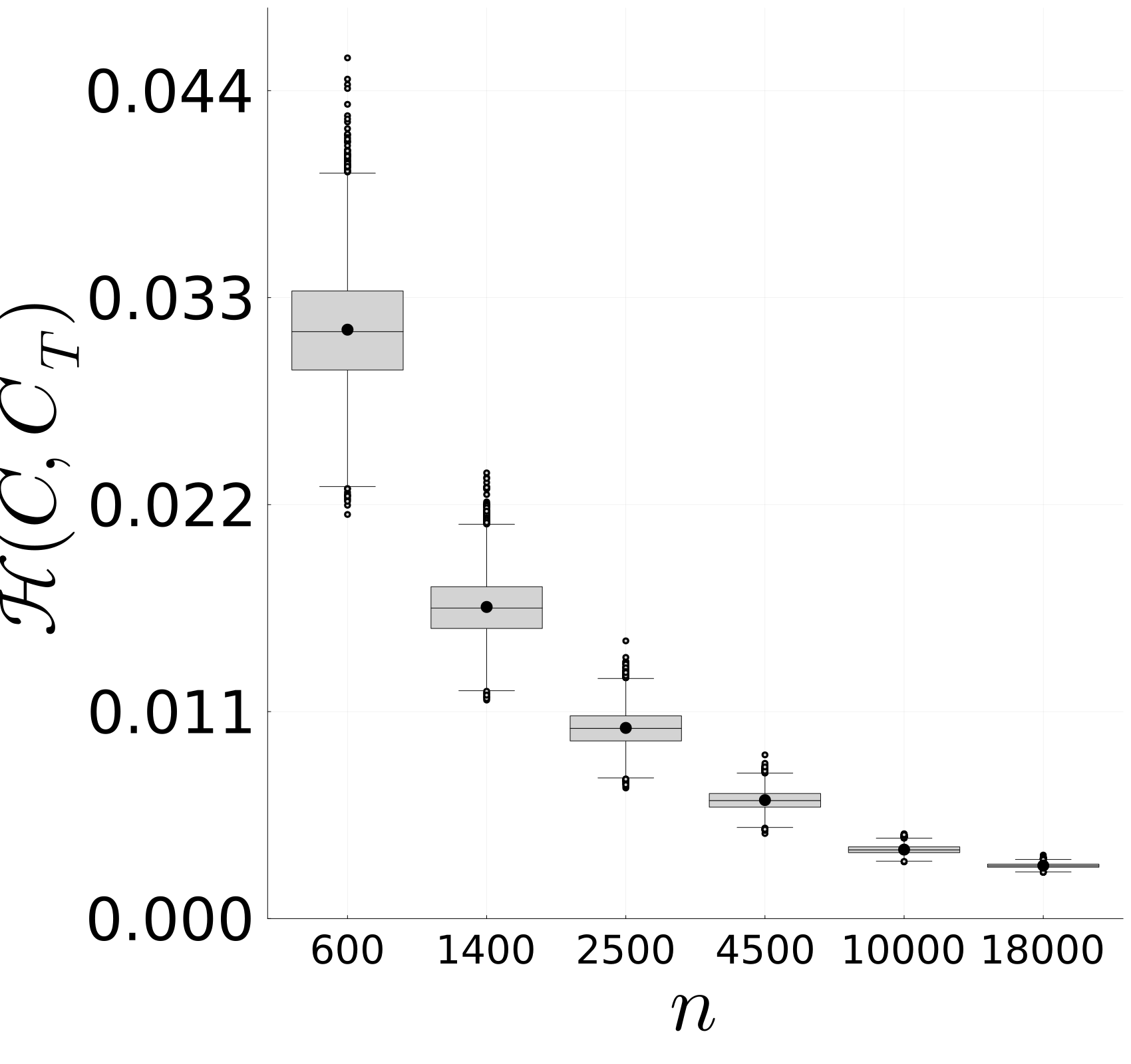}
}
\subfigure[M2 - $k$ = 10.]{%
\includegraphics[width=3.8cm]{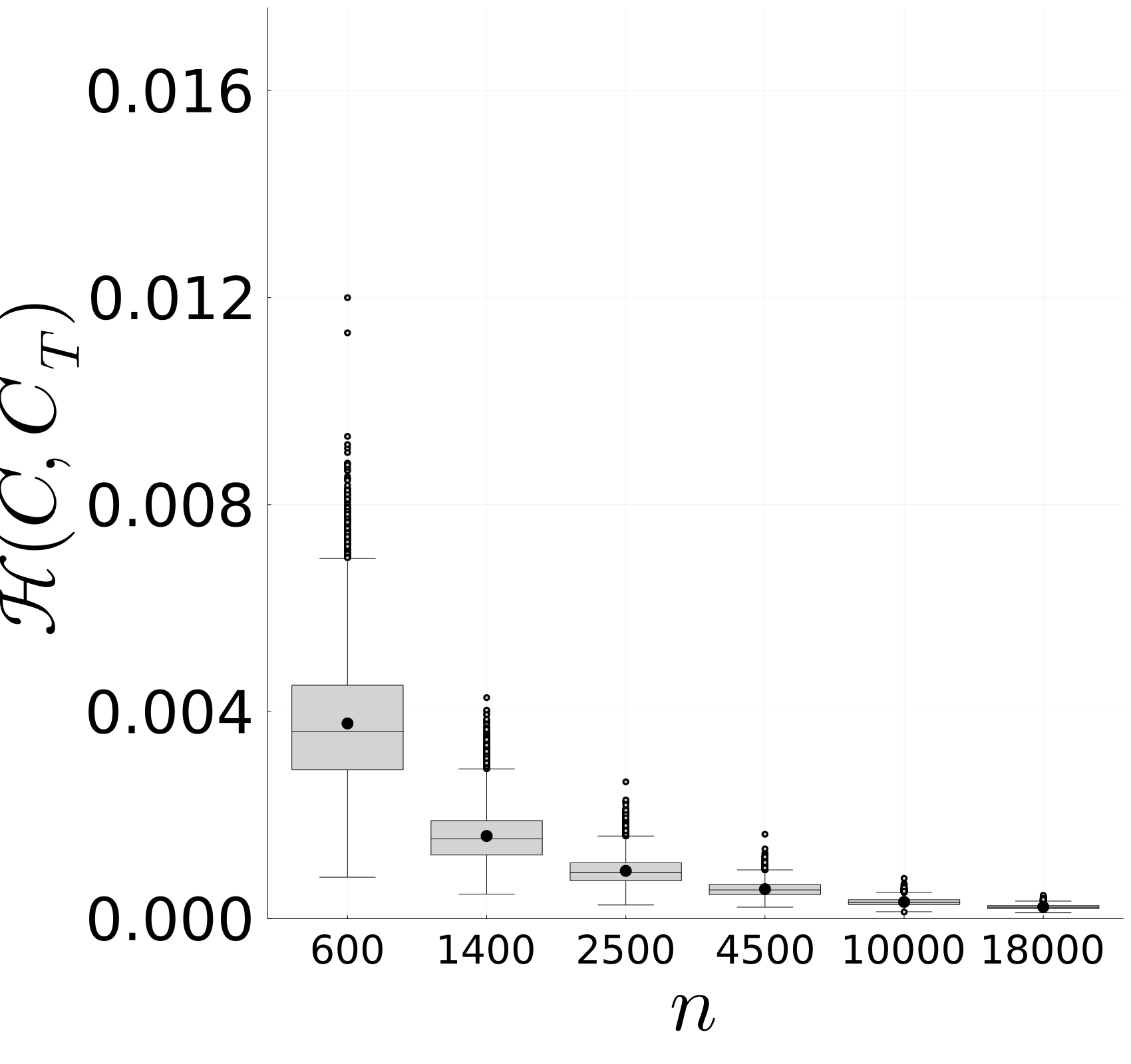} 
}      
\subfigure[M2 - $k$ = 20.]{%
\includegraphics[width=3.8cm]{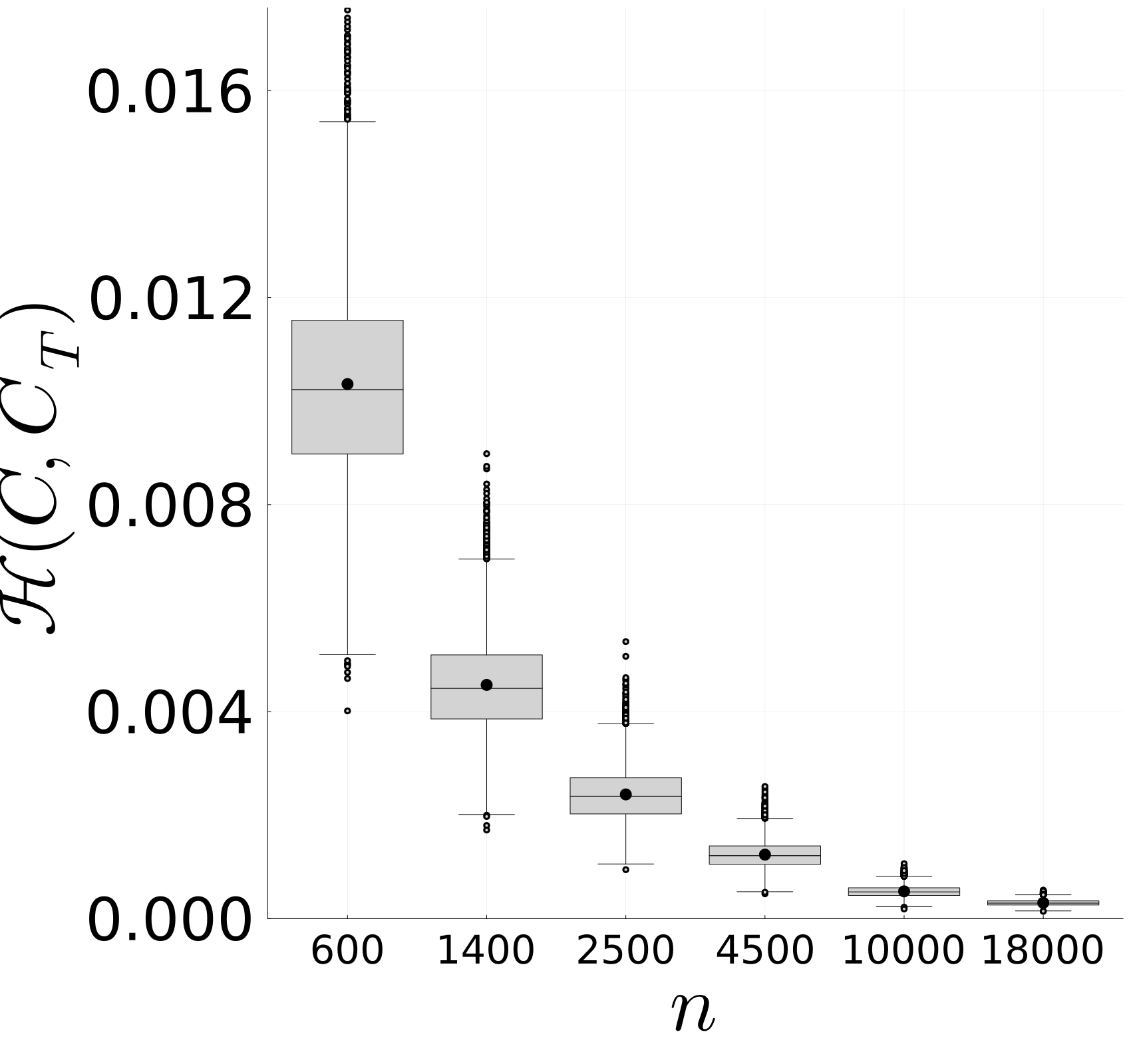}
}
\\ 
\subfigure[M3 - $k$ = 10.]{%
\includegraphics[width=3.8cm]{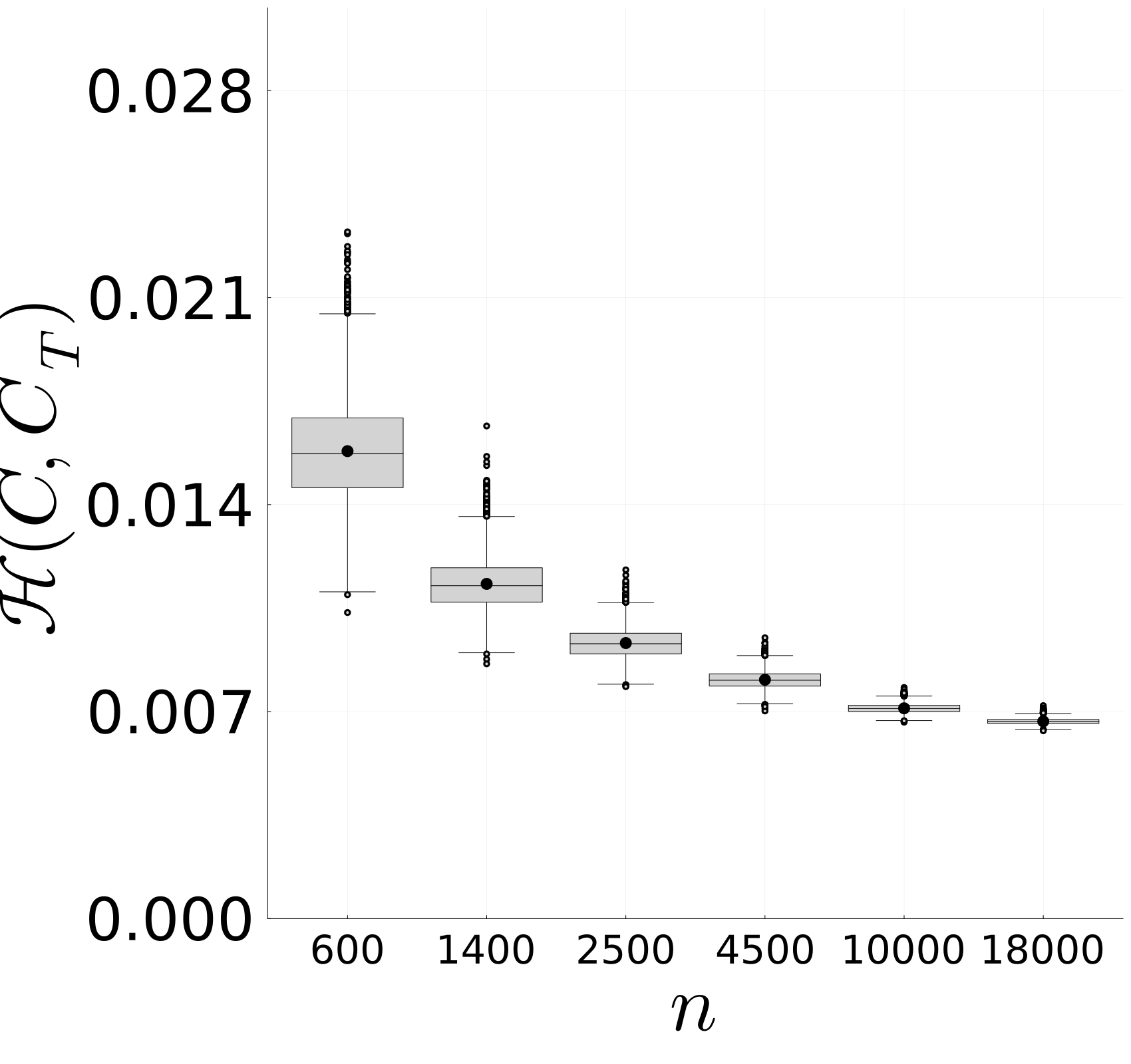} 
}      
\subfigure[M3 - $k$ = 20.]{%
\includegraphics[width=3.8cm]{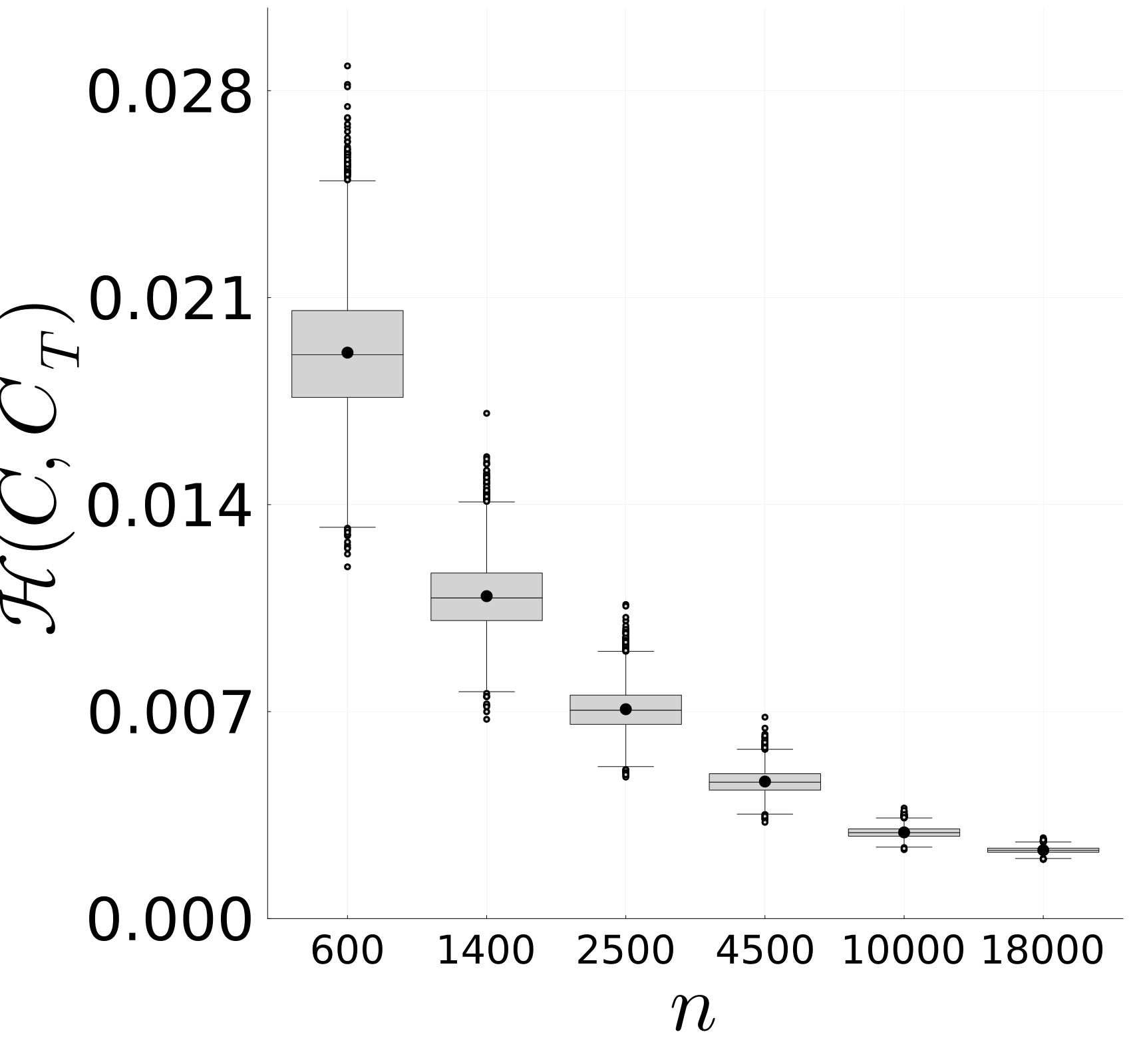}
}
\subfigure[M4 - $k$ = 10.]{%
\includegraphics[width=3.8cm]{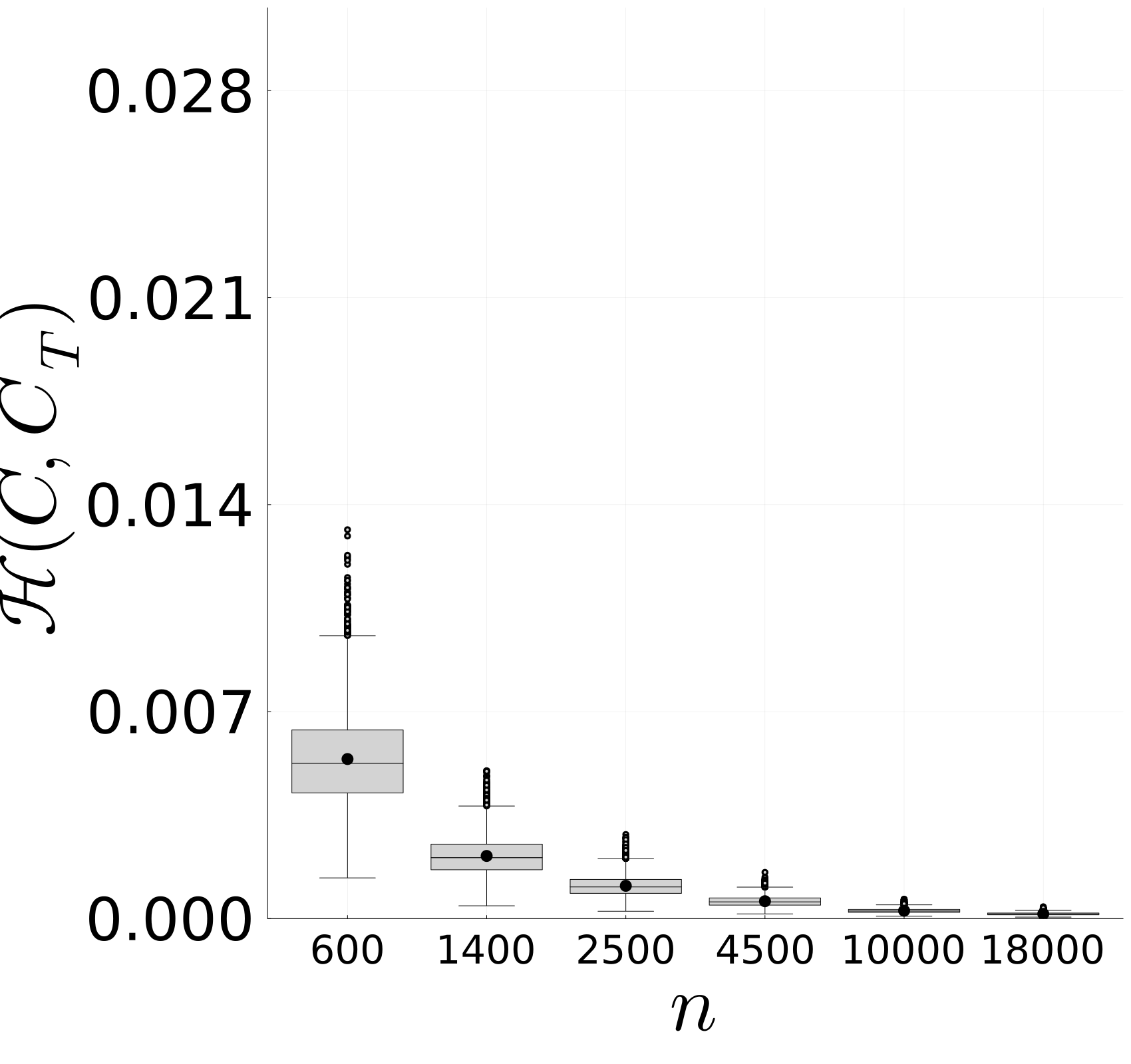} 
}      
\subfigure[M4 - $k$ = 20.]{%
\includegraphics[width=3.8cm]{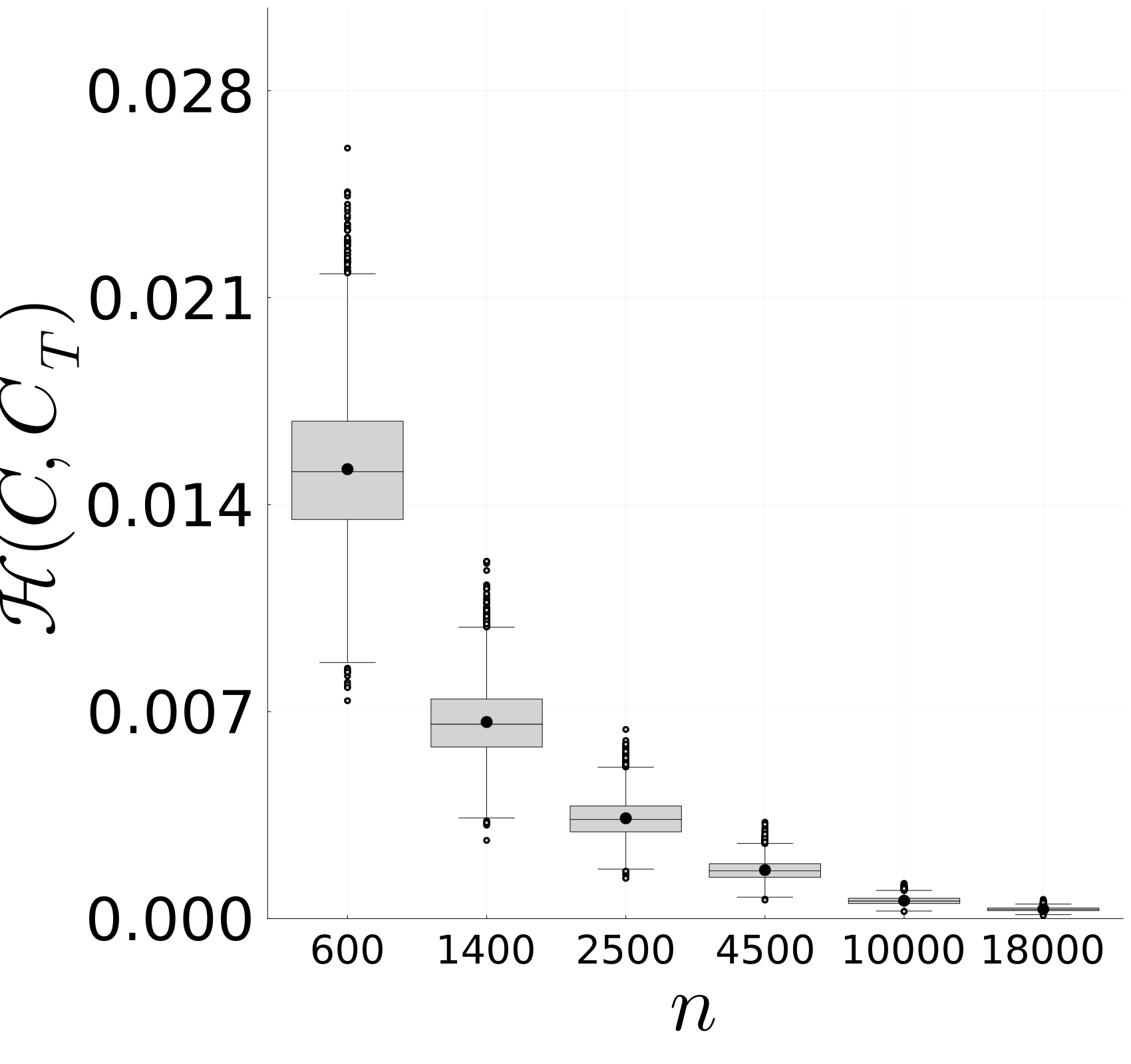}
}
\caption{Simulated data - known marginals: Box and whisker plot summarizing the posterior distribution of the Hellinger distance, $\mathcal{H}$, to the true copula model $C_T$, under the different true models, grid size, and sample size. Panels (a) and (b), (c) and (d), (e) and (f), and (g) and (h) display the results under true models 1 (M1), 2 (M2), 3 (M3) and 4 (M4), respectively. 
Panels (a), (c), (e), and (g)  display the results for $k$ = 10. Panels (b), (d), (f), and (h)  display the results for $k$ = 20. In all cases, the posterior mean is represented as a point.}
       \label{fig1:simulation2}
    \end{centering}
\end{figure}

\begin{figure}[!h]
\centering
\subfigure[$n = $ 600.] 
{
\includegraphics[width=3.5cm]{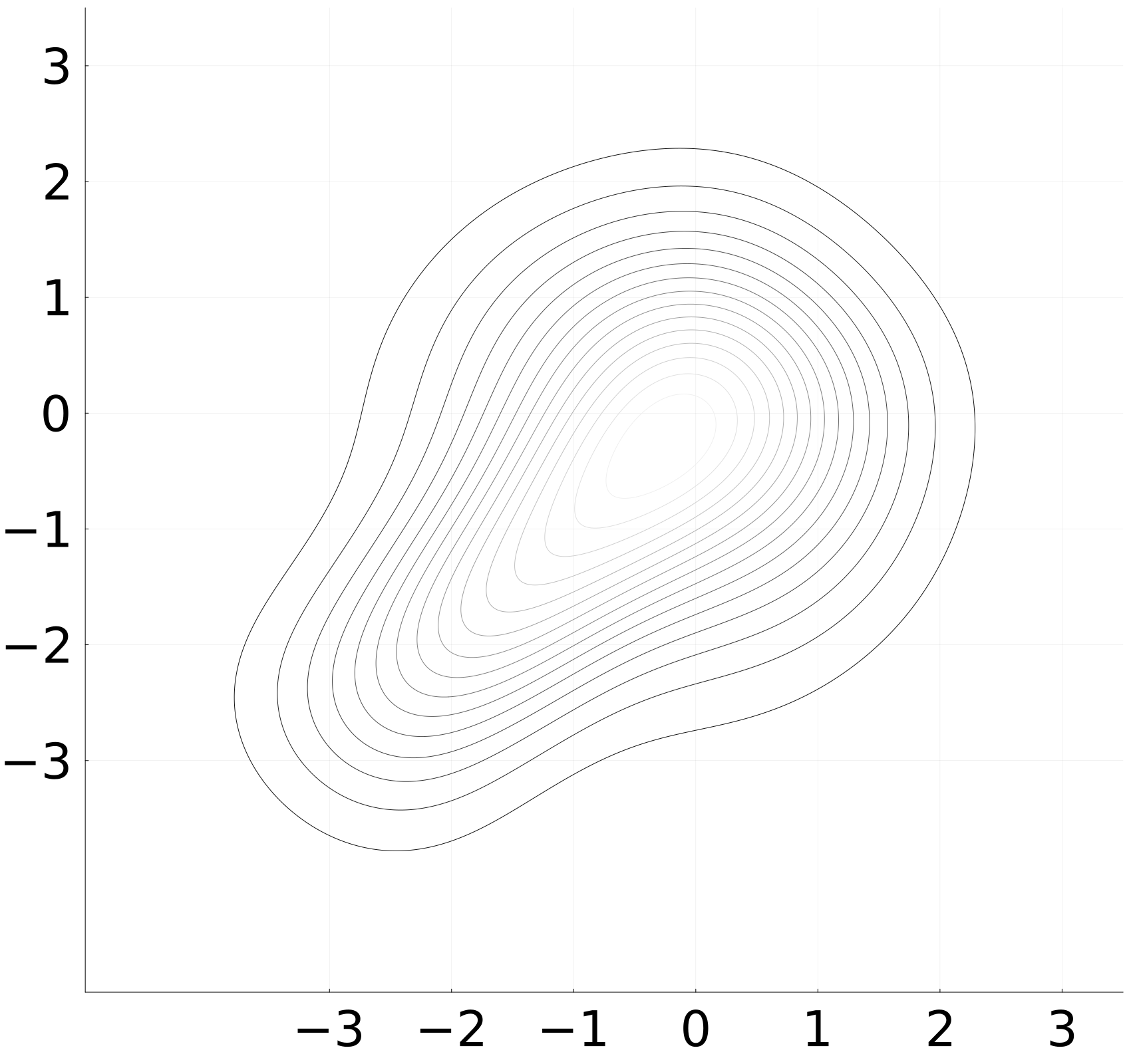}
 }      
 \subfigure[$n = $ 4,500.] 
{
\includegraphics[width=3.5cm]{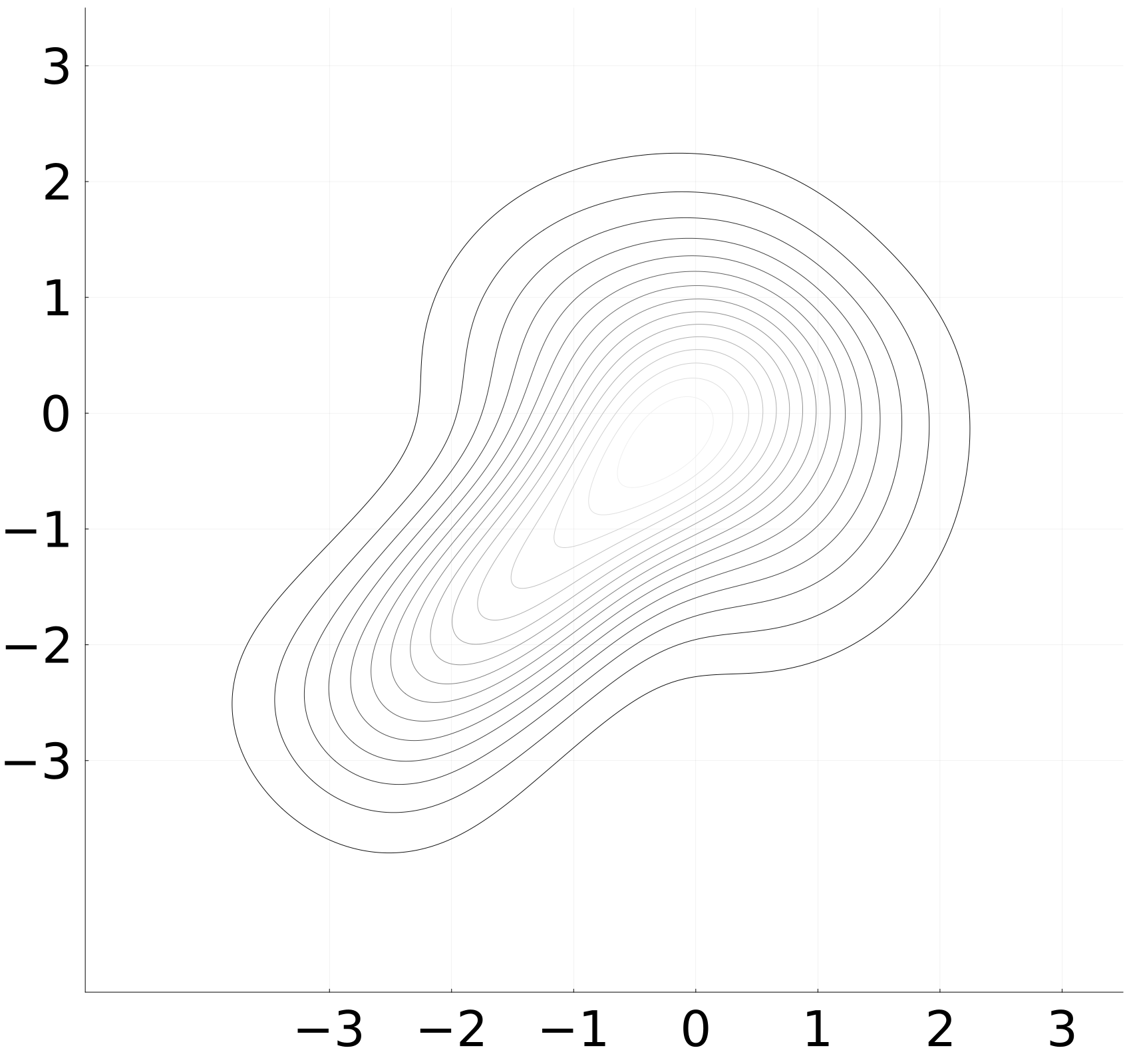}
 }      
\subfigure[$n = $ 18,000.] 
{
\includegraphics[width=3.5cm]{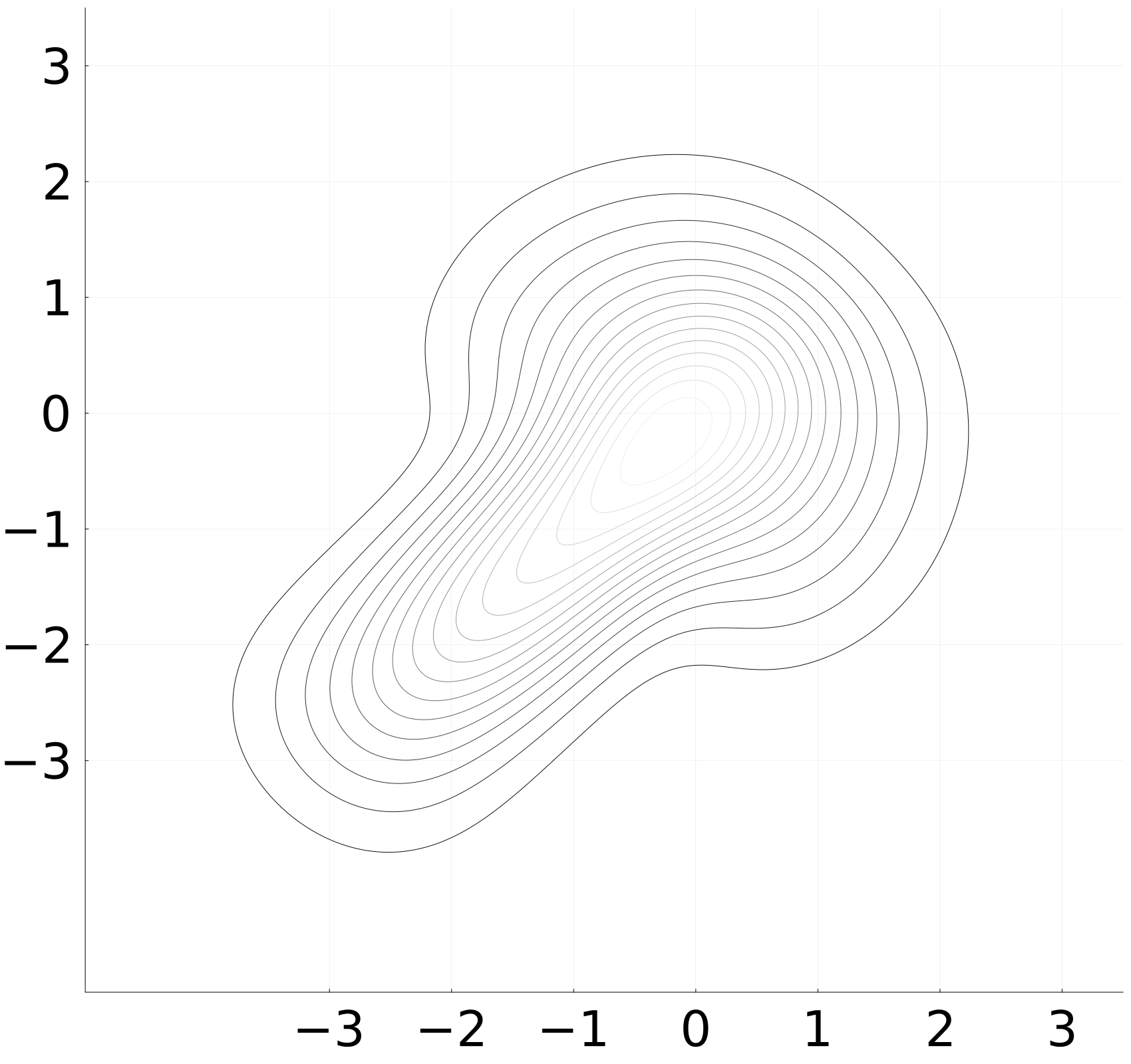}
 }      
\subfigure[True.] 
{
\includegraphics[width=3.5cm]{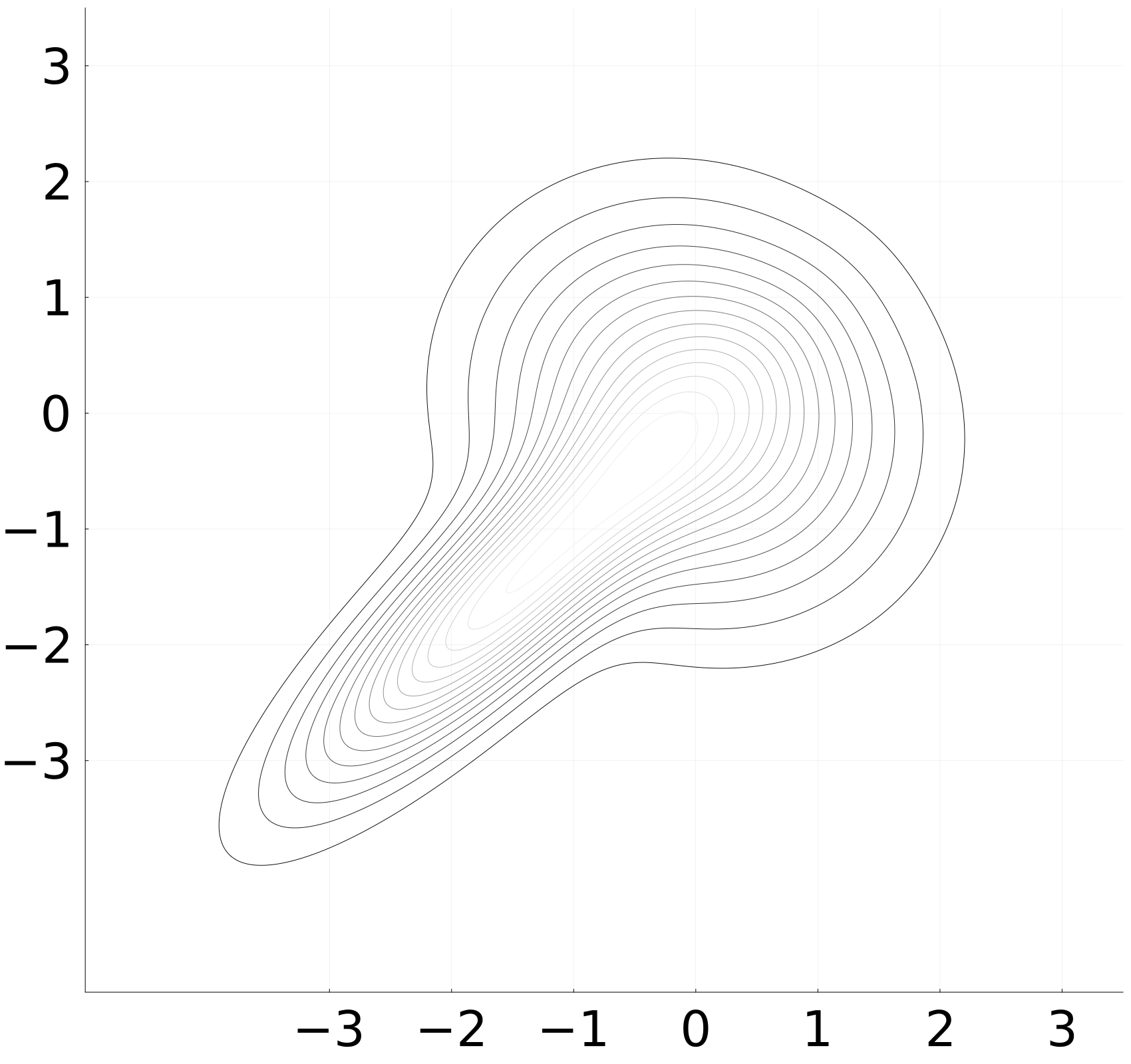}
 }      
\\ 
\subfigure[$n = $ 600.] 
{
\includegraphics[width=3.5cm]{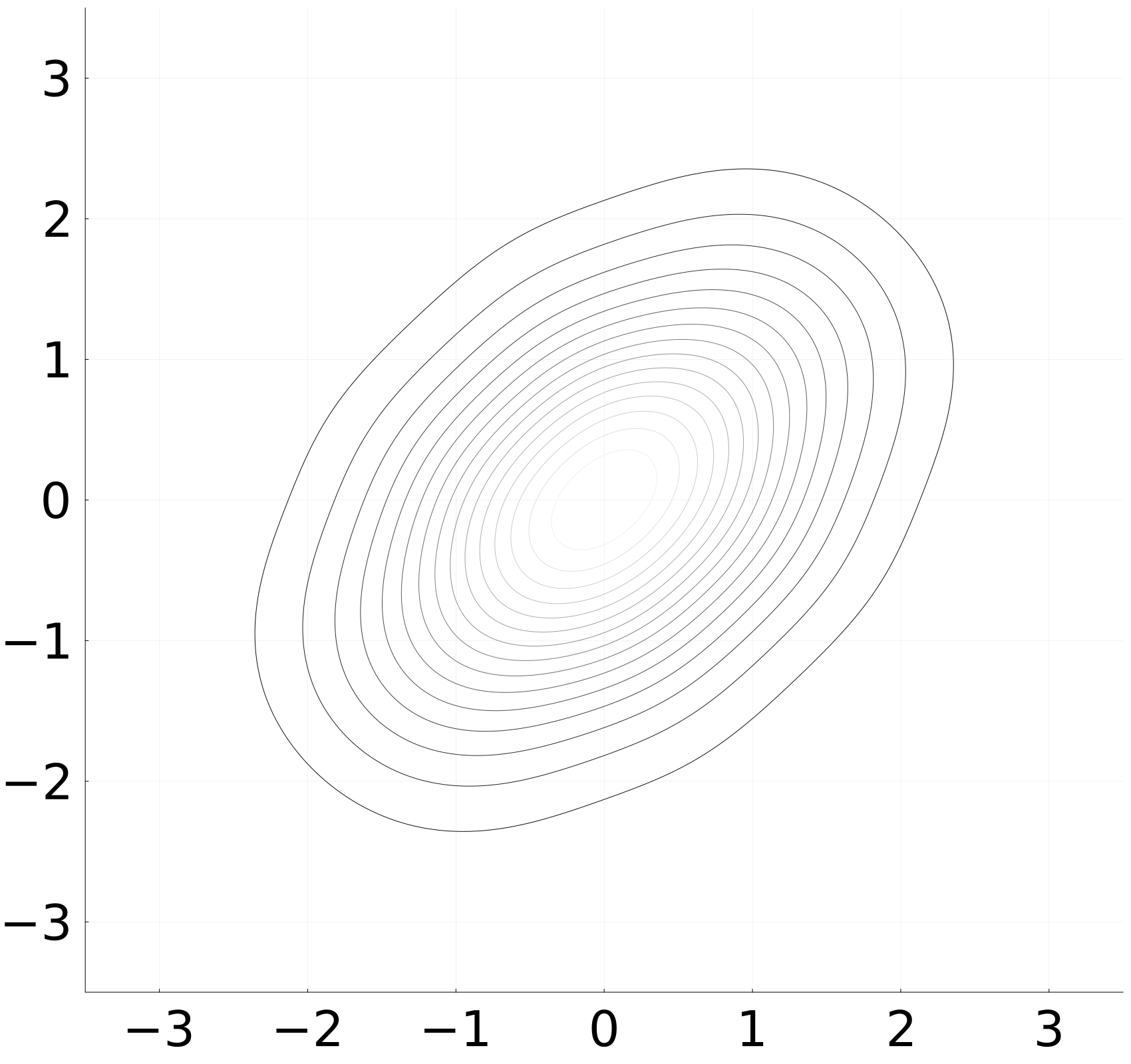}
 }
 \subfigure[$n = $ 4,500.] 
{
\includegraphics[width=3.5cm]{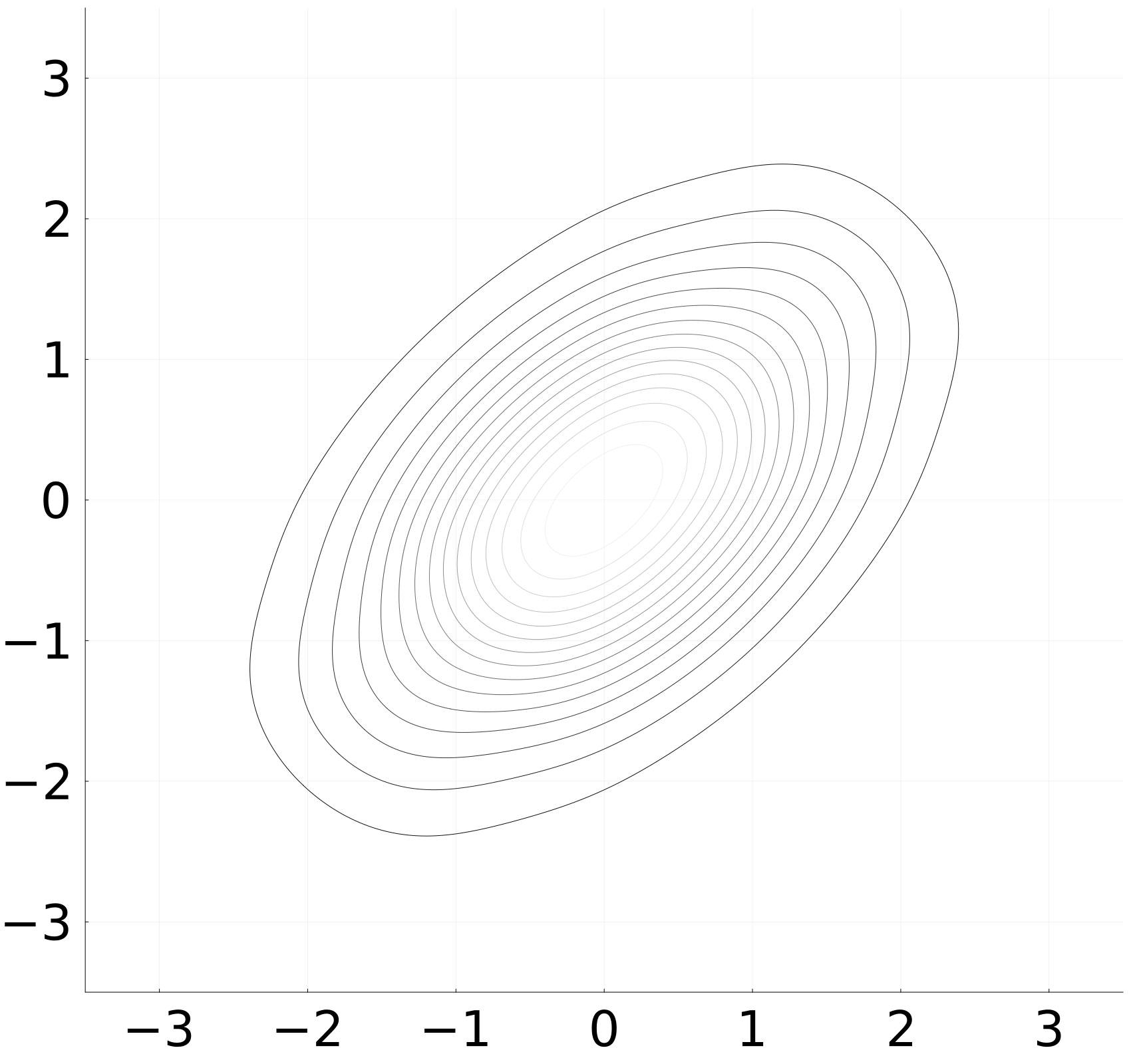}
 }      
\subfigure[$n = $ 18,000.] 
{
\includegraphics[width=3.5cm]{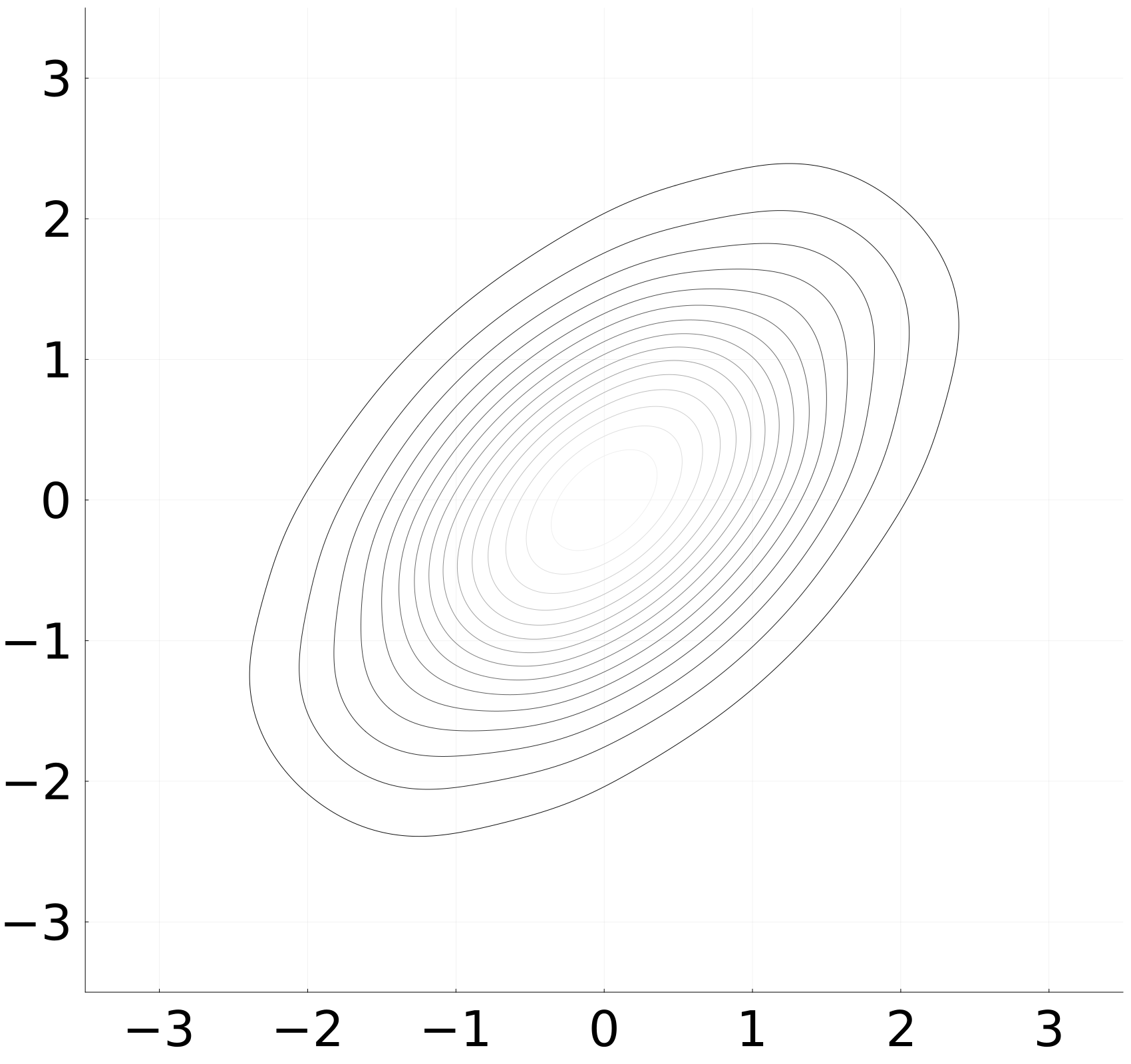}
 }      
\subfigure[True.] 
{
\includegraphics[width=3.5cm]{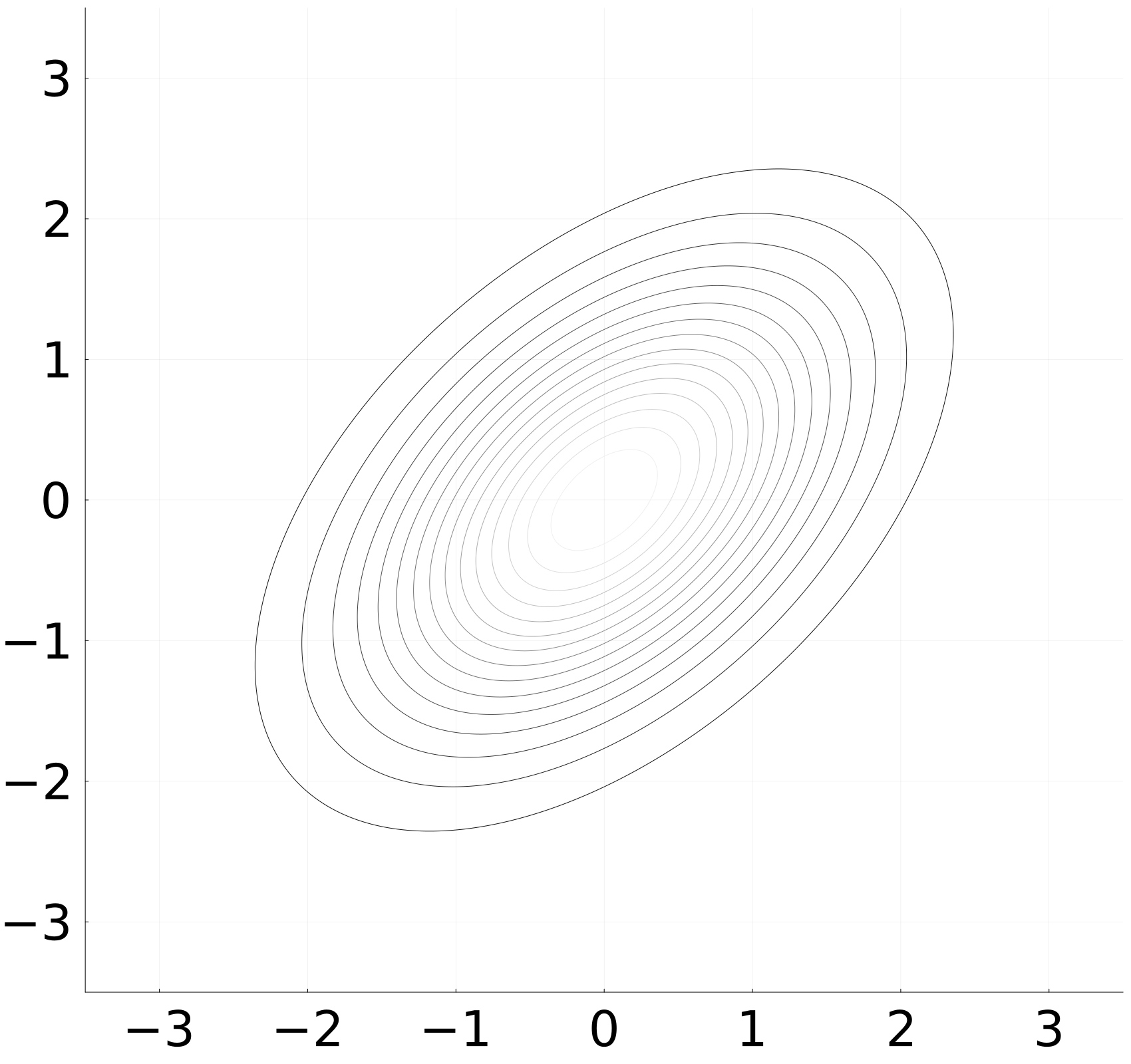}
 }      
\\ 
\subfigure[$n = $ 600.] 
{
\includegraphics[width=3.5cm]{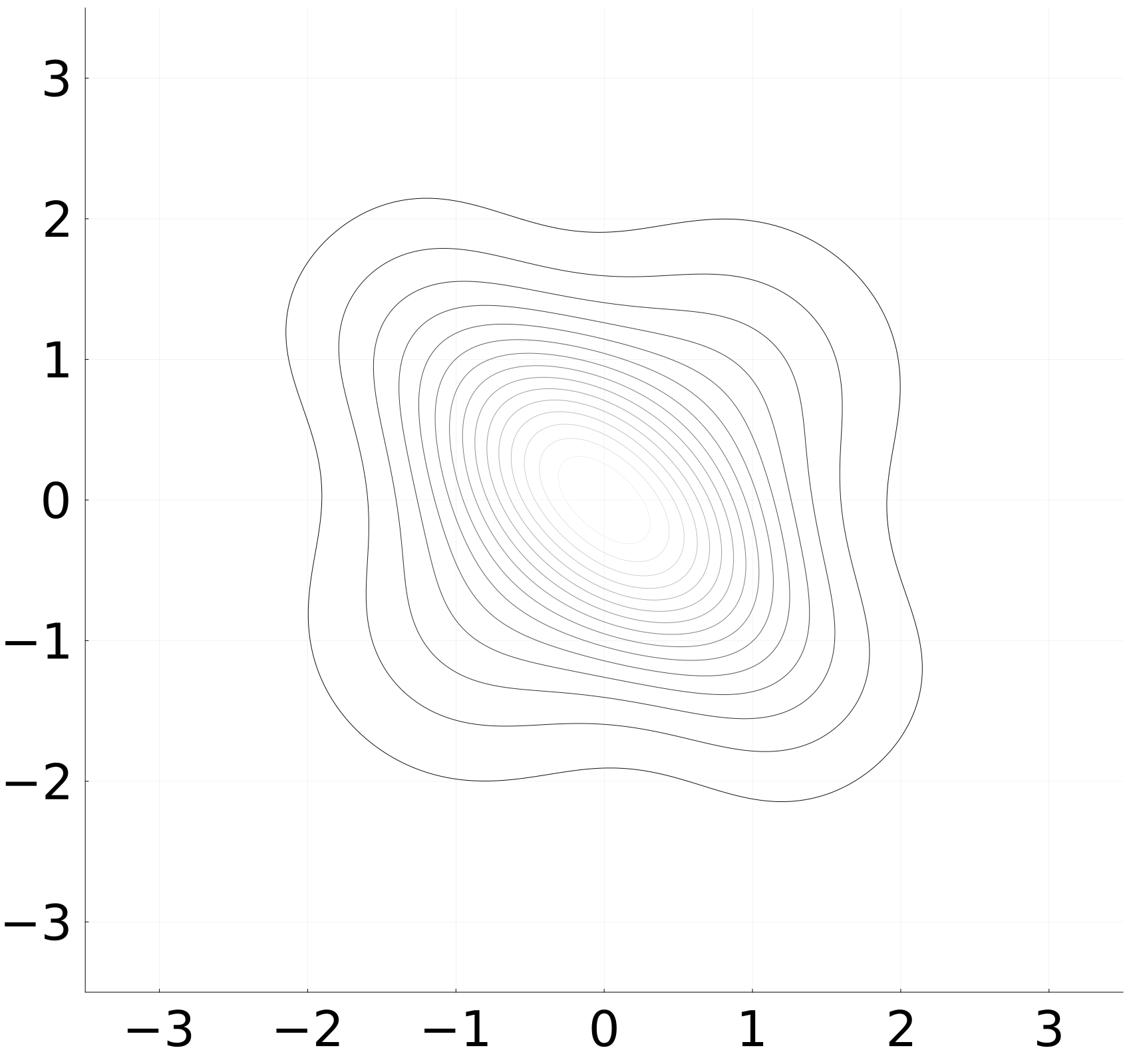}
 }
 \subfigure[$n = $ 4,500.] 
{
\includegraphics[width=3.5cm]{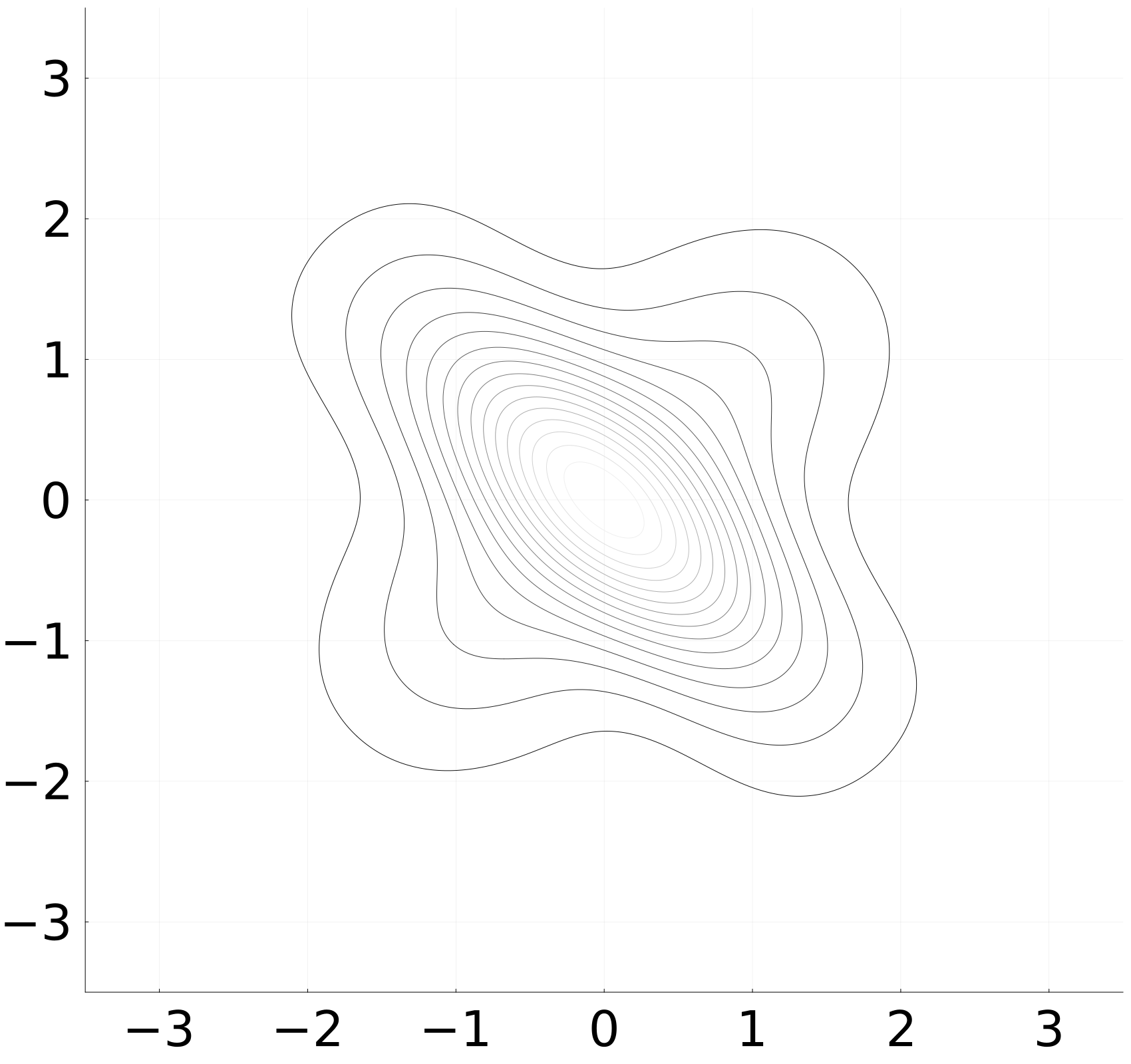}
 }      
\subfigure[$n = $ 18,000.] 
{
\includegraphics[width=3.5cm]{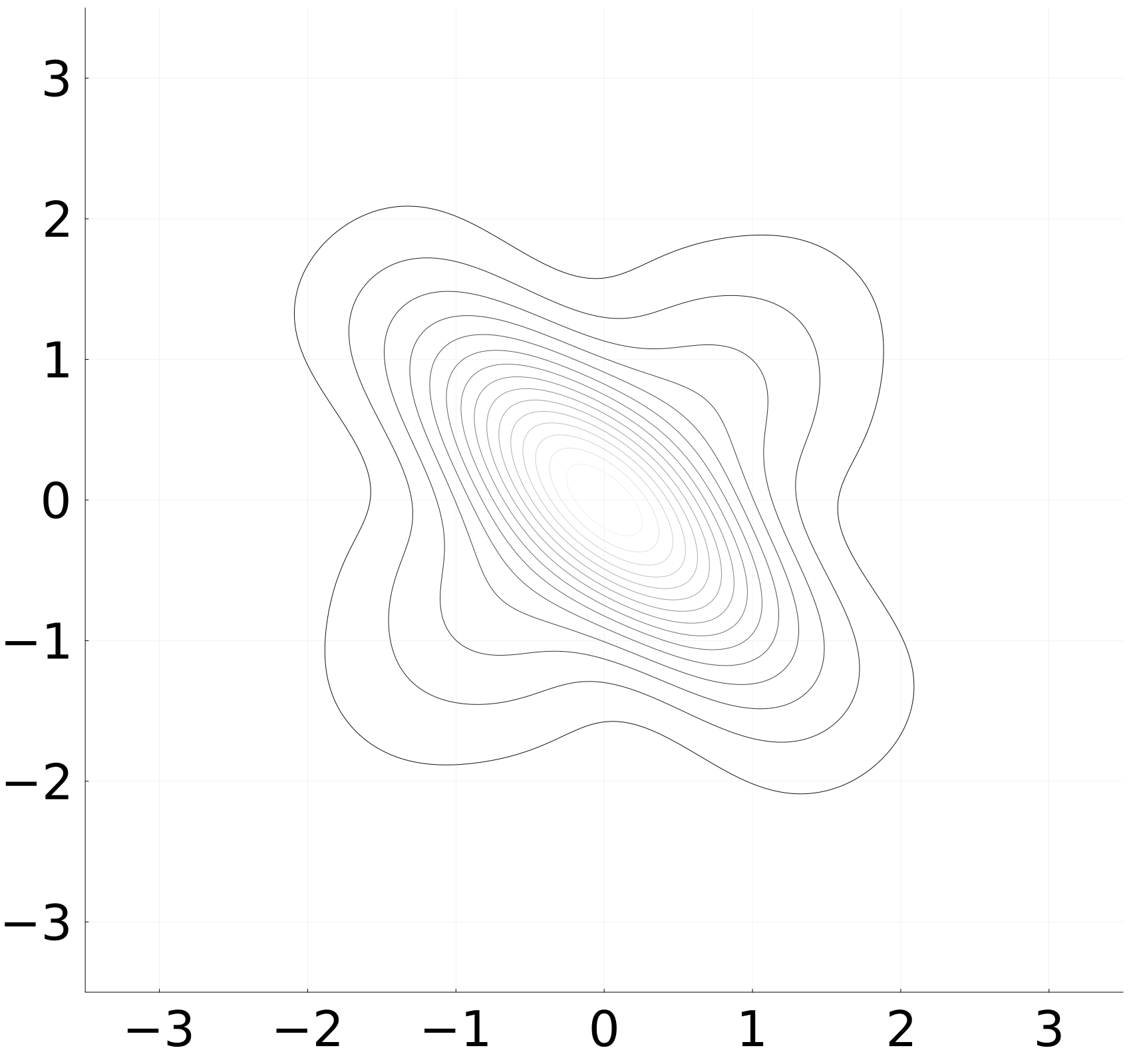}
 } 
 \subfigure[True.] 
{
\includegraphics[width=3.5cm]{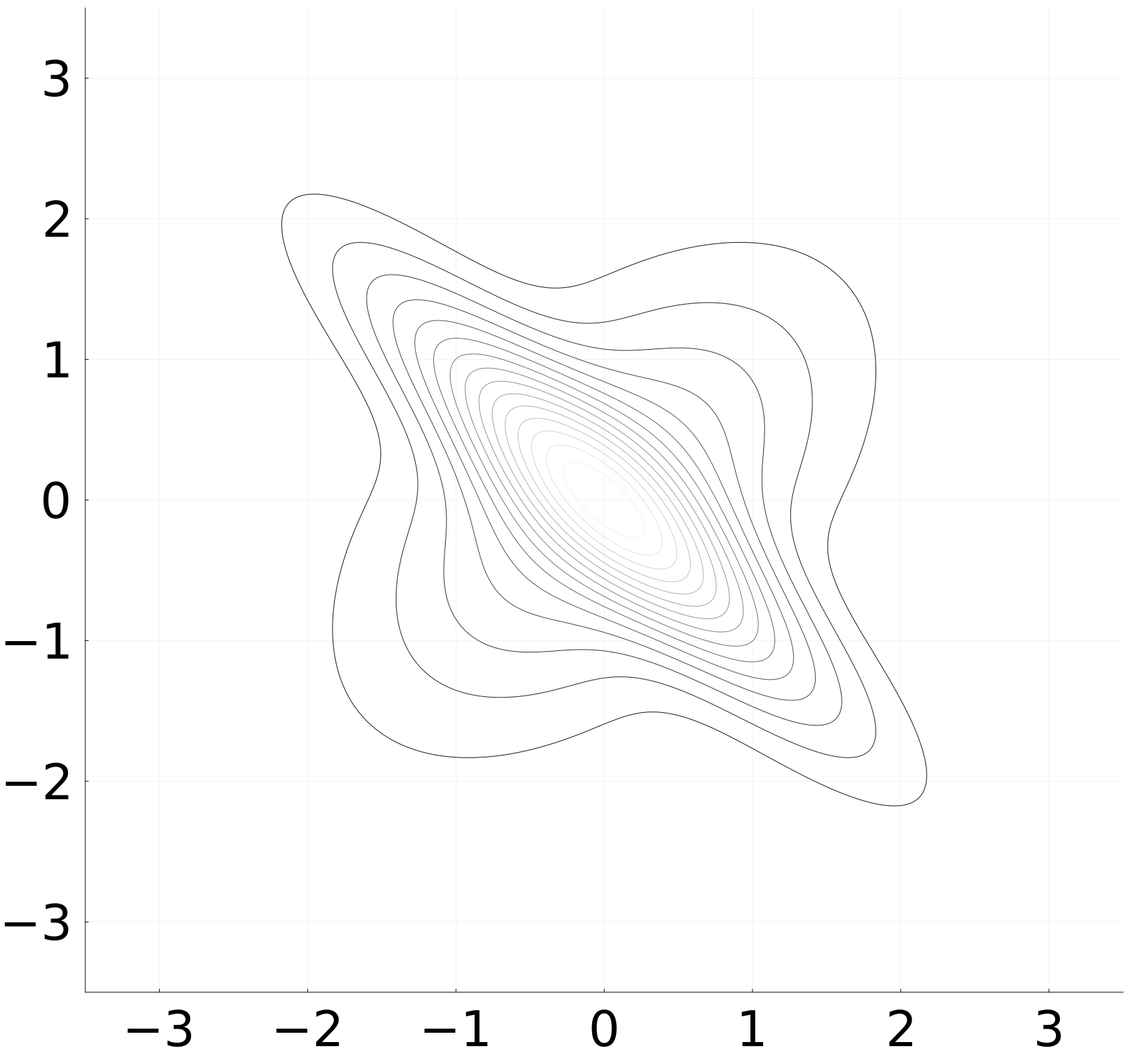}
 }      
 \\ 
\subfigure[$n = $ 600.] 
{
\includegraphics[width=3.5cm]{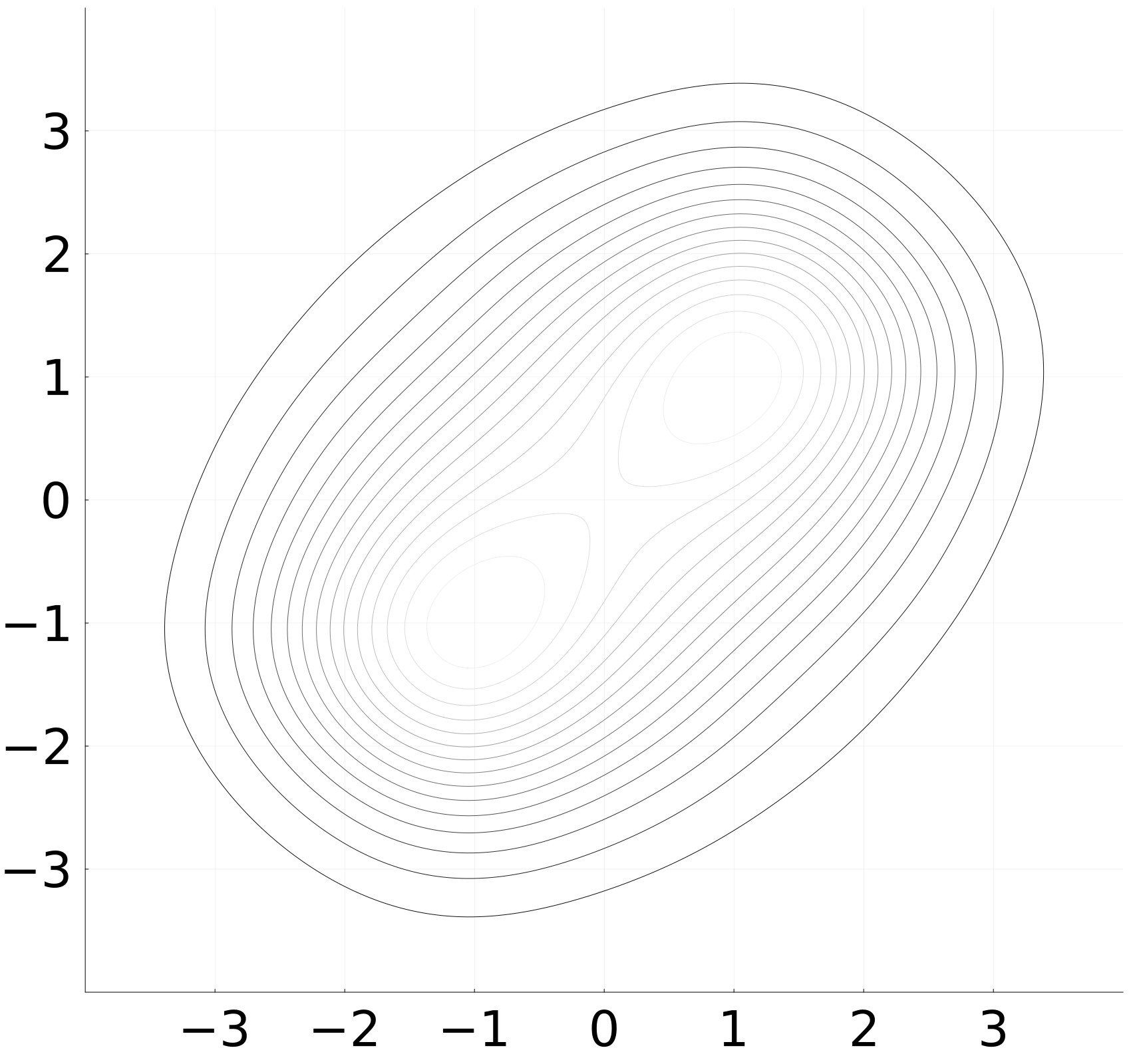}
 } 
\subfigure[$n = $ 4,500.] 
{
\includegraphics[width=3.5cm]{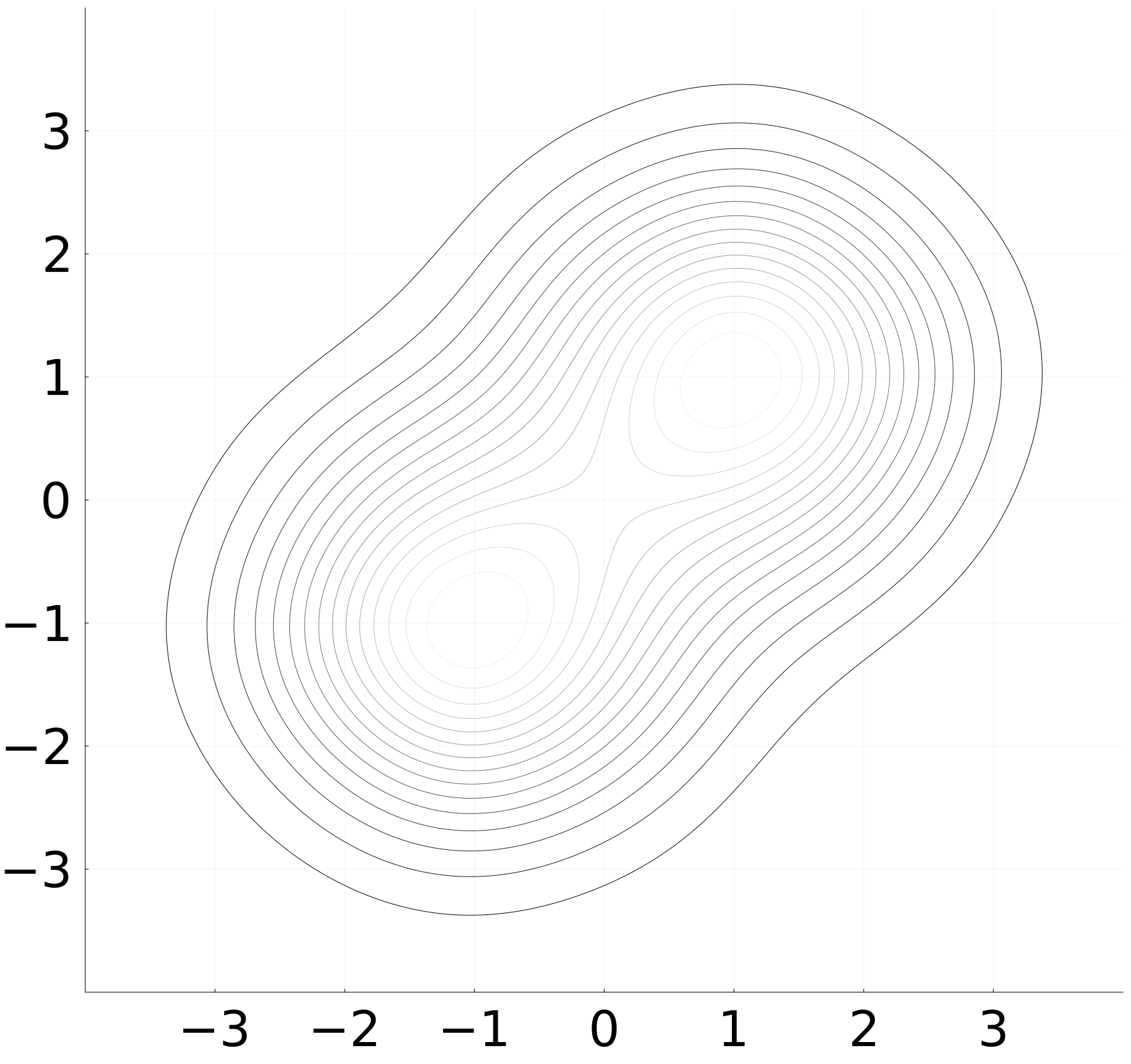}
 } 
\subfigure[$n = $ 18,000.] 
{
\includegraphics[width=3.5cm]{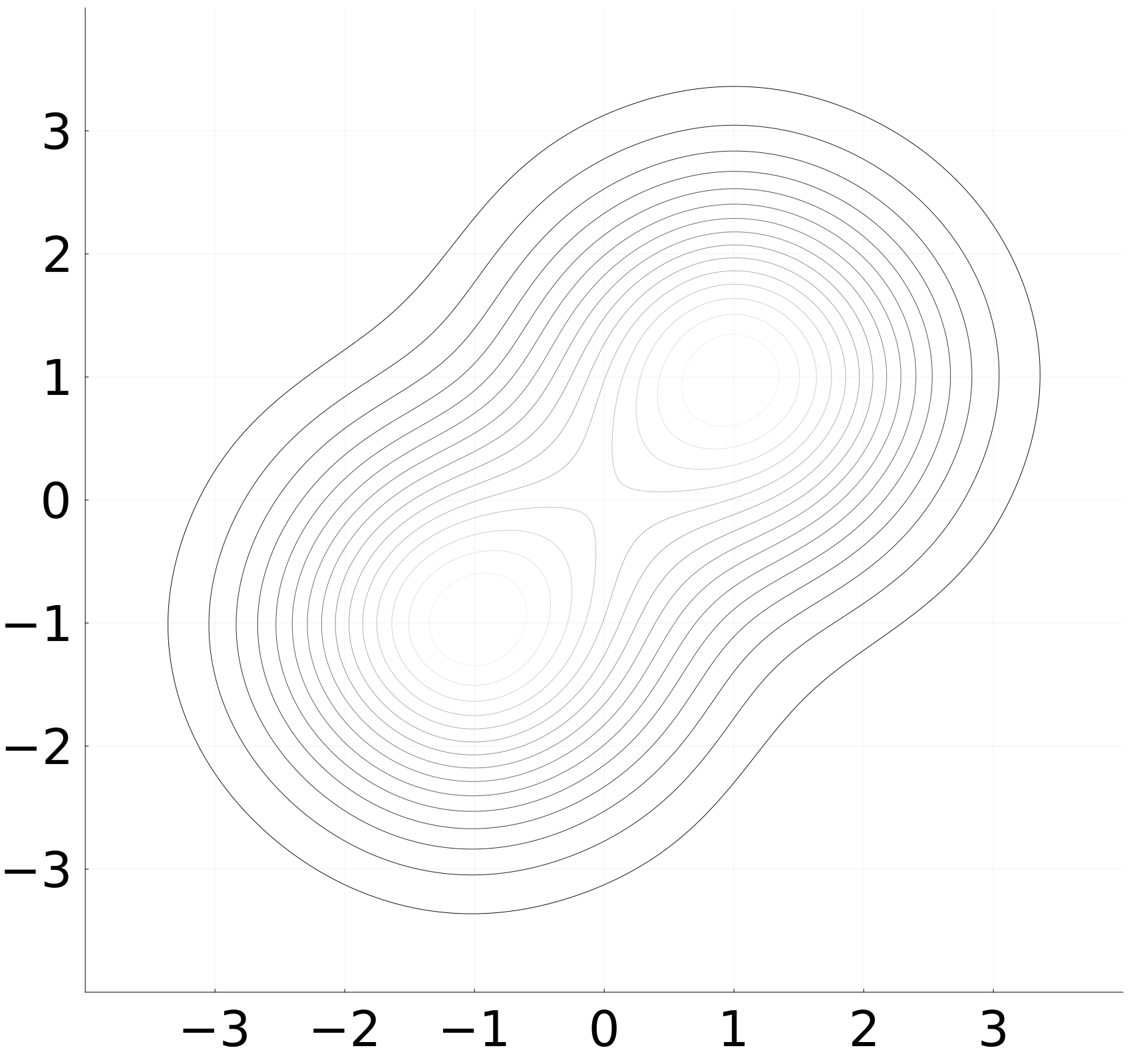}
 }
 \subfigure[True.] 
{
\includegraphics[width=3.5cm]{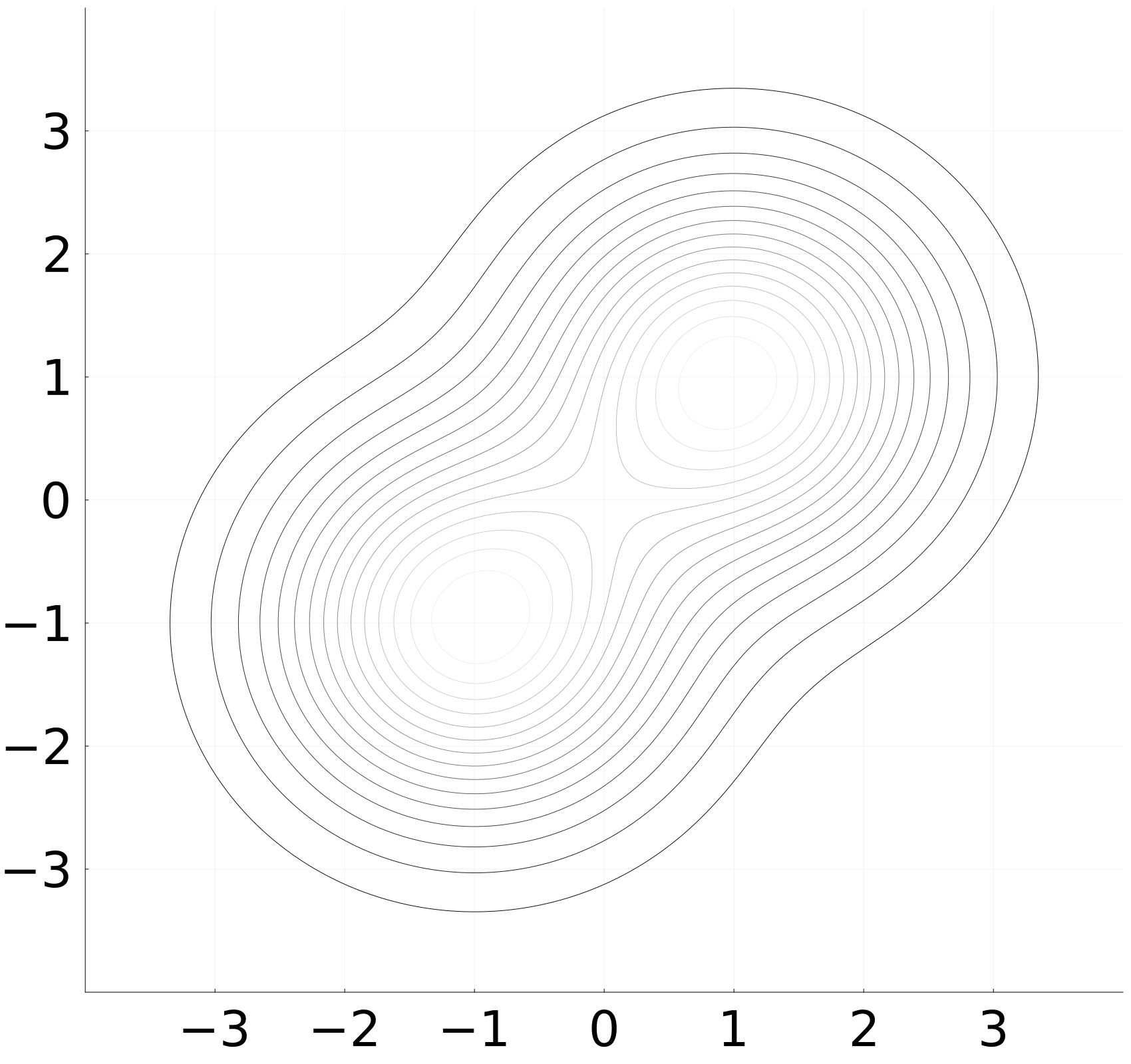}
 }      
 \caption{Simulated data  - known marginals - $k =$ 10: Posterior mean of the joint distribution under different models and sample sizes $n$. Panels (a) -- (c),
 show the results under Model 1.  Panels (e) -- (g),  show the results under Model 2. Panels (i) -- (k), show the results under Model 3. Panels (m) -- (o), show the results under Model 4. Panels (d), (h), (l), and (p) show the true joint models.}
\label{fig2:simulation2}
\end{figure}

\begin{figure}[!h]
\centering
\subfigure[$n = $ 600.] 
{
\includegraphics[width=3.5cm]{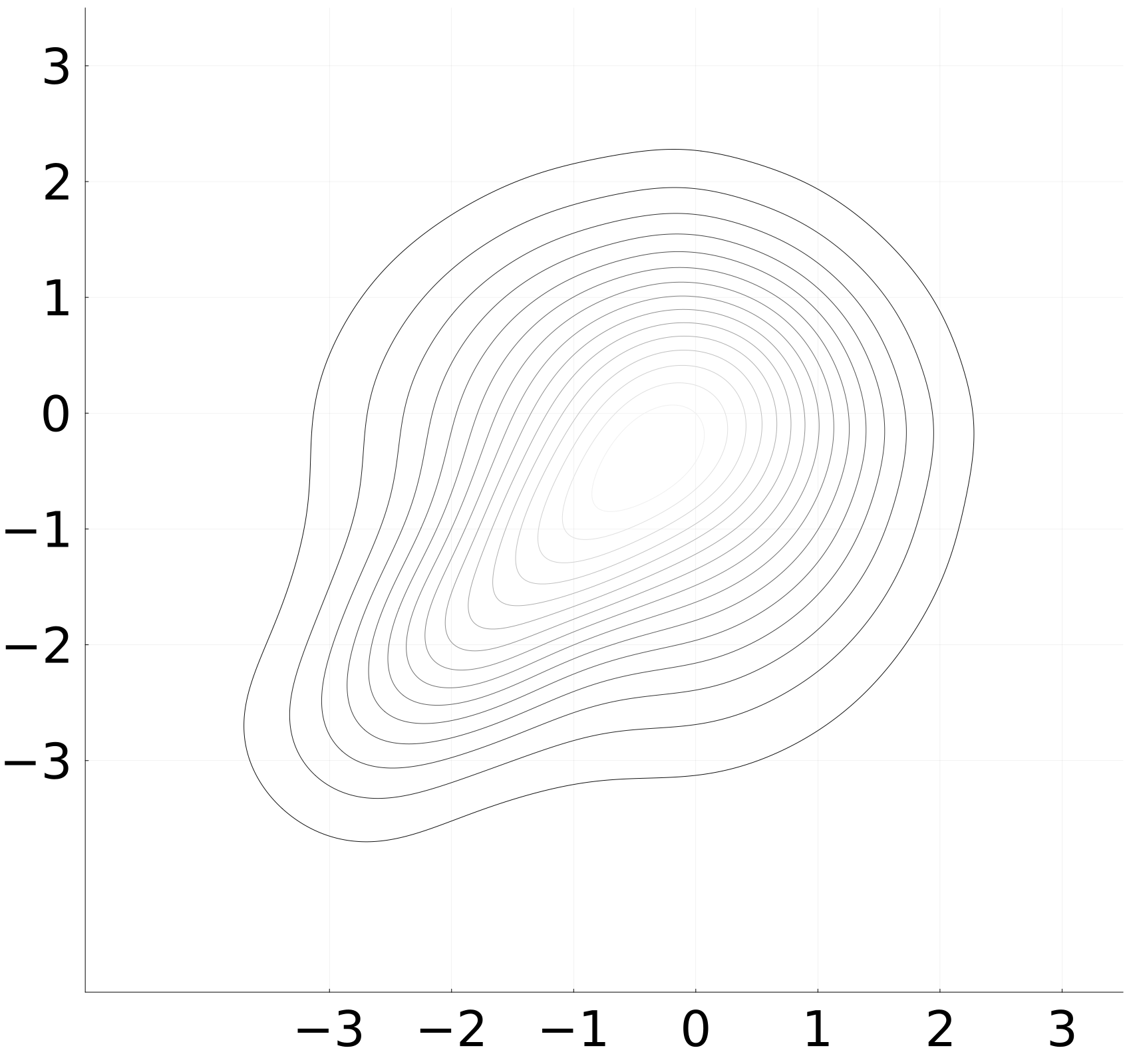}
 }      
 \subfigure[$n = $ 4,500.] 
{
\includegraphics[width=3.5cm]{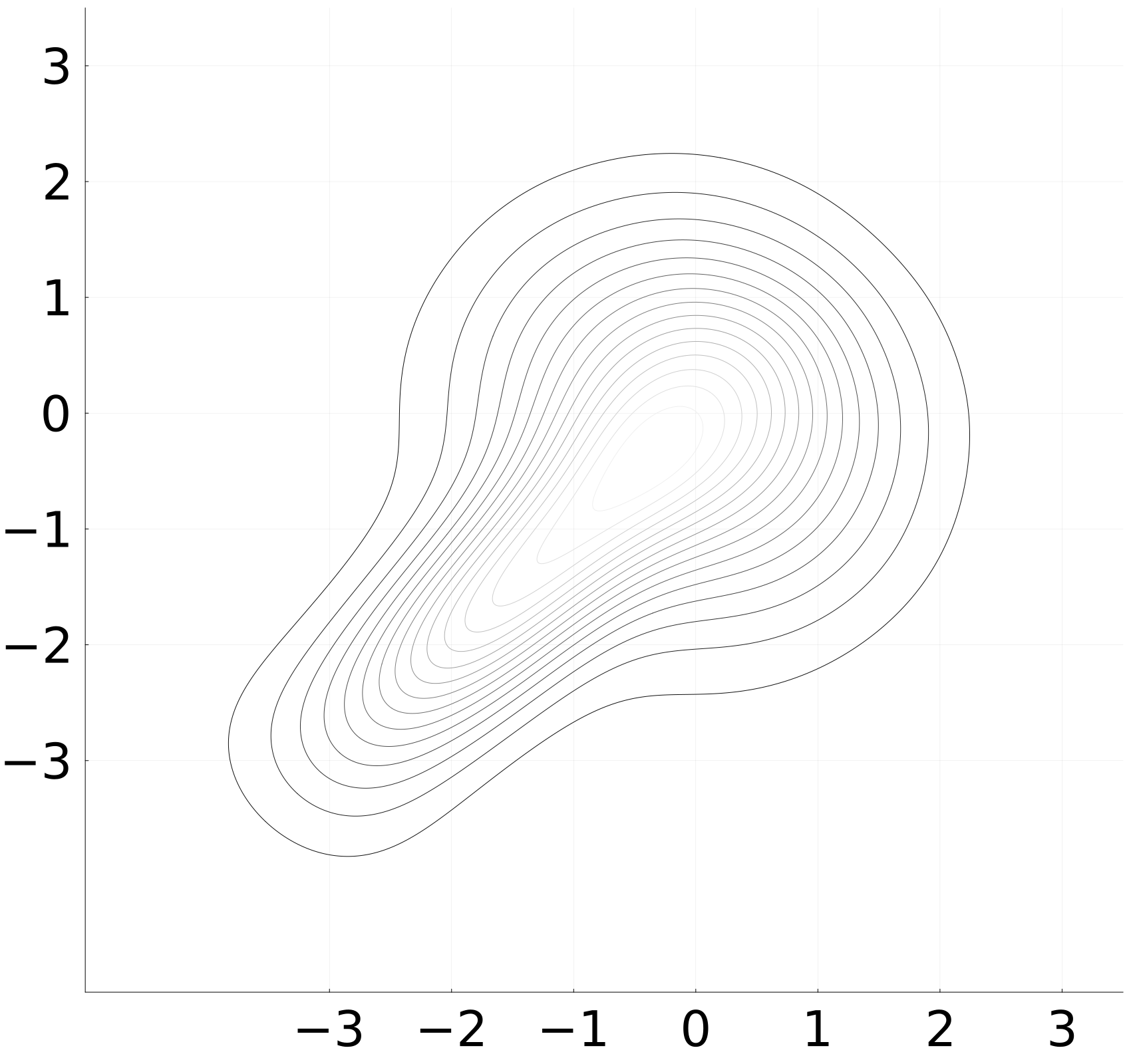}
 }      
\subfigure[$n = $ 18,000.] 
{
\includegraphics[width=3.5cm]{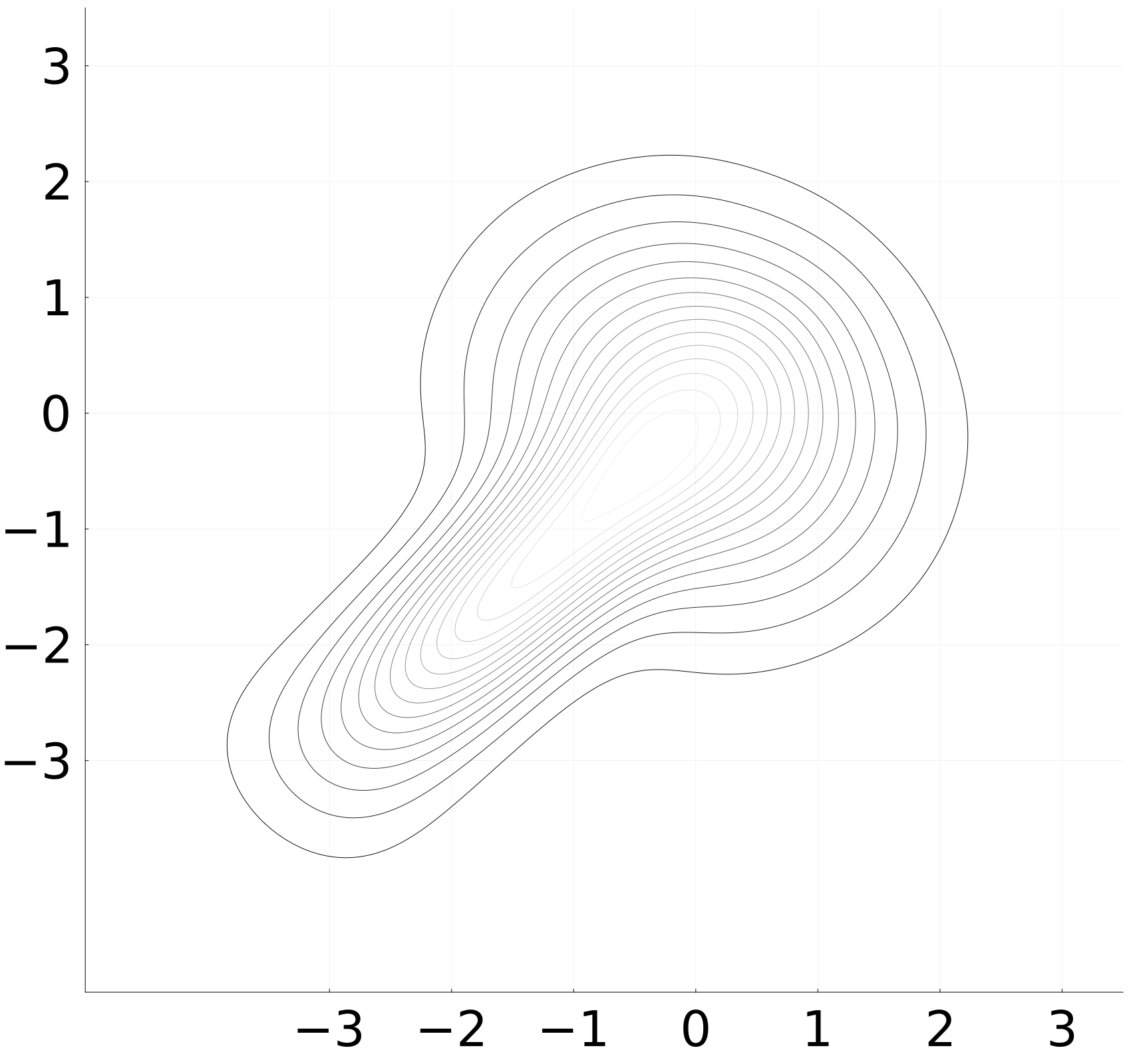}
 }
 \subfigure[True.] 
{
\includegraphics[width=3.5cm]{./figures/simulation2/modelopseudoclaytonverdad.png}
 }      
\\ 
\subfigure[$n = $ 600.] 
{
\includegraphics[width=3.5cm]{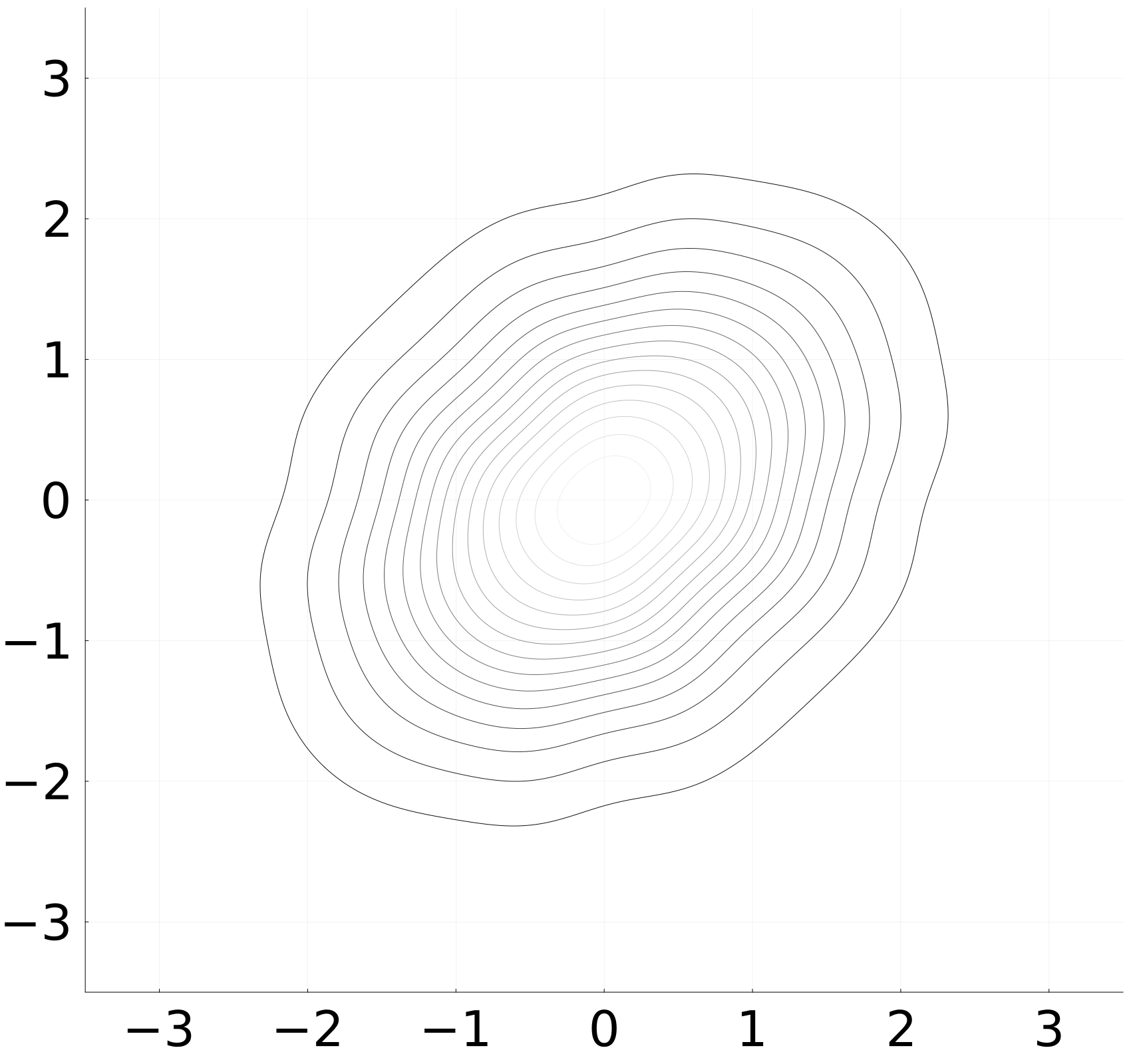}
 }
 \subfigure[$n = $ 4,500.] 
{
\includegraphics[width=3.5cm]{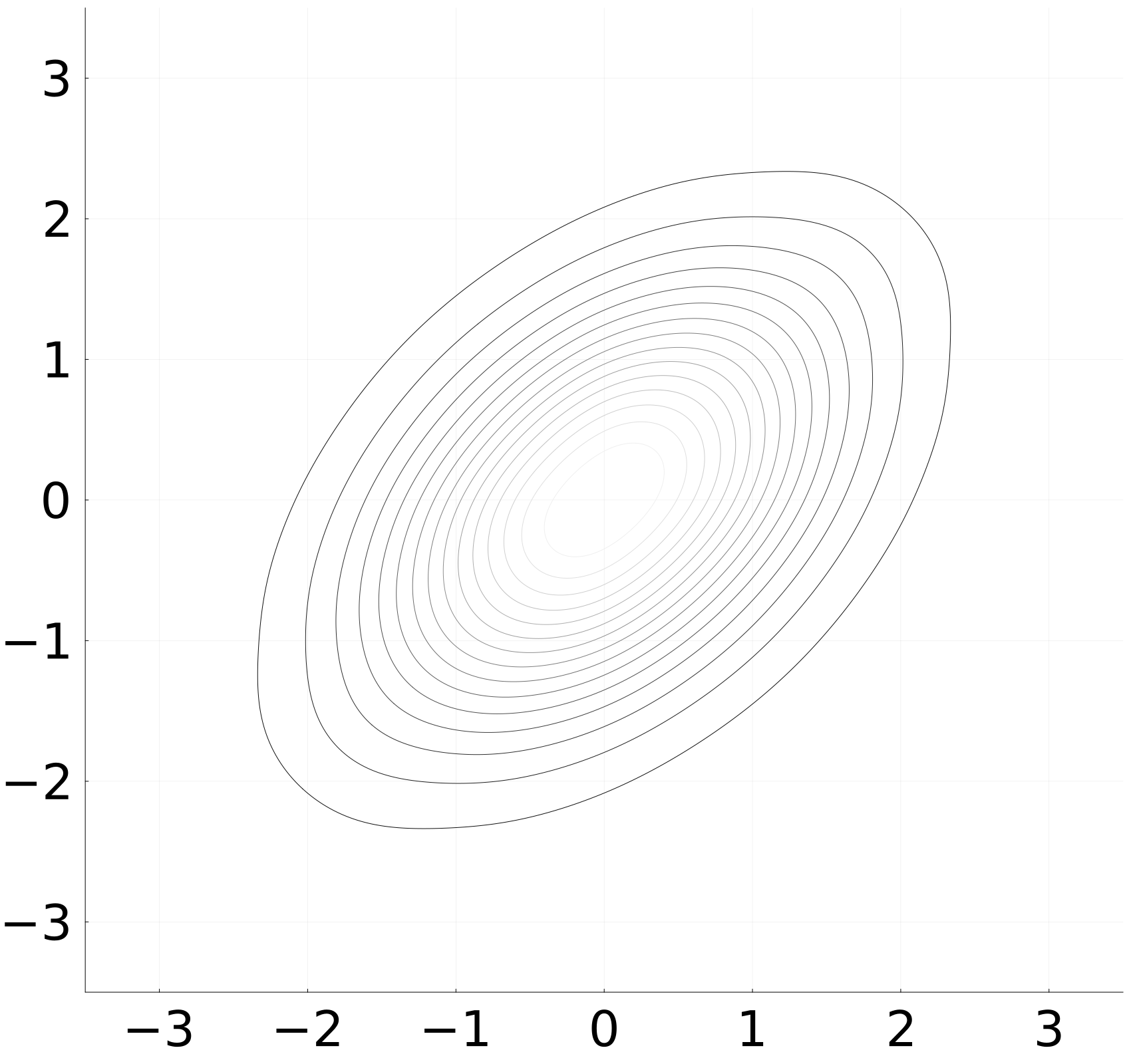}
 }      
\subfigure[$n = $ 18,000.] 
{
\includegraphics[width=3.5cm]{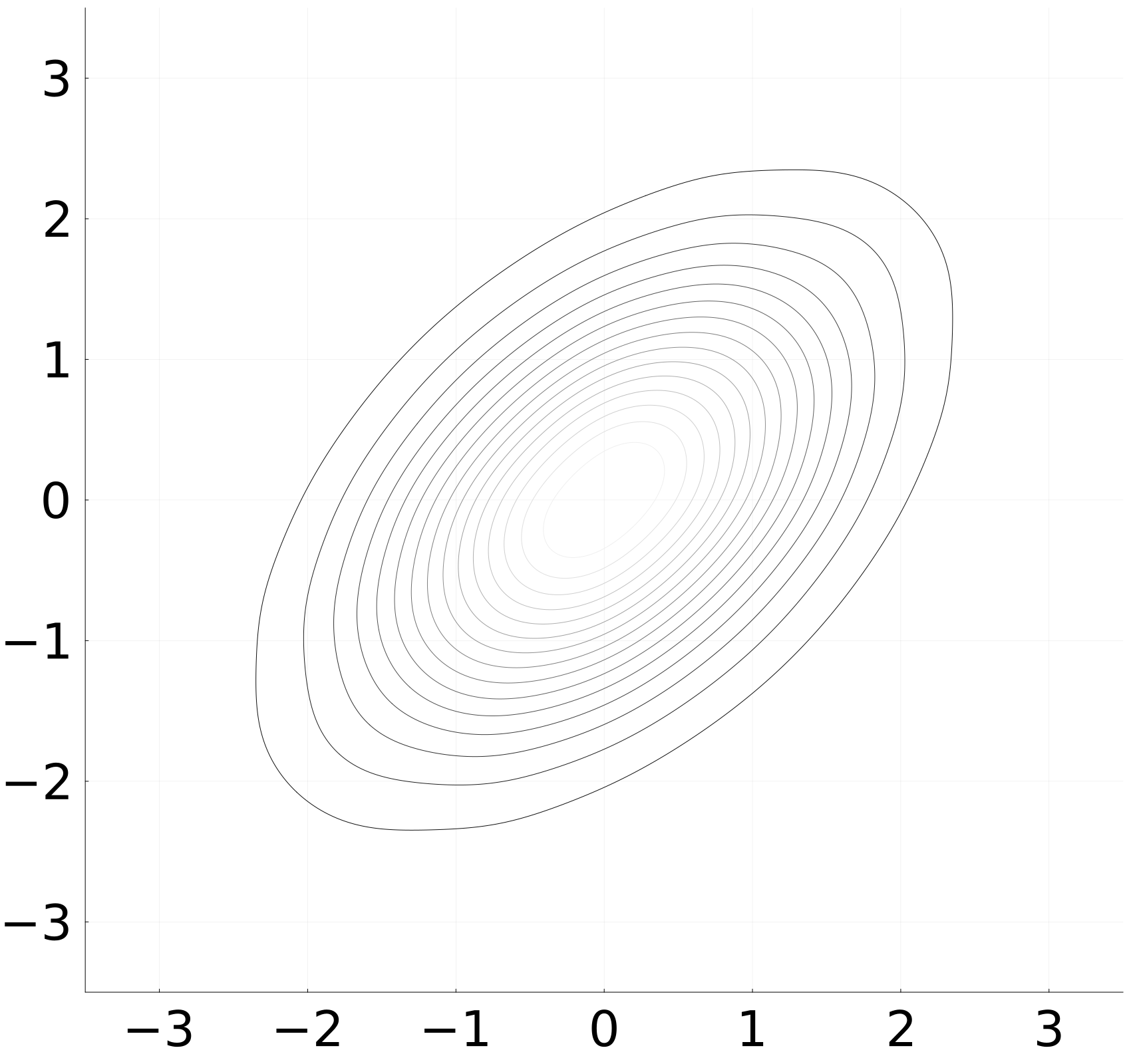}
 }
\subfigure[True.] 
{
\includegraphics[width=3.5cm]{./figures/simulation2/modelogausverdad.png}
 }      
\\ 
\subfigure[$n = $ 600.] 
{
\includegraphics[width=3.5cm]{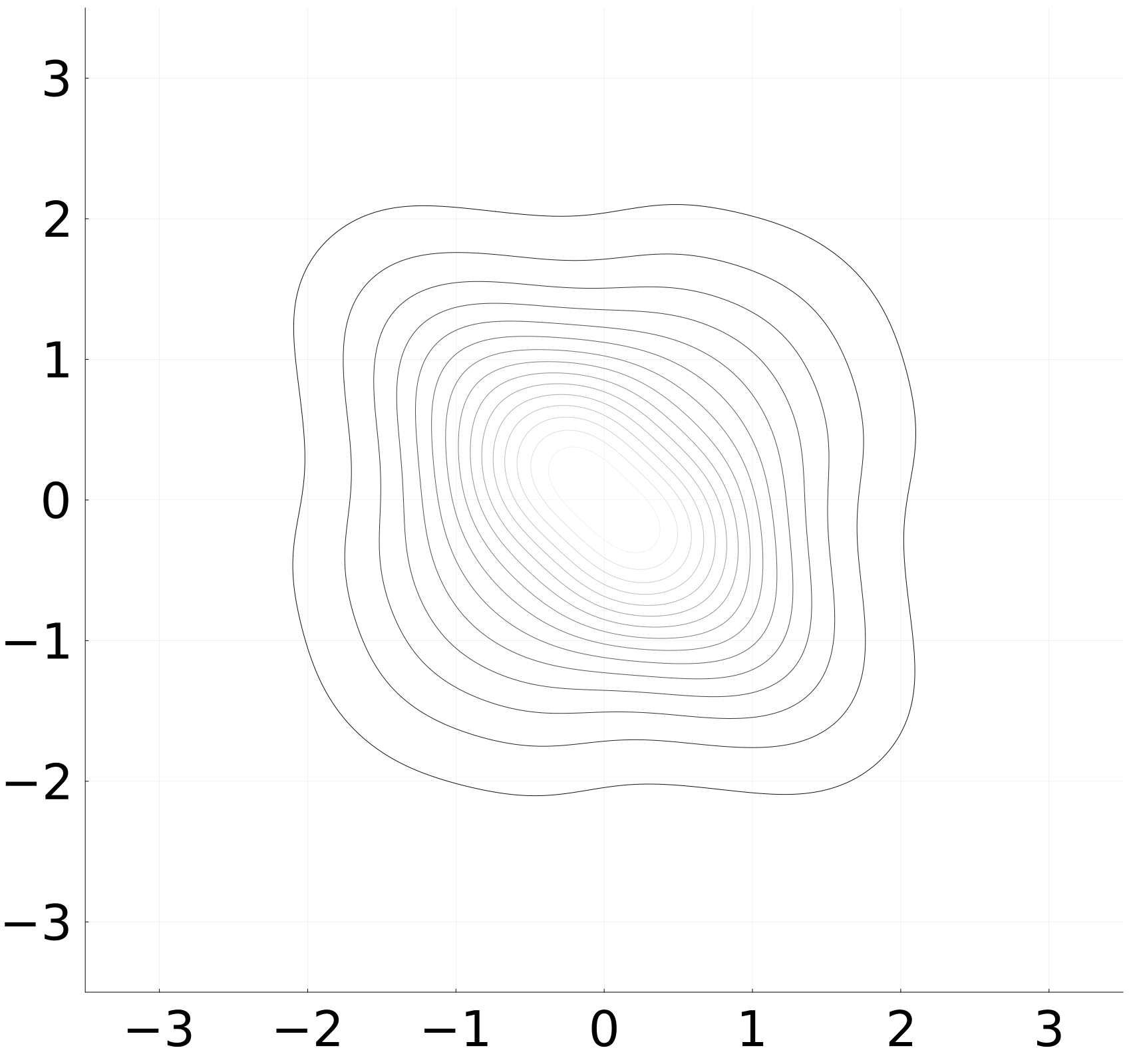}
 }
 \subfigure[$n = $ 4,500.] 
{
\includegraphics[width=3.5cm]{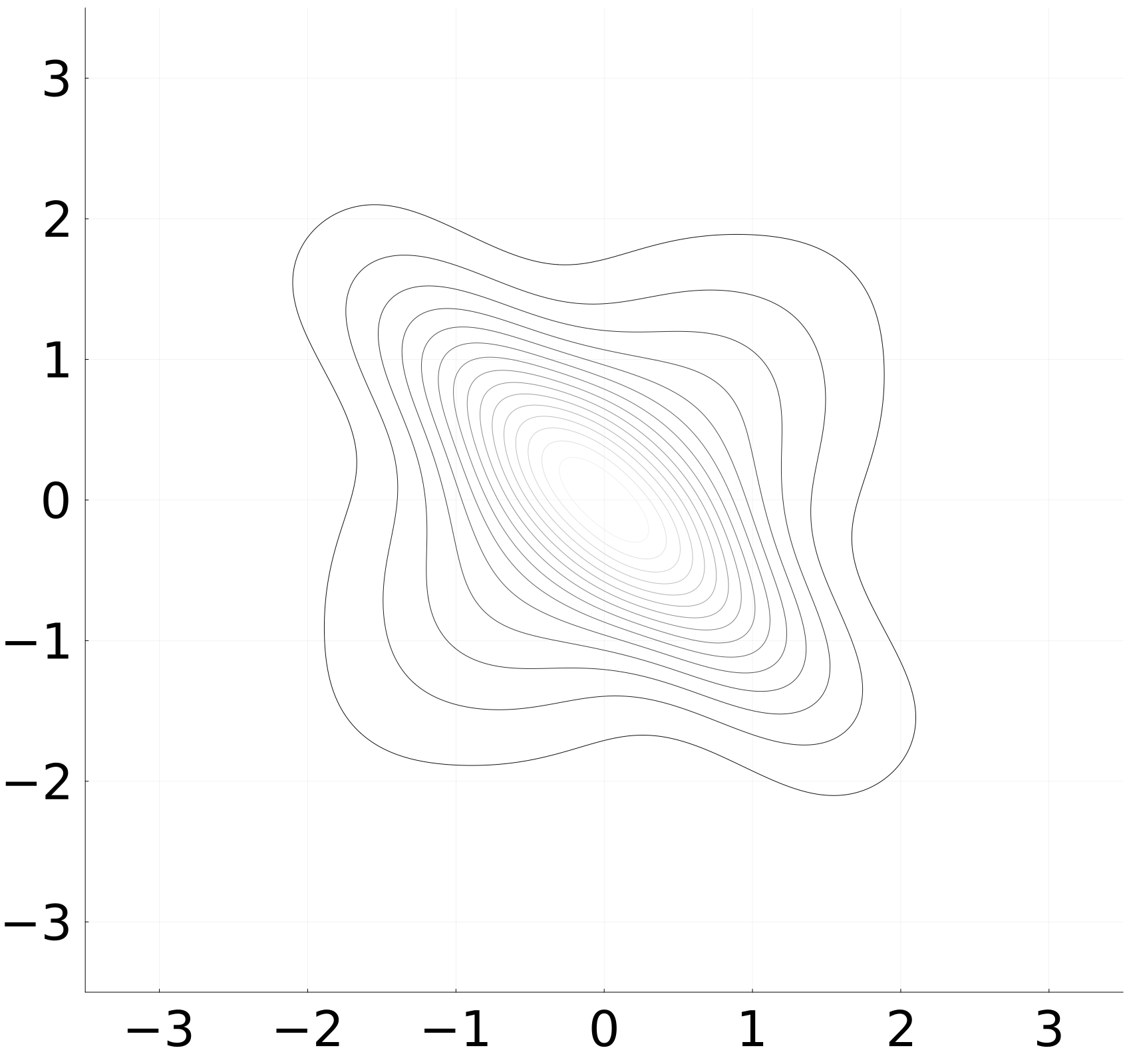}
 }      
\subfigure[$n = $ 18,000.] 
{
\includegraphics[width=3.5cm]{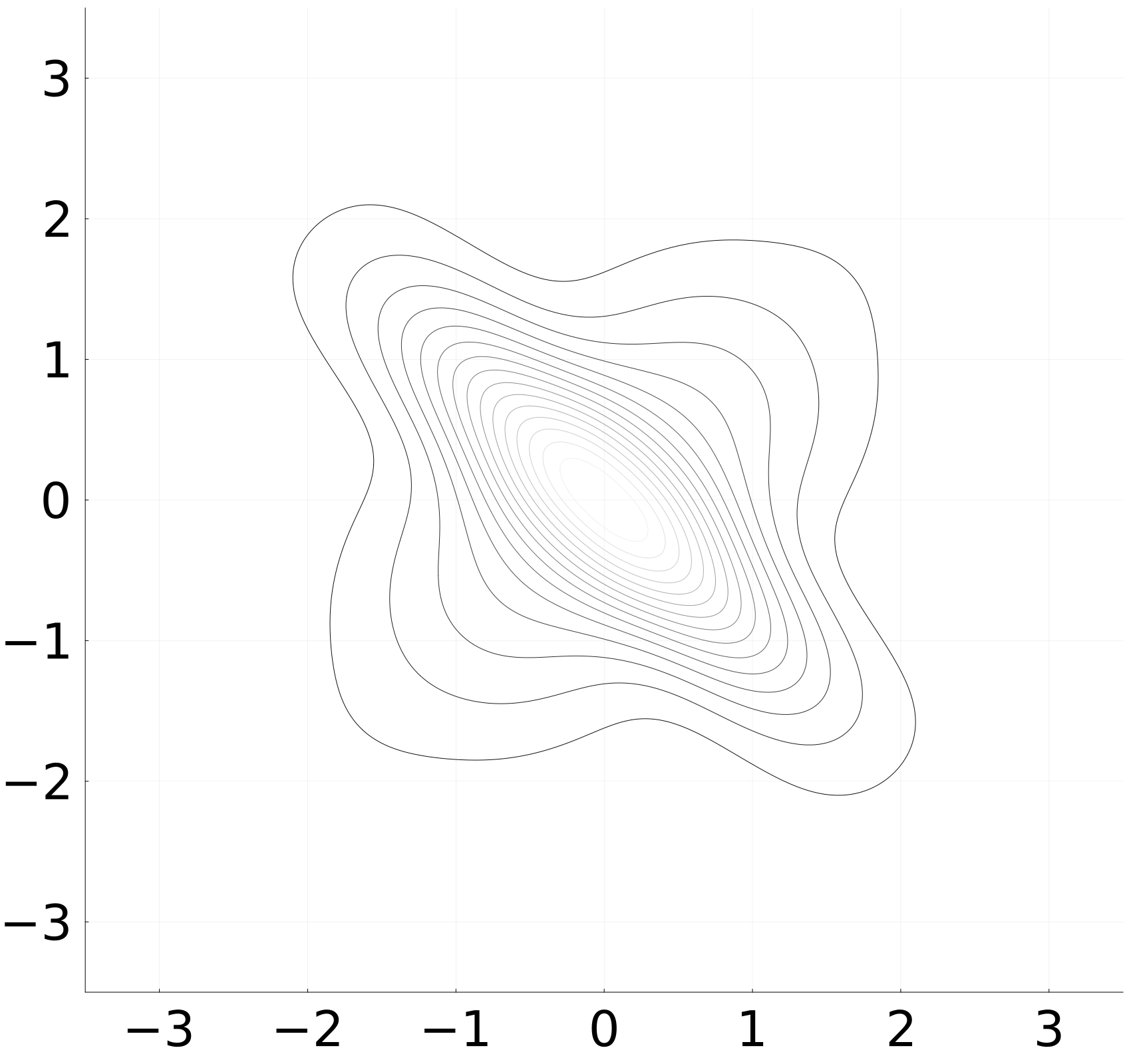}
 }
  \subfigure[True.] 
{
\includegraphics[width=3.5cm]{./figures/simulation2/modelomezclagausverdad.png}
 }        
 \\ 
\subfigure[$n = $ 600.] 
{
\includegraphics[width=3.5cm]{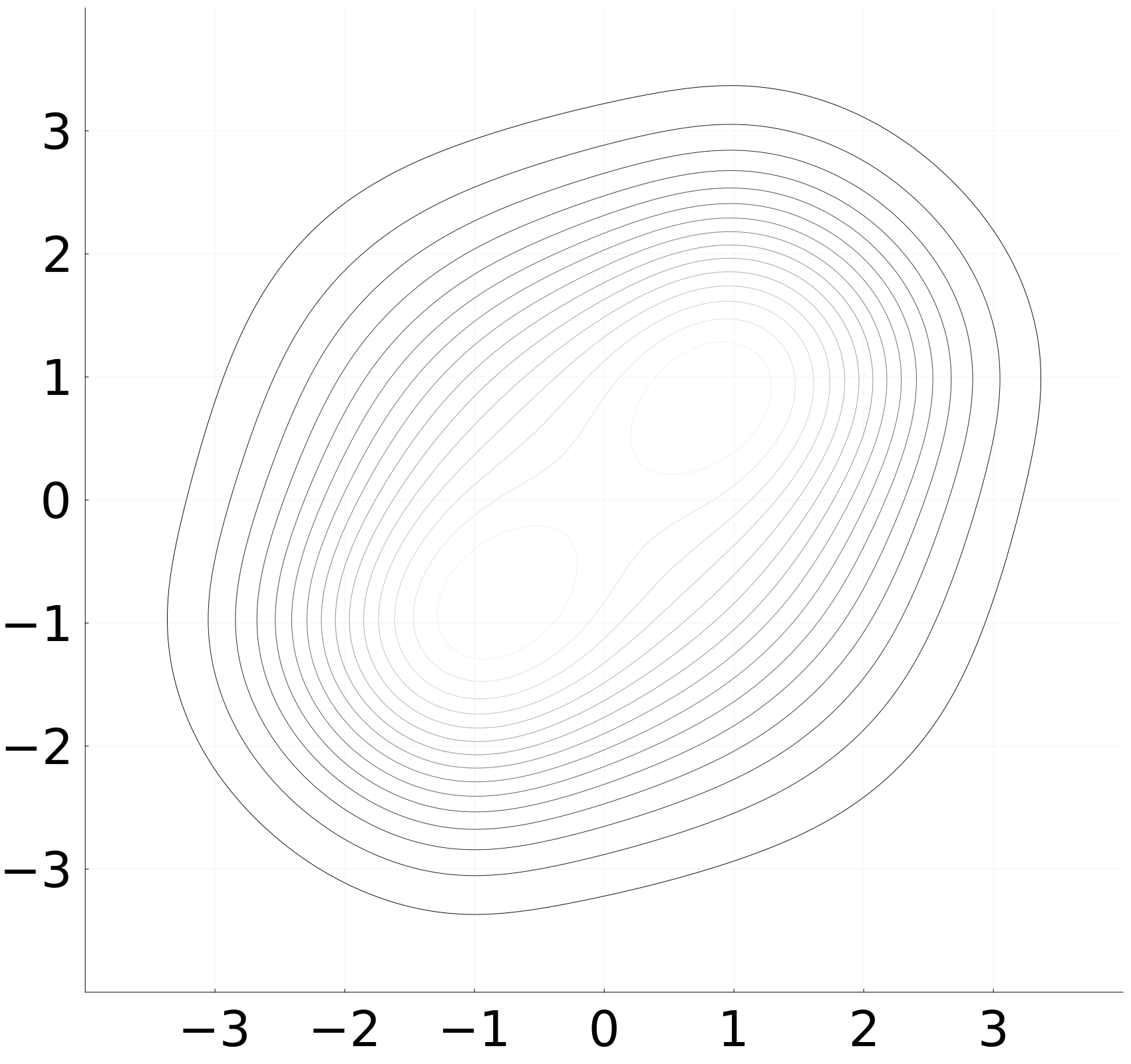}
 } 
\subfigure[$n = $ 4,500.] 
{
\includegraphics[width=3.5cm]{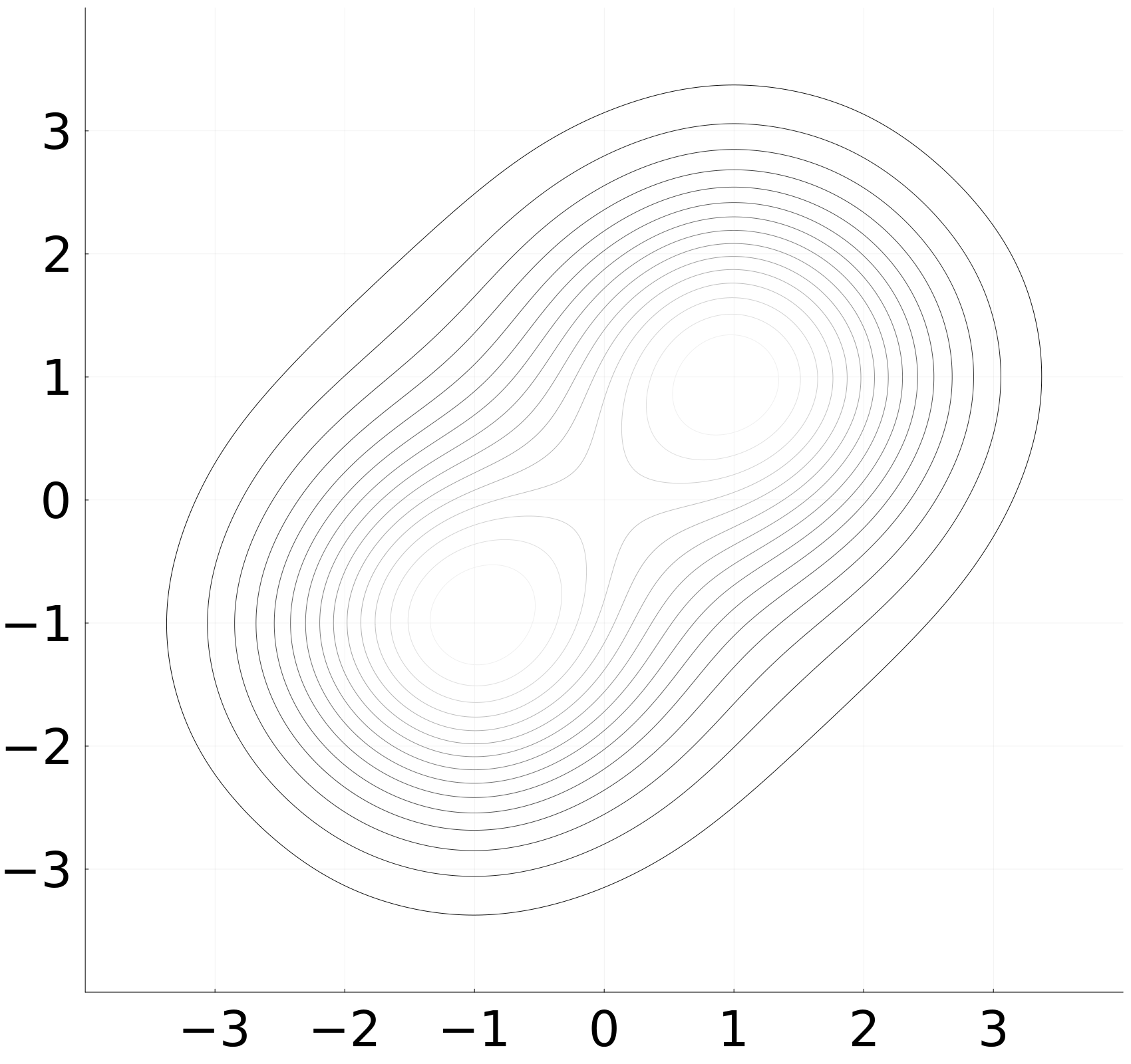}
 } 
\subfigure[$n = $ 18,000.] 
{
\includegraphics[width=3.5cm]{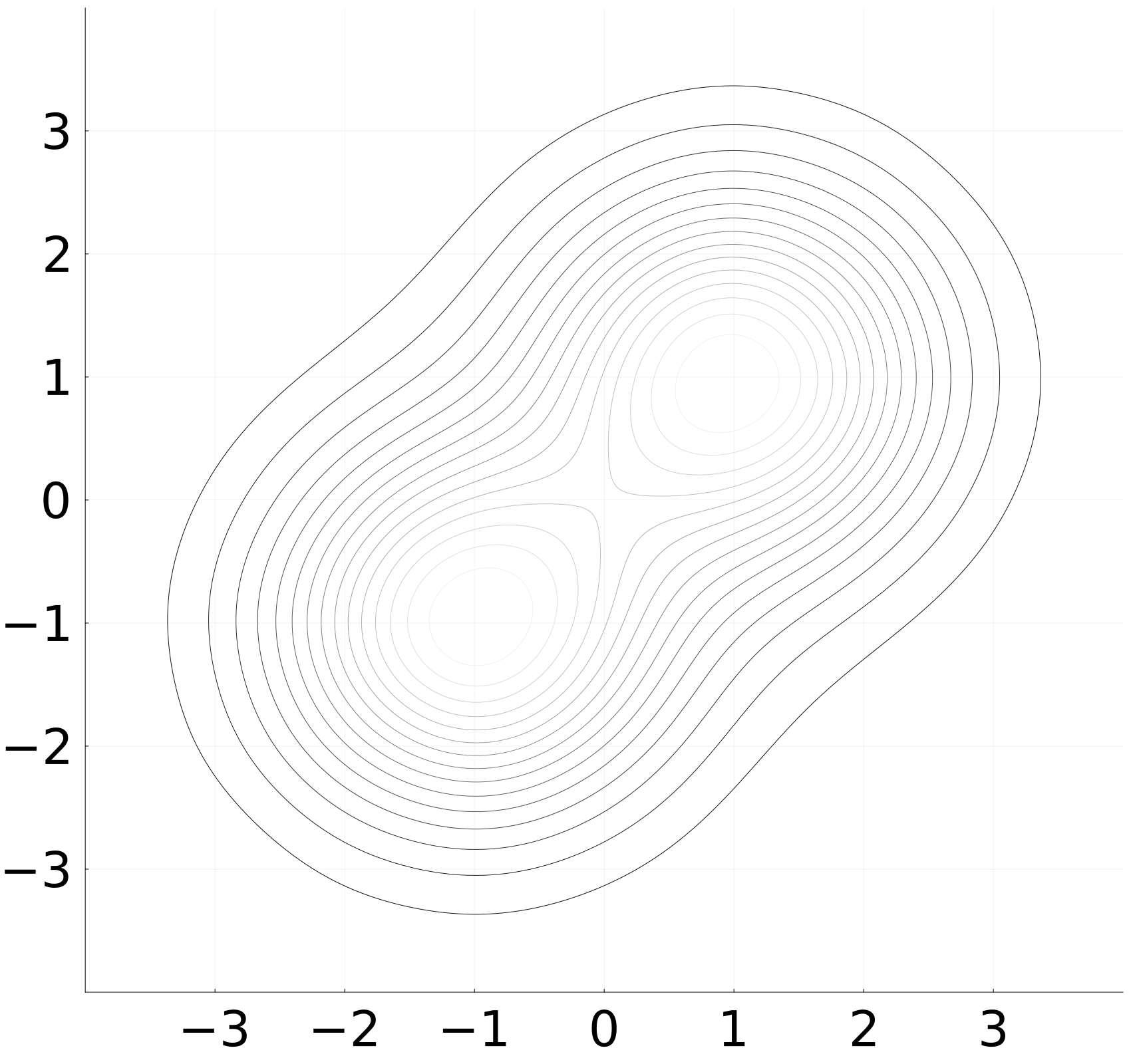}
 }
 \subfigure[True.] 
{
\includegraphics[width=3.5cm]{./figures/simulation2/modelobimodalverdad.png}
 }      
 \caption{Simulated data  - known marginals - $k =$ 20: Posterior mean of the joint distribution under different models and sample sizes $n$. Panels (a) -- (c),
 show the results under Model 1.  Panels (e) -- (g),  show the results under Model 2. Panels (i) -- (k), show the results under Model 3. Panels (m) -- (o), show the results under Model 4. Panels (d), (h), (l), and (p) show the true joint models.}
\label{fig2:simulation3}
\end{figure}

Figure~\ref{fig1:simulation2} summarizes the posterior distribution of the Hellinger distance to the true copula model under the different true models, degrees of the BYP prior, and sample sizes. Figures~\ref{fig2:simulation2} and~\ref{fig2:simulation3} show the posterior mean of the joint distribution under the different true models for $k =$ 10 and $k =$ 20, resectively.
\clearpage
The results show that adequate estimates for complex true models can be obtained, even for reduced sample sizes, and that when the copula model is simple, the proposed model does not overfit the data. The results also show that the posterior mean gets closer to the true model when the sample size increases and that the posterior distribution's mass concentrates around the true model as the sample size increases. 
\clearpage

\clearpage

\subsection{Estimation of a four dimensional copula with unknown marginals}

We illustrate the model under a more difficult set of circumstances. A random sample was generated from a four dimensional distribution with a Gaussian copula and correlation matrix reflecting different degrees of association, given by
{\small $$
\left(\begin{array}{cccc}
    1.0 & 0.4 & 0.6 & 0.7 \\
    0.4 & 1.0 & 0.7 & 0.3 \\
    0.6 & 0.7 & 1.0 & 0.2 \\
    0.7 & 0.3 & 0.2 & 1.0 \\
\end{array}\right).
$$}
We considered different supports and types for the marginal distributions. Specifically, for coordinate one, we considered mixture of Gaussian distributions given by $0.5\times N(0,1) + 0.5 \times N(3,1)$. For the second coordinate we considered a log-normal distribution with location and scale parameters set to 0 and 1, respectively. For the third coordinate we considered Gamma distribution with shape and scale parameter set to 5 and 9, respectively. Finally, for the fourth coordinate, we considered a Beta distribution with parameters 2 and 4. 

We considered three sample sizes: $n =$ 1,000, 5,000, and 10,000. For each simulated dataset, inference was performed under the ICAR prior and degree $k$ = 5. Posterior inference was performed using the  $t$-walk algorithm described previously, assuming the true marginal distributions with unknown parameters, and flat priors on them. We considered a burn-in period of 10,000 and the chain was run until it produced an effective posterior sample of length 1,000.  

\begin{figure}[!h]
\begin{centering}
\includegraphics[width=7cm]{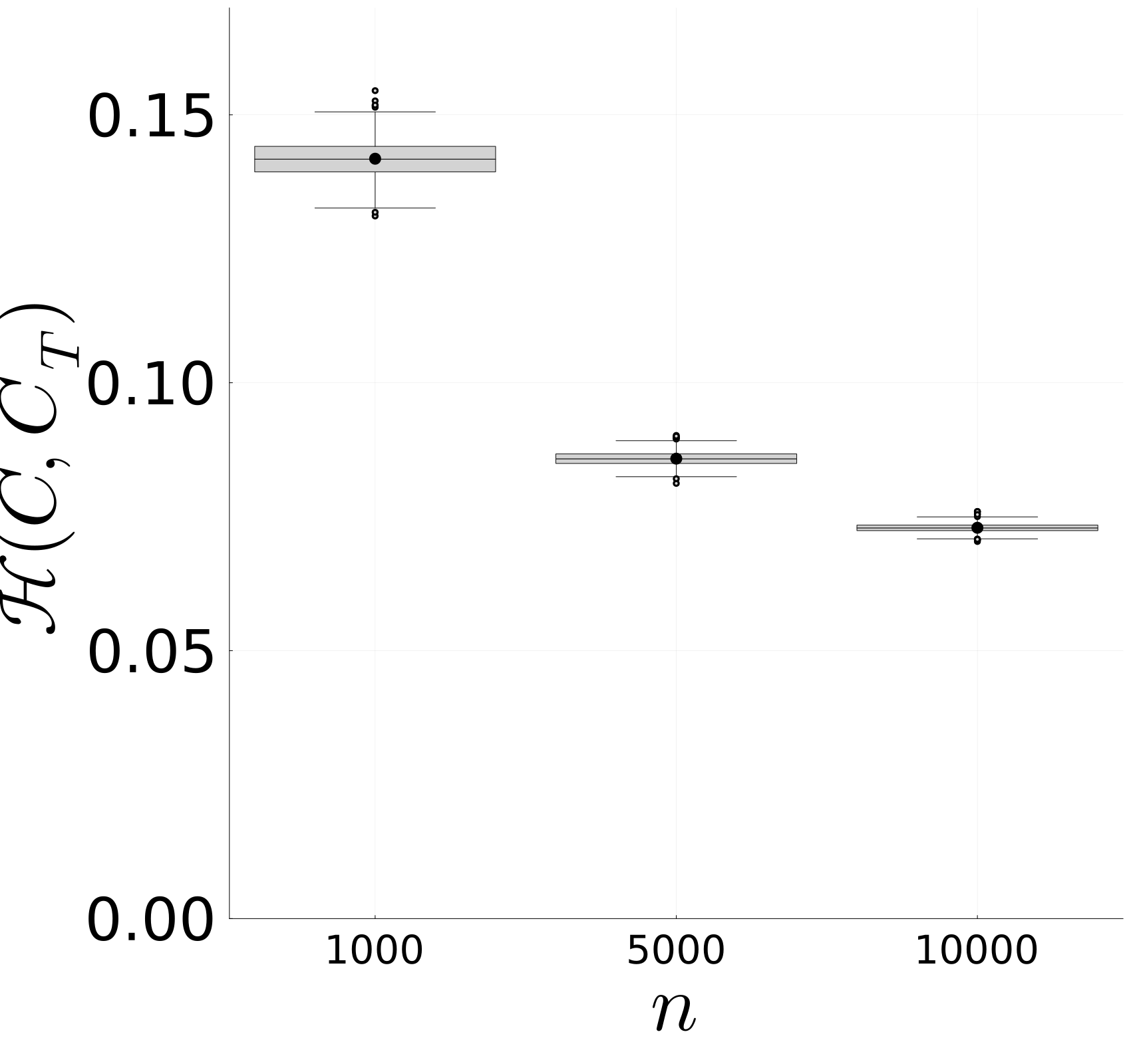}
\caption{Simulated data - unknown marginals: Box and whisker plot summarizing the posterior distribution of the Hellinger distance, $\mathcal{H}$, to the true copula model $C_T$, under different sample sizes for the four-dimensional joint model.}
       \label{fig1:simulation3}
    \end{centering}
\end{figure}

Figure  \ref{fig1:simulation3} shows summary statistics of the posterior distribution of the Hellinger distance to the true copula function for  the different sample sizes. As expected, we observe that posterior distribution concentrate the mass around the true copula function as the sample size increases. 

\begin{figure}
    \begin{centering}
     \subfigure[$n =$ 1,000]{%
            \includegraphics[scale=0.14]{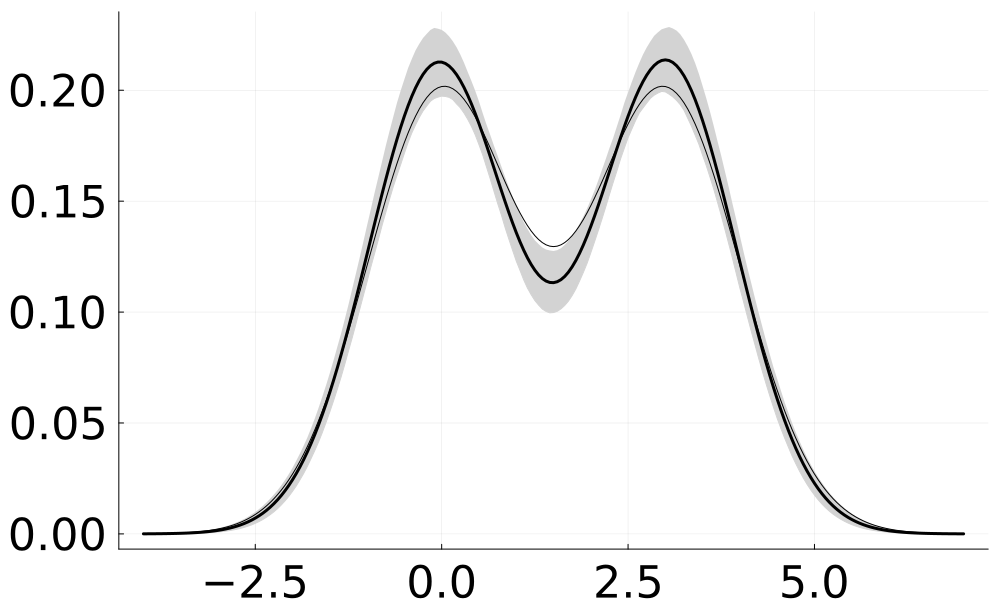} 
     }
     \subfigure[$n =$ 5,000]{%
            \includegraphics[scale=0.14]{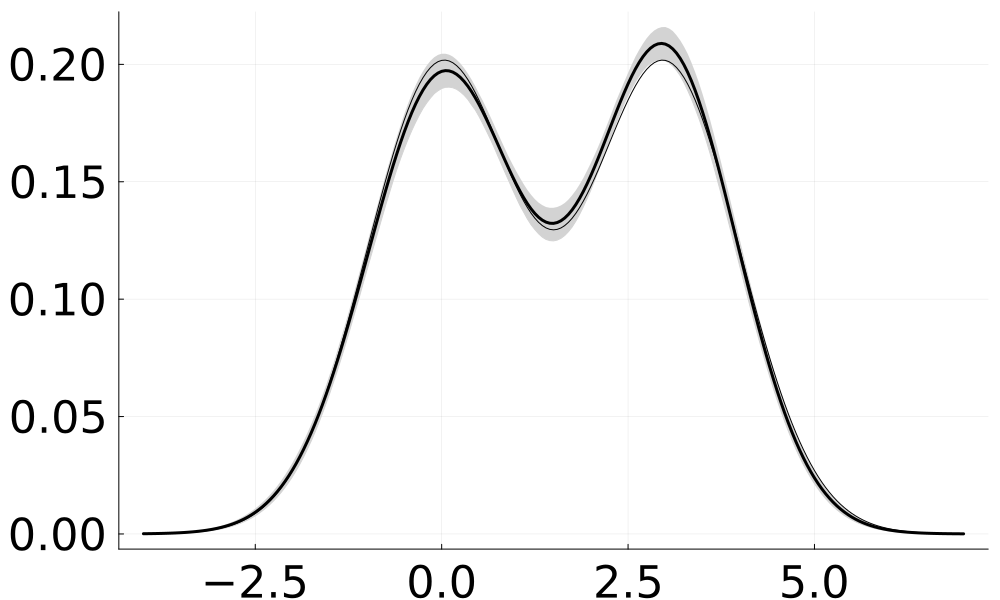} 
     }
     \subfigure[$n =$ 10,000]{%
            \includegraphics[scale=0.14]{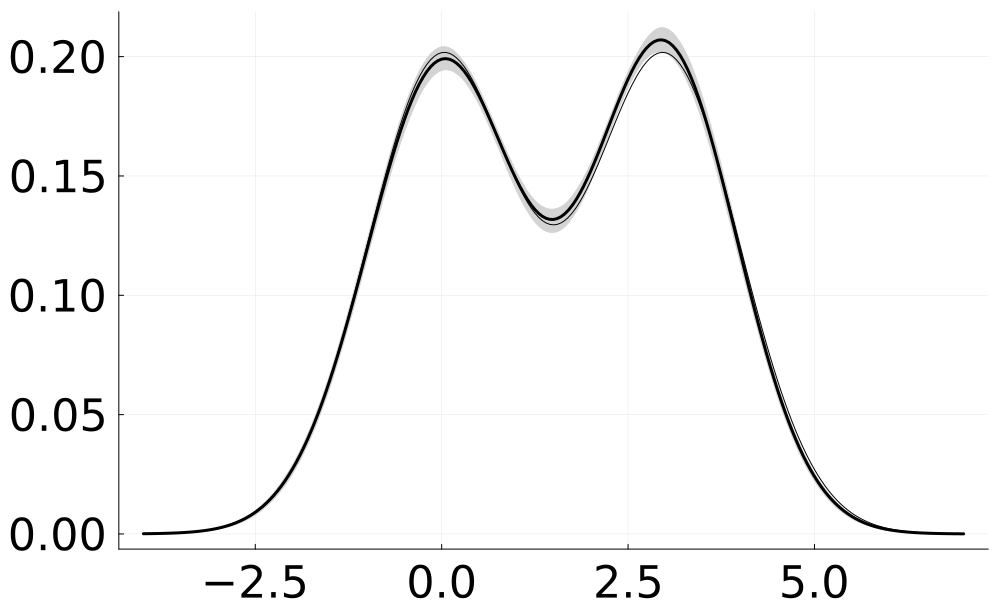} 
     }\\
     \subfigure[$n =$ 1,000]{%
            \includegraphics[scale=0.14]{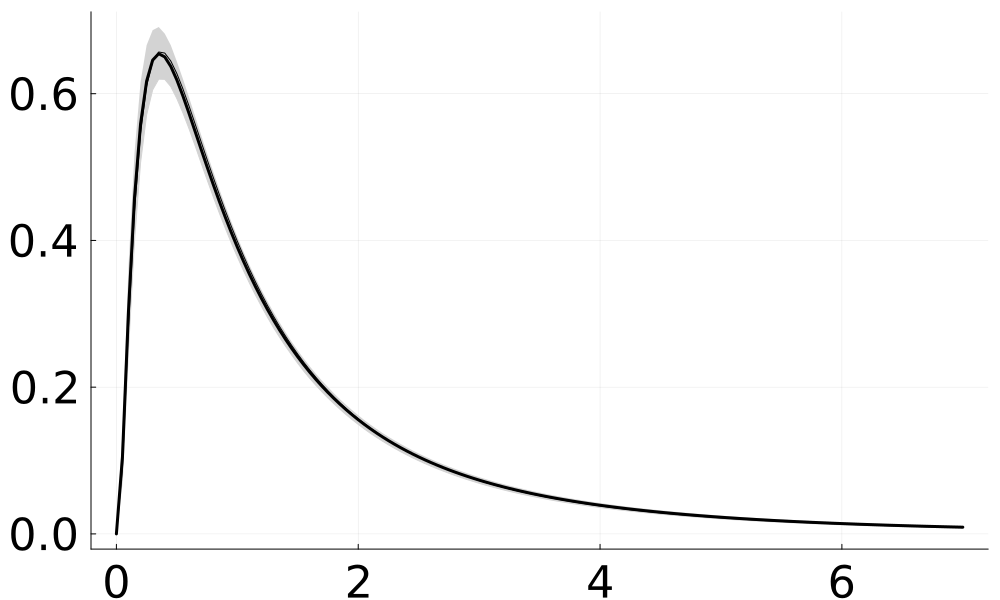}
     }
     \subfigure[$n =$ 5,000]{%
            \includegraphics[scale=0.14]{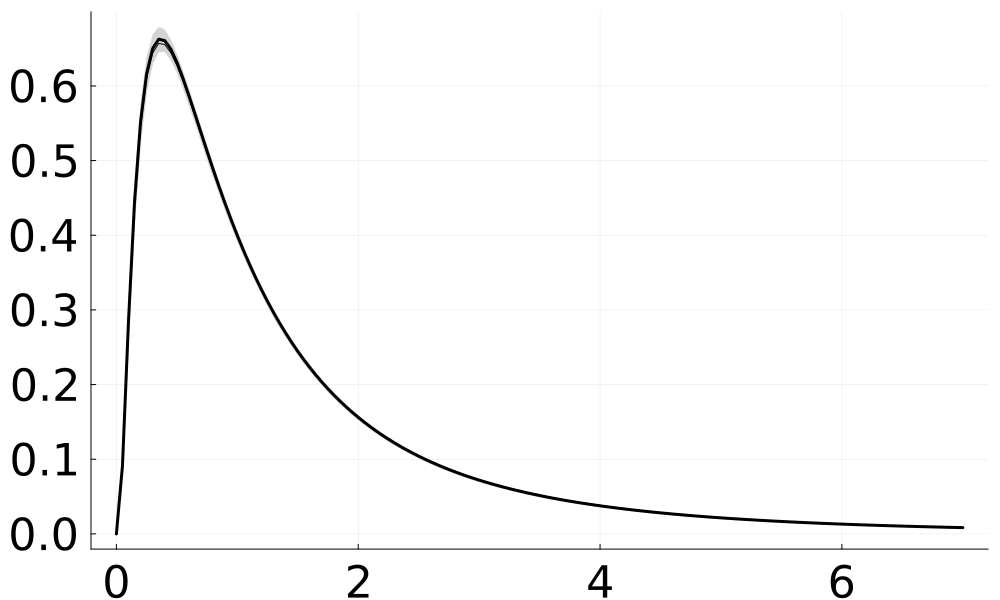}
     }
     \subfigure[$n =$ 10,000]{%
            \includegraphics[scale=0.14]{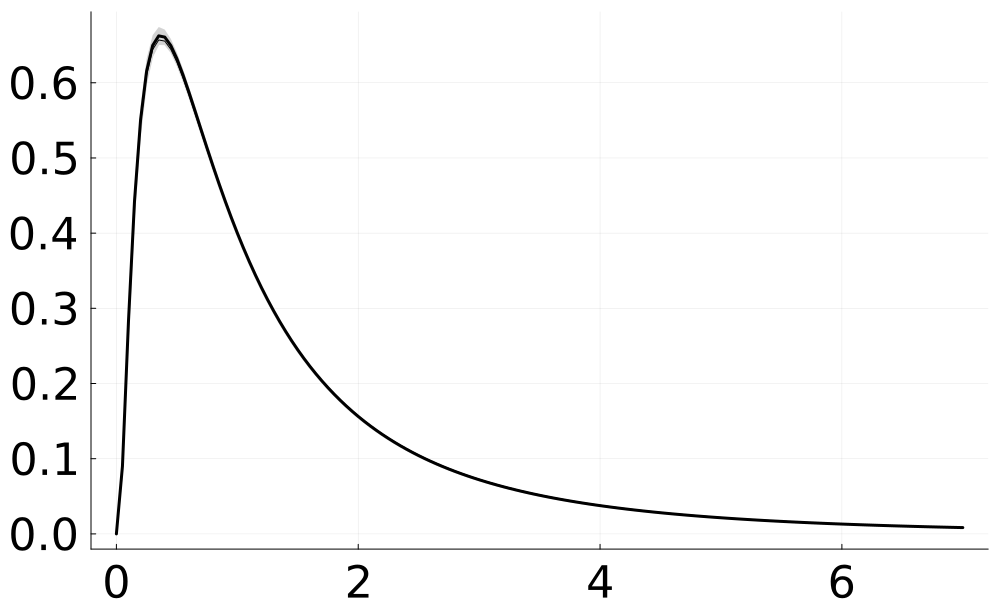}
     }\\
     \subfigure[$n =$ 1,000]{%
            \includegraphics[scale=0.14]{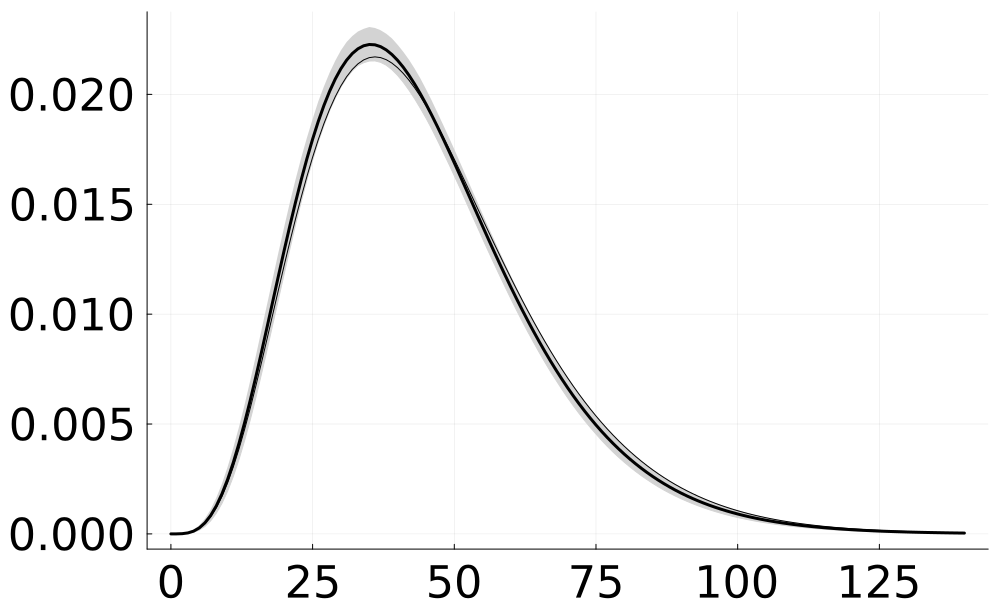}
     }       
     \subfigure[$n =$ 5,000]{%
            \includegraphics[scale=0.14]{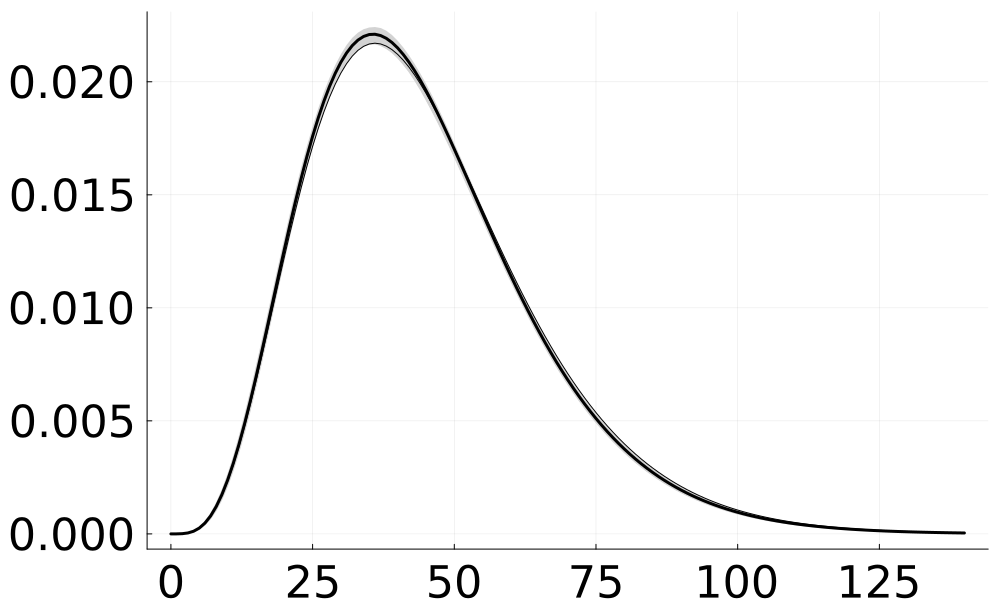}
     }       
     \subfigure[$n =$ 10,000]{%
            \includegraphics[scale=0.14]{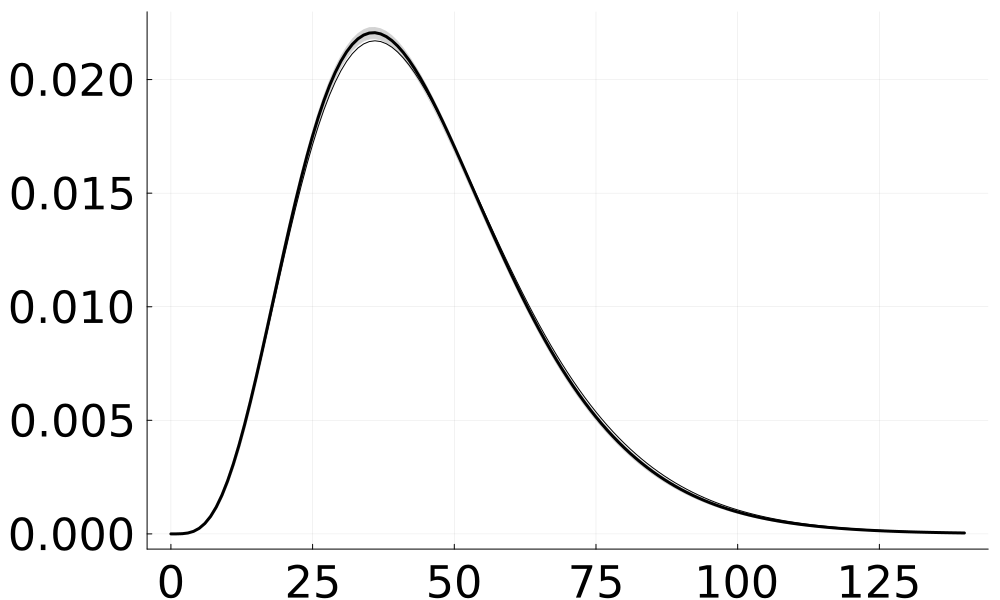}
     }\\       
     \subfigure[$n =$ 1,000]{%
            \includegraphics[scale=0.14]{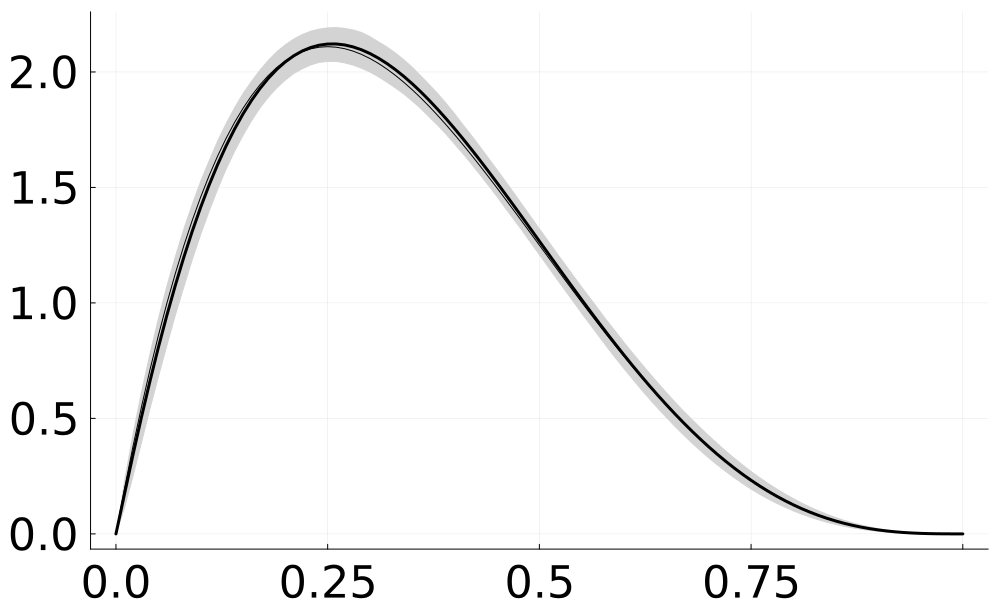}
     }       
     \subfigure[$n =$ 5,000]{%
            \includegraphics[scale=0.14]{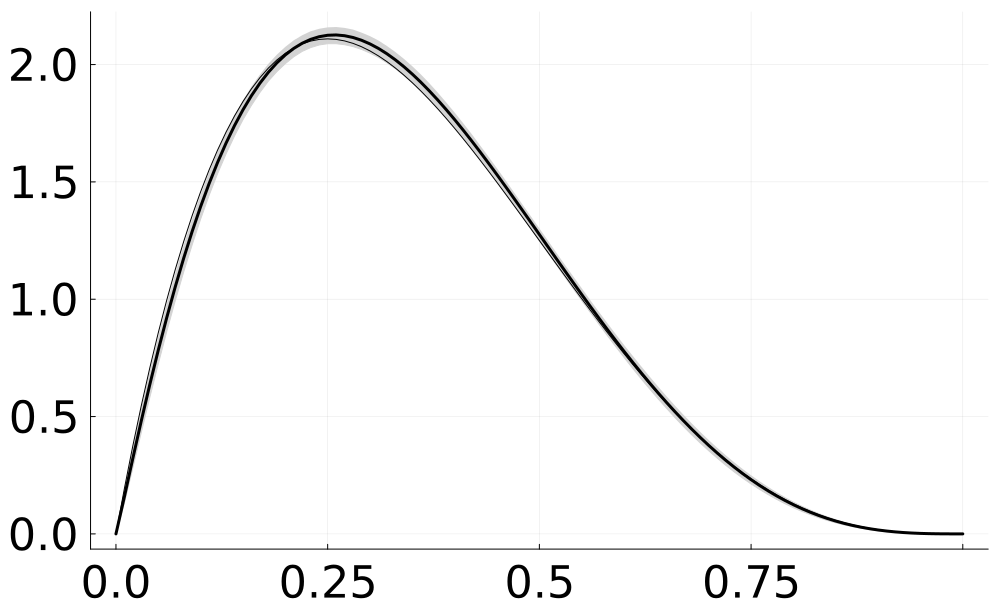}
     }       
     \subfigure[$n =$ 10,000]{%
            \includegraphics[scale=0.14]{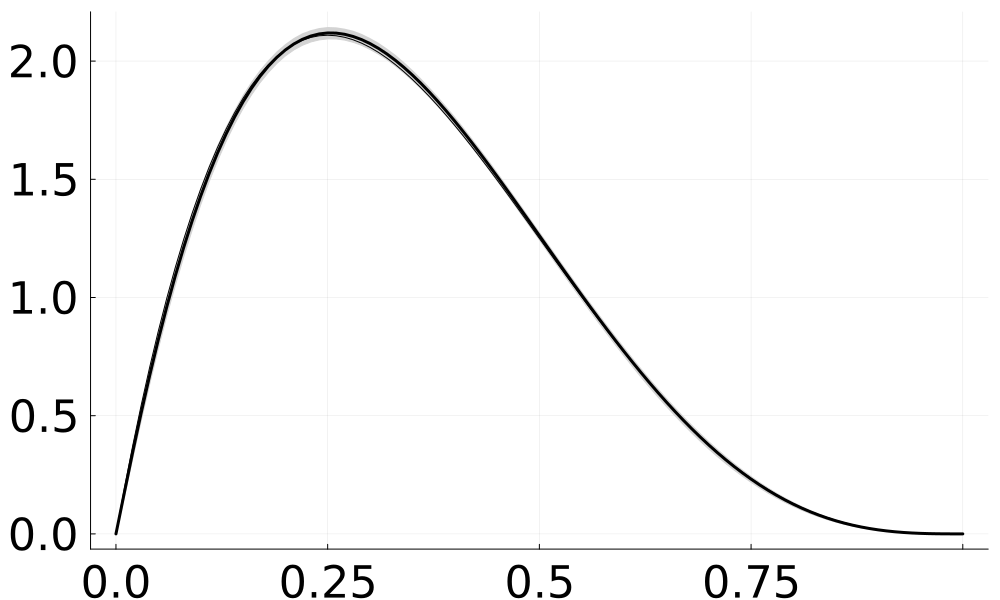}
     }       
        \caption{Four dimensional copula with unknown marginals.  Posterior mean (tick line) and 95\% point-wise credibility interval (grey band) for the marginal distribution. Panels (a)--(c), (d)--(f), (g)--(i), and (j)--(l) display the results for coordinate 1, 2, 3, and 4, respectively. Panels (a), (d), (g), and (j), (b), (e), (h), and (k), and (c), (f), (i), and (l) display the results for $n = $ 1,000, 5,000, and 10,000, respectively. In each figure, the true marginal distribution is shown as a thin line.}
        \label{marginalfig}
    \end{centering}
\end{figure}

Figure \ref{marginalfig} shows the posterior inferences on the marginal distributions for the different sample sizes. We observe that even under a complex marginal distribution, the marginal estimation is close to the truth in all cases and that the posterior distribution's mass concentrates around the true density function as the sample size increases.

\subsection{The Ames house price data}

\begin{figure}[!h]
\centering
\subfigure[Gaussian copula.] 
{
       \includegraphics[width=6cm]{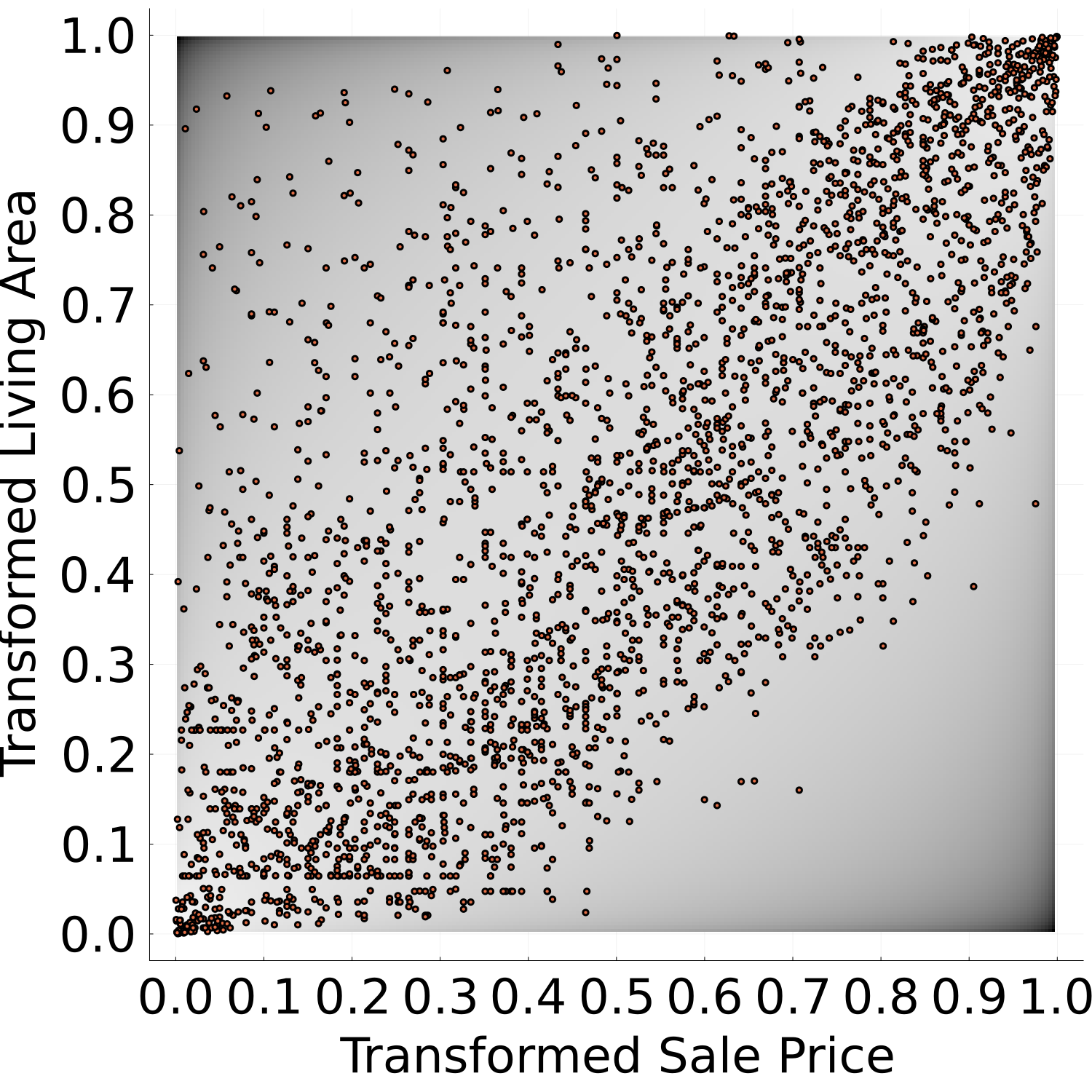}
 }      
\subfigure[BYU model with $k =$ 10.] 
{
    \includegraphics[width=6cm]{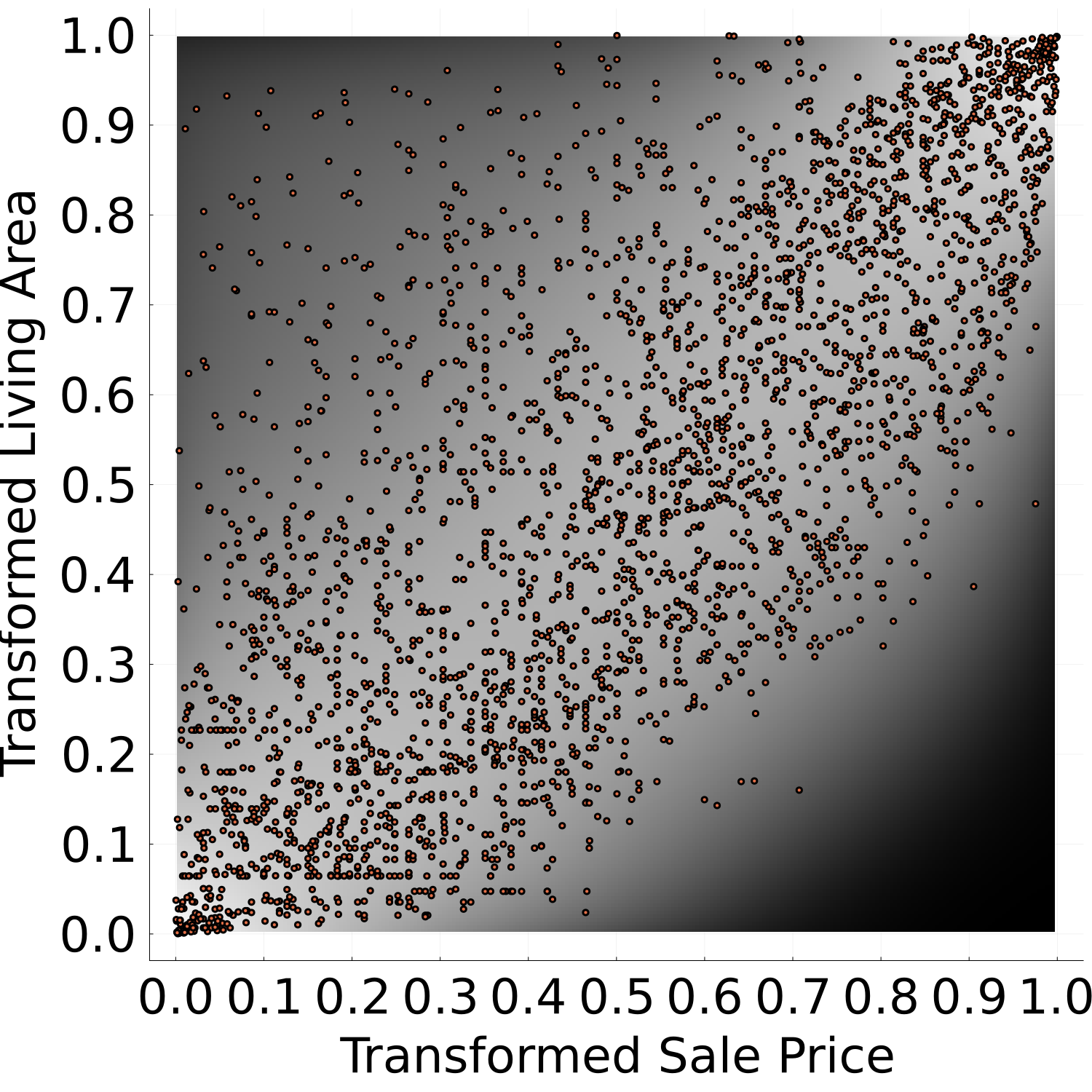}
}\\
\subfigure[BYU model with $k =$ 15] 
{
       \includegraphics[width=6cm]{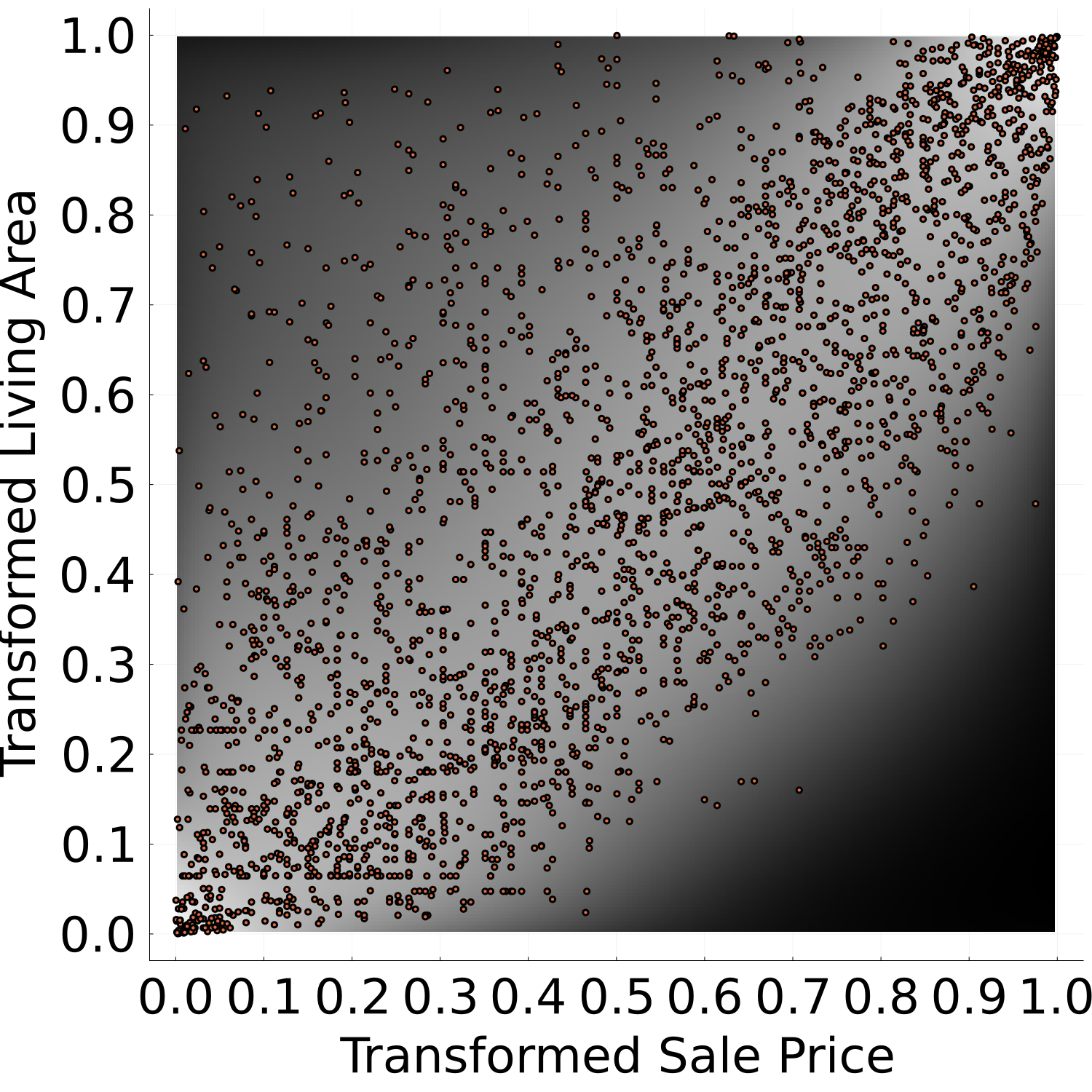}
 }      
\subfigure[BYU model with $k =$ 20] 
{
    \includegraphics[width=6cm]{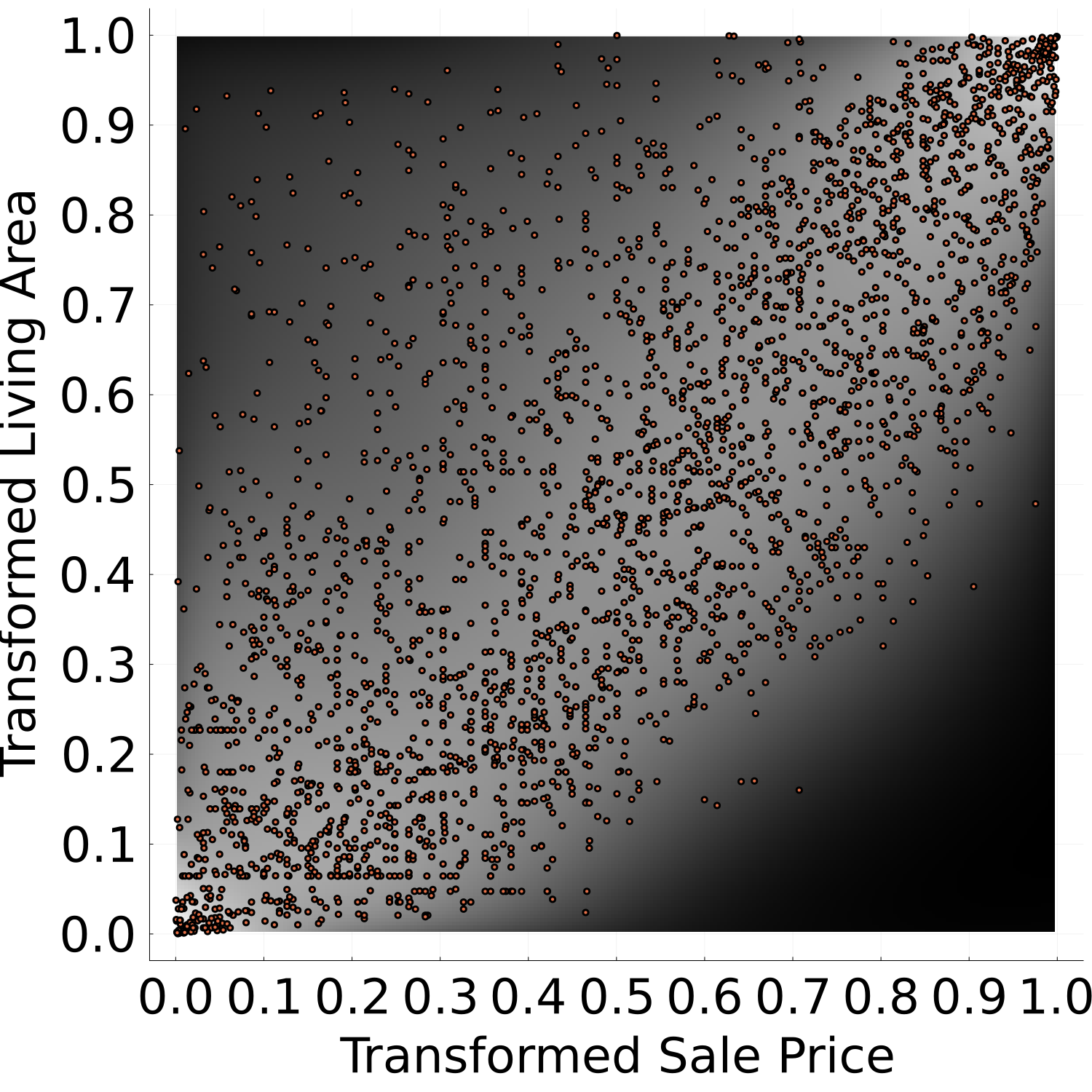}
}
 \caption{House price data: Panels (a), (b), (c), and (d) display the logarithm of the posterior mean of the copula function for sales price ($x$-coordinate) and above ground living area ($y$-coordinate) under the Gaussian copula model, and the BYU model with $k =$ 10, 15, and 20, respectively. In each plot the transformed observed data is also displayed.}
    \label{fig1:application1}
\end{figure}

We explore the relationship between two housing variables included in the Ames Housing dataset compiled by  \cite{houssing:data}. The data set describes the sale of individual residential property in Ames, Iowa from 2006 to 2010. The data set contains 2,930 observations and a large number of variables associated with home values. We considered two variables which have an interesting dependence relationship here;  the property's sale price in dollars and the above grade (ground) living area in square feet. Here we assume the marginal distributions to be known, which were estimated using an empirical distribution function, i.e., $\hat{F}_n(t) = \sum_{i=1}^{n} I(t \leq X_i)/(n+1)$. The data shown in Figure~\ref{fig1:application1}, 
shows that the sale price and the  above grade square footage have  an asymmetrical relationship. Specifically, there are homes with low sales prices and large above grade square footage. However the opposite is not observed.  The are no homes with high sales prices and small above grade square footage.  

Using the iterated rectangular exchanges MCMC algorithm, we create a Markov chain of size 1,020,000 to explore the posterior distribution under the proposed model by considering $k$ = 5, 10, and 20, the hierarchically centered prior with the ICAR correlation structure, and with $G_0$ being an independent copula. The full chain was subsampled every 10 steps after a burn-in period of 20,000 samples, to give a reduced chain of length 100,000. We also considered a Gaussian copula model. The acceptance rate of the MCMC was 23\%, 22\%, and
18\% for the BYU model with $k$ = 10, 15, and 20, respectively. 

Figure~\ref{fig1:application1} 
displays the logarithm of the posterior mean of the copula function under the Gaussian copula model and the different versions of the BYU model. The results illustrate that all versions of the proposed model are able to capture the asymmetrical relationship between the variables and that a parametric model, such as the Gaussian one, is not able to capture the deviation from the symmetrical behavior, assigning too much density to areas where no data points are actually observed.

We compare the models using the widely applicable Bayesian information criterion (WAIC) proposed by \cite{waic}. As expected, any of the considered versions of the BYU
model outperformed the parametric normal model from a goodness-of-fit point of view for these data. The WAIC under the Gaussian copula model was -2201.8. The WAIC under the BYU model with $k =$ 10, 15, and 20 was  -2280.6, -2308.0, and  -2280.2, respectively.

\section{Concluding remarks}\label{sec6}

Copula-based models provide a great deal of flexibility in modelling multivariate distributions, allowing for the specifications of models for the marginal distributions separately from the dependence structure (copula) that links them to form a joint distribution. Choosing a class of copula models is not a trivial task and its misspecification can lead to wrong conclusions. We have introduced a novel class of random Bernstein copula functions, which is dense in the space of all continuous copula functions in a Hellinger sense and has appealing smoothness, support and posterior consistency properties. The behavior of the Bayesian model was illustrated by means of simulated and real data.

We developed and compared alternative MCMC algorithms for exploring the corresponding posterior distribution for fix degree $\bk \in \mathbb{N}^d$. The user-specified degree may have an important influence in the resulting model in practice. Based on a set of possible degrees $\bk$, the Bayesian information criteria can be easily employed to select the best model. However, this approach does not provide a way to quantify the uncertainty on $\bk$. The study of strategies for the estimation under a random degree $\bk$ is the currently subject the subject of ongoing research.

Recent work suggests ways to implement efficient MCMC algorithms on convex polytopes \citep[see, e.g.,][]{karras2022dbsop,chen2018fast,yao2017walkr}.  However, copulas are only one type of convex polytope and are highly constrained, relative to most other convex polytopes.  The high degree of constraints on yett-uniform copulas, at present, make these and many other methods unsuitable for our use.  The study of ways to adapt those efficient MCMC algorithms to yett-uniform copulas is also subject of ongoing research.

The proposed model suffers from the curse of dimensionality. The implementation of the BYUP in high dimensions and with fine grids would result in an explosion of parameters that need to be updated, which makes the implementation of this approach practically impossible. The study of marginal versions of the model, where the copula probabilities are integrated out of the model is subject of ongoing research. The study of strategies for the estimation of the optimal size and location of the grid is also the subject of ongoing research. Finally, the extension of the model to handle mixed, discrete and continuous, variables and to copula regression problems would be interesting areas of future research.

\section*{Acknowledgements}
N. Kuschinski's  research was supported by Fondecyt 3210553 grant and ANID – Millennium Science Initiative Program – NCN17\_059. A. Jara's research was supported by Fondecyt 1220907 grant and ANID – Millennium Science Initiative Program – NCN17\_059.

\end{document}